\documentclass[10pt]{article}

\usepackage{setspace}
\usepackage{url}
\usepackage{fancyheadings}
\usepackage{reportcomments}
\usepackage{times}
\usepackage{natbib}
\usepackage{amsmath}
\usepackage{epsfig}
\usepackage{subfigure}
\usepackage{graphicx}
\usepackage{moreverb}

\begin{document}

\singlespacing

\title{\Huge CLAIRLIB Documentation\\ v1.03}

\author{\\ \\ \\ \\ \\ \\ \\ \\ \\ \\
        Dragomir Radev, University of Michigan \\
        Mark Hodges, University of Michigan \\
        Anthony Fader, University of Michigan \\
        Mark Joseph, University of Michigan \\
        Joshua Gerrish, University of Michigan \\
        Mark Schaller, University of Michigan \\
        Jonathan dePeri, Columbia University \\
        Bryan Gibson, University of Michigan \\
 \\
        \url{http://www.clairlib.org} \\ \\ }

\maketitle

\cleardoublepage

\tableofcontents

\cleardoublepage

\pagestyle{fancy}
\lhead {Clairlib}
\rhead {User Documentation}

\section{Introduction}

The University of Michigan CLAIR (Computational Linguistics and
Information Retrieval) group is happy to present version 1.03
of the Clair Library.

The Clair library is intended to simplify a number of generic tasks
in Natural Language Processing (NLP), Information Retrieval (IR), and
Network Analysis (NA).
Its architecture also allows for external software to be plugged in
with very little effort.

We are distributing the Clair library in two forms: Clairlib-core,
which has essential functionality and minimal dependence on external
software, and Clairlib-ext, which has extended functionality that
may be of interest to a smaller audience.  Depending on whether you
choose to install only Clairlib-core or both Clairlib-core and
Clairlib-ext, some of the content of this manual will not apply to
your installation.  Throughout this
document, for the sake of brevity, we will usually say ``the
Clair library'' or the more abbreviated ``Clairlib'' to refer to the
software we're distributing.

This work has been supported in part by National Institutes of Health
grants R01 LM008106 ``Representing and Acquiring Knowledge of Genome
Regulation'' and U54 DA021519 ``National center for integrative
bioinformatics,'' as well as by grants IDM 0329043 ``Probabilistic and
link-based Methods for Exploiting Very Large Textual Repositories,''
DHB 0527513 ``The Dynamics of Political Representation and Political
Rhetoric,'' 0534323 ``Collaborative Research: BlogoCenter - Infrastructure
for Collecting, Mining and Accessing Blogs,'' and 0527513 ``The Dynamics of
Political Representation and Political Rhetoric,'' from the National
Science Foundation.

\subsection{Functionality}

Much can be done using Clairlib on its own.  Some of the things that
Clairlib can do are listed below, in separate lists indicating whether
that functionality comes from within a particular distribution of
Clairlib, or is made available through Clairlib interfaces, but actually
is imported from another source, such as a CPAN module, or external
software.

\subsubsection{Native to Clairlib-core}
\begin{itemize}
\item Tokenization
\item Summarization
\item LexRank
\item Biased LexRank
\item Document Clustering
\item Document Indexing
\item PageRank
\item Biased Pagerank
\item Web Graph Analysis
\item Network Generation
\item Power Law Distribution Analysis
\item Network Analysis
        \begin{itemize}
        \item clustering coefficient
        \item degree distribution plotting
        \item average shortest path
        \item diameter
        \item triangles
        \item shortest path matrices
        \item connected components
        \end{itemize}
\item Cosine Similarity
\item Random Walks on Graphs
\item Statistics
        \begin {itemize}
        \item Distributions
        \item Tests
        \end{itemize}
\item Tf
\item Idf
\item Perceptron Learning and Classification
\item Phrase Based Retrieval and Fuzzy OR Queries
\end{itemize}

\subsubsection{Imported and available via Clairlib-core}
\begin{itemize}
\item Parsing
\item Stemming
\item Sentence Segmentation
\item Web Page Download
\item Web Crawling
\item XML Parsing
\item XML Tree Building
\item XML Writing
\end{itemize}

\subsection{Native to Clairlib-ext}
\begin{itemize}
\item Interfacing with Weka, a machine-learning Java toolkit
\item Latent Semantic Indexing
\item Sentence Segmentation using Adwait Ratnaparkhi's MxTerminator
\item Parsing using a Charniak Parser
\item Using the Automatic Link Extractor (ALE)
\item Using Google WebSearch
\end{itemize}

\subsection{Contributors}

Timothy Allison, Michael Dagitses, Jonathan DePeri, Aaron Elkiss,
Gunes Erkan, Bryan Gibson, Scott Gifford, Patrick Jordan, Mark Joseph, Jung-bae Kim,
Samuela Pollack, and Adam Winkel

\subsection{Changes}

\subsubsection*{1.03 August 2007\label{1_03_August_2007}\index{1.03 August 2007}}
\begin{itemize}

\item Added functionality to perform community finding within weighted, undirected networks
\item Added util/chunk\_document.pl to break documents into smaller files by word number
\item Added option to retain punctuation for idf and tf queries
\item Added option to print out full lists of idf and tf values for a corpus
\item LexRank moved from Clair::Network to Clair::Network::Centrality::LexRank
\item LexRank use now follows the same use pattern as the other centrality modules\end{itemize}
\subsubsection*{1.02 July 2007\label{1_02_July_2007}\index{1.02 July 2007}}
\begin{itemize}

\item Distribution reorganized in standard format
\item Improved and expanded installation documentation (INSTALL)
\item Improved POD (inline) documentation
\item Additional examples
\item Updated PDF documentation\end{itemize}
\subsubsection*{1.01 May 2007\label{1_01_May_2007}\index{1.01 May 2007}}
\begin{itemize}

\item Added Phrase-based Retrieval and Fuzzy OR Queries
\item Extended Clairlib-ext with interfaces for the Cluster class and the Document class to the Weka machine learning toolkit
\item Added LSI functionality
\item Extended parsing of strings / files into Documents
\item Added perceptron learning and classification for documents\end{itemize}
\subsubsection*{1.0 RC1 April 2007\label{1_0_RC1_April_2007}\index{1.0 RC1 April 2007}}
\begin{itemize}

\item Moved all Clair modules beneath the Clair::* namespace, updated documentation
\item Improved Network Analysis, added Clustering Coefficients code
\item Added Network Generation and Statistics modules\end{itemize}
\subsubsection*{0.955 March 2007\label{0_955_March_2007}\index{0.955 March 2007}}
\begin{itemize}

\item Made it possible to distribute clairlib in two distributions, one containing core code and another containing code that may be dependent on other resources
\item Cleaned up unit tests\end{itemize}
\subsubsection*{0.953 February 2007\label{0_953_February_2007}\index{0.953 February 2007}}
\begin{itemize}

\item Fixed bugs in Clair::Cluster, Clair::Document involving stemming
\item Cleaned up t/ and test/ directories
\item Created util/ directory
\item Added scripts to util/ directory to:\begin{itemize}

\item Run a Google query and save the returned URLs to a file
\item Download files from a URL and build a corpus
\item Segment a document into sentences and build a corpus of the sentences
\item Take all documents in a directory and create a corpus
\item Index the corpus (compute TF*IDF, etc.)
\item Compute cosine similarity measures between all documents in a corpus
\item Generate networks corresponding to various cosine thresholds
\item Print network statistics about a network file
\item Generate plots of degree distribution and cosine transitions\end{itemize}

\item New methods in Clair::Network:\begin{verbatim}
    print_network_info
    get_network_info_as_string
    get_cumulative_distribution
    cumulative_power_law_exponent
    find_components
    newman_clustering_coefficient
    linear_regression
\end{verbatim}
\end{itemize}

\section{Getting Started}

\subsection{Downloading}

Clairlib can be downloaded from http://www.clairlib.org/.

\subsection{Installing}

This guide explains how to install both Clairlib distributions, Clairlib-Core and Clairlib-Ext. To install Clairlib-core, follow the instructions in the section immediately below. To install Clairlib-Ext, first follow the instructions for installing
Clairlib-Core, then follow those for Clairlib-Ext itself. Clairlib-Ext requires an installed version of Clairlib-Core in order to run; it is not a stand-alone distribution.

\section{Install and Test Clairlib-Core\label{Install_and_Test_Clairlib-Core}\index{Install and Test Clairlib-Core}}
\subsection*{System Requirements\label{System_Requirements}\index{System Requirements}}

Clairlib-Core requires Perl 5.8.2 or greater. Before you proceed, confirm that the version of Perl you are running is at least this recent by entering

\begin{verbatim}
        perl -v
\end{verbatim}

at the shell prompt.

\subsection*{Install MEAD\label{Install_MEAD}\index{Install MEAD}}

Download MEAD 3.11 or later from \textsf{http://www.summarization.com/mead/}. The installation package is in \textbf{.tar.gz} ("tarball") format.  To install MEAD in, say, the directory \textbf{\$HOME/mead}, ensure that the installation package is located in \textbf{\$HOME}, and enter the following at the shell prompt:

\begin{verbatim}
        $ cd $HOME
        $ gunzip MEAD-3.11.tar.gz
        $ tar -xf MEAD-3.11.tar
        $ cd mead
        $ perl Install.PL
\end{verbatim}

Next, you will need to compile \textbf{tf2gen.cpp} to produce an executable required by MEAD. Enter the following:

\begin{verbatim}
        $ cd $HOME/mead/bin/feature-scripts
        $ g++ tf2gen.cpp -o tf2gen
\end{verbatim}
\subsection*{Install CPAN Libraries\label{Install_CPAN_Libraries}\index{Install CPAN Libraries}}

Clairlib-Core depends on access to the following Perl modules:

\begin{description}

\item[{BerkeleyDB}] \mbox{}
\item[{Carp}] \mbox{}
\item[{File::Type}] \mbox{}
\item[{Getopt::Long}] \mbox{}
\item[{Graph::Directed}] \mbox{}
\item[{Hash::Flatten}] \mbox{}
\item[{HTML::LinkExtractor}] \mbox{}
\item[{HTML::Parse}] \mbox{}
\item[{IO::File}] \mbox{}
\item[{IO::Handle}] \mbox{}
\item[{IO::Pipe}] \mbox{}
\item[{Lingua::Stem}] \mbox{}
\item[{Math::MatrixReal}] \mbox{}
\item[{Math::Random}] \mbox{}
\item[{MLDBM}] \mbox{}
\item[{PDL}] \mbox{}
\item[{POSIX}] \mbox{}
\item[{Scalar::Util}] \mbox{}
\item[{Statistics::ChisqIndep}] \mbox{}
\item[{Storable}] \mbox{}
\item[{Test::More}] \mbox{}
\item[{Text::Sentence}] \mbox{}
\item[{XML::Parser}] \mbox{}
\item[{XML::Simple}] \mbox{}\end{description}

There are multiple approaches to locating and installing these modules; using the automated CPAN installer, which is bundled with Perl, is perhaps the quickest and easiest. To do so, enter the following at the shell prompt:

\begin{verbatim}
        $ perl -MCPAN -e shell
\end{verbatim}

If you have not yet configured the CPAN installer, then you'll have to do so this one time. If you do not know the answer to any of the questions asked, simply hit enter, and the default options will likely suit your environment adequately. However, when asked about parameter options for the \texttt{perl Makefile.PL} command, users without root permissions or who otherwise wish to install Perl libraries within their personal \textbf{\$HOME} directory structure should enter the suggested path when prompted:

\begin{verbatim}
        Your choice:  ] PREFIX=~/perl
\end{verbatim}

This will cause the CPAN installer to install all modules it downloads and tests into \textbf{\$HOME/perl}, which means that all subdirectories of this directory that contain Perl modules will need to be added to Perl's \texttt{@INC} variable so that they will be found when needed (see section V below for further explanation).

As a side note, if you ever need to reconfigure the installer, type at the shell prompt:

\begin{verbatim}
        $ perl -MCPAN -e shell
        cpan>o conf init
\end{verbatim}

After configuration (if needed), return to the CPAN shell prompt,

\begin{verbatim}
        cpan>
\end{verbatim}

and type the following to upgrade the CPAN installer to the latest version:

\begin{verbatim}
        cpan>install Bundle::CPAN
        cpan>q
\end{verbatim}

If asked whether to prepend the installation of required libraries to the queue, hit return (or enter \texttt{yes}). After quitting the shell, type the following to install or upgrade \texttt{Module::Build} and make it the preferred installer:

\begin{verbatim}
        $ perl -MCPAN -e shell
        cpan>install Module::Build
        cpan>o conf prefer_installer MB
        cpan>o conf commit
        cpan>q
\end{verbatim}

Finally, install each of the following dependencies (if you are at all unsure whether the latest versions of each have already been installed) by entering the following at the shell prompt:

\begin{verbatim}
        $ perl -MCPAN -e shell
        cpan>install BerkeleyDB
        cpan>install Carp
        cpan>install File::Type
        cpan>install Getopt::Long
        cpan>install Graph::Directed
        cpan>install HTML::LinkExtractor
        cpan>install HTML::Parse
        cpan>install IO::File
        cpan>install IO::Handle
        cpan>install IO::Pipe
        cpan>install Lingua::Stem
        cpan>install Math::MatrixReal
        cpan>install Math::Random
        cpan>install MLDBM
        cpan>install PDL
        cpan>install POSIX
        cpan>install Scalar::Util
        cpan>install Statistics::ChisqIndep
        cpan>install Storable
        cpan>install Test::More
        cpan>install Text::Sentence
        cpan>install XML::Parser
        cpan>install XML::Simple
\end{verbatim}
\subsection*{Configure Clairlib-Core\label{Configure_Clairlib-Core}\index{Configure Clairlib-Core}}

Download the Clairlib-Core distribution (\textbf{clairlib-core.tar.gz}) into, say, the directory \textbf{\$HOME}. Then to install Clairlib-Core in \textbf{\$HOME/clairlib-core}, enter the following at the shell prompt:

\begin{verbatim}
        $ cd $HOME
        $ gunzip clairlib-core.tar.gz
        $ tar -xf clairlib-core.tar
        $ cd clairlib-core/lib/Clair
\end{verbatim}

Then edit Config.pm, which is located in \textbf{clairlib-core/lib/Clair}. You will see the following message at the top of the file:

\begin{verbatim}
        #################################
        # For Clairlib-core users:
        # 1. Edit the value assigned to $CLAIRLIB_HOME and give it the value
        #    of the path to your installation.
        # 2. Edit the value assigned to $MEAD_HOME and give it the value
        #    that points to your installation of MEAD.
        # 3. Edit the value assigned to $EMAIL and give it an appropriate
        #    value.
\end{verbatim}

Follow those instructions. In the case of our example, we would assign

\begin{verbatim}
        $CLAIRLIB_HOME=$HOME/clairlib-core
\end{verbatim}

and

\begin{verbatim}
        $MEAD_HOME=$HOME/mead
\end{verbatim}

where \textbf{\$HOME} must be replaced by an explicit path string such as \textbf{/home/username}. Also, note that the following MEAD variables reflect the structure of a standard MEAD installation and should typically be kept as assigned:

\begin{verbatim}
        $CIDR_HOME "$MEAD_HOME/bin/addons/cidr";
        $PRMAIN    "$MEAD_HOME/bin/feature-scripts/lexrank/prmain";
        $DBM_HOME  "$MEAD_HOME/etc";
\end{verbatim}
\subsection*{Test and Install the Clairlib-Core Modules\label{Test_and_Install_the_Clairlib-Core_Modules}\index{Test and Install the Clairlib-Core Modules}}

Before testing and installing the Clairlib-core modules, you'll need to modify Perl's \texttt{@INC} variable to ensure that it includes 1) paths to all Clairlib dependencies (the required libraries installed above), and 2) the path to Clairlib's own modules (in the case of our example, \textbf{\$HOME/clairlib-core/lib}). The simplest way to do this is by modifying the contents of your \texttt{PERL5LIB} environment variable from the shell prompt:

\begin{verbatim}
        $ export PERL5LIB=$HOME/clairlib-core/lib:$HOME/perl/lib     (*)
\end{verbatim}

Here \textbf{\$HOME/clairlib-core/lib} is the path to Clairlib's own modules, while \textbf{\$HOME/perl} is the path to Clairlib's required modules, installed above (assuming that path is their location). However, doing this requires that you export \texttt{PERL5LIB} each time you invoke the shell environment, so a better way to modify \texttt{@INC} is the following:

\begin{verbatim}
        $ cd $HOME
\end{verbatim}

Edit \textbf{.profile} or the appropriate configuration file for your shell environment, or create it if it does not already exist. Add \texttt{(*)} to to the file, or prepend the necessary paths using colons, as in \texttt{(*)}. Save the file and enter:

\begin{verbatim}
        $ . .profile
\end{verbatim}

This way you will not have to export \texttt{PERL5LIB} each time you invoke the
shell. Enter

\begin{verbatim}
        $ echo $PERL5LIB
\end{verbatim}

to confirm that your modifications have been applied.

Now you may test your Clairlib-Core installation. Enter its directory, in the case of our example:

\begin{verbatim}
        $ cd $HOME/clairlib-core
\end{verbatim}

Then enter the following commands to test the Clairlib-Core modules:

\begin{verbatim}
        $ perl Makefile.PL
        $ make
        $ make test
\end{verbatim}

If you would like to have the Clairlib-Core modules installed for you, and you have the necessary (root) permissions to do so, you may install them by entering the following command:

\begin{verbatim}
        $ make install
\end{verbatim}

If, on the other hand, you have only local permissions, but you have a personal perl library located at, say, \textbf{\$HOME/perl} (as described earlier), then you can install Clairlib-Core there by entering the commands:

\begin{verbatim}
        $ perl Makefile.PL PREFIX=~/perl
        $ make install
\end{verbatim}
\subsection*{Using the Clairlib-Core Modules\label{Using_the_Clairlib-Core_Modules}\index{Using the Clairlib-Core Modules}}

To use the Clairlib-Core modules in a Perl script, you must add a path to the modules to Perl's \texttt{@INC} variable. You may use either 1) \textbf{\$CLAIRLIB\_HOME/lib}, where \texttt{\$CLAIRLIB\_HOME} is the path defined in \textbf{Config.pm}, or 2) \textbf{\$INSTALL\_PATH}, where \texttt{\$INSTALL\_PATH} is a path to the location of the installed Clairlib-Core modules (if you installed them in section V, immediately above). Either of these paths can be added to \texttt{@INC} either by appending the path to the \texttt{PERL5LIB} environment variable or by putting a \texttt{use lib PATH} statement at the top of the script. See the beginning of section V above for a detailed explanation of how to modify the \texttt{PERL5LIB} variable.

\section{Install and Test Clairlib-Ext\label{Install_and_Test_Clairlib-Ext}\index{Install and Test Clairlib-Ext}}

The Clairlib-Ext distribution contains optional extensions to Clairlib-Core as well as functionality that depends on other software. The sections below explain how to configure different functionalities of Clairlib-Ext. As each is independent of the rest, you may configure as many or as few as you wish. Section VI provides instructions for the installation and testing of the Clairlib-ext modules itself.

\subsection*{Sentence Segmentation using Adwait Ratnaparkhi's MxTerminator\label{Sentence_Segmentation_using_Adwait_Ratnaparkhi_s_MxTerminator}\index{Sentence Segmentation using Adwait Ratnaparkhi's MxTerminator}}

To use MxTerminator for sentence segmentation, download the installation package from:

\begin{verbatim}
        L<ftp://ftp.cis.upenn.edu/pub/adwait/jmx/jmx.tar.gz>.
\end{verbatim}

Putting the tarball in, say, \textbf{\$HOME/jmx}, enter the following to unpack:

\begin{verbatim}
        $ cd $HOME/jmx
        $ gunzip jmx.tar.gz
        $ tar -xf .tar
\end{verbatim}

Uncomment and modify the following lines in \textbf{clairlib-core/lib/Clair/Config.pm}. Point \texttt{\$JMX\_HOME} to the top directory of your MxTerminator installation, and point \texttt{\$JMX\_MODEL\_PATH} to the location of your MxTerminator trained data, as for example

\begin{verbatim}
        # $JMX_HOME                "$HOME/jmx";
        # $SENTENCE_SEGMENTER_TYPE "MxTerminator";
        # $JMX_MODEL_PATH          "$HOME/jmx/eos.project";
\end{verbatim}

where \texttt{\$HOME} must be replaced by a literal path string such as \textbf{/home/username}. Note that the \textbf{/bin} directory of a Java installation must be located in your search path, or MxTerminator will not work.

\subsection*{Parsing using a Charniak Parser\label{Parsing_using_a_Charniak_Parser}\index{Parsing using a Charniak Parser}}

To use a Charniak parser with Clairlib, uncomment the following variables in \textbf{clairlib-core/lib/Clair/Config.pm} and point them to it, as for example:

\begin{verbatim}
        # Default parser and data paths for the Charniak parser for use in Parse.pm
        # (Note that CHARNIAK_DATA should end with a slash and that the other
        # paths include the executable)
        # $CHARNIAK_PATH      "/data0/tools/charniak/PARSE/parseIt";
        # $CHARNIAK_DATA_PATH "/data0/tools/charniak/DATA/EN/";
\end{verbatim}
\begin{verbatim}
        # Default path to Chunklink
        # $CHUNKLINK_PATH "/data2/tools/chunklink/chunklink.pl";
\end{verbatim}
\subsection*{Using the Weka Machine Learning Toolkit\label{Using_the_Weka_Machine_Learning_Toolkit}\index{Using the Weka Machine Learning Toolkit}}

To use the Weka Machine Learning Toolkit, a Java machine learning library, with Clairlib, download Weka from \textsf{http://www.cs.waikato.ac.nz/ml/weka/} and uncomment the following line in \textbf{clairlib-core/lib/Clair/Config.pm}. Point the variable to the location of Weka's \textbf{.jar} file, as for example:

\begin{verbatim}
        # $WEKA_JAR_PATH "$HOME/weka/weka-3-4-11/weka.jar"
\end{verbatim}

where \texttt{\$HOME} must be replaced by an explicit path string such as \textbf{/home/username}. Note that the \textbf{/bin} directory of a Java installation must be located in your search path, or MxTerminator will not work.

\subsection*{Using the Automatic Link Extractor (ALE)\label{Using_the_Automatic_Link_Extractor_ALE_}\index{Using the Automatic Link Extractor (ALE)}}

If you have MySQL installed and wish to use ALE, uncomment the following variables. Point \texttt{\$ALE\_PORT} at your MySQL socket, and provide the root password to your MySQL installation:

\begin{verbatim}
        # $ALE_PORT "/tmp/mysql.sock";
        # $ALE_DB_USER "root";
        # $ALE_DB_PASS "";
\end{verbatim}
\subsection*{Using Google WebSearch\label{Using_Google_WebSearch}\index{Using Google WebSearch}}

To use the Google WebSearch module, first install the CPAN module \texttt{Net::Google} (refer to section II of the Clairlib-Core installation instructions for further explanation) Then, uncomment the following line and provide a Google SOAP API key. Unfortunately, Google no longer gives out SOAP API keys but has moved to an AJAX Search API. If you have a SOAP API key, you can still use it, and WebSearch will still work.

\begin{verbatim}
        # $GOOGLE_DEFAULT_KEY "";
\end{verbatim}
\subsection*{Configure Clairlib-Ext\label{Configure_Clairlib-Ext}\index{Configure Clairlib-Ext}}

Download the Clairlib-Ext distribution (\textbf{clairlib-ext.tar.gz}) into, for example, the directory \textbf{\$HOME}. Then to install Clairlib-Ext in \textbf{\$HOME/clairlib-ext}, enter the following at the shell prompt:

\begin{verbatim}
        $ cd $HOME
        $ gunzip clairlib-ext.tar.gz
        $ tar -xf clairlib-ext.tar
        $ cd clairlib-ext
\end{verbatim}

To test the Clairlib-Ext modules, you must first have installed the Clairlib-Core modules. Confirm that you have, then enter the following:

\begin{verbatim}
        $ perl Makefile.PL
        $ make
        $ make test
\end{verbatim}

If you would like to have the Clairlib-Ext modules installed, and you have the necessary (root) permissions to do so, you may install them by entering:

\begin{verbatim}
        $ make install
\end{verbatim}

If, on the other hand, you have only local permissions, but you have a personal perl library located at, say, \textbf{\$HOME/perl} (as described earlier), then you can install Clairlib-Ext there by entering the commands:

\begin{verbatim}
        $ perl Makefile.PL PREFIX=~/perl
        $ make install
\end{verbatim}
\subsection*{Using the Clairlib-Ext Modules\label{Using_the_Clairlib-Ext_Modules}\index{Using the Clairlib-Ext Modules}}

To use the Clairlib-Ext modules in a script, you must add a path to the modules to Perl's \texttt{@INC} variable. You may use either 1) \textbf{\$CLAIRLIB\_EXT\_HOME/lib}, where \textbf{\$CLAIRLIB\_EXT\_HOME} is the path to the top directory of your Clairlib-Ext installation, or 2) \textbf{\$INSTALL\_PATH}, where \textbf{\$INSTALL\_PATH} is a path to the location of the installed Clairlib-Ext modules (if you installed them in section V, immediately above). Either of these paths can be added to \texttt{@INC} either by appending the path to the \texttt{PERL5LIB} environment variable or by putting a \texttt{use lib PATH} statement at the top of the script. See the beginning of section V of the Clairlib-Core installation instructions for a detailed explanation of how to modify the \texttt{PERL5LIB} variable.

\subsection*{Support and Documentation\label{Support_and_Documentation}\index{Support and Documentation}}

After installing Clairlib, you may access documentation for a module using the \texttt{perldoc} command, as for example:

\begin{verbatim}
        $ perldoc Clair::Document
\end{verbatim}

Each Clairlib distribution also includes a PDF tutorial. Online API documentation is available at:

\begin{verbatim}
        http://belobog.si.umich.edu/clair/clairlib/pdoc
\end{verbatim}

\section{Structure of the Clairlib Code}

The Clairlib code is divided into many modules, located in subdirectories within the \texttt{lib/Clair}
directory.  Some of the key functionality is in the \texttt{lib/Clair} directory itself:

\begin{itemize}
\item \texttt{Clair::Document} - Represents a single document
\item \texttt{Clair::Cluster} - Represents a collection of many documents
\item \texttt{Clair::Network} - Represents a network, like a graph.  The
nodes of the network may often be of type \texttt{Clair::Document}, but do
not have to be.
\item \texttt{Clair::Gen} - Works with Poisson and Power Law distributions
\item \texttt{Clair::Util} - Provides utility functions needed when using the Clair library
\item \texttt{Clair::Config} - Provides configurable constants needed by the Clair library (library paths, etc.)
\end{itemize}

Other modules in the top directory include the following:

\begin{itemize}
\item \texttt{Clair::Features} - Carry out feature selection using Chi-squared algorithm with Clair::GenericDoc
\item \texttt{Clair::Debug} - A simple class that Exports debugmsg and errmsg subs.
\item \texttt{Clair::Learn} - Implement various learning algorithms here. Default algorithm is Perceptron.
\item \texttt{Clair::Index} - Creates various indexes from supplied Clair::GenericDoc objects.
\item \texttt{Clair::Classify} - Take in the model file generated by Learn.pm and then carry out the classification
\item \texttt{Clair::StringManip} - Majority of the string manipulation routines required by other packages
\item \texttt{Clair::Centroid}
\item \texttt{Clair::Corpus} - Class for dealing with TREC corpus format data
\item \texttt{Clair::CIDR} - single pass document clustering
\item \texttt{Clair::SyntheticCollection} - Generate synthetic clusters of documents
\item \texttt{Clair::Extensions} - Versioning File for the Clairlib-ext distribution
\item \texttt{Clair::IDF} - Handle IDF databases
\item \texttt{Clair::SentenceFeatures} - a collection of sentence feature subroutines
\end{itemize}

Within the \texttt{lib/Clair/Utils/} directory, several modules are provided to work with
corpora:

\begin{itemize}
\item \texttt{Clair::Utils::CorpusDownload} - Download corpora from a list of URLs or from a single URL as a starting point, compute IDF and TF values
\item \texttt{Clair::Utils::Idf} - Retrieve IDF values calculated by CorpusDownload
\item \texttt{Clair::Utils::Tf} - Retrieve TF values calculated by CorpusDownload
\item \texttt{Clair::Utils::TFIDFTUtils} - Provides utility functions needed for the IDF/TF calculations
\item \texttt{Clair::Utils::Robot2} - configurable web traversal engine (for web robots \& agents)
\item \texttt{Clair::Utils::LinearAlgebra}
\item \texttt{Clair::Utils::Stem} - An implementation of a stemmer
\item \texttt{Clair::Utils::MxTerminator}
\item \texttt{Clair::Utils::ALE} - The Automatic Link Extrapolator
\end{itemize}

The Clairlib-ext distribution also contains the following modules in lib/Clair/Utils/:

\begin{itemize}
\item \texttt{Clair::Utils::WebSearch} - Performs Google searches and downloads files
\item \texttt{Clair::Utils::Parse} - Parse a file using the Charniak parser or use
the Chunklink tool.
\end{itemize}

Clairlib includes a large collection of network and graph processing
modules:

\begin{itemize}
\item \texttt{Clair::Network} - Network Class for the CLAIR Library
\item \texttt{Clair::NetworkWrapper} - A subclass of \texttt{Clair::Network} that wraps the C++ version of Lexrank.

\item \texttt{Clair::Network::Sample} - Network sampling algorithms
\begin{itemize}
\item \texttt{Clair::Network::Sample::RandomEdge} - Random edge sampling
\item \texttt{Clair::Network::Sample::RandomNode} - Random node sampling
\item \texttt{Clair::Network::Sample::ForestFire} - Random sampling using Forest Fire model
\item \texttt{Clair::Network::Sample::SampleBase} - Abstract class for
  network sampling
\end{itemize}

\item \texttt{Clair::Network::Reader} - Different network file type readers
\begin{itemize}
\item \texttt{Clair::Network::Reader} - Abstract class for reading in network formats
\item \texttt{Clair::Network::Reader::GraphML} - Class for reading in GraphML network files
\item \texttt{Clair::Network::Reader::Pajek} - Class for reading in Pajek network files
\item \texttt{Clair::Network::Reader::Edgelist} - Class for reading in edgelist network files
\end{itemize}

\item \texttt{Clair::Network::Generator} - Random network generators
\begin{itemize}
\item \texttt{Clair::Network::Generator::GeneratorBase} - Network generator abstract class
\item \texttt{Clair::Network::Generator::ErdosRenyi} - ErdosRenyi network generator abstract class
\end{itemize}

\item \texttt{Clair::Network::Writer} - Different network file type writers
\begin{itemize}
\item \texttt{Clair::Network::Writer} - Abstract class for exporting various Network formats
\item \texttt{Clair::Network::Writer::GraphML} - Class for writing GraphML network files
\item \texttt{Clair::Network::Writer::Pajek} - Class for writing Pajek network files
\item \texttt{Clair::Network::Writer::Edgelist} - Class for writing edge list network files
\end{itemize}

\item \texttt{Clair::Network::Centrality} - Network centrality measures
\begin{itemize}
\item \texttt{Clair::Network::Centrality} - Abstract class for computing network centrality
\item \texttt{Clair::Network::Centrality::Degree} - Class for computing degree
\item \texttt{Clair::Network::Centrality::Closeness} - Class for computing closeness
centrality
\item \texttt{Clair::Network::Centrality::Betweenness} - Class for computing betweenness
centrality
\end{itemize}

\item \texttt{Clair::Network::CFNetwork} - Class for performing community finding
\end{itemize} % Network modules

The Network modules uses the Graph CPAN module by default, but this
other graph libraries such as Boost can be used:

\begin{itemize}
\item \texttt{Clair::GraphWrapper} - Abstract class for underlying graphs
\item \texttt{Clair::GraphWrapper::Boost} - GraphWrapper class that provides an interface to the Boost graph library
\end{itemize}

There are also packages for dealing with discrete and continuous
distributions:

\begin{itemize}
\item \texttt{Clair::RandomDistribution::RandomDistributionBase} - base class for all distributions
\item \texttt{Clair::RandomDistribution::Gaussian}
\item \texttt{Clair::RandomDistribution::LogNormal}
\item \texttt{Clair::RandomDistribution::Poisson}
\item \texttt{Clair::RandomDistribution::RandomDistributionFromWeights}
\item \texttt{Clair::RandomDistribution::Zipfian}
\end{itemize}

\begin{itemize}
\item \texttt{Clair::Statistics::Distributions::TDist}
\item \texttt{Clair::Statistics::Distributions::DistBase}
\item \texttt{Clair::Statistics::Distributions::Geometric}
\end{itemize}

Here is a listing of the other modules in Clairlib:

\begin{itemize}
\item \texttt{Clair::ALE::Default::Tokenizer}
\item \texttt{Clair::ALE::Default::Stemmer} - ALE's default stemmer.
\item \texttt{Clair::ALE::Tokenizer}
\item \texttt{Clair::ALE::Stemmer} - Internal stemmer used by ALE

\item \texttt{Clair::ALE::Conn} - A connection between two pages, consisting of one or more links, created the the Automatic Link Extrapolator.
\item \texttt{Clair::ALE::Link} - A link between two URLs created by the Automatic Link Extrapolator.
\item \texttt{Clair::ALE::\_SQL} - Internal SQL adapter for use by ALE
\item \texttt{Clair::ALE::URL} - A URL created by the Automatic Link Extrapolator
\item \texttt{Clair::ALE::NormalizeURL}

% \item \texttt{Clair::Polisci} - Polisci modules
% \begin{itemize}
% \item \texttt{Clair::Polisci::AU::XMLHandler}
% \item \texttt{Clair::Polisci::US::XMLHandler}
% \item \texttt{Clair::Polisci::US::Connection} - Read records from the US polisci database
% \item \texttt{Clair::Polisci::Speaker} - An object representing a hansard speaker
% \item \texttt{Clair::Polisci::Record} - An object representing a hansard record

% \item \texttt{Clair::Polisci::Graf} - An object representing a hansard graf
% \item \texttt{Clair::Polisci::AustralianParser} - A class for parsing Australian hansard html.
% \end{itemize}

\item \texttt{Clair::MEAD::DocsentConverter} - Document => Mead Cluster converter
\item \texttt{Clair::MEAD::Summary} - access to a MEAD summary
\item \texttt{Clair::MEAD::Wrapper} - A perl module wrapper for MEAD

% \item \texttt{Clair::Bio} - CLAIR Bio utilities
% \begin{itemize}
% \item \texttt{Clair::Bio::EUtils::ESearchHandler} - an XML handler for parsing ESearch results
% \item \texttt{Clair::Bio::EUtils::ESearch} - a Perl interface to the ESearch utility
% \item \texttt{Clair::Bio::EUtils} - a base class for Bio::EUtils objects
% \item \texttt{Clair::Bio::Connection} - Connect to the Bio database using SOAP
% \item \texttt{Clair::Bio::GeneRIF} - Perl module for parsing GeneRIF files
% \end{itemize}

\item \texttt{Clair::LinkPolicy} - Different document linking policies
\begin{itemize}
\item \texttt{Clair::LinkPolicy::MenczerMacro} - Class implementing the Menczer Micro link model
\item \texttt{Clair::LinkPolicy::LinkPolicyBase} - Base class for creating corpora from collections
\item \texttt{Clair::LinkPolicy::RadevPAMixed} - Class implementing the RadevPAMixed  link model
\item \texttt{Clair::LinkPolicy::MenczerPAMixed} - Class implementing the MenczerPAMixed Micro link model
\item \texttt{Clair::LinkPolicy::RadevMicro} - Class implementing the Radev Micro link model
\item \texttt{Clair::LinkPolicy::BarabasiAlbert} - Class implementing the Barabasi Albert link model.
\item \texttt{Clair::LinkPolicy::WattsStrogatz} - Class implementing the Watts/Strogatz link model
\item \texttt{Clair::LinkPolicy::ErdosRenyi} - Class implementing the Erdos Renyi link model
\end{itemize}

\item \texttt{Clair::SentenceSegmenter} - Sentence segmentation
\begin{itemize}
\item \texttt{Clair::SentenceSegmenter::SentenceSegmenter}
\item \texttt{Clair::SentenceSegmenter::Text}
\item \texttt{Clair::SentenceSegmenter::MxTerminator}
\end{itemize}

\item \texttt{Clair::CIDR::Wrapper} - A wrapper script for the original cidr script
\item \texttt{Clair::Nutch::Search} - A class for performing simple Nutch searches.

\item \texttt{Clair::Interface::Weka}
\item \texttt{Clair::Index::mldbm} - A submodule that gets dynamically loaded by Index.pm.
\item \texttt{Clair::Index::dirfiles} - Builds the index into the filesystem namespace.

\item \texttt{Clair::Algorithm::LSI}
\item \texttt{Clair::Info::Query} - A module that implements different types of queries.
\item \texttt{Clair::Info::Stats}

\item \texttt{Clair::GenericDoc} - Generic document representations and parsing modules
\begin{itemize}
\item \texttt{Clair::GenericDoc} - A class to standardize and create generic representation of documents.
\item \texttt{Clair::GenericDoc::html} - a submodule that strips out html tags.
\item \texttt{Clair::GenericDoc::shakespear} - specialized to parse shakespear html files.
\item \texttt{Clair::GenericDoc::octet\_stream} - a submodule that parses xml and converts it into a hash
\item \texttt{Clair::GenericDoc::sports} - a specialized module for parsing docs for hw2
\item \texttt{Clair::GenericDoc::xml} - a submodule that parses xml and converts it into a hash
\item \texttt{Clair::GenericDoc::plain} - A submodule that returns the document as is.
\end{itemize}

\end{itemize}

Many of the above modules are described in more detail in the following section.

\section{Clairlib Network Processing Utilities Tutorial\label{Clairlib_Network_Processing_Utilities_Tutorial}\index{Clairlib Network Processing Utilities Tutorial}}

A tutorial explaining how to use the Clairlib library and tools to create a network from a group of files and process that network to extract information.

\subsection*{Introduction\label{Introduction}\index{Introduction}}

This tutorial will walk you through downloading files, creating a corpus from them, creating a network from the corpus, and extracting information along the way.  We'll be using utilities included in the Clairlib package to do the work.

Before beginning, install the clairlib package.  To do so, follow the instructions at:

\begin{verbatim}
 http://www.clairlib.org/mediawiki/index.php/Installation.
\end{verbatim}

The best way to use this document is to read all the way through as each command is explained.  The commands at the end of the tutorial in the code section.

\subsection*{Generating the corpus\label{Generating_the_corpus}\index{Generating the corpus}}

The first thing we will need is a corpus of files to run our tests against.  As an example we will be using a set of files extracted from Wikipedia.  We'll first download those files into a folder:

\begin{verbatim}
 mkdir corpus
\end{verbatim}

We'll use the 'wget' command to download the files.  The -r means to recursively get all of the files in the folder, -nd means don't create the directory path, and -nc means only get one copy of each file:

\begin{verbatim}
 cd corpus
 wget -r -nd -nc http://belobog.si.umich.edu/clair/corpora/chemical
 cd ..
\end{verbatim}

Now that we have our files, we can create the corpus.  To do this we'll use the 'directory\_to\_corpus.pl' utility.  The options used here are fairly consistent for all utilities:  --corpus, or -c, refers to the name of the corpus we are creating.  This should be something fairly simple, since we use it often and it is used to name several of the files we'll be creating.  In this case, we call our corpus 'chemical'.  --base, or -b, refers to the base directory of our corpus' data files.  A common practice is to use 'produced'.  Lastly --directory, or -d, refers to the directory where our files to be converted are located:

\begin{verbatim}
 directory_to_corpus.pl --corpus chemical --base produced \
  --directory corpus
\end{verbatim}

Now that our corpus has been organized, we'll index it so we can then start extacting data from it.  To do that we'll use 'index\_corpus.pl'.  Again, we'll specify the corpus name and the base directory where the index files should be produced:

\begin{verbatim}
 index_corpus.pl --corpus chemical --base produced
\end{verbatim}

We've now got our corpus and our indices and are ready to extract data.

\subsection*{Tfs and Idfs\label{Tfs_and_Idfs}\index{Tfs and Idfs}}

First we'll run a query for the term frequency of a single term.  To do this we'll use 'tf\_query.pl'.  Let's query 'health':

\begin{verbatim}
 tf_query.pl -c chemical -b produced -q health
\end{verbatim}

This outputs a list of the files in our corpus which contain the term 'health' and the number of times those terms occur in that file.  To get term frequencies for all terms in the corpus, pass the --all option:

\begin{verbatim}
 tf_query.pl -c chemical -b produced --all
\end{verbatim}

This returns a list of terms, their frequencies, and the number of documents each occurs in.

In order to see the full list of term frequencies for stemmed terms, pass the stemmed option:

\begin{verbatim}
 tf_query.pl -c chemical -b produced --stemmed --all
\end{verbatim}

Next we'll run a query for the inverse document frequency of a single term.  To do this we'll use 'idf\_query'.  Again, we'll query 'health':

\begin{verbatim}
 idf_query.pl -c chemical -b produced -q health
\end{verbatim}

We can also pass the --all option to idf\_query.pl to get a list of idf's for all terms in the corpus:

\begin{verbatim}
 idf_query.pl -c chemical -b produced --all
\end{verbatim}
\subsection*{Creating a Network\label{Creating_a_Network}\index{Creating a Network}}

We now have a corpus from which we can extract some data.  Next we'll create a network from this corpus.  To do this, we'll use 'corpus\_to\_network.pl'.  This command creates a network of hyperlinks from our corpus.  It produces a graph file with each line containing two linked nodes.  This command requires a specified output file which we'll call 'chemical.graph':

\begin{verbatim}
 corpus_to_network.pl -c chemical -b produced -o chemical.graph
\end{verbatim}

Now we can gather some data on this network.  To do that we'll run 'print\_network\_stats.pl' on our graph file.  This command can be used to produce many different types of data.  The easiest way to use it is with the --all option, which run all of its various tests.  We'll redirect its output to a file:

\begin{verbatim}
 print_network_stats.pl -i chemical.graph --all > chemical.graph.stats
\end{verbatim}

If we now look at 'chemical.graph.stats' we can see statistics for our network including numbers of nodes and edges, degree statistics, clustering coefficients, and path statistics.  This command also creates three centrality files (betweenness, closeness, and degree) which are lists of all terms and their centralities.

\subsection*{Conclusions\label{Conclusions}\index{Conclusions}}

With the tools described above you should be able to create a corpus from a set of files and extract statistics from that corpus.  For additional functionality or to get more information on the utilites used, go to

\begin{verbatim}
 http://www.clairlib.org/mediawiki/index.php/Documentation.
\end{verbatim}
\subsection*{CODE\label{CODE}\index{CODE}}

This is a list of all of the commands used in this tutorial:

\begin{verbatim}
 mkdir corpus
 cd corpus
 wget -r -nd -nc http://belobog.si.umich.edu/clair/corpora/chemical
 cd ..
 directory_to_corpus.pl --corpus chemical --base produced \
  --directory corpus
 index_corpus.pl --corpus chemical --base produced
 tf_query.pl -c chemical -b produced -q health
 tf_query.pl -c chemical -b produced --all
 idf_query.pl -c chemical -b produced -q health
 idf_query.pl -c chemical -b produced --all
 corpus_to_network.pl -c chemical -b produced -o chemical.graph
 print_network_stats.pl -i chemical.graph --all > chemical.graph.stats
\end{verbatim}
\section{Recipes}

In this section we will be using Clairlib utilities to create corpora, generate networks, extract plots and statistics, and demonstrate how to perform other useful tasks.  The chapter is organized into the following sections:

\begin{enumerate}
 \item Generating Corpora
 \item Gathering Corpora Statistics
 \item Generating Networks
 \item Gathering Network Statistics
 \item Other Useful Tools
\end{enumerate}

\subsection{Generating Corpora}
\subsubsection{Generate a corpus by downloading files}

\begin{boxedverbatim}

 output: indexed corpus

 mkdir corpus
 cd corpus
 wget -r -nd -nc \
   http://belobog.si.umich.edu/clair/corpora/chemical
 cd ..
 directory_to_corpus.pl -c chemical -b produced -d corpus
 index_corpus.pl -c chemical -b produced

\end{boxedverbatim}
\subsubsection{Generate a corpus by crawling a site}

\begin{boxedverbatim}
 output: indexed corpus

 crawl_url.pl -u http://www.asdg.com/ -o asdg.urls
 download_urls.pl -c asdg -i asdg.urls -b produced
 index_corpus.pl -c asdg -b produced
\end{boxedverbatim}
\subsubsection{Generate a corpus from a Google search}

\begin{boxedverbatim}
 output: indexed corpus

 search_to_url.pl -q bulgaria -n 10 > bulgaria.10.urls
 download_urls.pl -i bulgaria.10.urls -c bulgaria-10 -b produced
 index_corpus.pl -c bulgaria-10 -b produced
\end{boxedverbatim}
\subsubsection{Generate a corpus of sentences from a document}

\begin{boxedverbatim}
 input: collection of documents
 output: indexed corpus

 sentences_to_docs.pl -d $CLAIRLIB/corpora/1984/ -o docs
 directory_to_corpus.pl -c 1984sents -b produced -d docs
 index_corpus.pl -c 1984sents -b produced
\end{boxedverbatim}
\subsubsection{Generate a corpus using Zipfian distribution}

\begin{boxedverbatim}

 input: indexed corpus
 output: synthetic corpus

 make_synth_collection.pl --policy zipfian --alpha 1 -o synth \
   -d synth_out -c chemical -b produced --size 11 --verbose
\end{boxedverbatim}

\subsection{Gathering Corpus Statistics}

\subsubsection{Run IDF queries on a corpus}

\begin{boxedverbatim}
 input: indexed corpus
 output: idf query data
 
 idf_query.pl -c chemical -b produced -q health
 idf_query.pl -c chemical -b produced --all
\end{boxedverbatim}
\subsubsection{Run TF queries on a corpus}

\begin{boxedverbatim}
 input: indexed corpus
 output: tf query data

 tf_query.pl -c chemical -b produced -q health
 tf_query.pl -c chemical -b produced --all
 tf_query.pl -c chemical -b produced --stemmed --all
 tf_query.pl -c chemical -b produced -q "atomic number"
\end{boxedverbatim}

\subsection{Generating Networks}

\subsubsection{Generate a network from a corpus}

\begin{boxedverbatim}
 input: indexed corpus
 output: network graph

 corpus_to_network.pl -c chemical -b produced -o chemical.graph
\end{boxedverbatim}
\subsubsection{Generate synthetic network using Erdos/Renyi linking model}

\begin{boxedverbatim}

 output: synthetic graph

# With n nodes and m edges 

 generate_random_network.pl -o synthetic.graph \
   -t erdos-renyi-gnm -n 100 -m 88

# With n nodes and random edge with probability p

 generate_random_network.pl -o synthetic.graph \
   -t erdos-renyi-gnp -n 100 -p .1

# Based on another graph

 generate_random_network.pl -o synthetic.graph \
   -i $CLAIRLIB/corpora/david_copperfield/adjnoun.graph \
   -t erdos-renyi-gnp -p .1
\end{boxedverbatim}

\subsection{Gathering Network Statistics}

\subsubsection{Generate plots and statistics from a corpus}

\begin{boxedverbatim}
 input: indexed corpus
 output: plots and stats

 corpus_to_cos.pl -c chemical -o chemical.cos -b produced
 cos_to_cosplots.pl -i chemical.cos
 cos_to_histograms.pl -i chemical.cos
 cos_to_stats.pl -i chemical.cos -o chemical.stats
\end{boxedverbatim}
\subsubsection{Generate plots from a network}

\begin{boxedverbatim}
 input: network file 
 output: degree distribution plots 

 network_to_plots.pl -i chemical.cos --bins 100
\end{boxedverbatim}

\subsubsection{Putting it all together: plots and stats generated from a document}

\begin{boxedverbatim}
 input: sample news data
 output: plots and statistics
 optional: Matlab

 sentences_to_docs.pl -i \
   $CLAIRLIB/corpora/news-sample/lexrank-sample.txt \
   -o lexrank-sample
 directory_to_corpus.pl -c lexrank-sample  -b produced \
   -d lexrank-sample
 index_corpus.pl -c lexrank-sample -b produced
 corpus_to_cos.pl -c lexrank-sample -b produced \
    -o lexrank-sample.cos
 cos_to_histograms.pl -i lexrank-sample.cos
 cos_to_cosplots.pl -i lexrank-sample.cos
 cos_to_stats.pl --graphs -i lexrank-sample.cos \
   -o lexrank-sample.stats
 print_network_stats.pl --triangles -i lexrank-sample-0.26.graph
 stats2matlab.pl -i lexrank-sample.stats -o lexrank-sample.m
 network_growth.pl -c lexrank-sample -b produced
 stats2matlab.pl -i lexrank-sample.wordmodel.stats \
   -o lexrank-sample-wordmodel.m

# Now make the synthetic collection

 make_synth_collection.pl --policy zipfian --alpha 1 -o synth \
   -d synth_out -c lexrank-sample -b produced --size 11 --verbose
 link_synthetic_collection.pl -n synth -b produced -c synth \
   -d synth_out -l erdos -p 0.2
 index_corpus.pl -c synth -b produced
 corpus_to_cos.pl -c synth -b produced -o synth.cos
 cos_to_histograms.pl -i synth.cos
 cos_to_cosplots.pl -i synth.cos
 cos_to_stats.pl -i synth.cos -o synth.stats --graphs --all -v
 stats2matlab.pl -i synth.stats -o synth.m
 network_growth.pl -c synth -b produced
 stats2matlab.pl -i synth.wordmodel.stats -o synth-wordmodel.m

# If you are on a machine with MATLAB, 
# run the following to generate plots:

 mkdir plots
 mv *.m matlab
 matlab -nojvm -nosplash < lexrank-sample-cosine-cumulative.m
 matlab -nojvm -nosplash < lexrank-sample-cosine-hist.m
 matlab -nojvm -nosplash < lexrank-sample-distplots.m
 matlab -nojvm -nosplash < lexrank-sample.m
 matlab -nojvm -nosplash < lexrank-sample-wordmodel.m
 matlab -nojvm -nosplash < synth-cosine-cumulative.m
 matlab -nojvm -nosplash < synth-cosine-hist.m
 matlab -nojvm -nosplash < synth-distplots.m
 matlab -nojvm -nosplash < synth.m
 matlab -nojvm -nosplash < synth-wordmodel.m
\end{boxedverbatim}

\subsection{Other Useful Tools}

\subsubsection{Selecting a subset of a corpus for processing}

\begin{boxedverbatim}
 input: existing corpus
 output: directory containing subset of corpus
 
 corpus_to_cluster.pl -c bulgaria-10 -b produced \
  -f '^https://www.cia.gov/' \
  -f '^http://en.wikipedia.org/' -o filtered 
 directory_to_corpus.pl -c bulgaria-filtered -b produced \
   -d filtered
\end{boxedverbatim}
\subsubsection{Convert a network from one format to another}

\begin{boxedverbatim}

 input: gml file (or pajek file)
 output: edgelist file

 convert_network.pl -v \
   -input $CLAIRLIB/corpora/david_copperfield/adjnoun.gml \
   --input-format gml --output ./adjnoun.graph \
   --output-format edgelist
 print_network_stats.pl -i ./adjnoun.graph --undirected

\end{boxedverbatim}
\subsubsection{Extract ngrams from document and create network}

\begin{boxedverbatim}
 input: document
 output: stats

 extract_ngrams.pl -r "$CLAIRLIB/corpora/1984/1984.txt" \
   -f text -w 1984.2gram -N 2 -sort -v
 print_network_stats -i 1984.2gram -v --all --sample 100 \
   --sample-type forestfire > 1984.2gram.stats
\end{boxedverbatim}
\subsubsection{Generate statistics for word growth model from a corpus}

\begin{boxedverbatim}
 input: indexed corpus
 output: stats
 required: Matlab

 network_growth.pl -c chemical -b produced
 stats2matlab.pl -i chemical.wordmodel.stats -o wordmodel.m
 matlab -nojvm -nosplash < wordmodel.m
\end{boxedverbatim}
\section{Modules}

\subsection{Clair::Document}

Clairlib's Document class can be used to perform some basic analysis
and perform some calculations on a single document.

Documents have three types of values: `html', `text', and `stem'.  A
document must be created as one of the three types.  It can then be
converted from html to text and from text to stem.  Performing a
conversion does not cause the old information to be lost.  For
example, if a document starts as html, and is converted to text, the
html is not forgotten, the document now holds an html version and a
text version of the original html document.

Creating a new document: A new document can be created either from a
string or from a file.  To create a document from a string, the
string parameter should be specified, while the file parameter
should be specified with the filename to load the document from.  It
is an error if both are specified.

The initial type of a document must be specified.  This is done by
setting the type parameter to `html', `text', or `stem'.
Additionally, an id must be specified for the document.  Care should
be taken to keep ids of documents unique, as putting documents with
the same id into a Cluster or Network can cause problems.

Finally, the language of the document may be specified by passing a
value as the language parameter.
\\
\\
\begin{boxedverbatim}

my $doc = new Clair::Document(file => 'doc.html', id => 'doc1',
                              type => 'html');

\end{boxedverbatim}
\\

Using a single Document: strip\_html and stem convert an html
version of the document to text and a text version to stem,
respectively.

The html, text, or stem version of the document can be retrieved
using get\_html, get\_text, and get\_stem respectively.  For these
methods and all those used by document, the programmer is expected
to ensure that any time a particular type of a document is used,
that type is valid.  That is, a document that is created as an html
document and is never converted to a text document should never have
get\_text called or save (described later) called with type
specified as anything but `html'.
\\
\\
\begin{boxedverbatim}

# We start off with the html version
my $html = $doc->get_html;

# But can now get the text version
my $text = $doc->strip_html();
die if ($text ne $doc->get_text);

# And then the stemmed version
my $stem = $doc->stem();
die if $stem ne $doc->get_stem;

# Note that the html version is unchanged:
die if $html ne $doc->get_html;

\end{boxedverbatim}
\\

Several different operations can be performed on a document.  It can
be split into lines, sentences, or words using split\_into\_lines,
split\_into\_sentences, and split\_into\_words which return an array
of the text of the document separated appropriately.
split\_into\_lines and split\_into\_sentences can only be performed
on the text version of the document, but split\_into\_words can be
performed on any type of document.  It defaults to text, but this
can be overridden by specifying the type parameter.

A document can be saved to a file using the save method.  The method
requires the type to be saved be specified.

Documents may have parent documents as well.  This can be used to
track the original source of a document.  For example, a new
document could be created for each sentence of an original document.
By using set\_parent\_document and get\_parent\_document, each new
document can point to the document it was created from.

\subsection{Clair::Cluster}

Clairlib makes analyzing relationships beween documents very simple.
Generally, for simplicity, documents should be loaded as a cluster,
then converted to a network, but documents can be added directly to
a network.

Creating a Cluster: Documents can be added individually or loaded
collectively into a cluster.  To add documents individually, the
insert function is provided, taking the id and the document, in that
order.  It is not a requirement that the id of the document in the
cluster match the id of the document, but it is recommended for
simplicity.

Several functions are provided to load many documents quickly.
load\_file\_list\_array adds each file from the provided array as a
document and adds it to the cluster.  load\_file\_list\_from\_file
does the same for a list of documents that are given in a provided
file. load\_documents does the same for each document that matches
the expression passed along as a parameter.

Each of these functions must assign a type to each document created.
`text' is the default, but this may be changed by specifying a type
parameter.  Files can be loaded by filename or by `count', an index
that is incremented for each file.  Using the filename is the
default, but specifying a parameter count\_id of 1 changes that.  To
allow the load functions to be called repeatedly, a start\_count
parameter may be specified to have the counts started at a higher
number (to avoid repeated ids).  Each load function returns the next
safe count (note that if start\_count is not specified, this is the
number of documents loaded).

load\_lines\_from\_file loads each line from a file as an individual
document and adds it to the cluster.  It behaves very similarly to
the other load functions except that ids must be based on the count.
\\
\\
\begin{boxedverbatim}

my $cluster = Clair::Cluster->new();

$cluster->load_documents("directory/*.txt", type => 'text');

\end{boxedverbatim}
\\

\subsubsection{Working with Documents Collectively}
The functions
strip\_all\_documents, stem\_all\_documents, and
save\_documents\_to\_directory act on every document in the cluster,
stripping the html, stemming the text, or saving the documents.
\\
\\
\begin{boxedverbatim}

$cluster->stem_all_documents();

\end{boxedverbatim}
\\

\subsubsection{Analyzing a Cluster}
The documents in a cluster can be analyzed in
two ways.  The first is that an IDF database can be built from the
documents in the cluster with build\_idf.  The second is analyzing
the similarity between documents in the cluster.  First,
compute\_cosine\_matrix is provided which computes the similarity
between every pair of documents in the cluster.  These values are
returned in a hash, but are also saved with the cluster.
compute\_binary\_cosine returns a hash of cosine values that are
above the threshold.  It can be provided a cosine hash or can use a
previously computed hash stored with the cluster.
get\_largest\_cosine returns the largest cosine value, and the two
keys that produced it in a hash.  It also can be passed a cosine
hash or can use a hash stored with the cluster.
\\
\\
\begin{boxedverbatim}

my %cos_hash = $cluster->compute_cosine_matrix();

my %bin_hash = $cluster->compute_binary_cosine(0.2);

\end{boxedverbatim}
\\

\subsection{Clair::Network}

\subsubsection{Creating a Network}

There are three ways to create a network from a cluster, based on
what statistics are desired from the network.  For statistics based
on the similarity relationships, create\_network creates a network
based on a cosine hash.  Any two documents with a positive cosine
relationship will have an edge between them in the network.
Optionally, all documents can have an edge by specifying parameter
include\_zeros as 1.  The transition values to compute lexrank are
also set, although the values can be saved to a different attribute
name by specifying a property parameter.

For statistics based on hyperlink relationship,
create\_hyperlink\_network\_from\_array and \linebreak
create\_hyperlink\_network\_from\_file creates a network with edges
between pairs of documents in an array or on lines of a file,
respectively.

create\_sentence\_based\_network creates a network with a node for
every sentence in every document.  The cosine between each sentence
is then computed and, if a threshold is specified, the binary cosine
is computed.  The edges are created based on the similarity values
as with create\_network.
\\
\\
\begin{boxedverbatim}

my $network = $cluster->create_network(cosine_matrix => %bin_hash);

\end{boxedverbatim}
\\

\subsubsection{Importing a Network}

Networks can also be read in from various cross-platform graph
formats.  Currently, the following formats are supported:

\begin{itemize}
\item Edgelist
\item GraphML
\item Pajek
\end{itemize}

To read in a network, create a Clair::Network::Reader object of the
appropiate type and call the read\_network method with a filename.  A
new Clair::Network object will be returned.

Example of reading a Pajek file:
\\
\\
\begin{boxedverbatim}

use Clair::Network::Reader::Pajek;

my $reader = Clair::Network::Reader::Pajek->new();
my $net = $reader->read_network("example.net");

\end{boxedverbatim}

\subsubsection{Exporting a Network}

You can also export a Network to any of the above formats with the
Writer classes.

Example of writing a Pajek format network:
\\
\\
\begin{boxedverbatim}

use Clair::Network::Writer::Pajek;

my $export = Clair::Network::Writer::Pajek->new();
$export->set_name("networkname");
$export->write_nework($net, "example.net");

\end{boxedverbatim}

\subsubsection{Analyzing a Network}

Once a network has been created, much more analysis is possible.
Basic statistics like the number of nodes and edges are available
from num\_nodes and num\_links.  The average and maximum diameters
can be ascertained from diameter, specifying either a max parameter
as 1 or an avg parameter as 1 (max is the default).  The average in
degree, out degree, and total degree can be computed with
avg\_in\_degree, avg\_out\_degree, and avg\_total\_degree
respectively.

\textbf{Shortest Path Length}

Clairlib can compute the shortest path between all pairs of vertices.
It returns the results as a hash of hashes of the shortest path
matrix.
\\
\\
\begin{boxedverbatim}

use Clair::Network;

my $net = new Clair::Network();
my $sp_matrix = $net->get_shortest_path_matrix();

\end{boxedverbatim}
\\

\textbf{Average Shortest Path Length}

Clairlib can compute the average shortest path length between all
pairs of vertices.  See the examples for usage.

\textbf{Clustering Coefficient}

Clairlib supports two clustering coefficient functions.  The
Watts-Strogatz clustering coefficient and the Newman clustering
coefficient.

\textbf{Assortativity}

Clairlib can compute degree assortativity.  It returns a global
measure of network assortativity, the degree assortativity
coefficient.
\\
\\
\begin{boxedverbatim}

my $sp_matrix = $net->degree_assortativity_coefficient();

\end{boxedverbatim}
\\

\textbf{Centrality}

Clairlib supports several network centrality measures.  These measures
assign a value to each vertex depending on how ``central'' that vertex
is.

The Centrality modules are in namespace Clair::Network::Centrality.
Each module has two centrality member functions, which both return a
hash of vertices and their corresponding centrality.  The first
function returns the base centrality measure.  The second returns a
centrality normalized to between 0 and 1.

\textbf{Degree Centrality}

Ranking each vertex by vertex degree is the simplest measure of
network centrality.  This is called degree centrality.  For undirected
networks, it is simply the degree of each vertex.  For directed
networks, it is the total degree divided by two.

\textbf{Closeness Centrality}

Closeness centrality measures how close each vertex is to the other
vertices.  This is found by measuring the length from the target
vertex to every other reachable vertex along the shortest path.  The
reciprocal of this is the closeness centrality.

\textbf{Betweenness Centrality}

Betweenness centrality measures how many shortest paths the target
vertex is between.  The betweenness index is the sum of the number of
shortest paths between two actors through the target actor, divided by
the total number of shortest paths between the two actors.

\textbf{LexRank Centrality}

To compute the lexrank from a network built from a cluster using
create\_network or \linebreak
create\_sentence\_based\_network,
compute\_lexrank is provided.  Initial values or bias values can
also be loaded from a file using
read\_lexrank\_initial\_distribution and read\_lexrank\_bias (the
default for both is uniform).  If the network was not created from a
cluster appropriately (or to change the values), transition values
can also be loaded from a file using
read\_lexrank\_probabilities\_from\_file.
\\
\\
\begin{boxedverbatim}

my %lex_hash = $network->compute_lexrank();

\end{boxedverbatim}
\\

\textbf{PageRank Centrality}

Similarly, the pagerank can be computed with compute\_pagerank.
Transition values are already set for a network created with one of
the create\_hyperlink\_network functions, but can be read from a
file using \linebreak
read\_pagerank\_probabilities\_from\_file otherwise.
Initial distribution and personalization values can be read from
files using read\_pagerank\_initial\_distribution and
read\_pagerank\_personalization.

The results of these computations are returned by compute\_lexrank
and compute\_pagerank, but can also be saved to a file using
save\_current\_lexrank\_distribution or printed to standard out
using \linebreak
print\_current\_lexrank\_distribution (for pagerank,
save\_current\_pagerank\_distribution and \linebreak
print\_current\_pagerank\_distribution, respectively).
\\
\\
\begin{boxedverbatim}

$network->print_current_lexrank_distribution();
$network->save_current_lexrank_distribution('lex_out');

\end{boxedverbatim}
\\

Many other network based statistics can be computed.  For examples
of what can be computed, please see test\_network\_stat.pl in the
test directory.

\subsubsection{Network Generation}

Random networks can also be generated with the
Clair::Network::Generator package.  Currently, this includes
generation of Erd\H{o}s-R\'{e}nyi random graphs.

\textbf{Clair::Network::Generator::ErdosRenyi}

Two models of Erd\H{o}s-R\'{e}nyi random networks can be generated.
One includes a set number of nodes and edges.  The other type includes
a set number of nodes with an edge existing between two nodes with a
probability $p$.

Example:
\\
\\
\begin{boxedverbatim}

use Clair::Network::Generator::ErdosRenyi;
my $generator = Clair::Network::Generator::ErdosRenyi->new();
my $set_edges = $generator->generate(10, 20, type => "gnm");
my $random_number_edges = $generator->generate(10, 0.2, type => "gnp");

\end{boxedverbatim}

\subsubsection{Network Sampling}

Sometimes a network may be too large to process in its entirety.
Sampling can be used to extract a representive subset of the network
for analysis.  Different methods preserve different network
properties.  Clairlib provides several network sampling algorithms.

\begin{itemize}
\item{Clair::Network::Sample::RandomNode}

  Random node sampling simply chooses a number of nodes from the
  original graph, choosing nodes uniformly at random.  If there is an
  edge between two nodes that have been selected in the original
  network, that edge will be included in the sampled network.

\item{Clair::Network::Sample::RandomEdge}

  Random edge sampling chooses edges randomly from the original
  network, and includes the two incident nodes.

\item{Clair::Network::Sample::ForestFire}

  ForestFire sampling chooses an initial random node, and performs a
  probabilistic recursive breadth-first search from that initial node.
  If the "fire" dies out, it will restart at another random node.

\end{itemize}

Example:
\\
\\
\begin{boxedverbatim}

use Clair::Network::Sample::ForestFire;

my $fire = new Clair::Network::Sample::ForestFire($net);
print "Sampling 3 nodes using Forest Fire model\n";
$new_net = $fire->sample(3, 0.9);

\end{boxedverbatim}

\subsection{Clair::Statistics}

Clairlib provides several statistical tools for analyzing and
generating distributions.  New distributions include Geometric,
Gaussian, LogNormal, Zipfian and students T-distribution.  There is
also experimental support for statistical inference.  These
distribution and test modules are included under the Clair::Statistics
namespace.  See the test\_statistics.pl recipe for more information.
The older Clair::Gen will be folded into this in the next release.

\subsection{Clair::Gen}

Clair::Gen is for use when working with distributions.  It can
produce expected Power Law and Poisson distributions, or analyze
observed distributions.  The read\_from\_file method reads an
observed distribution from a file.

The plEstimate function accepts a distribution as input and
produces the best-fitting $\hat{c}$ and $\hat{\alpha}$ values.
genPl does the opposite--using $\hat{c}$ and $\hat{\alpha}$ as
input, it produces the expected distribution.

For Poisson distributions, poisEstimate and genPois are provided
which mirror the functionality of plEstimate and genPl.  plEstimate
is currently just a stub function, however.

To compare estimated and actual distributions, compareChiSquare is
included in the package.  This returns the number of degrees of
freedom and the p-value.
\\
\\
\begin{boxedverbatim}

my $g = new Clair::Gen;

$g->read_from_file("trial1.dist");
my @observed = $g->distribution;

my ($c_hat, $alpha_hat) = g->plEstimate(\@observed);
my @expected = g->genPL($c_hat, $alpha_hat);

my ($df, $pv) = $g->compareChiSquare(\@observed, \@expected, 2);

\end{boxedverbatim}
\\

\subsection{Clair::Util}

Clair::Util provides several different methods that are useful but
do not fit in other modules.  For example, build\_IDF\_database reads
a list of files and writes the IDF values from those files to a
database (Berkeley DB).  build\_idf\_by\_line can also be used to
build an IDF database, in this case, using text pass to it and
treating each line as a different document and computing the IDF
from those.  read\_idf opens a database and returns the
hash from it.  This is particularly useful for examining the contents
of an IDF database, but can be easily used for many other tasks as well.
\\
\\
\begin{boxedverbatim}

Clair::Util::build_idf_by_line("This is a test.\n" .
             "This is considered another document.\n" .
             "A third sample document.",
             "test_dbm_file");

my %idf = Clair::Util::read_idf("test_dbm_file");

print "The idf of 'this' is: ", $idf{this}, "\n";

\end{boxedverbatim}
\\

\subsection{Clair::Utils::CorpusDownload}

\subsubsection{Creating a Corpus}

The CorpusDownload module is provided to create a
corpus.  Create a CorpusDownload object using new().  A corpus name
must be provided, and a rootdir is optional, but strongly
recommended since the default is `/data0/projects/tfidf'.  The
rootdir must be an absolute path, rather than a relative path.  The
root directory is where the corpus files will be placed.  Many
corpora can be made with the same root directory, as long as the
corpusname is different for each.

Two functions are provided to create a corpus.  buildCorpus is used
to download files to form the corpus, while buildCorpusFromFiles is
used to form a corpus with files already on the computer.  Both
require a reference to an array with either the urls or absolute
paths to the files for buildCorpus and buildCorpusFromFiles,
respectively.  These files will then be copied to the root directory
provided and a corpus created from them in TREC format.

Because CorpusDownload was designed to use a downloaded corpus,
results from a corpus build with buildCorpusFromFiles will list
files with ``http://'' at the beginning, then the full path of the
file.

To use a base URL and find files based on links from that file, the
function poach is provided as an interface to `poacher.'  This
returns an array with URLs that can be passed to buildCorpus.
\\
\\
\begin{boxedverbatim}

$corpus = Clair::Utils::CorpusDownload->new(corpusname => 'new_corpus',
              rootdir => '/usr/username/');

$corpus->buildCorpus(urlsref => $@urls);

\end{boxedverbatim}
\\

\subsubsection{Computing IDF and TF Values}

To compute the IDF and TF values for
the corpus, buildIdf and buildTf are provided.  Both accept stemmed
as a parameter which can be set to 1 to compute the stemmed values
or 0 (the default) to compute the unstemmed values.  Before buildTf
can be called, build\_docno\_dbm must be called.
\\
\\
\begin{boxedverbatim}

$corpus->buildIdf(stemmed => 0);
$corpus->buildIdf(stemmed => 1);

$corpus->build_docno_dbm();

$corpus->buildTf(stemmed => 0);
$corpus->buildTf(stemmed => 1);

\end{boxedverbatim}
\\

\subsection{Clair::Utils::TF and Clair::Utils::IDF}

Once IDF values have been
computed, they can be accessed by creating an Idf object.  In the
constructor, rootdir and corpusname parameters should be supplied
that match the CorpusDownload parameters, along with a stemmed
parameter depending on whether stemmed or unstemmed values are
desired (1 and 0 respectively).  To get the IDF for a word, then,
use the method getIdfForWord, supplying the desired word.

A Tf object is created with the same parameters passed to the
constructor.  The function getFreq returns the number of times a
word appears in the corpus, getNumDocsWithWord returns the number of
documents it appears in, and getDocs returns the array of documents
it appears in.
\\
\\
\begin{boxedverbatim}

my $idf = Clair::Utils::Idf->new( rootdir=> '/usr/username/',
          corpusname =>'new\_corpus', stemmed => 0);

print "The idf of 'and' is ", $idf->getIdfForWord("and"), "\n";

my $tf = Clair::Utils::Idf->new( rootdir=> '/usr/username/',
         corpusname =>'new_corpus', stemmed => 0);

print $tf->getNumDocsWithWord("and"), " docs have 'and' in them\n";
print "'and' appears ", $tf->getFreq("and"), "times.\n";

print "The documents are:\n" my @docs = $tf->getDocs("and");
foreach my $doc (@docs) {
  print "$doc\n";
}

\end{boxedverbatim}
\\

\subsection{Clair::Utils::WebSearch}

\emph{This applies only to users of Clairlib-ext!}

The WebSearch module is used to perform Google searches.  A key must
be obtained from Google in order to do this.  Follow the instructions
in the section "Installing the Clair Library" to obtain a key and
have the WebSearch module use it.

Once the key has been obtained and the appropriate variables are set,
use the googleGet method to obtain a list of results to a Google query.
The following code gets the top 20 results to a search for the
"University of Michigan," and then prints the results to the screen.
\\
\\
\begin{boxedverbatim}

my @results = @{Clair::Utils::WebSearch::googleGet("University of \
Michigan", 20)};

foreach my $r (@results) {
  print "$r\n\n";
}

\end{boxedverbatim}
\\

The WebSearch module also provides the ability to download a single page
as a URI::URL-escaped file using the downloadUrl method.  This method
needs two parameters: the URL to download and the filename where the
downloaded page should be saved.
\\
\\
\begin{boxedverbatim}

Clair::Utils::WebSearch::downloadUrl("http://www.mgoblue.com/", \
"mgoblue_home.htm");

\end{boxedverbatim}
\\

\subsection{Clair::Utils::Parse}

\emph{This applies only to users of Clairlib-ext!}

The Parse module provides a wrapper for the Charniak parser and the
chunklink tool.

\subsubsection{Preparing a File for the Charniak Parser}

To be parsed by the Charniak parser, a file must be formatted in
a specific way, with sentences on separate lines, placed inside
$<$s$></$s$>$ tags.  For example:
\\
\\
\begin{boxedverbatim}

<s>This is one sentence.</s>
<s>This is another sentence.</s>

\end{boxedverbatim}
\\

To make this formatting easier, the the prepare\_for\_parse function
is provided.  This function will read a file, split it into sentences
(using Clair::Document::split\_into\_sentences), then put each sentence on
its own line, surrounded by $<$s$></$s$>$ tags, in a new file.
\\
\\
\begin{boxedverbatim}

Clair::Utils::Parse::prepare_for_parse("input.txt", "output.txt");

\end{boxedverbatim}
\\

If a file is already correctly formatted, this step should not be performed.

\subsubsection{Charniak Parser}

The parse function runs a file through the Charniak
parser.  The result of parsing will be returned from
the function as a string, and may optionally be written to a file
by specifying an output file.

Note that a file must be correctly
formatted to be parsed.  See the previous section, ``Preparing a File
for the Charniak Parser'' for more information.
\\
\\
\begin{boxedverbatim}

my $parse_output = Clair::Utils::Parse::parse("to_be_parsed.txt",
                          output_file => "output.txt");

\end{boxedverbatim}
\\

\subsubsection{Chunklink}

Chunklink is a very useful tool to analyze file from the Penn Treebank.
The Parse module also provides a wrapper to it, with the function
Parse::chunklink.  This function takes an input file and returns the
result as a string, and may optionally also write the results to a file.
\\
\\
\begin{boxedverbatim}

my $chunkout = Clair::Utils::Parse::chunklink("WSJ_0021.MRG",
                      output_file => "output.txt");

\end{boxedverbatim}
\\

\subsection{Clair::Utils::Stem}

This is an implementation of a stemmer, to take one word at a time
and return the stem of it.  There are only two functions: new and
stem.  Creating an object with new initializes the stemmer.
Subsequent calls to stem will return the stemmed version of a word.
Note that this is not the same stemmer that is used by
Document::stem.
\\
\\
\begin{boxedverbatim}

my $stemmer = new Clair::Utils::Stem;

print "'testing' stemmed is: ", $stemmer->stem("testing"), "\n";

\end{boxedverbatim}
\\
\section{Sample Code Example}

Several code examples are provided with Clairlib, in the `test' directory
and also in the next section of the tutorial.  This section takes
a thorough look at one of these, `test\_mega.pl.'  This script 
combines many pieces of functionality in Clairlib, so it serves as
a good example.

We now walk through this example section by section:
\\
\\
\begin{boxedverbatim}

# script: test_mega.pl
# functionality: Downloads documents using CorpusDownload, then makes IDFs,
# functionality: TFs, builds a cluster from them, a network based on a
# functionality: binary cosine, and tests the network for a couple of
# functionality: properties

use strict;
use warnings;
use FindBin;
use Clair::Utils::CorpusDownload;
use Clair::Utils::Idf;
use Clair::Utils::Tf;
use Clair::Document;
use Clair::Cluster;
use Clair::Network;

\end{boxedverbatim}
\\

We start by declaring the packages we will use.  We use FindBin
to make the example system independent, because we know the
relative location of the library to the script, rather than
the more typical situation of knowing the absolute path of the
library.  Typically, scripts
are more likely to change relative paths to the library than the
library is to move, so simply hard-coding the path here may be best
in most situations.

Next, we determine the ``base directory'' (where the script is located)
and remember the directory where we will put all produced files.  We
then create a CorpusDownload object, giving it a corpus name of ``testhtml''
and specifying the produced files directory as the root directory for the
corpus.  Note that we are specifying an absolute path, not a relative pass
for the rootdir parameter (otherwise, some CorpusDownload functions may not
work correctly).
\\
\\
\begin{boxedverbatim}

my $basedir = $FindBin::Bin;
my $gen_dir = "$basedir/produced/mega";

my $corpusref = Clair::Utils::CorpusDownload->new(corpusname => "testhtml",
                rootdir => $gen_dir);

\end{boxedverbatim}
\\

We use CorpusDownload::poach to start with a single URL and follow links
on that page, then links on those pages, etc. and return those URLs in
an array reference.  We iterate through those URLs and print them out
to the screen.  Finally, we pass those URLs to CorpusRef::buildCorpus
to download the URLs and create a corpus in TREC format.
\\
\\
\begin{boxedverbatim}

# Get the list of urls that we want to download
my $uref =                                                            \
$corpusref->poach("http://tangra.si.umich.edu/clair/testhtml/index.ht \
ml", error_file => "$gen_dir/errors.txt");

my @urls = @$uref;

foreach my $v (@urls) {
    print "URL: $v\n";
}

# Build the corpus using the list of urls
# This will index and convert to TREC format
$corpusref->buildCorpus(urlsref => $uref);


\end{boxedverbatim}
\\

Our next step is to build the IDF and TF files.  This computes the
IDF and TF values for every word, then stores them in files from which
those values can be easily retrieved.  We build the unstemmed IDF,
then the stemmed IDF first.  Next, we must build the DOCNO/URL
database before we build the TF files.  Again, we build the
unstemmed, and then the stemmed files (this order is not important
for either calculation).
\\
\\
\begin{boxedverbatim}

# -------------------------------------------------------------------
#  This is how to build the IDF.  First we build the unstemmed IDF,
#  then the stemmed one.
# -------------------------------------------------------------------
$corpusref->buildIdf(stemmed => 0);
$corpusref->buildIdf(stemmed => 1);

# -------------------------------------------------------------------
#  This is how to build the TF.  First we build the DOCNO/URL
#  database, which is necessary to build the TFs.  Then we build
#  unstemmed and stemmed TFs.
# -------------------------------------------------------------------
$corpusref->build_docno_dbm();
$corpusref->buildTf(stemmed => 0);
$corpusref->buildTf(stemmed => 1);

\end{boxedverbatim}
\\

Now that we have build these values, we want to be able to see what
the values are for specific words.  We create an Idf object, giving
it the same rootdir and corpusname as our CorpusDownload object.  We
choose whether we want the IDFs for the stemmed or unstemmed versions,
choosing unstemmed in this example.  We then get and print the IDF
values for several words: `have,' `and', and `zimbabwe.'  Note that
these words should be in lowercase.
\\
\\
\begin{boxedverbatim}

# -------------------------------------------------------------------
#  Here is how to use a IDF.  The constructor (new) opens the
#  unstemmed IDF.  Then we ask for IDFs for the words "have"
#  "and" and "zimbabwe."
# -------------------------------------------------------------------
my $idfref = Clair::Utils::Idf->new( rootdir => $gen_dir,
                       corpusname => "testhtml" ,
                       stemmed => 0 );

my $result = $idfref->getIdfForWord("have");
print "IDF(have) = $result\n";
$result = $idfref->getIdfForWord("and");
print "IDF(and) = $result\n";
$result = $idfref->getIdfForWord("zimbabwe");
print "IDF(zimbabwe) = $result\n";

\end{boxedverbatim}
\\

We now compute the TF values similarly.  We create a Tf object,
again using the same rootdir and corpusname as we did for
CorpusDownload, and again choosing whether we want the stemmed or
unstemmed information.  Now that we have our Tf object, we can
call getNumDocsWithWord to get the number of unique documents that
have a word, getFreq to get the number of times a word is in the
corpus, and getDocs to get all the URLs of all the documents that
have that word in them.  We do this with `washington', `and,' and
`zimbabwe.'
\\
\\
\begin{boxedverbatim}

# -------------------------------------------------------------------
#  Here is how to use a TF.  The constructor (new) opens the
#  unstemmed TF.  Then we ask for information about the
#  word "have":
#
#  1 first, we get the number of documents in the corpus with
#    the word "Washington"
#  2 then, we get the total number of occurrences of the word         \
"Washington"
#  3 then, we print a list of URLs of the documents that have the
#    word "Washington"
# -------------------------------------------------------------------
my $tfref = Clair::Utils::Tf->new( rootdir => $gen_dir,
                     corpusname => "testhtml" ,
                     stemmed => 0 );

$result = $tfref->getNumDocsWithWord("washington");
my $freq   = $tfref->getFreq("washington");
@urls = $tfref->getDocs("washington");
print "TF(washington) = $freq total in $result docs\n";
print "Documents with \"washington\"\n";
foreach my $url (@urls)  {  print "  $url\n";  }
print "\n";

# -------------------------------------------------------------------
#  Then we do 1-2 with the word "and"
# -------------------------------------------------------------------
$result = $tfref->getNumDocsWithWord("and");
$freq   = $tfref->getFreq("and");
@urls = $tfref->getDocs("and");
print "TF(and) = $freq total in $result docs\n";

# -------------------------------------------------------------------
#  Then we do 1-3 with the word "zimbabwe"
# -------------------------------------------------------------------
$result = $tfref->getNumDocsWithWord("zimbabwe");
$freq   = $tfref->getFreq("zimbabwe");
@urls = $tfref->getDocs("zimbabwe");
print "TF(zimbabwe) = $freq total in $result docs\n";
print "Documents with \"zimbabwe\"\n";
foreach my $url (@urls)  {  print "  $url\n";  }
print "\n";

\end{boxedverbatim}
\\

We now change direction, using the fact that CorpusDownload has
downloaded all of the html files to a specific directory.  The
directory location depends upon the root directory, the corpusname
and the url of each downloaded file.  In this case, all the
downloaded files are from the same host and same path in the URL,
so they are all in the same folder.

We create a new Clair::Cluster and use load\_documents to get all
the files from that directory.  We give a type of `html' so that
every Clair::Document that is created has type `html.'  Once we
have loaded the documents, we display a message saying how many
we have, then strip and stem all the documents.
\\
\\
\begin{boxedverbatim}

# Create a cluster with the documents
my $c = new Clair::Cluster;

$c->load_documents("$gen_dir/download/testhtml/tangra.si.umich.edu/cl \
air/testhtml/*", type => 'html');

print "Loaded ", $c->count_elements, " documents.\n";

$c->strip_all_documents;
$c->stem_all_documents;

print "I'm done stripping and stemming\n";


\end{boxedverbatim}
\\

In order to shorten the computation for the rest of the example,
we only want to look at 40 of the documents.  To do this, we
simply use a foreach loop that inserts the first 40 documents
into a new cluster.  Which 40 documents are inserted will vary
from system to system (and possibly from run to run) since they
are not specified or explicitly ordered in any way.
\\
\\
\begin{boxedverbatim}

my $count = 0;
my $c2 = new Clair::Cluster;
foreach my $doc (values %{ $c->documents} ) {
    $count++;

    if ($count <= 40) {
        $c2->insert($doc->get_id, $doc);
    }
}

\end{boxedverbatim}
\\

We now compute the cosine matrix for the new cluster.  This will
return a hash.  By indexing into the hash using a pair of documents,
we can get the cosine similarity of those two documents.  We next
compute the binary cosine using a threshold of 0.15.  We could
specify the cosine matrix, but not specifying it will result in the use of the
cosine matrix from the last compute\_cosine\_matrix.  This returns
a hash with the same format as that returned by compute\_cosine\_matrix.

Next, we create a network based on the binary cosine.  Every document
with at least one edge (explained next) will become a vertex
in the network, and every pair of documents
with a non-zero cosine matrix will have an edge between their
corresponding vertices.

Using this network, we compute a few statistics, getting the number
of documents in the network (remember, this will probably be less
than the 40 we started with because it is the number of documents
with at least one edge).  We also print out the average and
maximum diameter of the network we created.
\\
\\
\begin{boxedverbatim}

my %cm = $c2->compute_cosine_matrix();
my %bin_cos = $c2->compute_binary_cosine(0.15);
my $network = $c2->create_network(cosine_matrix => \%bin_cos);

print "Number of documents in network: ", $network->num_documents,    \
"\n";

print "Average diameter: ", $network->diameter(avg => 1), "\n";
print "Maximum diameter: ", $network->diameter(), "\n";

\end{boxedverbatim}
\\

\section{All Code Examples}

This section contains many different scripts which can
help understand Clairlib, and can be used as a starting point for
many common tasks.  It includes all unit tests, all stand-alone tests,
and all utilities distributed in both Clairlib-core and Clairlib-ext.

\let\oldboxedverbatim=\boxedverbatim
\renewcommand{\boxedverbatim}{\footnotesize\oldboxedverbatim}

\subsection{List of Recipes}

\begin{itemize}
\item Unit Tests

\begin{itemize}
  \item test\_cidrwrapper.t

 Using CIDR::Wrapper, add a document cluster and verify
 clustering 
  \item test\_corpus\_download.t

 Test CorpusDownload, downloading a corpus and checking the
 produced TF and IDF against expected results 
  \item test\_gen.t

 Test some statistical computations using Clair::Gen 
  \item test\_meadwrapper.t

 Test basic Clair::MEAD::Wrapper functions, such as
 summarization, varying compression ratios, feature sorting,
 etc., having assumed the use of Text::Sentence as a sentence
 splitting tool 
  \item test\_network.t

 Test basic Network functionality, such as node/edge addition
 and removal, path generation, statistics, matlab graphics
 generation, etc. 
  \item test\_networkwrapper\_docs.t

 Test the NetworkWrapper's lexrank generation for a small
 cluster of documents 
  \item test\_networkwrapper\_sents.t

 Test the NetworkWrapper's lexrank generation for a small        
 cluster of documents built from an array of sentences 
  \item test\_sentence\_combiner.t

 Test a variety of sentence-oriented Document functions, such
 as sentence scoring, and combining sentence feature scores 
  \item test\_sentence\_features\_cluster.t

 Test the propagation of feature scores between sentences
 related to each other through clusters. 
  \item test\_sentence\_features\_subs.t

 Test the assignment of standard features, such as length,
 position, and centroid, to sentences in a small Document 
  \item test\_sentence\_features.t

 Using a short document, test many sentence feature functions 
  \item test\_aleextract.t

 Using ALE, extract a corpus in a DB and perform several
 searches on it 
  \item test\_alesearch.t

 From a small set of documents, build an ALE DB and do some
 searches 
  \item test\_lexrank\_large\_mxt.t

 Test lexrank calculation on a network having used MxTerminator
 as the tool to split sentences. 
  \item test\_meadwrapper\_mxt.t

 Test basic Clair::MEAD::Wrapper functions, such as
 summarization, varying compression ratios, feature sorting,
 etc., having assumed the use of MxTerminator as a sentence
 splitting tool 
  \item test\_web\_search.t

 Test Clair::Utils::WebSearch and its use of the Google
 search API for returning varying numbers of webpages
 in response to queries 
\end{itemize}

\item Example Tests

\begin{itemize}
  \item biased\_lexrank.pl

 Computes the lexrank value of a network given bias sentences
  \item cidr.pl

 Creates a CIDR from input files and writes sample
 centroid files 
  \item classify.pl

 Classifies the test documents using the perceptron parameters
 calculated previously; requires that learn.pl has been run
  \item cluster.pl

 Creates a cluster, a sentence-based network from it,
 calculates a binary cosine and builds a network based
 on the cosine, then exports it to Pajek 
  \item compare\_idf.pl

 Compares results of Clair::Util idf calculations with
 those performed by the build\_idf script 
  \item corpusdownload\_hyperlink.pl

 Downloads a corpus and creates a network based on the
 hyperlinks between the webpages
  \item corpusdownload\_list.pl

 Downloads a corpus and makes stemmed and unstemmed IDFs
  and TFs
  \item corpusdownload.pl

 Downloads a corpus from a file containing URLs;
  makes IDFs and TFs
  \item document\_idf.pl

 Loads documents from an input dir; strips and stems them,
 and then builds an IDF from them 
  \item document.pl

 Creates Documents from strings, files, strips and stems them,
 splits them into lines, sentences, counts words, saves them 
  \item features\_io.pl

 Same as features.pl BUT, outputs the train data set as
 document and feature vectors in svm\_light format, reads
 the svm\_light formatted file and converts it to perl hash
  \item features.pl

 Reads docs from input/features/train, calculates chi-squared
 values for all extracted features, shows ways to retrieve
 those features
  \item features\_traintest.pl

 Builds the feature vector for training and testing datasets,
 and is a prerequisite for learn.pl and classify.pl
  \item genericdoc.pl

 Tests parsing of simple text/html file/string, conversion
 into xml file, instantiation via constructor and morph()
  \item html.pl

 Tests the html stripping functionality in Documents 
  \item hyperlink.pl

 Makes and populates a cluster, builds a network from
  hyperlinks between them; then tests making a subset
  \item idf.pl

 Creates a cluster from some input files, then builds an idf
 from the lines of the documents 
  \item index\_dirfiles\_incremental.pl

 Tests index update using Index/dirfiles.pm; requires
 index\_dirfiles.pl to be run previously
  \item index\_dirfiles.pl

 Tests index update using Index/dirfiles.pm, index is created
 in produces/index\_dirfiles, complementary to index\_mldbm.pl
  \item index\_mldbm\_incremental.pl

 Tests index update using Index/mldbm.pm; requires that
 index\_mldbm.pl was run previously
  \item index\_mldbm.pl

 Tests index creation using Index/mldbm.pm, outputs stats,
 uses input/index/Shakespear, creates produces/index\_mldbm
  \item ir.pl

 Builds a corpus from some text files, then makes an IDF, a
 TF, and outputs some information from them
  \item learn.pl

 Uses feature vectors in the svm\_light format and calculates
 and saves perceptron parameters; needs features\_traintest.pl
  \item lexrank2.pl

 Computes lexrank from a stemmed line-based cluster
  \item lexrank3.pl

 Computes lexrank from line-based, stripped and stemmed
 cluster
  \item lexrank4.pl

 Based on an interactive script, this test builds a sentence-
 based cluster, then a network, computes lexrank, and then
 runs MMR on it 
  \item lexrank\_large.pl

 Builds a cluster from a set of files, computes a cosine matrix
 and then lexrank, then creates a network and a cluster using
 a lexrank-based threshold of 0.2
  \item lexrank.pl

 Computes lexrank on a small network
  \item linear\_algebra.pl

 A variety of arithmetic tests of the linear algebra module 
  \item mead\_summary.pl

 Tests MEAD's summarizer on a cluster of two documents,
 prints features for each sentence of the summary 
  \item mega.pl

 Downloads documents using CorpusDownload, then makes IDFs,
 TFs, builds a cluster from them, a network based on a
 binary cosine, and tests the network for a couple of
 properties
  \item mmr.pl

 Tests the lexrank reranker on a network  
  \item networkstat.pl

 Generates a network, then computes and displays a large
 number of network statistics 
  \item pagerank.pl

 Creates a small cluster and runs pagerank, displaying
 the pagerank distribution
  \item query.pl

 Requires indexes to be built via index\_*.pl scripts, shows
 queries implemented in Clair::Info::Query, single-word and
 phrase queries, meta-data retrieval methods
  \item random\_walk.pl

 Creates a network, assigns initial probabilities and tests
 taking single steps and calculating stationary distribution 
  \item read\_dirfiles.pl

 Requires index\_*.pl scripts to have been run, shows how to
 access the document\_index and the inverted\_index, how to
 use common access API to retrieve information
  \item sampling.pl

 Exercises network sampling using RandomNode and ForestFire 
  \item statistics.pl

 Tests linear regression and T test code 
  \item stem.pl

 Tests the Clair::Utils::Stem stemmer
  \item summary.pl

 Test the cluster summarization ability using various features 
  \item wordcount\_dir.pl

 Counts the words in each file of a directory; outputs report 
  \item wordcount.pl

 Using Cluster and Document, counts the words in each file
 of a directory 
  \item xmldoc.pl

 Tests the XML to text function of Document 
  \item classify\_weka.pl

 Extracts bag-of-words features from each document
 in a training corpus of baseball and hockey documents,
 then trains and evaluates a Weka decision tree classifier,
 saving its output to files
  \item lsi.pl

 Constructs a latent semantic index from a corpus of
 baseball and hockey documents, then uses that index
 to map terms, queries, and documents to latent semantic
 space. The position vectors of documents in that space
 are then used to train and evaluate a SVM classifier
 using the Weka interface provided in Clair::Interface::Weka
  \item parse.pl

 Parses an input file and then runs chunklink on it 
\end{itemize}

\item Utilities

\begin{itemize}
  \item chunk\_document.pl

 Breaks a text file into multiple files of a given word length
  \item corpus\_to\_cos.pl

 Calculates cosine similarity for a corpus
  \item corpus\_to\_cos-threaded.pl

 Calculates cosine similarity using multiple threads
  \item corpus\_to\_lexical\_network.pl

 Generates a lexical network for a corpus
  \item corpus\_to\_network.pl

 Generates a hyperlink network from corpus HTML files
  \item cos\_to\_cosplots.pl

 Generates cosine distribution plots, creating a
 histogram in log-log space, and a cumulative cosine plot
 histogram in log-log space
  \item cos\_to\_histograms.pl

 Generates degree distribution histograms from
 degree distribution data
  \item cos\_to\_networks.pl

 Generate series of networks by incrementing through cosine cutoffs
  \item cos\_to\_stats.pl

 Generates a table of network statistics for networks by
 incrementing through cosine cutoffs
  \item crawl\_url.pl

 Crawls from a starting URL, returning a list of URLs
  \item directory\_to\_corpus.pl

 Generates a clairlib Corpus from a directory of documents
  \item download\_urls.pl

 Downloads a set of URLs
  \item generate\_random\_network.pl

 Generates a random network
  \item idf\_query.pl

 Looks up idf values for terms in a corpus
  \item index\_corpus.pl

 Builds the TF and IDF indices for a corpus 
 as well as several other support indices
  \item link\_synthetic\_collection.pl

 Links a collection using a certain network generator
  \item make\_synth\_collection.pl

 Makes a synthetic document set
  \item network\_growth.pl

 Generates graphs for queries in web search engine
 query logs and measures network statistics
  \item network\_to\_plots.pl

 Generates degree distribution plots, creating a
 histogram in log-log space, and a cumulative degree
 distribution histogram in log-log space.
  \item print\_network\_stats.pl

 Prints various network statistics
  \item sentences\_to\_docs.pl

 Converts a document with sentences into a set of
 documents with one sentence per document
  \item tf\_query.pl

 Looks up tf values for terms in a corpus
  \item search\_to\_url.pl

 Searches on a Google query and prints a list of URLs
  \item wordnet\_to\_network.pl

 Generates a synonym network from WordNet
\end{itemize}

\end{itemize}

\subsection{Unit Tests}

This section contains the unit tests included with Clairlib.

\subsubsection{test\_cidrwrapper.t}
\begin{boxedverbatim}

# script: test_cidrwrapper.t
# functionality: Using CIDR::Wrapper, add a document cluster and verify
# functionality: clustering 

use strict;
use warnings;
use FindBin;
use Test::More;
use Clair::Config;
use DB_File;

if (not defined $CIDR_HOME or not -d $CIDR_HOME) {
    plan( skip_all => 
        '$CIDR_HOME not defined or doesn\'t exist in Clair::Config' );
} else {
    plan( tests => 6 );
}

use_ok("Clair::CIDR::Wrapper");
use_ok("Clair::Cluster");

my $cidr = Clair::CIDR::Wrapper->new( 
    cidr_home => $CIDR_HOME, 
    dest => "$FindBin::Bin/produced/cidrwrapper/temp.cidr"
);

my $cluster = Clair::Cluster->new();
$cluster->load_documents("$FindBin::Bin/input/cidrwrapper/*");
$cidr->add_cluster($cluster);

my @results = $cidr->run_cidr();
is(@results, 2, "Two clusters");

foreach my $map(@results) {
    my $cluster = $map->{cluster};
    my $docs = $cluster->documents();
    if ($cluster->count_elements() == 2) {
        ok(exists $docs->{"fed1.txt"}, "fed1.txt exists");
        ok(exists $docs->{"fed2.txt"}, "fed2 txt exists");
    } else {
        ok(exists $docs->{"41.docsent"}, "41.docsent exists");
    }
}

\end{boxedverbatim}

\subsubsection{test\_corpus\_download.t}
\begin{boxedverbatim}

# script: test_corpus_download.t
# functionality: Test CorpusDownload, downloading a corpus and checking the
# functionality: produced TF and IDF against expected results 

$ENV{ALECACHE} = "/tmp";
use strict;
use warnings;
use FindBin;
use Test::More tests => 11;

use_ok('Clair::Utils::CorpusDownload');
use_ok('Clair::Util');

my $base_dir = $FindBin::Bin;
my $input_dir = "$base_dir/input/corpus_download";
my $root_dir = "$base_dir/produced/corpus_download";

my $corpus_name = "download_test";
my $corpusref = Clair::Utils::CorpusDownload->new(corpusname => $corpus_name,
    rootdir => $root_dir);

# Make sure we read in the correct number of URLs
my $uref = $corpusref->readUrlsFile("$base_dir/input/corpus_download/t.urls");
is(scalar @$uref, 6, "Number of url refs");

# Build the corpus
$corpusref->buildCorpus(urlsref => $uref);

# Now check to make sure the correct files have been downloaded
foreach my $url (@$uref) {
    my $remote_path = $url;
    $remote_path =~ s{^http://}{}g;
    if ($remote_path =~ m{/([^/]+)$}) {
        my $file_name = $1;
        ok( cd_compare("download/$corpus_name/$remote_path", $file_name), 
            "downloaded $file_name" );
    } else {
        fail("Bad URL: $url, check input dir $input_dir");
    }
}

$corpusref->buildIdf(stemmed => 1);
$corpusref->buildIdf(stemmed => 0);
$corpusref->build_docno_dbm();
$corpusref->buildTf(stemmed => 1);
$corpusref->buildTf(stemmed => 0);

ok( cd_compare("corpus-data/$corpus_name-tf/a/ab/abused.tf", "abused.tf"),
    "abused.tf" );
ok( cd_compare("corpus-data/$corpus_name-tf-s/a/ab/abus.tf", "abus.tf"),
    "abus.tf" );

sub cd_compare {
    my ($file1, $file2) = @_;
    return Clair::Util::compare_files(
        "$base_dir/produced/corpus_download/$file1",
        "$base_dir/expected/corpus_download/$file2"
    );
}

\end{boxedverbatim}

\subsubsection{test\_gen.t}
\begin{boxedverbatim}

# script: test_gen.t
# functionality: Test some statistical computations using Clair::Gen 

use strict;
use warnings;
use FindBin;
use Test::More tests=> 9;

use_ok('Clair::Gen');

my $file_input_dir = "$FindBin::Bin/input/gen";

my $g = new Clair::Gen;
$g->read_from_file("$file_input_dir/j.dist");
my $n = $g->count;

is($n, 8, "count");

my @expected_dist = (7, 4, 1, 0, 0, 0, 0, 3);
my @observed = $g->distribution;
is_deeply(\@observed, \@expected_dist, "distribution");

my ($c_hat, $alpha_hat) = $g->plEstimate(\@observed);
cmp_ok( abs($c_hat - 4.7265), '<', 0.0005, "plEstimate c_hat" );
cmp_ok( abs($alpha_hat + 0.465), '<', 0.005, "plEstimate alpha_hat" );

my @expected = $g->genPL($c_hat, $alpha_hat, $n);
my ($df, $pv) = $g->compareChiSquare(\@observed, \@expected, 2);
is($df, 5, "compareChiSquare df");
cmp_ok( abs($pv - 0.0895), '<', 0.0005, "compareChiSquare pv" );

# lambda = 8, nsamples = 20
my $lambda = 8;
my $n_samples = 20;
my @samples = $g->genPois($lambda, $n_samples);

is(scalar @samples, $n_samples, "genPois number of samples");
my $all_pos = 1;
for (@samples) {
    last and $all_pos = 0 if $_ <= 0;
}
ok($all_pos, "genPois positive samples");


\end{boxedverbatim}

\subsubsection{test\_meadwrapper.t}
\begin{boxedverbatim}

# script: test_meadwrapper.t
# functionality: Test basic Clair::MEAD::Wrapper functions, such as
# functionality: summarization, varying compression ratios, feature sorting,
# functionality: etc., having assumed the use of Text::Sentence as a sentence
# functionality: splitting tool 

use strict;
use warnings;
use FindBin;
use Clair::Config;
use Test::More;

use vars qw($SENTENCE_SEGMENTER_TYPE);
my $old_SENTENCE_SEGMENTER_TYPE = $SENTENCE_SEGMENTER_TYPE;
$SENTENCE_SEGMENTER_TYPE = "Text";

if (not defined $MEAD_HOME or not -d $MEAD_HOME) {
    plan( skip_all => 
        '$MEAD_HOME not defined in Clair::Config or doesn\'t exist' );
} else {
    plan( tests => 15 );
}

use_ok("Clair::MEAD::Wrapper");
use_ok("Clair::Cluster");
use_ok("Clair::Document");

my $cluster_dir = "$FindBin::Bin/produced/meadwrapper";
my $cluster = Clair::Cluster->new();
$cluster->load_documents("$FindBin::Bin/input/meadwrapper/*");

my $mead = Clair::MEAD::Wrapper->new( 
    mead_home => $MEAD_HOME,
    cluster => $cluster,
    cluster_dir => $cluster_dir
);

my %files = ( "fed1.txt" => 1, "fed2.txt" => 1, "41" => 1);
my @dids = $mead->get_dids();
for (@dids) {
    ok(exists $files{$_}, "listing dids: $_ exists");
}


map { delete $ENV{$_} } keys %ENV;

my @summary1 = $mead->run_mead();
is(@summary1, 13, "Generic summary");

$mead->add_option("-s -p 100");
my @summary2 = $mead->run_mead();
# This test is appropriate for MxTerminator.  Eventually this will be smart
# enough to know which sentence splitter is in use.
#is(@summary2, 64, "No compression");
# This test is appropriate for Text::Sentence.
# Furthermore, this unit test is now intended to only test the Text             \
SentenceSegmenter.
is(@summary2, 61, "No compression");

my @expected_features = sort ("Centroid", "Length", "Position");
my @features = sort $mead->get_feature_names();

is(scalar @features, scalar @expected_features, "Feature names");

for (my $i = 0; $i < @features; $i++) {
    ok($features[$i] eq $expected_features[$i], 
        "Feature names: $features[$i]");

\end{boxedverbatim}

\begin{boxedverbatim}

}

my %features = $mead->get_feature("Centroid");
my $centroid_41 = scalar @{ $features{"41"} };
my $centroid_fed1 = scalar @{ $features{"fed1.txt"} };
my $centroid_fed2 = scalar @{ $features{"fed2.txt"} };

is($centroid_41, 26, "Centroid scores: 41");
# See above comments re: MxTerminator/Text::Sentence.
#is($centroid_fed1, 21, "Centroid scores: fed1.txt");
#is($centroid_fed2, 18, "Centroid scores: fed2.txt");
is($centroid_fed1, 19, "Centroid scores: fed1.txt");
is($centroid_fed2, 16, "Centroid scores: fed2.txt");

$SENTENCE_SEGMENTER_TYPE = $old_SENTENCE_SEGMENTER_TYPE;

\end{boxedverbatim}

\subsubsection{test\_network.t}
\begin{boxedverbatim}

# script: test_network.t
# functionality: Test basic Network functionality, such as node/edge addition
# functionality: and removal, path generation, statistics, matlab graphics
# functionality: generation, etc. 

use strict;
use warnings;
use FindBin;
use Test::More tests => 64;

use_ok('Clair::Network');
use_ok('Clair::Network::Writer::Pajek');
use_ok('Clair::Network::Writer::Edgelist');
use_ok('Clair::Util');

my $file_gen_dir = "$FindBin::Bin/produced/network";
my $file_doc_dir = "$FindBin::Bin/input/network";
my $file_exp_dir = "$FindBin::Bin/expected/network";

my $g1 = Clair::Network->new();
$g1->add_node(1, text => "Random sentence");
$g1->add_node(2, text => "unique");
$g1->add_node(3, text => "mark hodges");
$g1->add_node(4, text => "mark liffiton");
$g1->add_node(5, text => "dragomir radev");
$g1->add_node(6, text => "mike dagitses");

$g1->add_edge(1, 2);
$g1->add_edge(1, 3);
$g1->add_edge(2, 4);
$g1->add_edge(4, 5);
$g1->add_edge(5, 6);
$g1->add_edge(4, 6);

#is($g1->diameter(filename => "$file_gen_dir/graph.diameter"), 3, "diameter");
#ok(compare_sorted_proper_files("graph.diameter"), "diameter files");

is($g1->diameter(), 3, "diameter"); 

is($g1->diameter(), 3, "diameter"); 
$g1->remove_edge(4, 6);
is($g1->diameter(), 4, "diameter"); 

$g1->add_node(7, text => "");
$g1->add_edge(1, 7);
$g1->add_edge(7, 6);

my @path = $g1->find_path(1, 6);
my $path_length = @path;
is($path_length, 3, "find_path");

$g1->set_node_weight(7, 20);
is($g1->get_node_weight(7), 20, "get_node_weight");

$g1->remove_node(7);

@path = $g1->find_path(1, 6);
$path_length = @path;
is($path_length, 5, "find_path");

# Test Pajek writing and reading
my $export = Clair::Network::Writer::Pajek->new();
$export->set_name('test_graph');
$export->write_network($g1, "$file_gen_dir/graph.pajek");

my $reader = Clair::Network::Reader::Pajek->new();
my $pajek_net = $reader->read_network("$file_gen_dir/graph.pajek");

\end{boxedverbatim}

\begin{boxedverbatim}

ok($pajek_net->{graph} eq $g1->{graph}, "Pajek reading and writing");

is($g1->num_documents(), 6, "num_documents");
is($g1->num_pairs(), 15, "num_pairs");
is($g1->num_links(), 5, "num_links");
my $graph = $g1->{graph};

$g1->add_node('EX8', text => 'an external node');
$g1->add_edge('EX8', 4);
$g1->add_edge(5, 'EX8');

is($g1->num_links(), 5, "num_links");
is($g1->num_links(external => 1), 2, "num_links external => 1");

my %deg_hist = $g1->compute_in_link_histogram();
is($deg_hist{1}, 5, "compute_in_link_histogram");

%deg_hist = $g1->compute_out_link_histogram();
is($deg_hist{1}, 3, "compute_out_link_histogram");

my $avg_deg = $g1->avg_total_degree();
is($avg_deg, 2, "avg_total_degree");

%deg_hist = $g1->compute_total_link_histogram();
is($deg_hist{1}, 2, "compute_total_link_histogram");

my $retString = $g1->power_law_out_link_distribution();
like($retString, qr/y = 3 x\^-0\.5849\d+/, "power_law_out_link_distribution");

$retString = $g1->power_law_in_link_distribution();
like($retString, qr/y = 5 x\^-2\.3219\d+/, "power_law_in_link_distribution");

$retString = $g1->power_law_total_link_distribution();
like($retString, qr/y = 2\.204\d+ x\^0\.0629\d+/, 
    "power_law_total_link_distribution");

is($g1->diameter(), 4, "diameter");
is($g1->diameter(undirected => 1), 5, "diameter undirected");

my $diameter = $g1->diameter(avg => 1);
cmp_ok(abs($diameter - 2.055), "<", 0.005, "diameter avg");
 
$diameter = $g1->diameter(avg => 1, undirected => 1);
cmp_ok(abs($diameter - 2.285), "<", 0.005, "diameter undirected avg");

# Test average shortest path
my $asp = $g1->average_shortest_path();
cmp_ok(abs($asp - 1.535), '<', 0.005, "average_shortest_path");

# Test Newman's power law exponent formula
my @npl = $g1->newman_power_law_exponent(\%deg_hist, 1);
cmp_ok(abs($npl[0] - 2.635), '<', 0.005, "newman_power_law_exponent");

# Test finding largest component
my $largest_component = $g1->find_largest_component("weakly");
is($largest_component->num_nodes(), 7, "find_largest_component");

$export = Clair::Network::Writer::Edgelist->new();
$export->write_network($g1, "$file_gen_dir/graph.links");
ok(compare_sorted_proper_files("graph.links"), "write_links");

$g1->write_nodes("$file_gen_dir/graph.nodes");
ok(compare_sorted_proper_files("graph.nodes"), "write_nodes");

my $wscc = $g1->Watts_Strogatz_clus_coeff;
cmp_ok(abs($wscc - 0.235), '<', 0.005, "Watts_Strogatz_clus_coeff");


\end{boxedverbatim}

\begin{boxedverbatim}

my $newman_cc = $g1->newman_clustering_coefficient();
cmp_ok($newman_cc, "=", 0.375, "newman_clustering_coefficient");

my @triangles = $g1->get_triangles();
cmp_ok($triangles[0][0], "eq", "4-5-EX8", "get_triangles");

my $spl = $g1->get_shortest_path_length("1", "4");
cmp_ok($spl, "=", 2, "shortest_path_length");

my %dist = $g1->get_shortest_paths_lengths("1");
cmp_ok($dist{5}, "=", 3, "shortest_paths_lengths");

$g1->write_db("$file_gen_dir/graph.db");
ok(-e "$file_gen_dir/graph.db", "write_db");

$g1->write_db("$file_gen_dir/xpose.db", transpose => 1);
ok(-e "$file_gen_dir/xpose.db", "write_db transpose");

$g1->find_scc("$file_gen_dir/graph.db", "$file_gen_dir/xpose.db", 
    "$file_gen_dir/graph-scc-db.fin");
ok(compare_sorted_proper_files("graph-scc-db.fin"), "find_scc");

$g1->get_scc("$file_gen_dir/graph-scc-db.fin", "$file_doc_dir/link_map", 
    "$file_gen_dir/graph.scc");
ok(compare_sorted_proper_files("graph.scc"), "get_scc");

my %in_hist = $g1->compute_in_link_histogram();
$g1->write_link_matlab(\%in_hist, "$file_gen_dir/graph_in.m", 'graph');
ok(compare_proper_files("graph_in.m"), "write_link_matlab");

$g1->write_link_dist(\%in_hist, "$file_gen_dir/graph-inLinks");
ok(compare_sorted_proper_files("graph-inLinks"), "write_link_dist");

my %cos = ();
$cos{1} = ();
$cos{1}{2} = .1;
$cos{1}{3} = .3;
$cos{1}{4} = .6;
$cos{2} = ();
$cos{2}{1} = .1;
$cos{2}{3} = .4;
$cos{2}{4} = .1;
$cos{3} = ();
$cos{3}{1} = .3;
$cos{3}{2} = .4;
$cos{3}{4} = .2;
$cos{4} = ();
$cos{4}{1} = .6;
$cos{4}{2} = .1;
$cos{4}{3} = .2;

my ($la, $nla) = $g1->average_cosines(\%cos);
cmp_ok(abs($la - 0.1665), "<", 0.0005, "average_cosines la");
cmp_ok(abs($nla - 0.3225), "<", 0.0005, "average_cosines nla");

my ($lb_ref, $nlb_ref) = $g1->cosine_histograms(\%cos);
my @lb = @$lb_ref;
my @nlb = @$nlb_ref;
is($lb[10], 2, "cosine_histograms lb");
is($nlb[10], 2, "cosine_histograms nlb");

$g1->write_histogram_matlab($lb_ref, $nlb_ref, "$file_gen_dir/graph", 
    "test_network");
ok(compare_sorted_proper_files("graph_linked_hist.m"),                          \
"write_histogram_matlab");
ok(compare_sorted_proper_files("graph_linked_cumulative.m"),                    \
"write_histogram_matlab");

\end{boxedverbatim}

\begin{boxedverbatim}

ok (compare_sorted_proper_files("graph_not_linked_hist.m"),                     \
"write_histogram_matlab");

my $hist_as_string = $g1->get_histogram_as_string($lb_ref, $nlb_ref);
open (HIST_FILE, "> $file_gen_dir/graph.hist")  
    or die "Couldn't open $file_gen_dir/graph.hist: $!";
print HIST_FILE $hist_as_string;
close(HIST_FILE);
ok(compare_sorted_proper_files("graph.hist"), "get_histogram_as_string");

$g1->create_cosine_dat_files('graph', \%cos, directory => "$file_gen_dir");
ok(compare_sorted_proper_files("graph-point-one-all.dat"), 
    "create_cosine_dat_files graph-point-one-all.dat");
ok(compare_sorted_proper_files("graph-all-cosine"), "...                        \
graph-all-cosine.dat");
ok(compare_sorted_proper_files("graph-0-1.dat"), "... graph-0-1.dat");
ok(compare_sorted_proper_files("graph-0-2.dat"), "... graph-0-2.dat");
ok(compare_sorted_proper_files("graph-0-3.dat"), "... graph-0-3.dat");
ok(compare_sorted_proper_files("graph-0-4.dat"), "... graph-0-4.dat");
ok(compare_sorted_proper_files("graph-0-5.dat"), "... graph-0-5.dat");
ok(compare_sorted_proper_files("graph-0-6.dat"), "... graph-0-6.dat");
ok(compare_sorted_proper_files("graph-0-7.dat"), "... graph-0-7.dat");
ok(compare_sorted_proper_files("graph-0-8.dat"), "... graph-0-8.dat");
ok(compare_sorted_proper_files("graph-0-9.dat"), "... graph-0-9.dat");
ok(compare_sorted_proper_files("graph-0.dat"), "... graph-0.dat");

my $network = Clair::Network->new();
open DEBUG, "$file_exp_dir/debug.graph";
while (<DEBUG>) {
    chomp;
    my ($from, $to) = split / /, $_;
    $network->add_edge($from, $to);
}
close DEBUG;
is($network->avg_in_degree(), $network->avg_out_degree(), "avg deg on graph");

# Compares two files named filename
# from the t/docs/expected directory and
# from the t/docs/produced directory
sub compare_proper_files {
	my $filename = shift;
	return Clair::Util::compare_files("$file_exp_dir/$filename", 
        "$file_gen_dir/$filename");
}

sub compare_sorted_proper_files {
	my $filename = shift;
	return Clair::Util::compare_sorted_files("$file_exp_dir/$filename", 
        "$file_gen_dir/$filename");
}

\end{boxedverbatim}

\subsubsection{test\_networkwrapper\_docs.t}
\begin{boxedverbatim}

# script: test_networkwrapper_docs.t
# functionality: Test the NetworkWrapper's lexrank generation for a small
# functionality: cluster of documents 

use strict;
use warnings;
use FindBin;
use Clair::Config;
use Test::More;

if (not defined $PRMAIN or -d $PRMAIN) {
    plan( skip_all => 
        '$PRMAIN not defined in Clair::Config or doesn\'t exist' );
} else {
    plan( tests => 7 );
}

use_ok("Clair::Cluster");
use_ok("Clair::Document");
use_ok("Clair::NetworkWrapper");
use_ok("Clair::Network::Centrality::CPPLexRank");

my @files = grep { /^[^\.]/ }                                                   \
glob("$FindBin::Bin/input/networkwrapper_docs/*");
my @expected_scores = ( [0.38, 0.40], [0.15, 0.17], [0.42, 0.44] );

my $cluster = Clair::Cluster->new();
my $i = 1;
for (@files) {
    chomp;
    my $doc = Clair::Document->new(
        file => $_,
        type => "text",
    );
    $doc->stem();
    $cluster->insert($i, $doc);
    $i++;
}

my %matrix = $cluster->compute_cosine_matrix();
my $network = $cluster->create_network(
    cosine_matrix => \%matrix, 
    include_zeros => 1
);
my $wrapped_network = Clair::NetworkWrapper->new(
    prmain => $PRMAIN,
    network => $network,
    clean => 1
);
my $cent = Clair::Network::Centrality::CPPLexRank->new($network);
$cent->centrality();

my @vertices = $wrapped_network->{graph}->vertices();
my $vector = $wrapped_network->get_property_vector(\@vertices, 
    "lexrank_value");

my @actual_scores;
for (my $i = 0; $i < ($vector->dim())[0]; $i++) {
    push @actual_scores, $vector->element($i + 1, 1);
}

for (my $i = 0; $i < @files; $i++) {
    ok($expected_scores[$i]->[0] <= $actual_scores[$i] &&
       $actual_scores[$i] <= $expected_scores[$i]->[1], "File: $files[$i]");
}

\end{boxedverbatim}

\subsubsection{test\_networkwrapper\_sents.t}
\begin{boxedverbatim}

# script: test_networkwrapper_sents.t
# functionality: Test the NetworkWrapper's lexrank generation for a small        
# functionality: cluster of documents built from an array of sentences 

use strict;
use FindBin;
use Clair::Config;
use Test::More;

if (not defined $PRMAIN or -d $PRMAIN) {
    plan( skip_all => 
        '$PRMAIN not defined in Clair::Config or doesn\'t exist' );
} else {
    plan( tests => 7 );
}

use_ok("Clair::Cluster");
use_ok("Clair::Document");
use_ok("Clair::NetworkWrapper");
use_ok("Clair::Network::Centrality::CPPLexRank");


my @sents           = ( "foo bar",    "bar baz",    "baz foo"    );
my @expected_scores = ( [0.30, 0.32], [0.41, 0.43], [0.24, 0.26] );

my $cluster = Clair::Cluster->new();
my $i = 1;
for (@sents) {
    chomp;
    my $doc = Clair::Document->new(
        string => $_,
        type => "text",
    );
    $doc->stem();
    $cluster->insert($i, $doc);
    $i++;
}

my %matrix = $cluster->compute_cosine_matrix();
my $network = $cluster->create_network(
    cosine_matrix => \%matrix, 
    include_zeros => 1
);
my $wrapped_network = Clair::NetworkWrapper->new(
    prmain => $PRMAIN,
    network => $network,
    clean => 1
);
my $cent = Clair::Network::Centrality::CPPLexRank->new($network);
$cent->centrality();

my @vertices = $wrapped_network->{graph}->vertices();
my $vector = $wrapped_network->get_property_vector(\@vertices, 
    "lexrank_value");

my @actual_scores;
for (my $i = 0; $i < ($vector->dim())[0]; $i++) {
    push @actual_scores, $vector->element($i + 1, 1);
}

for (my $i = 0; $i < @sents; $i++) {
    ok($expected_scores[$i]->[0] <= $actual_scores[$i] &&
       $actual_scores[$i] <= $expected_scores[$i]->[1], "Sentence:              \
$sents[$i]");
}

\end{boxedverbatim}

\subsubsection{test\_sentence\_combiner.t}
\begin{boxedverbatim}

# script: test_sentence_combiner.t
# functionality: Test a variety of sentence-oriented Document functions, such
# functionality: as sentence scoring, and combining sentence feature scores 

# mjschal edited this file.
# I removed the one test that generates a warning message in order to not have
# warnings cluttering up the screen when an installation of clairlib-core is
# being tested by an end-user.

use strict;
use Test::More tests => 15;
use Clair::Document;

my $text = "The first sentence ends with a period. Does the second sentence? "
         . "Last sentence here!";
my $doc = Clair::Document->new( string => $text, did => "doc", type => "text"   \
);

# Make sure that scores are undefined at the beginning
is($doc->get_sentence_score(0), undef, "can't get uncomputed scores");

# Compute some simple test features. This assumes that the tests for that
# part of the code have already passed.
$doc->compute_sentence_feature( name => "has_q_mark", feature => \&has_q_mark   \
);
$doc->compute_sentence_feature( name => "char_length", 
    feature => \&char_length );

# Get a basic combiner that does a linear combination.
my $combiner = linear_combiner( has_q_mark => 10, char_length => 1 );

# Score the sentences and normalize them
$doc->score_sentences( combiner => $combiner );
my @expected = (1, 16/19, 0);
scores_ok($doc, \@expected, "score_sentences");

# Test the default weight method
$doc->score_sentences( weights => { has_q_mark => 10, char_length => 1} );
scores_ok($doc, \@expected, "score_sentences with default weights");

# Score the sentences, but don't normalize
$doc->score_sentences( combiner => $combiner, normalize => 0 );
@expected = (39, 36, 20);
scores_ok($doc, \@expected, "score_sentences without normalizing");

# A one sentence document should just output its score as 1 (normalized)
my $unit_doc = Clair::Document->new( string => "One sent.", type => "text", 
    did => "unit" );
$unit_doc->compute_sentence_feature( name => "char_length", 
    feature => \&char_length );
$unit_doc->score_sentences( combiner => $combiner );
@expected = (1);
scores_ok($unit_doc, \@expected, "score_sentences with only one sentence");

# Case when score isn't normalized
$unit_doc->score_sentences( combiner => $combiner, normalize => 0 );
@expected = (10);
scores_ok($unit_doc, \@expected, "score_sentences one sent no normalize");

# Give all sentences the same feature, and the resulting scores should be 1
my $doc2 = Clair::Document->new( string => $text, type => "text" );
$doc2->compute_sentence_feature( name => "uniform", feature => \&uniform );
$doc2->score_sentences( combiner => linear_combiner( uniform => 1 ) );
@expected = (1, 1, 1);
scores_ok($doc2, \@expected, "score_sentences uniform feature");



\end{boxedverbatim}

\begin{boxedverbatim}

# The following test has been removed because it (intentionally) generates
# a warning message.

# A combiner should always return a real number
# my $doc3 = Clair::Document->new( string => $text, type => "text" );
# $doc3->compute_sentence_feature( name => "uniform", feature => \&uniform );
# my $ret = $doc3->score_sentences( combiner => \&bad_combiner );
# is($ret, undef, "Combiner should always return a real number");


sub scores_ok {
    my $doc = shift;
    my $expected = shift;
    foreach my $i ( 0 .. ($doc->sentence_count() - 1) ) {
        is($doc->get_sentence_score($i), $expected->[$i], "score $i ok");
    }
}

sub has_q_mark {
    my %params = @_;
    chomp $params{sentence};
    if ($params{sentence} =~ /\?/) {
        return 1;
    } else {
        return 0;
    }
}

sub char_length {
    my %params = @_;
    return length($params{sentence});
}

sub uniform {
    return 0;
}

sub linear_combiner {
    my %weights = @_;
    my $combiner = sub {
        my %features = @_;
        my $score = 0;
        foreach my $name (keys %weights) {
            if ($features{$name}) {
                $score += $weights{$name} * $features{$name};
            }
        }
        return $score;
    };
}

sub bad_combiner {
    return "text";
}

\end{boxedverbatim}

\subsubsection{test\_sentence\_features\_cluster.t}
\begin{boxedverbatim}

# script: test_sentence_features_cluster.t
# functionality: Test the propagation of feature scores between sentences
# functionality: related to each other through clusters. 

use strict;
use Test::More tests => 25;
use Clair::Cluster;
use Clair::Document;

my $text1 = "First sentence from doc1. The second sent from doc1.";
my $text2 = "First sentence from doc2. The second sent from doc2.";

my $doc1 = Clair::Document->new(string => $text1, id => 1);
my $doc2 = Clair::Document->new(string => $text2, id => 2);

my $cluster = Clair::Cluster->new(id => "cluster");
$cluster->insert(1, $doc1);
$cluster->insert(2, $doc2);

$cluster->compute_sentence_feature(name => "cid", feature => \&cid_feat);
$cluster->compute_sentence_feature(name => "did", feature => \&did_feat);

foreach my $did (1, 2) {
    foreach my $i (0, 1) {
        my $cvalue = $cluster->get_sentence_feature($did, $i, "cid");
        my $dvalue = $cluster->get_sentence_feature($did, $i, "did");
        is($cvalue, "cluster", "individ feature score ok");
        is($dvalue, $did, "individ feature score ok");
    }
}

$cluster->remove_sentence_features();

# Test cluster-wide normalization
$cluster->set_sentence_feature(1, 0, feat => 1); # did, sno, feature => value
$cluster->set_sentence_feature(1, 1, feat => 2);
$cluster->set_sentence_feature(2, 0, feat => 3);
$cluster->set_sentence_feature(2, 1, feat => 4);

$cluster->score_sentences( weights => { feat => 1 } );

is( $cluster->get_sentence_score(1, 0), 0, "sent 1" );
is( $cluster->get_sentence_score(1, 1), 1/3, "sent 2" );
is( $cluster->get_sentence_score(2, 0), 2/3, "sent 3" );
is( $cluster->get_sentence_score(2, 1), 1, "sent 4" );

my %scores = ( 1 => [0, 1/3],  2 => [2/3, 1] );
my %got_scores = $cluster->get_sentence_scores();
is_deeply(\%got_scores, \%scores, "hash of scores ok");

$cluster->remove_sentence_features();
$cluster->compute_sentence_feature( name => "state", feature => \&state_feat );
is( $cluster->get_sentence_feature(1, 0, "state"), 1, "state 1.0");
is( $cluster->get_sentence_feature(1, 1, "state"), 2, "state 1.1");
is( $cluster->get_sentence_feature(2, 0, "state"), 3, "state 2.0");
is( $cluster->get_sentence_feature(2, 1, "state"), 4, "state 2.1");

$cluster->remove_sentence_features();
$cluster->compute_sentence_feature( name => "state", feature => \&state_feat,
    normalize => 1);
is( $cluster->get_sentence_feature(1, 0, "state"), 0, "normalized 1.0");
is( $cluster->get_sentence_feature(1, 1, "state"), 1/3, "normalized 1.1");
is( $cluster->get_sentence_feature(2, 0, "state"), 2/3, "normalized 2.0");
is( $cluster->get_sentence_feature(2, 1, "state"), 1, "normalized 2.1");

$cluster->compute_sentence_feature( name => "unif", 
    feature => sub { return 0 }, normalize => 1);

\end{boxedverbatim}

\begin{boxedverbatim}

is( $cluster->get_sentence_feature(1, 0, "unif"), 1, "unif 1.0");
is( $cluster->get_sentence_feature(1, 1, "unif"), 1, "unif 1.1");
is( $cluster->get_sentence_feature(2, 0, "unif"), 1, "unif 2.0");
is( $cluster->get_sentence_feature(2, 1, "unif"), 1, "unif 2.1");

sub cid_feat {
    my %params = @_;
    return $params{cluster}->get_id();
}

sub did_feat {
    my %params = @_;
    return $params{document}->get_id();
}

sub state_feat {
    my %params = @_;

    unless (defined $params{state}->{feats}) {
        $params{state}->{feats} = { 1 => [1, 2], 2 => [3, 4] };
    }

    my $did = $params{document}->get_id();
    my $index = $params{sentence_index};

    return $params{state}->{feats}->{$did}->[$index];

}


\end{boxedverbatim}

\subsubsection{test\_sentence\_features\_subs.t}
\begin{boxedverbatim}

# script: test_sentence_features_subs.t
# functionality: Test the assignment of standard features, such as length,
# functionality: position, and centroid, to sentences in a small Document 

use strict;
use Test::More tests => 8;
use Clair::Document;
use Clair::SentenceFeatures qw(length_feature position_feature                  \
centroid_feature);

my $text = "Roses are red. Violets are blue. Sugar is sweet. This is the        \
longest sentence.";
my $doc = Clair::Document->new(string => $text);

my %feats = ( 
    lf => \&length_feature, 
    pf => \&position_feature,
#    cf => \&centroid_feature
);

my %expected = (
    lf => [3, 3, 3, 5],
    pf => [1, 3/4, 2/4, 1/4]
);

$doc->compute_sentence_features(%feats);

features_ok($doc, "lf", $expected{lf});
features_ok($doc, "pf", $expected{pf});

sub features_ok {
    my $doc = shift;
    my $name = shift;
    my $expected = shift;
    for (my $i = 0; $i < @$expected; $i++) {
        my $feat = $doc->get_sentence_feature($i, $name);
        is($expected->[$i], $feat, "$name for $i ok");
    }
}


\end{boxedverbatim}

\subsubsection{test\_sentence\_features.t}
\begin{boxedverbatim}

# script: test_sentence_features.t
# functionality: Using a short document, test many sentence feature functions 

# mjschal edited this file.
# I removed a test that intentionally and correctly generated a warning.  This  \
is
# to prevent warning messages from cluttering up the screen for an enduser of 
# Clairlib-core who is testing his or her installation.

use strict;
use Test::More tests => 34;
use Clair::Document;
use Clair::Cluster;

my $text = "This is the first sentence. This is short. So is this. But perhaps  \
the longest sentence of all is the last sentence.";
my $doc = Clair::Document->new( string => $text, type => "text", id => "doc" );


##########################
# Sentence feature tests #
##########################

# Check to make sure the sentences are being split correctly
is($doc->sentence_count(), 4, "Correct # of sents");

# Shouldn't be able to set sentence features for sentences out of range
my $ret = $doc->set_sentence_feature(4, test_feature => 100);
is(undef, $ret, "Can't set out of range features");

# Should be able to set and get sentence features
$ret = $doc->set_sentence_feature(0, test_feature => 100);
ok($ret, "Set in range freatures");
is($doc->get_sentence_feature(0, "test_feature"), 100,
    "Can get sent feat back");

# should return undef if feature doesn't exist
is($doc->get_sentence_feature(1, "test_feature"), undef,
    "Undefined feature returns undef");

# Return undef after feature has been removed
$doc->remove_sentence_feature(0, "test_feature");
is($doc->get_sentence_feature(0, "test_feature"), undef,
    "Undefined after removed feature");

# Set many features at once
my %s0_feats = ( feature1 => 1, feature2 => 2, feature3 => 3);
$doc->set_sentence_feature(0, %s0_feats);
my %got_s0_feats = $doc->get_sentence_features(0);
is_deeply(\%s0_feats, \%got_s0_feats, "Can set/get list of features");

# Compute a simple feature that counts how many ts or Ts there are 
$doc->compute_sentence_feature( name => "count_t", feature => \&count_t );
my @e_feats = (4, 2, 1, 7);
features_ok($doc, "count_t", \@e_feats);

# Compute a feature that copies the document id to check that a reference
# to the document is actually getting passed to the sentence feature
# sub.
$doc->compute_sentence_feature( name => "did", feature => \&did_feat );
@e_feats = ("doc", "doc", "doc", "doc");
features_ok($doc, "did", \@e_feats);

# Compute a feature that returns the index of the document to check that
# this argument is passed to the feature sub.
$doc->compute_sentence_feature( name => "index", feature => \&index_feat );
@e_feats = (0, 1, 2, 3);

\end{boxedverbatim}

\begin{boxedverbatim}

features_ok($doc, "index", \@e_feats);


# This next test has been removed because it (intentionally) generates warning
# messages.

# Compute a feature that just dies in order to make sure that a feature 
# calculation can't crash the system.
#eval {
#    no warnings;
#    $doc->compute_sentence_feature( name => "bad", feature => \&bad_feat );
#};
#is("", $@, "stopped from feature dying");
#features_ok($doc, "bad", [undef, undef, undef, undef]);


# See if we can pass state between calls to the feature subroutine
$doc->remove_sentence_features();
$doc->compute_sentence_feature( name => "state", feature => \&state_feat );
features_ok($doc, "state", [0, 1, 2, 3]);

# Make sure that we can normalize sentence features
$doc->remove_sentence_features();
$doc->compute_sentence_feature( name => "count_t", feature => \&count_t, 
    normalize => 1 );
features_ok($doc, "count_t", [1/2, 1/6, 0, 1]);

# Make sure that normalizes correctly with uniform scores
$doc->remove_sentence_features();
$doc->compute_sentence_feature( name => "unif", feature => \&unif, 
    normalize => 1 );
features_ok($doc, "unif", [1, 1, 1, 1]);

$doc->remove_sentence_features();
$doc->compute_sentence_feature( name => "did", feature => \&did_feat );
$doc->compute_sentence_feature( name => "unif", feature => \&unif );
is($doc->is_numeric_feature("did"), 0, "did not numeric feature" );
ok( $doc->is_numeric_feature("unif"), "unif numeric feature" );
$doc->set_sentence_feature(0, mixed => 1);
$doc->set_sentence_feature(1, mixed => 1);
$doc->set_sentence_feature(2, mixed => 1);
$doc->set_sentence_feature(2, mixed => "string");
is( $doc->is_numeric_feature("mixed"), 0, "mixed not numeric" );


sub features_ok {
    my $doc = shift;
    my $name = shift;
    my $expected = shift;
    for (my $i = 0; $i < @$expected; $i++) {
        my $feat = $doc->get_sentence_feature($i, $name);
        is($feat, $expected->[$i], "$name for $i ok");
    }
}

sub count_t {
    my %params = @_;
    my $doc = $params{document};
    my $sent = $params{sentence};
    $sent =~ s/[^tT]//g;
    return length($sent);
}

sub did_feat {
    my %params = @_;
    my $doc = $params{document};
    return $doc->get_id();

\end{boxedverbatim}

\begin{boxedverbatim}

}

sub index_feat {
    my %params = @_;
    return $params{sentence_index};
}

sub char_length {
    my %params = @_;
    return length($params{sentence});
}

sub bad_feat {
    die;
}

sub unif {
    return 0;
}

sub state_feat {
    my %params = @_;

    if (defined $params{state}->{count}) {
        $params{state}->{count} = $params{state}->{count} + 1;
    } else {
        $params{state}->{count} = 0;
    }

    return $params{state}->{count};
}

\end{boxedverbatim}

\subsubsection{test\_aleextract.t}
\begin{boxedverbatim}

# script: test_aleextract.t
# functionality: Using ALE, extract a corpus in a DB and perform several
# functionality: searches on it 

use warnings;
use strict;
use Clair::Config qw($ALE_PORT $ALE_DB_USER $ALE_DB_PASS);
use FindBin;
use Test::More;

if (not defined $ALE_PORT or not -e $ALE_PORT) {
    plan(skip_all => "ALE_PORT not defined in Clair::Config or doesn't exist");
} else {
    plan(tests => 10);
}

use_ok("Clair::ALE::Extract");
use_ok("Clair::ALE::Search");
use Clair::Utils::ALE qw(%ALE_ENV);

# Set up the ALE environment
my $doc_dir = "$FindBin::Bin/input/ale";
$ENV{MYSQL_UNIX_PORT} = $ALE_PORT;
$ALE_ENV{ALESPACE} = "test_extract";
$ALE_ENV{ALECACHE} = $doc_dir;
if (defined $ALE_DB_USER) {
    $ALE_ENV{ALE_DB_USER} = $ALE_DB_USER;
}
if (defined $ALE_DB_PASS) {
    $ALE_ENV{ALE_DB_PASS} = $ALE_DB_PASS;
}

# Extract the links
my $e = Clair::ALE::Extract->new();
my @files = glob("$doc_dir/tangra.si.umich.edu/clair/testhtml/*.html");
$e->extract( drop_tables => 1, files => \@files );

# TEST 1 - total pages
my $search = Clair::ALE::Search->new(
    limit => 200,
);
is(count_results($search), 107, "Total links indexed");

# TEST 2 - just from index.html
$search = Clair::ALE::Search->new(
    limit => 100,
    source_url => "http://tangra.si.umich.edu/clair/testhtml"
);
is(count_results($search), 3, "From index.html");

# TEST 3 - just to google
$search = Clair::ALE::Search->new(
    limit => 100,
    dest_url => "http://www.google.com"
);
is(count_results($search), 1, "To google.com");

# TEST 4 - "search the web"
$search = Clair::ALE::Search->new(
    limit => 100,
    link1_text => "Search the web"
);
is(count_results($search), 1, "With text \"Search the web\"");

# TEST 5,6 - "search the web" urls
$search = Clair::ALE::Search->new(
    limit => 100,

\end{boxedverbatim}

\begin{boxedverbatim}

    link1_text => "Search the web"
);
my $conn = $search->queryresult();
my $link = $conn->{links}->[0];
is($link->{from}->{url}, 
    "http://tangra.si.umich.edu/clair/testhtml", "link from");
is($link->{to}->{url}, "http://www.google.com", "link to");

# Clean up
$e->drop_tables();

# TEST 7,8 - from CorpusDownload style corpus 
$e = Clair::ALE::Extract->new();
my $old_space = $ALE_ENV{ALESPACE};
$e->extract(
    corpusname => "myCorpus",
    rootdir => "$FindBin::Bin/input/ale/corpus"
);
is($ALE_ENV{ALESPACE}, $old_space, "extract doesn't change ALESPACE");
$ALE_ENV{ALESPACE} = "myCorpus";
$search = Clair::ALE::Search->new();
is(count_results($search), 5, "Total links");
#$e->drop_tables();


# Helper
sub count_results {
    my $search = shift;
    my $total = 0;
    $total++ while $search->queryresult();
    return $total;
}

\end{boxedverbatim}

\subsubsection{test\_alesearch.t}
\begin{boxedverbatim}

# script: test_alesearch.t
# functionality: From a small set of documents, build an ALE DB and do some
# functionality: searches 

use warnings;
use strict;
use Clair::Config;
use FindBin;
use Test::More;

if (not defined $ALE_PORT or not -e $ALE_PORT) {
    plan(skip_all => "ALE_PORT not defined in Clair::Config or doesn't exist");
} else {
    plan(tests => 7);
}

use_ok("Clair::ALE::Extract");
use_ok("Clair::ALE::Search");
use Clair::Utils::ALE qw(%ALE_ENV);

# Set up the ALE environment
my $doc_dir = "$FindBin::Bin/input/ale";
$ENV{MYSQL_UNIX_PORT} = $ALE_PORT;
$ALE_ENV{ALESPACE} = "test_search";
$ALE_ENV{ALECACHE} = $doc_dir;
if (defined $ALE_DB_USER) {
    $ALE_ENV{ALE_DB_USER} = $ALE_DB_USER;
}
if (defined $ALE_DB_PASS) {
    $ALE_ENV{ALE_DB_PASS} = $ALE_DB_PASS;
}


my $extract = Clair::ALE::Extract->new();
my @files = glob("$doc_dir/foo.com/*.html");
$extract->extract(files => \@files);

# TEST 1 - total links
my $search = Clair::ALE::Search->new();
is(count_results($search), 5, "Total links");

# TEST 2 - links to self
$search  = Clair::ALE::Search->new(link1_word => "self");
is(count_results($search), 2, "Self links");

# TEST 3 - limit the results
$search = Clair::ALE::Search->new(limit => 1);
is(count_results($search), 1, "limit results");

# TEST 4 - case shouldn't matter
$search = Clair::ALE::Search->new(link1_word => "self");
my $search2 = Clair::ALE::Search->new(link1_word => "SeLF");
is(count_results($search), count_results($search2), "case");

# TEST 5 - mulltilink testing
$search = Clair::ALE::Search->new( link2_word => "web", link1_word => "self" );
is(count_results($search), 1, "multilink search");

# Clean up
$extract->drop_tables();

sub count_results {
    my $search = shift;
    my $total = 0;
    $total++ while $search->queryresult();
    return $total;
}

\end{boxedverbatim}

\subsubsection{test\_lexrank\_large\_mxt.t}
\begin{boxedverbatim}

# script: test_lexrank_large_mxt.t
# functionality: Test lexrank calculation on a network having used MxTerminator
# functionality: as the tool to split sentences. 

use strict;
use warnings;
use FindBin;
use Test::More;

use Clair::Config;

use vars qw($SENTENCE_SEGMENTER_TYPE $JMX_HOME);
my $old_SENTENCE_SEGMENTER_TYPE = $SENTENCE_SEGMENTER_TYPE;

if (defined $JMX_HOME) {
    $SENTENCE_SEGMENTER_TYPE = "MxTerminator";
    plan(tests => 10);
} else {
    plan(skip_all => "No path assigned to Clair::Config::JMX_HOME.  Test        \
skipped.");
}

use_ok('Clair::Network');
use_ok('Clair::Network::Centrality::LexRank');
use_ok('Clair::Cluster');
use_ok('Clair::Document');
use_ok('Clair::Util');


my $file_gen_dir = "$FindBin::Bin/produced/lexrank_large";
my $file_input_dir = "$FindBin::Bin/input/lexrank_large";
my $file_exp_dir = "$FindBin::Bin/expected/lexrank_large";

my $c = new Clair::Cluster();

$c->load_documents("$file_input_dir/*", type => 'html', count_id => 1);
$c->strip_all_documents();
$c->stem_all_documents();

is($c->count_elements, 3, "count_elements");
my $sent_n = $c->create_sentence_based_network;

is($sent_n->num_nodes(), 44, "num_nodes");
# is($sent_n->num_nodes(), 25, "num_nodes");

my %cos_matrix = $c->compute_cosine_matrix(text_type => 'stem');

my $n = $c->create_network(cosine_matrix => \%cos_matrix);

my $cent = Clair::Network::Centrality::LexRank->new($n);
$cent->centrality();
$cent->save_lexrank_probabilities_to_file("$file_gen_dir/lex1_prob");
ok(compare_proper_files("lex1_prob"), "save_lexrank_probabilities_to_file" );

my $lex_network = $n->create_network_from_lexrank(0.33);
is($lex_network->num_nodes, 2, "num_nodes");

my $lex_cluster = $n->create_cluster_from_lexrank(0.33);
is($lex_cluster->count_elements(), 2, "count_elements");

$SENTENCE_SEGMENTER_TYPE = $old_SENTENCE_SEGMENTER_TYPE;

# Compares two files named filename
# from the t/docs/expected directory and
# from the t/docs/produced directory
sub compare_proper_files {
  my $filename = shift;

\end{boxedverbatim}

\begin{boxedverbatim}


  return Clair::Util::compare_files("$file_exp_dir/$filename",                  \
"$file_gen_dir/$filename");
}


\end{boxedverbatim}

\subsubsection{test\_meadwrapper\_mxt.t}
\begin{boxedverbatim}

# script: test_meadwrapper_mxt.t
# functionality: Test basic Clair::MEAD::Wrapper functions, such as
# functionality: summarization, varying compression ratios, feature sorting,
# functionality: etc., having assumed the use of MxTerminator as a sentence
# functionality: splitting tool 

use strict;
use warnings;
use FindBin;
use Clair::Config;
use Test::More;

use vars qw($SENTENCE_SEGMENTER_TYPE $JMX_HOME);

if (not defined $MEAD_HOME or not -d $MEAD_HOME) {
    plan( skip_all => 
        '$MEAD_HOME not defined in Clair::Config or doesn\'t exist' );
} else {
    if (not defined $JMX_HOME) {
        plan( skip_all => '$JMX_HOME not defined in Clair::Config.' );
    } else {
        plan( tests => 15 );
    }
}

my $old_SENTENCE_SEGMENTER_TYPE = $SENTENCE_SEGMENTER_TYPE;
$SENTENCE_SEGMENTER_TYPE = "MxTerminator";

use_ok("Clair::MEAD::Wrapper");
use_ok("Clair::Cluster");
use_ok("Clair::Document");

my $cluster_dir = "$FindBin::Bin/produced/meadwrapper";
my $cluster = Clair::Cluster->new();
$cluster->load_documents("$FindBin::Bin/input/meadwrapper/*");

my $mead = Clair::MEAD::Wrapper->new( 
    mead_home => $MEAD_HOME,
    cluster => $cluster,
    cluster_dir => $cluster_dir
);

my %files = ( "fed1.txt" => 1, "fed2.txt" => 1, "41" => 1);
my @dids = $mead->get_dids();
for (@dids) {
    ok(exists $files{$_}, "listing dids: $_ exists");
}


map { delete $ENV{$_} } keys %ENV;

my @summary1 = $mead->run_mead();
is(@summary1, 13, "Generic summary");

$mead->add_option("-s -p 100");
my @summary2 = $mead->run_mead();
is(@summary2, 64, "No compression");
# This test is only appropriate for Text::Sentence.
#is(@summary2, 61, "No compression");

my @expected_features = sort ("Centroid", "Length", "Position");
my @features = sort $mead->get_feature_names();

is(scalar @features, scalar @expected_features, "Feature names");

for (my $i = 0; $i < @features; $i++) {
    ok($features[$i] eq $expected_features[$i], 

\end{boxedverbatim}

\begin{boxedverbatim}

        "Feature names: $features[$i]");
}

my %features = $mead->get_feature("Centroid");
my $centroid_41 = scalar @{ $features{"41"} };
my $centroid_fed1 = scalar @{ $features{"fed1.txt"} };
my $centroid_fed2 = scalar @{ $features{"fed2.txt"} };

is($centroid_41, 26, "Centroid scores: 41");
is($centroid_fed1, 21, "Centroid scores: fed1.txt");
is($centroid_fed2, 18, "Centroid scores: fed2.txt");

$SENTENCE_SEGMENTER_TYPE = $old_SENTENCE_SEGMENTER_TYPE;

\end{boxedverbatim}

\subsubsection{test\_web\_search.t}
\begin{boxedverbatim}

# script: test_web_search.t
# functionality: Test Clair::Utils::WebSearch and its use of the Google
# functionality: search API for returning varying numbers of webpages
# functionality: in response to queries 

use strict;
use warnings;
use FindBin;
use Clair::Config;
use Test::More;

if (not defined $GOOGLE_DEFAULT_KEY) {
    plan(skip_all => "GOOGLE_DEFAULT_KEY not defined in Clair::Config");
} else {
    plan(tests => 5);
}

use_ok('Clair::Utils::WebSearch');
use_ok('Clair::Util');

my $file_gen_dir = "$FindBin::Bin/produced/web_search";
my $file_exp_dir = "$FindBin::Bin/expected/web_search";

Clair::Utils::WebSearch::download("http://tangra.si.umich.edu/", 
   "$file_gen_dir/tangrapage");
ok(compare_proper_files("tangrapage"), "WebSearch::download" );

my @results = @{Clair::Utils::WebSearch::googleGet("Westminster Abbey", 15)};
# We cannot be sure what the results will be, but we can be pretty safe
# that there will be at least 15

is(scalar @results, 15, "googleGet 1");

@results = @{Clair::Utils::WebSearch::googleGet("Arwad Island", 25)};
# Again, we don't know how what the results will be, but this call should
# return exactly 25
is(scalar @results, 25, "googleGet 2");


# Compares two files named filename
# from the t/docs/expected directory and
# from the t/docs/produced directory
sub compare_proper_files {
  my $filename = shift;

  return Clair::Util::compare_files("$file_exp_dir/$filename",                  \
"$file_gen_dir/$filename");
}


\end{boxedverbatim}

\subsection{Example tests}

This section contains the different sample programs that show off the features included in Clairlib.

\subsubsection{biased\_lexrank.pl}
\begin{boxedverbatim}

#!/usr/local/bin/perl

# script: test_biased_lexrank.pl
# functionality: Computes the lexrank value of a network given bias sentences

use strict;
use warnings;
use FindBin;
use Clair::Config;
use Clair::Cluster;
use Clair::Document;
use Clair::NetworkWrapper;

my @sents = ("The president's neck is missing", 
             "The human torch was denied a bank loan today",
             "The verdict was mail fraud");
my @bias = ("The president's neck is missing",
             "The president was given a bank loan");

print "Sentences:\n";
map { print "\t$_\n" } @sents;
print "\nBias sentences:\n";
map { print "\t$_\n" } @bias;

my $cluster = Clair::Cluster->new();
my $i = 1;

for (@sents) {
    chomp;
    my $doc = Clair::Document->new(
        string => $_,
        type => "text",
    );
    $doc->stem();
    $cluster->insert($i, $doc);
    $i++;
}

my %matrix = $cluster->compute_cosine_matrix();
my $network = $cluster->create_network(
    cosine_matrix => \%matrix, 
    include_zeros => 1
);
my $wn = Clair::NetworkWrapper->new(
    prmain => $PRMAIN,
    network => $network
);

my @verts = $wn->{graph}->vertices();

my $lr = Clair::Network::Centrality::LexRank->new($network);

my $lrv = $lr->compute_lexrank_from_bias_sents( bias_sents=>\@bias );

for (my $i =0; $i < @verts; $i++) {
    print "$sents[$i]\t", $lrv->element($i + 1, 1), "\n";
}

\end{boxedverbatim}

\subsubsection{cidr.pl}
\begin{boxedverbatim}

#!/usr/local/bin/perl

# script: test_cidr.pl
# functionality: Creates a CIDR from input files and writes sample
# functionality: centroid files 

use warnings;
use strict;
use FindBin;
use Clair::Cluster;
use Clair::CIDR;
use Getopt::Long;

my $input_dir = "$FindBin::Bin/input/cidr";
my $output_dir = "$FindBin::Bin/produced/cidr";

unless (-d $output_dir) {
    mkdir $output_dir or die "Couldn't mkdir $output_dir: $!";
}

opendir INPUT, $input_dir or die "Couldn't opendir $input_dir: $!";
my @files = map { "$input_dir/$_" } grep { /\.txt$/ } readdir INPUT;
closedir INPUT;

my $cluster = Clair::Cluster->new();
$cluster->load_file_list_array(\@files, type => "text");

my $cidr = Clair::CIDR->new();
my @results = $cidr->cluster($cluster);

chdir $output_dir or die "Couldn't chdir to $output_dir: $!";
foreach my $result (@results) {

    my $cluster = $result->{cluster};
    my $centroid = $result->{centroid};

    my @words= sort { $centroid->{$b} <=> $centroid->{$a} } keys %$centroid;
    my $docs = $cluster->documents();

    my $str = "$words[0]_$words[1]_$words[2]";
    mkdir "$str" or die "Couldn't mkdir $output_dir/$str: $!";

    open CENTROID, "> $str/centroid.txt" 
        or die "Couldn't open $str/centroid.txt: $!";
    foreach my $word (@words) {
        print CENTROID "$word\t$centroid->{$word}\n";
    }
    close CENTROID; 

    $cluster->save_documents_to_directory($str, "text");

    print "cluster: $str\n";
    map { print "\t$_\n" } keys %{ $cluster->documents() };
    print "\n";

}

\end{boxedverbatim}

\subsubsection{classify.pl}
\begin{boxedverbatim}

#!/usr/local/bin/perl

# script: test_classify.pl
# functionality: Classifies the test documents using the perceptron parameters
# functionality: calculated previously; requires that learn.pl has been run

use strict;
use FindBin;
# use lib "$FindBin::Bin/../lib";
# use lib "$FindBin::Bin/lib"; # if you are outside of bin path.. just in case
use vars qw/$DEBUG/;

use Benchmark;
use Clair::Classify;
use Data::Dumper;
use File::Find;

$DEBUG = 0;

my $results_root = "$FindBin::Bin/produced/features";
mkpath($results_root, 0, 0777) unless(-d $results_root);

my $output = "feature_vectors";
my $test = "$results_root/$output.test";
my $model = "$results_root/model";
my $output = "$results_root/classify.results";


unless(-f $test)
{
	print "The test file is required. Make sure learn.pl has been run.\n";
	exit;
}


my $t0;
my $t1;

#
# Finding files
#
$t0 = new Benchmark;

my $cla = new Clair::Classify(DEBUG => $DEBUG, test => $test, model => $model);

my ($result, $correct_count, $total_count) = $cla->classify();

my $percent = sprintf("%.4f", ( $correct_count / $total_count ) * 100 );
# print Dumper(\@return);
print "accuracy: ( $correct_count / $total_count ) * 100 = $percent\n";


$cla->debugmsg($result, 1);

# save the output
open M, "> $output" or $cla->errmsg("cannot open file '$output': $!", 1);
for my $aref (@$result)
{
	my $line = join " ", @$aref;
	print M "$line\n";
}
close M;


$t1 = new Benchmark;
my $timediff_find = timestr(timediff($t1, $t0));


\end{boxedverbatim}

\subsubsection{cluster.pl}
\begin{boxedverbatim}

#!/usr/local/bin/perl

# script: test_cluster.pl
# functionality: Creates a cluster, a sentence-based network from it,
# functionality: calculates a binary cosine and builds a network based
# functionality: on the cosine, then exports it to Pajek 

# Note: Make sure java is in your path, it is used by the splitter.

use strict;
use warnings;
use FindBin;
use lib "$FindBin::Bin/../lib";

use Clair::Document;
use Clair::Cluster;
use Clair::Network;

my $basedir = $FindBin::Bin;
my $input_dir = "$basedir/input/cluster";
my $gen_dir = "$basedir/produced/cluster";

# Create a cluster
my $c = new Clair::Cluster;

my $count = 0;

# Read every document from the the 'text' directory
# And insert it into the cluster
# Convert from HTML to text, then stem as we do so
while ( <$input_dir/*> ) {
	my $file = $_;

	my $doc = new Clair::Document(type => 'html', file => $file, id => ++$count);
	$doc->strip_html;
	$doc->stem;

	$c->insert($count, $doc);
}

print "Loaded ", $c->count_elements, " documents.\n";

print "Creating sentence based network.\n";
my $n = $c->create_sentence_based_network();
print "Created sentence based network with: ", $n->num_nodes(), " documents and \
", $n->num_links, " edges.\n";

# Compute the cosine matrix
my %cos_matrix = $c->compute_cosine_matrix;

# Find the largest cosine
my %largest_cosine = $c->get_largest_cosine;
print "The largest cosine is ", $largest_cosine{'value'}, " produced by ",
      $largest_cosine{'key1'}, " and ", $largest_cosine{'key2'}, ".\n";

# Compute the binary cosine using threshold = 0.15, 
# then write it to file 'docs/produced/text.cosine'
my %bin_cosine = $c->compute_binary_cosine(0.15);
$c->write_cos("$gen_dir/text.cosine", cosine_matrix => \%bin_cosine);

# Create a network using the binary cosine,
# then export the network to Pajek
$n = $c->create_network(cosine_matrix => \%bin_cosine);
my $export = Clair::Network::Writer::Pajek->new();
$export->set_name('cosine_network');
$export->write_network($n, "$gen_dir/test.pajek");


\end{boxedverbatim}

\begin{boxedverbatim}

$c->save_documents_to_directory($gen_dir, 'text');

\end{boxedverbatim}

\subsubsection{compare\_idf.pl}
\begin{boxedverbatim}

#!/usr/local/bin/perl

# script: test_compare_idf.pl
# functionality: Compares results of Clair::Util idf calculations with
# functionality: those performed by the build_idf script 

# This is used to compare the results of the idf calculations in Clair::Util
# to the ones performed by the build_idf script
# Input should be a single file that has already been stemmed
use strict;
use warnings;
use FindBin;
use Clair::Util;
use Clair::Cluster;
use Clair::Document;
use DB_File;

# This file has been stemmed.
my $input_file = "$FindBin::Bin/input/compare_idf/speech.txt";
my $output_dir = "$FindBin::Bin/produced/compare_idf";

# Create cluster
my %documents = ();
my $c = Clair::Cluster->new(documents => \%documents);



# Create each document, stem it, and insert it into the cluster
# Add the stemmed text to the $text variable
my $doc = Clair::Document->new(type => 'text', file => $input_file, id =>       \
$input_file);
$c->insert(document => $doc, id => $input_file);
my $text .= $doc->get_text() . " ";

# Take off the last newline like the other build_idf does (for comparison)
$text = substr($text, 0, length($text) - 1);

# Make the produced directory unless it exists
unless (-d $output_dir) {
    mkdir $output_dir or die "Couldn't create $output_dir: $!";
}

Clair::Util::build_idf_by_line($text, "$output_dir/dbm2");

my %idf = Clair::Util::read_idf("$output_dir/dbm2");
my $l;
my $r;
my $ct = 0;

while (($l, $r) = each %idf) {
  $ct++;
  print "$ct\t$l\t*$r*\n";
}

\end{boxedverbatim}

\subsubsection{corpusdownload\_hyperlink.pl}
\begin{boxedverbatim}

#!/usr/local/bin/perl

# script: test_corpusdownload_hyperlink.pl
# functionality: Downloads a corpus and creates a network based on the
# functionality: hyperlinks between the webpages

use strict;
use warnings;
# -------------------------------------------------------------------
#   This is a sample driver for the TF/IDF CLAIR library modules
# -------------------------------------------------------------------

# -------------------------------------------------------------------
#  * Use CorpusDownload.pm to download and build a new corpus, or
#      to build a TF or IDF.
#  * Use Idf (Tf) to use an already-built Idf (Tf)
# -------------------------------------------------------------------
use DB_File;
use FindBin;
use Clair::Utils::CorpusDownload;
use Clair::Utils::Idf;
use Clair::Utils::Tf;
use Clair::Network;
use Clair::Network::Centrality::PageRank;

my $basedir = $FindBin::Bin;
my $input_dir = "$basedir/input/corpusdownload_hyperlink";

my $gen_dir = "$basedir/produced/corpusdownload_hyperlink";
unless (-d $gen_dir) {
    mkdir $gen_dir or die "Couldn't mkdir $gen_dir: $!";
}
unless (-d "$gen_dir/corpora") {
    mkdir "$gen_dir/corpora" or die "Couldn't mkdir $gen_dir/corpora: $!";
}

# -------------------------------------------------------------------
#  This is the constructor.  It simply stores the directory
#  and name of the corpus.  It must be called prior to
#  any other routine.
# -------------------------------------------------------------------
my $corpus_name = "test-hyper";
my $corpusref = Clair::Utils::CorpusDownload->new(corpusname => $corpus_name,
                rootdir => $gen_dir);

# -------------------------------------------------------------------
#  Here's how to build a corpus.  An array @urls needs to be
#  built somehow.  (Here, we read the URLs from a file
#  $corpusname.urls.)  Then, the corpus will be built in
#  the directory $rootdir/$corpusname
# -------------------------------------------------------------------
my $uref = $corpusref->readUrlsFile("$input_dir/t.urls");
$corpusref->buildIdf(stemmed => 0, rootdir => $gen_dir );
$corpusref->buildIdf(stemmed => 1, rootdir => $gen_dir );
$corpusref->buildCorpus(urlsref => $uref, rootdir => $gen_dir );
$corpusref->build_docno_dbm( rootdir => $gen_dir );

# -------------------------------------------------------------------
# Compute the file listing the links
# -------------------------------------------------------------------
$corpusref->write_links( rootdir => $gen_dir );

# -------------------------------------------------------------------
# Create the network based on the links
# -------------------------------------------------------------------
my $linkfile = "$gen_dir/corpus-data/$corpus_name/$corpus_name.links";
my $doc_to_file =                                                               \

\end{boxedverbatim}

\begin{boxedverbatim}

"$gen_dir/corpus-data/$corpus_name/$corpus_name-docid-to-file";
my $compress_dbm =                                                              \
"$gen_dir/corpus-data/$corpus_name/$corpus_name-compress-docid";

my $network = Clair::Network->new_hyperlink_network($linkfile,                  \
docid_to_file_dbm => $doc_to_file, compress_docid => $compress_dbm);
my $networkEX = Clair::Network->new_hyperlink_network($linkfile, ignore_EX =>   \
0, docid_to_file_dbm => $doc_to_file, compress_docid => $compress_dbm);

# -------------------------------------------------------------------
# Create the network based on the links
# -------------------------------------------------------------------
print "Diameter without EX: ", $network->diameter(max => 1), "\n";
print "Avg diameter without EX: ", $network->diameter(avg => 1), "\n";

print "Diameter with EX: ", $networkEX->diameter(max => 1), "\n";
print "Avg diameter with EX: ", $networkEX->diameter(avg => 1), "\n";


my $cent = Clair::Network::Centrality::LexRank->new($network);

$network->centrality();

print "Pagerank results:\n";
$network->print_current_distribution();

$cent = Clair::Network::Centrality::LexRank->new($network);

$cent->centrality();
print "Pagerank results with EX:\n";
$cent->print_current_distribution();


\end{boxedverbatim}

\subsubsection{corpusdownload\_list.pl}
\begin{boxedverbatim}

#!/usr/local/bin/perl

# script: test_corpusdownload_list.pl
# functionality: Downloads a corpus and makes stemmed and unstemmed IDFs
# functionality:  and TFs

use strict;
use warnings;
use DB_File;
use FindBin;
use Clair::Utils::CorpusDownload;
use Clair::Utils::Idf;
use Clair::Utils::Tf;

my $basedir = $FindBin::Bin;
my $input_dir = "$basedir/input/corpusdownload_list";

my $gen_dir = "$basedir/produced/corpusdownload_list";
unless (-d $gen_dir) {
    mkdir $gen_dir or die "Couldn't mkdir $gen_dir: $!";
}
unless (-d "$gen_dir/corpora") {
    mkdir "$gen_dir/corpora" or die "Couldn't mkdir $gen_dir/corpora: $!";
}

# -------------------------------------------------------------------
#  This is the constructor.  It simply stores the directory
#  and name of the corpus.  It must be called prior to
#  any other routine.
# -------------------------------------------------------------------
my $corpus_name = "test-files";
my $corpusref = Clair::Utils::CorpusDownload->new(corpusname => $corpus_name,
                rootdir => "$gen_dir");

# -------------------------------------------------------------------
#  Here's how to build a corpus.  An array @urls needs to be
#  built somehow.  (Here, we read the URLs from a file
#  $corpusname.urls.)  Then, the corpus will be built in
#  the directory $rootdir/$corpusname
# -------------------------------------------------------------------
my $uref = $corpusref->readUrlsFile("$input_dir/files.list");
foreach my $url (@$uref) {
	$url = "$input_dir/" . $url;
}

foreach my $url (@$uref) {
	print "URL: $url\n";
}

print "Read ", scalar @$uref, " filenames.\n";
$corpusref->buildCorpusFromFiles(filesref => $uref, cleanup => 0);

# -------------------------------------------------------------------
#  This is how to build the IDF.  First we build the unstemmed IDF,
#  then the stemmed one.
# -------------------------------------------------------------------
$corpusref->buildIdf(stemmed => 0, rootdir => "$gen_dir/corpora");
$corpusref->buildIdf(stemmed => 1, rootdir => "$gen_dir/corpora");

# -------------------------------------------------------------------
#  This is how to build the TF.  First we build the DOCNO/URL
#  database, which is necessary to build the TFs.  Then we build
#  unstemmed and stemmed TFs.
# -------------------------------------------------------------------
$corpusref->build_docno_dbm( rootdir => "$gen_dir/corpora");
$corpusref->buildTf(stemmed => 0, rootdir => "$gen_dir/corpora");
$corpusref->buildTf(stemmed => 1, rootdir => "$gen_dir/corpora");

\end{boxedverbatim}

\begin{boxedverbatim}


# -------------------------------------------------------------------
#  Here is how to use a IDF.  The constructor (new) opens the
#  unstemmed IDF.  Then we ask for IDFs for the words "have"
#  "and" and "zimbabwe."
# -------------------------------------------------------------------
my $idfref = Clair::Utils::Idf->new( rootdir => "$gen_dir",
                       corpusname => $corpus_name ,
                       stemmed => 0 );

my $result = $idfref->getIdfForWord("have");
print "IDF(have) = $result\n";
$result = $idfref->getIdfForWord("and");
print "IDF(and) = $result\n";
$result = $idfref->getIdfForWord("zimbabwe");
print "IDF(zimbabwe) = $result\n";

# -------------------------------------------------------------------
#  Here is how to use a TF for term queries.  The constructor (new)
#  opens the unstemmed TF.  Then we ask for information about the
#  word "have":
#
#  1 first, we get the number of documents in the corpus with
#    the word "have"
#  2 then, we get the total number of occurrences of the word "have"
#  3 then, we print a list of URLs of the documents that have the
#    word "have" and the number of times each occurs in the document
# -------------------------------------------------------------------
my $tfref = Clair::Utils::Tf->new( rootdir => "$gen_dir",
                     corpusname => $corpus_name ,
                     stemmed => 0 );

print "\n\n---Direct term queries (unstemmed):---\n";
$result  = $tfref->getNumDocsWithWord("have");
my $freq = $tfref->getFreq("have");
my @urls = $tfref->getDocs("have");
print "\n";

print "TF(have) = $freq total in $result docs\n";
print "Documents with \"have\"\n";
foreach my $url (@urls)  {
    my $url_freq = $tfref->getFreqInDocument("have", url => $url);
    print "  $url: $url_freq\n";
}
print "\n";

# -------------------------------------------------------------------
#  Then we do 1-3 with the word "and"
# -------------------------------------------------------------------
$result = $tfref->getNumDocsWithWord("and");
$freq   = $tfref->getFreq("and");
@urls   = $tfref->getDocs("and");
print "TF(a) = $freq total in $result docs\n";
print "Documents with \"and\"\n";
foreach my $url (@urls)  {
    my $url_freq = $tfref->getFreqInDocument("and", url => $url);
    print "  $url: $url_freq\n";
}
print "\n";


# -------------------------------------------------------------------
#  Then we do 1-3 with the word "zimbabwe"
#  And also print out the number of times zimbabwe is used in each
#  document
# -------------------------------------------------------------------
$result = $tfref->getNumDocsWithWord("zimbabwe");

\end{boxedverbatim}

\begin{boxedverbatim}

$freq   = $tfref->getFreq("zimbabwe");
@urls = $tfref->getDocs("zimbabwe");
print "TF(zimbabwe) = $freq total in $result docs\n";
print "Documents with \"zimbabwe\"\n";
foreach my $url (@urls)  {
    my $url_freq = $tfref->getFreqInDocument("zimbabwe", url => $url);
    print "  $url: $url_freq\n";
}
print "\n";


# -------------------------------------------------------------------
#  Here is how to use a TF for phrase queries.  The constructor (new)
#  opens the stemmed TF.  Then we ask for information about the
#  phrase "result in":
#
#  1 first, we get the number of documents in the corpus with
#    the phrase "result in"
#  2 then, we get the total number of occurrences of the phrase
#    "result in"
#  3 then, we print a list of URLs of the documents that have the
#    word "result in" and the number of times each occurs in the
#    document, as well as the position in the document of the initial
#    term ("result") in each occurrence of the phrase
#  4 finally, using a different method, we print the number of times
#    "result in" occurs in each document in which it occurs (from 3),
#    as well as the position(s) of its occurrence (as in 3)
# -------------------------------------------------------------------
$tfref = Clair::Utils::Tf->new( rootdir => "$gen_dir",
                     corpusname => $corpus_name ,
                     stemmed => 1 );

print "\n---Direct phrase queries (stemmed):---\n";
my @phrase = ("result", "in");
$result = $tfref->getNumDocsWithPhrase(@phrase);
$freq   = $tfref->getPhraseFreq(@phrase);
my $positionsByUrlsRef = $tfref->getDocsWithPhrase(@phrase);
print "freq(\"result in\") = $freq total in $result docs\n";
print "Documents with \"result in\"\n";
foreach my $url (keys %$positionsByUrlsRef)  {
    my $url_freq = scalar keys %{$positionsByUrlsRef->{$url}};
    print "  $url:\n";
    print "      freq      = $url_freq\n";
    print "      positions = " . join(" ", reverse sort keys                    \
%{$positionsByUrlsRef->{$url}}) . "\n";
}
print "\n";

print "The following should be identical to the previous:\n";
foreach my $url (keys %$positionsByUrlsRef) {
    my ($url_freq, $url_positions_ref) =                                        \
$tfref->getPhraseFreqInDocument(\@phrase, url => $url);
    print "  $url:\n";
    print "      freq      = $url_freq\n";
    print "      positions = " . join(" ", reverse sort keys                    \
%$url_positions_ref) . "\n";
}
print "\n\n";


# -------------------------------------------------------------------
#  Then we do 1-4 with the phrase "resulting in"
#  And also print out the number of times "resulting in" is used in each
#  document
#  Because of stemming, the results this time should be the
#  same as those from last time (see directly above)
# -------------------------------------------------------------------

\end{boxedverbatim}

\begin{boxedverbatim}


@phrase = ("resulting", "in");
$result = $tfref->getNumDocsWithPhrase(@phrase);
$freq   = $tfref->getPhraseFreq(@phrase);
$positionsByUrlsRef = $tfref->getDocsWithPhrase(@phrase);
print "freq(\"result in\") = $freq total in $result docs\n";
print "Documents with \"resulting in\" (should be the same as for \"result      \
in\")\n";
foreach my $url (keys %$positionsByUrlsRef)  {
    my $url_freq = scalar keys %{$positionsByUrlsRef->{$url}};
    print "  $url:\n";
    print "      freq      = $url_freq\n";
    print "      positions = " . join(" ", reverse sort keys                    \
%{$positionsByUrlsRef->{$url}}) . "\n";
}
print "\n";

print "The following should be identical to the previous:\n";
foreach my $url (keys %$positionsByUrlsRef) {
    my ($url_freq, $url_positions_ref) =                                        \
$tfref->getPhraseFreqInDocument(\@phrase, url => $url);
    print "  $url:\n";
    print "      freq      = $url_freq\n";
    print "      positions = " . join(" ", reverse sort keys                    \
%$url_positions_ref) . "\n";
}
print "\n";


# -------------------------------------------------------------------
#  Here is how to use a TF for fuzzy OR queries.  We query the
#  (stemmed index of the) corpus as follows:
#
#  1 first, we get the number and scores of documents in the corpus
#    matching a query over the negated term !"thisisnotaword" (# = N),
#    then try the same query formulated as a negated phrase
#  2 then, we get the number and scores of documents in the corpus
#    matching a query over the term "result" (# = A),
#    then try the same query formulated as a phrase
#  3 then, we get the number and scores of documents in the corpus
#    matching a query over the term "in" (# = B)
#  4 then, we get the number and scores of documents in the corpus
#    matching a query over terms "result", "in" (# = C <= A + B)
#  5 then, we get the number and scores of documents in the corpus
#    matching the phrase query "result in" (# = D <= A, B)
#  6 then, we get the number and scores of documents in the corpus
#    matching a query over the negated phrase !"result in" (# = E = N - D)
#  7 finally, we get the number and scores of documents in the corpus
#    matching a query over the phrases "due to", "according to"
# -------------------------------------------------------------------

print "\n---Fuzzy OR Queries (stemmed):---\n";
#1a
    print "Query 1a: !\"thisisnotaword\" (negated term query)\n";
    my ($pTerms, $pNegTerms, $pPhrasePtrs, $pNegPhrasePtrs) = ([],              \
["thisisnotaword"], [], []);
    my %docScores = $tfref->getDocsMatchingFuzzyORQuery($pTerms, $pNegTerms,    \
$pPhrasePtrs, $pNegPhrasePtrs);
    my $N = scalar keys %docScores;
    my @scores = sort {$b <=> $a} values %docScores;
    print "    # docs matching: N = $N\n";
    print "             scores: " . join(" ", @scores) . "\n";
#1b
    print "Query 1b: !\"thisisnotaword\" (negated phrase query)\n";
    ($pTerms, $pNegTerms, $pPhrasePtrs, $pNegPhrasePtrs) = ([], [], [],         \
[["thisisnotaword"]]);
    %docScores = $tfref->getDocsMatchingFuzzyORQuery($pTerms, $pNegTerms,       \

\end{boxedverbatim}

\begin{boxedverbatim}

$pPhrasePtrs, $pNegPhrasePtrs);
    $N = scalar keys %docScores;
    @scores = sort {$b <=> $a} values %docScores;
    print "    # docs matching: N = $N\n";
    print "             scores: " . join(" ", @scores) . "\n\n";


#2a
    print "Query 2a: \"result\" (term query)\n";
    ($pTerms, $pNegTerms, $pPhrasePtrs, $pNegPhrasePtrs) = (["result"], [], [], \
[]);
    %docScores = $tfref->getDocsMatchingFuzzyORQuery($pTerms, $pNegTerms,       \
$pPhrasePtrs, $pNegPhrasePtrs);
    my $A = scalar keys %docScores;
    @scores = sort {$b <=> $a} values %docScores;
    print "    # docs matching: A = $A\n";
    print "             scores: " . join(" ", @scores) . "\n";
#2b
    print "Query 2b: \"result\" (phrase query)\n";
    ($pTerms, $pNegTerms, $pPhrasePtrs, $pNegPhrasePtrs) = ([], [],             \
[["result"]], []);
    %docScores = $tfref->getDocsMatchingFuzzyORQuery($pTerms, $pNegTerms,       \
$pPhrasePtrs, $pNegPhrasePtrs);
    $A = scalar keys %docScores;
    @scores = sort {$b <=> $a} values %docScores;
    print "    # docs matching: A = $A\n";
    print "             scores: " . join(" ", @scores) . "\n\n";
#3
    print "Query 3: \"in\"\n";
    ($pTerms, $pNegTerms, $pPhrasePtrs, $pNegPhrasePtrs) = (["in"], [], [],     \
[]);
    %docScores = $tfref->getDocsMatchingFuzzyORQuery($pTerms, $pNegTerms,       \
$pPhrasePtrs, $pNegPhrasePtrs);
    my $B = scalar keys %docScores;
    @scores = sort {$b <=> $a} values %docScores;
    print "    # docs matching: B = $B\n";
    print "             scores: " . join(" ", @scores) . "\n\n";
#4
    print "Query 4: \"result\", \"in\"\n";
    ($pTerms, $pNegTerms, $pPhrasePtrs, $pNegPhrasePtrs) = (["in"], [], [],     \
[]);
    %docScores = $tfref->getDocsMatchingFuzzyORQuery($pTerms, $pNegTerms,       \
$pPhrasePtrs, $pNegPhrasePtrs);
    my $C = scalar keys %docScores;
    @scores = sort {$b <=> $a} values %docScores;
    print "    # docs matching: C = $C <= A + B = " . ($A + $B) . "\n";
    print "             scores: " . join(" ", @scores) . "\n\n";
#5
    print "Query 5: \"result in\"\n";
    ($pTerms, $pNegTerms, $pPhrasePtrs, $pNegPhrasePtrs) = ([], [], [["result", \
"in"]], []);
    %docScores = $tfref->getDocsMatchingFuzzyORQuery($pTerms, $pNegTerms,       \
$pPhrasePtrs, $pNegPhrasePtrs);
    my $D = scalar keys %docScores;
    @scores = sort {$b <=> $a} values %docScores;
    print "    # docs matching: D = $D <= min{A, B}\n";
    print "             scores: " . join(" ", @scores) . "\n\n";
#6
    print "Query 6: !\"result in\"\n";
    ($pTerms, $pNegTerms, $pPhrasePtrs, $pNegPhrasePtrs) = ([], [], [],         \
[["result", "in"]]);
    %docScores = $tfref->getDocsMatchingFuzzyORQuery($pTerms, $pNegTerms,       \
$pPhrasePtrs, $pNegPhrasePtrs);
    my $E = scalar keys %docScores;
    @scores = sort {$b <=> $a} values %docScores;
    print "    # docs matching: E = $E = N - D = " . ($N - $D) . "\n";
    print "             scores: " . join(" ", @scores) . "\n\n";

\end{boxedverbatim}

\begin{boxedverbatim}

#7
    print "Query 7: \"due to\", \"according to\"\n";
    ($pTerms, $pNegTerms, $pPhrasePtrs, $pNegPhrasePtrs) = ([], [],             \
[["due","to"], ["according","to"]], []);
    %docScores = $tfref->getDocsMatchingFuzzyORQuery($pTerms, $pNegTerms,       \
$pPhrasePtrs, $pNegPhrasePtrs);
    my $F = scalar keys %docScores;
    @scores = sort {$b <=> $a} values %docScores;
    print "    # docs matching: F = $F\n";
    print "             scores: " . join(" ", @scores) . "\n\n";


# -------------------------------------------------------------------
#  Finally, we tell the user to have a nice day.
# -------------------------------------------------------------------
print "\nHave a nice day!\n";

\end{boxedverbatim}

\subsubsection{corpusdownload.pl}
\begin{boxedverbatim}

#!/usr/local/bin/perl

# script: test_corpusdownload.pl
# functionality: Downloads a corpus from a file containing URLs;
# functionality:  makes IDFs and TFs

use strict;
use warnings;
use FindBin;

use Clair::Utils::CorpusDownload;
use Clair::Utils::Idf;
use Clair::Utils::Tf;
use DB_File;

my $basedir = $FindBin::Bin;
my $gen_dir = "$basedir/produced/corpusdownload";
my $input_dir = "$basedir/input/corpusdownload";

# -------------------------------------------------------------------
#  This is the constructor.  It simply stores the directory
#  and name of the corpus.  It must be called prior to
#  any other routine.
# -------------------------------------------------------------------
my $corpusref = Clair::Utils::CorpusDownload->new(corpusname => "t2",
                rootdir => "$gen_dir");

# -------------------------------------------------------------------
#  Here's how to build a corpus.  An array @urls needs to be
#  built somehow.  (Here, we read the URLs from a file
#  $corpusname.urls.)  Then, the corpus will be built in
#  the directory $rootdir/$corpusname
# -------------------------------------------------------------------
my $uref = $corpusref->readUrlsFile("$input_dir/t.urls");

$corpusref->buildCorpus(urlsref => $uref, cleanup => 0);


# -------------------------------------------------------------------
#  This is how to build the IDF.  First we build the unstemmed IDF,
#  then the stemmed one.
# -------------------------------------------------------------------
$corpusref->buildIdf(stemmed => 0);
$corpusref->buildIdf(stemmed => 1);

# -------------------------------------------------------------------
#  This is how to build the TF.  First we build the DOCNO/URL
#  database, which is necessary to build the TFs.  Then we build
#  unstemmed and stemmed TFs.
# -------------------------------------------------------------------
$corpusref->build_docno_dbm();
$corpusref->buildTf(stemmed => 0);
$corpusref->buildTf(stemmed => 1);

# -------------------------------------------------------------------
#  Here is how to use a IDF.  The constructor (new) opens the
#  unstemmed IDF.  Then we ask for IDFs for the words "have"
#  "and" and "zimbabwe."
# -------------------------------------------------------------------
my $idfref = Clair::Utils::Idf->new( rootdir => "$gen_dir",
                       corpusname => "t2" ,
                       stemmed => 0 );

my $result = $idfref->getIdfForWord("have");
print "IDF(have) = $result\n";
$result = $idfref->getIdfForWord("and");
print "IDF(and) = $result\n";

\end{boxedverbatim}

\begin{boxedverbatim}

$result = $idfref->getIdfForWord("zimbabwe");
print "IDF(zimbabwe) = $result\n";

# -------------------------------------------------------------------
#  Here is how to use a TF for term queries.  The constructor (new)
#  opens the unstemmed TF.  Then we ask for information about the
#  word "have":
#
#  1 first, we get the number of documents in the corpus with
#    the word "have"
#  2 then, we get the total number of occurrences of the word "have"
#  3 then, we print a list of URLs of the documents that have the
#    word "have" and the number of times each occurs in the document
# -------------------------------------------------------------------
my $tfref = Clair::Utils::Tf->new( rootdir => "$gen_dir",
                     corpusname => "t2" ,
                     stemmed => 0 );

print "\n\n---Direct term queries (unstemmed):---\n";
$result = $tfref->getNumDocsWithWord("have");
my $freq   = $tfref->getFreq("have");
my @urls = $tfref->getDocs("have");
print "TF(have) = $freq total in $result docs\n";
print "Documents with \"have\"\n";
foreach my $url (@urls)  {
	my $url_freq = $tfref->getFreqInDocument("have", url => $url);
	print "  $url: $url_freq\n";
}
print "\n";

# -------------------------------------------------------------------
#  Then we do 1-3 with the word "and"
# -------------------------------------------------------------------
$result = $tfref->getNumDocsWithWord("and");
$freq   = $tfref->getFreq("and");
@urls = $tfref->getDocs("and");
print "TF(and) = $freq total in $result docs\n";
print "Documents with \"and\"\n";
foreach my $url (@urls)  {
	my $url_freq = $tfref->getFreqInDocument("and", url => $url);
	print "  $url: $url_freq\n";
}
print "\n";


# -------------------------------------------------------------------
#  Then we do 1-3 with the word "zimbabwe"
#  And also print out the number of times zimbabwe is used in each
#  document
# -------------------------------------------------------------------
$result = $tfref->getNumDocsWithWord("zimbabwe");
$freq   = $tfref->getFreq("zimbabwe");
@urls = $tfref->getDocs("zimbabwe");
print "TF(zimbabwe) = $freq total in $result docs\n";
print "Documents with \"zimbabwe\"\n";
foreach my $url (@urls)  {
	my $url_freq = $tfref->getFreqInDocument("zimbabwe", url => $url);
	print "  $url: $url_freq\n";
}
print "\n";


# -------------------------------------------------------------------
#  Here is how to use a TF for phrase queries.  The constructor (new)
#  opens the stemmed TF.  Then we ask for information about the
#  phrase "result in":
#

\end{boxedverbatim}

\begin{boxedverbatim}

#  1 first, we get the number of documents in the corpus with
#    the phrase "result in"
#  2 then, we get the total number of occurrences of the phrase
#    "result in"
#  3 then, we print a list of URLs of the documents that have the
#    word "result in" and the number of times each occurs in the
#    document, as well as the position in the document of the initial
#    term ("result") in each occurrence of the phrase
#  4 finally, using a different method, we print the number of times
#    "result in" occurs in each document in which it occurs (from 3),
#    as well as the position(s) of its occurrence (as in 3)
# -------------------------------------------------------------------
$tfref = Clair::Utils::Tf->new( rootdir => "$gen_dir",
                     corpusname => "t2" ,
                     stemmed => 1 );

print "\n---Direct phrase queries (stemmed):---\n";
my @phrase = ("result", "in");
$result = $tfref->getNumDocsWithPhrase(@phrase);
$freq   = $tfref->getPhraseFreq(@phrase);
my $positionsByUrlsRef = $tfref->getDocsWithPhrase(@phrase);
print "freq(\"result in\") = $freq total in $result docs\n";
print "Documents with \"result in\"\n";
foreach my $url (keys %$positionsByUrlsRef)  {
    my $url_freq = scalar keys %{$positionsByUrlsRef->{$url}};
    print "  $url:\n";
    print "      freq      = $url_freq\n";
    print "      positions = " . join(" ", reverse sort keys                    \
%{$positionsByUrlsRef->{$url}}) . "\n";
}
print "\n";

print "The following should be identical to the previous:\n";
foreach my $url (keys %$positionsByUrlsRef) {
    my ($url_freq, $url_positions_ref) =                                        \
$tfref->getPhraseFreqInDocument(\@phrase, url => $url);
    print "  $url:\n";
    print "      freq      = $url_freq\n";
    print "      positions = " . join(" ", reverse sort keys                    \
%$url_positions_ref) . "\n";
}
print "\n\n";


# -------------------------------------------------------------------
#  Then we do 1-4 with the phrase "resulting in"
#  And also print out the number of times "resulting in" is used in each
#  document
#  Because of stemming, the results this time should be the
#  same as those from last time (see directly above)
# -------------------------------------------------------------------

@phrase = ("resulting", "in");
$result = $tfref->getNumDocsWithPhrase(@phrase);
$freq   = $tfref->getPhraseFreq(@phrase);
$positionsByUrlsRef = $tfref->getDocsWithPhrase(@phrase);
print "freq(\"result in\") = $freq total in $result docs\n";
print "Documents with \"resulting in\" (should be the same as for \"result      \
in\")\n";
foreach my $url (keys %$positionsByUrlsRef)  {
    my $url_freq = scalar keys %{$positionsByUrlsRef->{$url}};
    print "  $url:\n";
    print "      freq      = $url_freq\n";
    print "      positions = " . join(" ", reverse sort keys                    \
%{$positionsByUrlsRef->{$url}}) . "\n";
}
print "\n";

\end{boxedverbatim}

\begin{boxedverbatim}


print "The following should be identical to the previous:\n";
foreach my $url (keys %$positionsByUrlsRef) {
    my ($url_freq, $url_positions_ref) =                                        \
$tfref->getPhraseFreqInDocument(\@phrase, url => $url);
    print "  $url:\n";
    print "      freq      = $url_freq\n";
    print "      positions = " . join(" ", reverse sort keys                    \
%$url_positions_ref) . "\n";
}
print "\n";


# -------------------------------------------------------------------
#  Here is how to use a TF for fuzzy OR queries.  We query the
#  (stemmed index of the) corpus as follows:
#
#  1 first, we get the number and scores of documents in the corpus
#    matching a query over the negated term !"thisisnotaword" (# = N),
#    then try the same query formulated as a negated phrase
#  2 then, we get the number and scores of documents in the corpus
#    matching a query over the term "result" (# = A),
#    then try the same query formulated as a phrase
#  3 then, we get the number and scores of documents in the corpus
#    matching a query over the term "in" (# = B)
#  4 then, we get the number and scores of documents in the corpus
#    matching a query over terms "result", "in" (# = C <= A + B)
#  5 then, we get the number and scores of documents in the corpus
#    matching the phrase query "result in" (# = D <= A, B)
#  6 then, we get the number and scores of documents in the corpus
#    matching a query over the negated phrase !"result in" (# = E = N - D)
#  7 finally, we get the number and scores of documents in the corpus
#    matching a query over the phrases "due to", "according to"
# -------------------------------------------------------------------

print "\n---Fuzzy OR Queries (stemmed):---\n";
#1a
    print "Query 1a: !\"thisisnotaword\" (negated term query)\n";
    my ($pTerms, $pNegTerms, $pPhrasePtrs, $pNegPhrasePtrs) = ([],              \
["thisisnotaword"], [], []);
    my %docScores = $tfref->getDocsMatchingFuzzyORQuery($pTerms, $pNegTerms,    \
$pPhrasePtrs, $pNegPhrasePtrs);
    my $N = scalar keys %docScores;
    my @scores = sort {$b <=> $a} values %docScores;
    print "    # docs matching: N = $N\n";
    print "             scores: " . join(" ", @scores) . "\n";
#1b
    print "Query 1b: !\"thisisnotaword\" (negated phrase query)\n";
    ($pTerms, $pNegTerms, $pPhrasePtrs, $pNegPhrasePtrs) = ([], [], [],         \
[["thisisnotaword"]]);
    %docScores = $tfref->getDocsMatchingFuzzyORQuery($pTerms, $pNegTerms,       \
$pPhrasePtrs, $pNegPhrasePtrs);
    $N = scalar keys %docScores;
    @scores = sort {$b <=> $a} values %docScores;
    print "    # docs matching: N = $N\n";
    print "             scores: " . join(" ", @scores) . "\n\n";


#2a
    print "Query 2a: \"result\" (term query)\n";
    ($pTerms, $pNegTerms, $pPhrasePtrs, $pNegPhrasePtrs) = (["result"], [], [], \
[]);
    %docScores = $tfref->getDocsMatchingFuzzyORQuery($pTerms, $pNegTerms,       \
$pPhrasePtrs, $pNegPhrasePtrs);
    my $A = scalar keys %docScores;
    @scores = sort {$b <=> $a} values %docScores;
    print "    # docs matching: A = $A\n";

\end{boxedverbatim}

\begin{boxedverbatim}

    print "             scores: " . join(" ", @scores) . "\n";
#2b
    print "Query 2b: \"result\" (phrase query)\n";
    ($pTerms, $pNegTerms, $pPhrasePtrs, $pNegPhrasePtrs) = ([], [],             \
[["result"]], []);
    %docScores = $tfref->getDocsMatchingFuzzyORQuery($pTerms, $pNegTerms,       \
$pPhrasePtrs, $pNegPhrasePtrs);
    $A = scalar keys %docScores;
    @scores = sort {$b <=> $a} values %docScores;
    print "    # docs matching: A = $A\n";
    print "             scores: " . join(" ", @scores) . "\n\n";
#3
    print "Query 3: \"in\"\n";
    ($pTerms, $pNegTerms, $pPhrasePtrs, $pNegPhrasePtrs) = (["in"], [], [],     \
[]);
    %docScores = $tfref->getDocsMatchingFuzzyORQuery($pTerms, $pNegTerms,       \
$pPhrasePtrs, $pNegPhrasePtrs);
    my $B = scalar keys %docScores;
    @scores = sort {$b <=> $a} values %docScores;
    print "    # docs matching: B = $B\n";
    print "             scores: " . join(" ", @scores) . "\n\n";
#4
    print "Query 4: \"result\", \"in\"\n";
    ($pTerms, $pNegTerms, $pPhrasePtrs, $pNegPhrasePtrs) = (["in"], [], [],     \
[]);
    %docScores = $tfref->getDocsMatchingFuzzyORQuery($pTerms, $pNegTerms,       \
$pPhrasePtrs, $pNegPhrasePtrs);
    my $C = scalar keys %docScores;
    @scores = sort {$b <=> $a} values %docScores;
    print "    # docs matching: C = $C <= A + B = " . ($A + $B) . "\n";
    print "             scores: " . join(" ", @scores) . "\n\n";
#5
    print "Query 5: \"result in\"\n";
    ($pTerms, $pNegTerms, $pPhrasePtrs, $pNegPhrasePtrs) = ([], [], [["result", \
"in"]], []);
    %docScores = $tfref->getDocsMatchingFuzzyORQuery($pTerms, $pNegTerms,       \
$pPhrasePtrs, $pNegPhrasePtrs);
    my $D = scalar keys %docScores;
    @scores = sort {$b <=> $a} values %docScores;
    print "    # docs matching: D = $D <= min{A, B}\n";
    print "             scores: " . join(" ", @scores) . "\n\n";
#6
    print "Query 6: !\"result in\"\n";
    ($pTerms, $pNegTerms, $pPhrasePtrs, $pNegPhrasePtrs) = ([], [], [],         \
[["result", "in"]]);
    %docScores = $tfref->getDocsMatchingFuzzyORQuery($pTerms, $pNegTerms,       \
$pPhrasePtrs, $pNegPhrasePtrs);
    my $E = scalar keys %docScores;
    @scores = sort {$b <=> $a} values %docScores;
    print "    # docs matching: E = $E = N - D = " . ($N - $D) . "\n";
    print "             scores: " . join(" ", @scores) . "\n\n";
#7
    print "Query 7: \"due to\", \"according to\"\n";
    ($pTerms, $pNegTerms, $pPhrasePtrs, $pNegPhrasePtrs) = ([], [],             \
[["due","to"], ["according","to"]], []);
    %docScores = $tfref->getDocsMatchingFuzzyORQuery($pTerms, $pNegTerms,       \
$pPhrasePtrs, $pNegPhrasePtrs);
    my $F = scalar keys %docScores;
    @scores = sort {$b <=> $a} values %docScores;
    print "    # docs matching: F = $F\n";
    print "             scores: " . join(" ", @scores) . "\n\n";


# -------------------------------------------------------------------
#  Finally, we tell the user to have a nice day.
# -------------------------------------------------------------------
print "\nHave a nice day!\n";

\end{boxedverbatim}

\subsubsection{document\_idf.pl}
\begin{boxedverbatim}

#!/usr/local/bin/perl

# script: test_document_idf.pl
# functionality: Loads documents from an input dir; strips and stems them,
# functionality: and then builds an IDF from them 

use strict;
use warnings;
use FindBin;
use DB_File;
use Clair::Document;
use Clair::Cluster;

my $basedir = $FindBin::Bin;
my $input_dir = "$basedir/input/document_idf";
my $gen_dir = "$basedir/produced/document_idf";

my $c = Clair::Cluster::->new();

$c->load_documents("$input_dir/*.txt", type => 'html');

$c->strip_all_documents();
$c->stem_all_documents();

my %idf_hash = $c->build_idf("$gen_dir/idf-dbm", type => 'text');

foreach my $k (keys %idf_hash) {
	print "$k\t", $idf_hash{$k}, "\n";
}

\end{boxedverbatim}

\subsubsection{document.pl}
\begin{boxedverbatim}

#!/usr/local/bin/perl

# script: test_document.pl
# functionality: Creates Documents from strings, files, strips and stems them,
# functionality: splits them into lines, sentences, counts words, saves them 

use strict;
use warnings;
use FindBin;
use Clair::Document;

my $basedir = $FindBin::Bin;
my $input_dir = "$basedir/input/document";
my $gen_dir = "$basedir/produced/document";

# Create a text document specifying the text directly
my $doc1 = new Clair::Document(string => 'She sees the facts with instruments   \
happily with embarassements.',
		               type => 'text', id => 'doc1');

# Create a text document by specifying the file to open
my $doc2 = new Clair::Document(file => "$input_dir/test.txt",
	                       type => 'text', id => 'doc2');

# Create an HTML document
my $doc3 = new Clair::Document(string => '<html><body><p>This is the HTML</p>'
	                       . '<p>She sees the facts with instruments happily with  \
embarassements.</p></body></html>',
			       type => 'html', id => 'doc3');

# Compute the text from the HTML
my $doc3_text = $doc3->strip_html;
print "The text from document 3:\n$doc3_text\n\n";

# Stem the text of the document
my $doc3_stem = $doc3->stem;
print "The stemmed text from document 3:\n$doc3_stem\n\n";

# Split the document into lines and sentences
# (Note that split_into_sentences uses MxTerminator which requires
# Perl 5.8)
my @doc3_lines = $doc3->split_into_lines;
my @doc3_sentences = $doc3->split_into_sentences;
print "\nDocument 3 has ", scalar @doc3_sentences, " sentences.\n\n"; 

# Count the number of words in each document
my $doc1_words = $doc1->count_words;
my $doc2_words = $doc2->count_words;
my $doc3_words = $doc3->count_words;
print ("Document 1 has $doc1_words words, Document2 has $doc2_words, and        \
Document 3 has $doc3_words.\n");

# Print the text version to the screen, then saved the stemmed version to disk
print "The text from document 3 is:\n";
$doc3->print(type => 'text');
print "\n";
$doc3->save(file => "$gen_dir/document_output.stem", type => 'stem');

\end{boxedverbatim}

\subsubsection{features\_io.pl}
\begin{boxedverbatim}

#!/usr/local/bin/perl

# script: features_io.pl
# functionality: Same as features.pl BUT, outputs the train data set as
# functionality: document and feature vectors in svm_light format, reads
# functionality: the svm_light formatted file and converts it to perl hash

use strict;
use FindBin;
# use lib "$FindBin::Bin/../lib";
# use lib "$FindBin::Bin/lib"; # if you are outside of bin path.. just in case
use vars qw/$DEBUG/;

use Clair::Features;
use Clair::GenericDoc;
use Data::Dumper;
use File::Find;
use File::Path;

# globals
$DEBUG = 0;
my %args;
my @train_files = (); # list of train files we will analyze
my @test_files = (); # list of test files we will analyze
my %container = (); # container for our file arrays.
my $results_root = "$FindBin::Bin/produced/features";

mkpath($results_root, 0, 0777) unless(-d $results_root);

my $n = $args{n} || 0;
my $train_root = "$FindBin::Bin/input/features/train";
my $test_root = "$FindBin::Bin/input/features/test";
my $output = "test_output";
my $feature_opt = $args{feature} if($args{feature});
my $filter = $args{filter} || '.*';


my $t0;
my $t1;

#
# Finding files
#
sub wanted_train
{
  return if( ! -f $File::Find::name );
  push @train_files, $File::Find::name;
}
find(\&wanted_train, ( $train_root ));
@train_files = grep { -f $_ && /$filter/ } @train_files;


#
# Processing documents
#
my $files = \@train_files;
my $files_count = scalar @train_files;


# we can limit the number of document per class
my $fea2 = new Clair::Features(
	DEBUG => $DEBUG,
	document_limit => 100, ## NOTICE THIS FLAG ##
	mode => "train", # train data
	# features_file => "$results_root/.features_lookup"
);
$fea2->debugmsg("registering $files_count documents with 100 limit per class",  \

\end{boxedverbatim}

\begin{boxedverbatim}

0);

# register each document into the Clair::Features object
for my $f (@$files)
{
	my $gdoc = new Clair::GenericDoc(
		DEBUG => $DEBUG,
		content => $f,
		stem => 1,
		lowercase => 1,
		use_parser_module => "sports" # the test data is formatted in pseudo xml.
	);
		
	$fea2->register($gdoc);
	undef $gdoc; # memory conscious
}


my $top10 = $fea2->select(20);
$fea2->debugmsg("top 20 features with 100 docs:\n" . Dumper($top10), 0);

# you can also get the feature chi-squared values for binary classified         \
documents.
$fea2->debugmsg("running \$fea2->chi_squared();", 0);

$fea2->{DEBUG} = 1; # to show more info
my $chisq_values = $fea2->chi_squared();
print Dumper($chisq_values);


# save the feature vectors in svm_light format
$fea2->output("$results_root/$output.train");
$fea2->debugmsg("feature vectors saved here: $results_root/$output.train", 0);

# feature and its associated id is saved here
# print Dumper($fea2->{features_map});

$fea2->debugmsg("retrieving feature vectors and converting to perl data         \
structure", 0);
my $vectors = $fea2->input("$results_root/$output.train");
print Dumper($vectors);

\end{boxedverbatim}

\subsubsection{features.pl}
\begin{boxedverbatim}

#!/usr/local/bin/perl

# script: test_features.pl
# functionality: Reads docs from input/features/train, calculates chi-squared
# functionality: values for all extracted features, shows ways to retrieve
# functionality: those features

use strict;
use FindBin;
# use lib "$FindBin::Bin/../lib";
# use lib "$FindBin::Bin/lib"; # if you are outside of bin path.. just in case
use vars qw/$DEBUG/;

use Clair::Features;
use Clair::GenericDoc;
use Data::Dumper;
use File::Find;
use File::Path;

# globals
$DEBUG = 0;
my %args;
my @train_files = (); # list of train files we will analyze
my @test_files = (); # list of test files we will analyze
my %container = (); # container for our file arrays.
my $results_root = "$FindBin::Bin/produced/features";

mkpath($results_root, 0, 0777) unless(-d $results_root);

my $n = $args{n} || 0;
my $train_root = "$FindBin::Bin/input/features/train";
my $test_root = "$FindBin::Bin/input/features/test";
my $output = "test_output";
my $feature_opt = $args{feature} if($args{feature});
my $filter = $args{filter} || '.*';


my $t0;
my $t1;

#
# Finding files
#
sub wanted_train
{
  return if( ! -f $File::Find::name );
  push @train_files, $File::Find::name;
}
find(\&wanted_train, ( $train_root ));
@train_files = grep { -f $_ && /$filter/ } @train_files;


#
# Processing documents
#
my $files = \@train_files;
my $files_count = scalar @train_files;

my $fea = new Clair::Features(
	DEBUG => $DEBUG,
	document_limit => $n,
	mode => "train", # train data
	# features_file => "$results_root/.features_lookup"
);


$fea->debugmsg("registering $files_count documents", 0);

\end{boxedverbatim}

\begin{boxedverbatim}

# register each document into the Clair::Features object
for my $f (@$files)
{
	my $gdoc = new Clair::GenericDoc(
		DEBUG => $DEBUG,
		content => $f,
		stem => 1,
		lowercase => 1,
		use_parser_module => "sports" # the test data is formatted in pseudo xml.
	);
		
	$fea->register($gdoc);
	undef $gdoc; # memory conscious
}

# print Dumper($fea->{features_global}); exit;

my $all = $fea->select();
$fea->debugmsg("feature counts: " . scalar @$all, 0);

my $top10 = $fea->select(10);
$fea->debugmsg("top 10 features:\n" . Dumper($top10), 0);

my $top50 = $fea->select(50);
$fea->debugmsg("top 50 features:\n" . Dumper($top50), 0);

# you can also get the feature chi-squared values for binary classified         \
documents.
$fea->debugmsg("running \$fea2->chi_squared();", 0);
$fea->{DEBUG} = 1; # to show more info
my $chisq_values = $fea->chi_squared();
print Dumper($chisq_values);

# save the classified data into a file in the svm_light format.
$fea->output("$results_root/$output.train");

$fea->debugmsg("feature vectors saved here: $results_root/$output.train", 0);

# print Dumper($fea->{features_global});
# print Dumper($fea->{feature_scores});


\end{boxedverbatim}

\subsubsection{features\_traintest.pl}
\begin{boxedverbatim}

#!/usr/local/bin/perl

# script: test_features_traintest.pl
# functionality: Builds the feature vector for training and testing datasets,
# functionality: and is a prerequisite for learn.pl and classify.pl

use strict;
use FindBin;
# use lib "$FindBin::Bin/../lib";
# use lib "$FindBin::Bin/lib"; # if you are outside of bin path.. just in case
use vars qw/$DEBUG/;

use Benchmark;
use Clair::Features;
use Clair::GenericDoc;
use Data::Dumper;
use File::Find;
use File::Path;

$DEBUG = 0;
my %args;
my @train_files = (); # list of train files we will analyze
my @test_files = (); # list of test files we will analyze
my %container = (); # container for our file arrays.
my $results_root = "$FindBin::Bin/produced/features";

mkpath($results_root, 0, 0777) unless(-d $results_root);

my $n = $args{n} || 0;
my $train_root = "$FindBin::Bin/input/features/train";
my $test_root = "$FindBin::Bin/input/features/test";
my $output = "feature_vectors";
my $feature_opt = $args{feature} if($args{feature});
my $filter = $args{filter} || '.*';


my $t0;
my $t1;

#
# Finding files
#
$t0 = new Benchmark;


sub wanted_train
{
  return if( ! -f $File::Find::name );
  push @train_files, $File::Find::name;
}
find(\&wanted_train, ( $train_root ));
@train_files = grep { -f $_ && /$filter/ } @train_files;


sub wanted_test
{
  return if( ! -f $File::Find::name );
  push @test_files, $File::Find::name;
}
find(\&wanted_test, ( $test_root ));
@test_files = grep { -f $_ && /$filter/ } @test_files;


$t1 = new Benchmark;
my $timediff_find = timestr(timediff($t1, $t0));



\end{boxedverbatim}

\begin{boxedverbatim}


#
# Processing documents
#
$t0 = new Benchmark;

$container{train} = \@train_files;
$container{test} = \@test_files;

# train the data first and then test
# this illustrates how you first use the train data to produce the feature      \
vectors
# and then use the test data to build the feature vectors with matching id's.

for my $dataset (qw/train test/)
{
	my $files = $container{$dataset};
	
	my $fea = new Clair::Features(
		DEBUG => $DEBUG,
		features_file => "$results_root/feature_lookup_map",
		# document_limit => $n,
		mode => $dataset,
		# features_file => "$results_root/.features_lookup"
	);
	$fea->debugmsg("building $dataset feature vectors", 0);

	for my $f (@$files)
	{
		my $gdoc = new Clair::GenericDoc(
			DEBUG => $DEBUG,
			content => $f,
			stem => 1,
			use_parser_module => "sports"
		);
		
		$fea->register($gdoc);
		undef $gdoc;
	}

	# you need to run $fea->select() in order to retain the feature id's across    \
the datasets.
	$fea->debugmsg("ordering features and saving the map for $dataset", 0)         \
if($dataset eq "train");
	$fea->select();
	# $fea->input("$output.$dataset");
	$fea->debugmsg("saving $dataset feature vectors:                               \
$results_root/$output.$dataset", 0);
	$fea->output("$results_root/$output.$dataset");
}


$t1 = new Benchmark;
my $timediff_prep = timestr(timediff($t1, $t0));


\end{boxedverbatim}

\subsubsection{genericdoc.pl}
\begin{boxedverbatim}

#!/usr/local/bin/perl

# script: genericdoc.pl
# functionality: Tests parsing of simple text/html file/string, conversion
# functionality: into xml file, instantiation via constructor and morph()

use strict;
use FindBin;
use Data::Dumper;
use Clair::GenericDoc;

my $DEBUG = 0;
my $basedir = $FindBin::Bin;
my $input_dir = "$basedir/input/document";
my $output_dir = "$basedir/produced/genericdoc";
my $testtxt = "$input_dir/test.txt";
my $testhtml = "$input_dir/test.html";


my $doc = new Clair::GenericDoc(
 content => $testtxt,
 use_system_file_cmd => 1,
 DEBUG => $DEBUG,
);

$doc->debugmsg("testing with $testtxt", 0);

  my $type = $doc->document_type($testtxt);

$doc->debugmsg("OK - document type is: $type", 0) if $type;


$doc->debugmsg("extracting content of $testtxt", 0);

  my $result = $doc->extract();

$doc->debugmsg("OK - content:\n". Dumper($result), 0) if $result;


$doc->debugmsg("converting to xml", 0);

  my $xml = $doc->to_xml($result->[0]);
  $doc->save_xml($xml, "$output_dir/test.xml");

$doc->debugmsg("saving to: $output_dir/test.xml", 0);
$doc->debugmsg("OK - output exists $output_dir/test.xml", 0) if -f              \
"$output_dir/test.xml";


$doc->debugmsg("reading from xml", 0);

  my $hash = $doc->from_xml("$output_dir/test.xml");

$doc->debugmsg("OK - content:\n". Dumper($hash), 0) if scalar keys %$hash;



$doc->debugmsg("testing with $testhtml", 0);

  my $type2 = $doc->document_type($testhtml);

$doc->debugmsg("OK - document type is: $type2", 0) if $type2;


$doc->debugmsg("extracting content of $testhtml", 0);

	$doc->{content} = $testhtml;

\end{boxedverbatim}

\begin{boxedverbatim}

	$doc->{stem} = 0; # suppress stemming
	$doc->{lowercase} = 0; # suppress lowercasing
  my $result2 = $doc->extract();

$doc->debugmsg("OK - content:\n". Dumper($result2), 0) if $result2;


$doc->debugmsg("using the shakespear parser module", 0);
# by supplying "use_parser_module", you can force the system to use 
# a specific parsing module.
my $doc2 = new Clair::GenericDoc(
 use_parser_module => "shakespear",
 content => $testhtml,
 # use_system_file_cmd => 1,
 DEBUG => $DEBUG,
);

my $result3 = $doc2->extract();

$doc->debugmsg("content:\n". Dumper($result3), 0);



my $doc3 = new Clair::GenericDoc(
 use_parser_module => "shakespear",
 content => $testhtml,
 # use_system_file_cmd => 1,
 DEBUG => $DEBUG,
 cast => 1, # we want the return object to be Clair::Document
);

print "Notice the Clair::Genericdoc gives you the ability to dynamically        \
instantiate Clair::Document\n";
$doc->debugmsg("OK - properly converted:\n" . Dumper($doc3)) if                 \
UNIVERSAL::isa($doc3, "Clair::Document");

$doc3->strip_html();
my $count = $doc3->count_words();
print "The Clair::Document object has text:\n". $doc3->{text} . "\n";
print "The Clair::Document object has $count words\n";



my $doc4 = $doc->morph();
print "What happens when you 'morph()' the existing Clair::Genericdoc           \
object?\n";
print Dumper($doc4);

\end{boxedverbatim}

\subsubsection{html.pl}
\begin{boxedverbatim}

#!/usr/local/bin/perl

# script: test_html.pl
# functionality: Tests the html stripping functionality in Documents 

use strict;
use warnings;
use FindBin;
use Clair::Document;

my $input_dir = "$FindBin::Bin/input/html";

#Take in a single file and parse the html, then document output the file

my $doc = new Clair::Document(type=>'html',file=>"$input_dir/test.html");
print "HTML version:\n";
my $html = $doc->get_html();
print "$html\n";

print "Stripped version:\n";
my $stripped = $doc->strip_html();
print "$stripped\n";

\end{boxedverbatim}

\subsubsection{hyperlink.pl}
\begin{boxedverbatim}

#!/usr/local/bin/perl

# script: test_hyperlink.pl
# functionality: Makes and populates a cluster, builds a network from
# functionality:  hyperlinks between them; then tests making a subset


use strict;
use warnings;
use FindBin;
use Clair::Network;
use Clair::Cluster;
use Clair::Document;

my $basedir = $FindBin::Bin;
my $input_dir = "$basedir/input/hyperlink";

my $c = new Clair::Cluster();
my $d1 = new Clair::Document(id => 1, type => 'text', string => 'Document 1');
$c->insert(1, $d1);
my $d2 = new Clair::Document(id => 2, type => 'text', string => 'Document 2');
$c->insert(2, $d2);
my $d3 = new Clair::Document(id => 3, type => 'text', string => 'Document 3');
$c->insert(3, $d3);
my $d4 = new Clair::Document(id => 4, type => 'text', string => 'Document 4');
$c->insert(4, $d4);

my $n = $c->create_hyperlink_network_from_file("$input_dir/t06.links");

print "Original edges:\n";
$n->print_hyperlink_edges();
my $n2 = $n->create_subset_network_from_file("$input_dir/t06.subset");
print "\nNew edges:\n";
$n2->print_hyperlink_edges();

\end{boxedverbatim}

\subsubsection{idf.pl}
\begin{boxedverbatim}

#!/usr/local/bin/perl

# script: test_idf.pl
# functionality: Creates a cluster from some input files, then builds an idf
# functionality: from the lines of the documents 

use strict;
use warnings;
use FindBin;
use Clair::Util;
use Clair::Cluster;
use Clair::Document;

my $basedir = $FindBin::Bin;
my $input_dir = "$basedir/input/idf";
my $gen_dir = "$basedir/produced/idf";

# Create cluster
my %documents = ();
my $c = Clair::Cluster->new(documents => \%documents);

my $text = "";
# Create each document, stem it, and insert it into the cluster
# Add the stemmed text to the $text variable
while ( <$input_dir/*> )
{
  my $file = $_;

  my $dl = Clair::Document->new(type => 'text', file => $file, id => $file);

  $c->insert(document => $dl, id => $file);

  # Get the number of lines in the text (because the stemmed version loses      \
them)
  my @lines = split("\n", $dl->{text});

  $dl->stem_keep_newlines();

  $text .= $dl->{stem} . " ";

  # print "Document: $dl->{stem}\n";
}

$text = substr($text, 0, length($text) - 1);

Clair::Util::build_idf_by_line($text, "$gen_dir/dbm2");

my %idf = Clair::Util::read_idf("$gen_dir/dbm2");
my $l;
my $r;
my $ct = 0;

while (($l, $r) = each %idf) {
  $ct++;
  print "$ct\t$l\t*$r*\n";
}


\end{boxedverbatim}

\subsubsection{index\_dirfiles\_incremental.pl}
\begin{boxedverbatim}

#!/usr/local/bin/perl

# script: test_index_dirfiles_incremental.pl
# functionality: Tests index update using Index/dirfiles.pm; requires
# functionality: index_dirfiles.pl to be run previously

use strict;
use FindBin;
use vars qw/$DEBUG/;

use Benchmark;
use Clair::GenericDoc;
use Clair::Index;
use Data::Dumper;
use File::Find;


$DEBUG = 0;
my %args;
my @files = ();
my $corpus_root = "$FindBin::Bin/input/index/Shakespear";
my $incremental_root = "$FindBin::Bin/input/index/incremental";
my $index_root = "$FindBin::Bin/produced/index_dirfiles",
my $stop_word_list = "$FindBin::Bin/input/index/stopwords.txt";
my $filter = "\.html";

# instantiate the index object
my $idx = new Clair::Index(
	DEBUG => $DEBUG,
	stop_word_list => $stop_word_list,
	index_root => $index_root,
	index_file_format => "dirfiles",
);

$idx->debugmsg("using stop word list: $stop_word_list", 0) if(-f                \
$stop_word_list);

my $t0;
my $t1;


# let's try incremental adding of index.
@files = ();
find(\&wanted, ( $incremental_root ));
@files = grep { /$filter/ } @files if($filter);
# print Dumper(\@files);


$t0 = new Benchmark;

# insert, build, and sync
for my $f (@files)
{
	my $gdoc = new Clair::GenericDoc(
		DEBUG => 1,
		# module_root => $module_root,
		content => $f,
		stem => 1,
		use_parser_module => "shakespear"
	);

	# insert the document into the index object
	$idx->insert($gdoc);
}
$idx->build();
$idx->sync();


\end{boxedverbatim}

\begin{boxedverbatim}


$t1 = new Benchmark;
my $timediff = timestr(timediff($t1, $t0));
$idx->debugmsg("incremental index update took : " . $timediff, 0);


my $doc2 = $idx->index_read($idx->{index_file_format}, "document_meta_index",   \
"all");

$idx->debugmsg("total documents   : " . scalar keys %$doc2, 0);
$idx->debugmsg($doc2, 1);


# to find all the shakespear html files by scenes
sub wanted
{
	return if(-d $File::Find::name || $File::Find::name =~                         \
/full\.html|index\.html|news\.html|^\./);
	push @files, $File::Find::name;
}

\end{boxedverbatim}

\subsubsection{index\_dirfiles.pl}
\begin{boxedverbatim}

#!/usr/local/bin/perl

# script: test_index_dirfiles.pl
# functionality: Tests index update using Index/dirfiles.pm, index is created
# functionality: in produces/index_dirfiles, complementary to index_mldbm.pl

use strict;
use FindBin;
use lib "$FindBin::Bin/../lib";
use lib "$FindBin::Bin/lib"; # if you are outside of bin path.. just in case
use vars qw/$DEBUG/;

use Benchmark;
use Clair::GenericDoc;
use Clair::Index;
use Data::Dumper;
use File::Find;
use Getopt::Long;
use Pod::Usage;


$DEBUG = 0;
my %args;
my @files = ();
my $corpus_root = "$FindBin::Bin/input/index/Shakespear";
my $incremental_root = "$FindBin::Bin/input/index/incremental";
my $index_root = "$FindBin::Bin/produced/index_dirfiles",
my $stop_word_list = "$FindBin::Bin/input/index/stopwords.txt";
my $filter = "\.html";

# Determine the GenericDoc module root here
# my @libpaths = grep { -d $_ && $_ =~ /GenericDoc/ } @INC;
# my $module_root = shift @libpaths;

# @libpaths = grep { -d $_ && $_ =~ /Index/ } @INC;
# my $rw_module_root = shift @libpaths;

GetOptions(\%args, 'help', 'man', 'debug=i', 'datadir=s', 'listfile=s',         \
'filter=s', 'stop_word_list=s') or pod2usage(2);
pod2usage(1) if($args{help});
pod2usage(-exitstatus => 0, -verbose => 2) if($args{man});
$corpus_root = $args{datadir} if($args{datadir});
$DEBUG = $args{debug} if($args{debug});
$stop_word_list = $args{stop_word_list} if($args{stop_word_list});



# instantiate the index object
my $idx = new Clair::Index(
	DEBUG => 1,
	stop_word_list => $stop_word_list,
	index_root => $index_root,
	index_file_format => "dirfiles",
);

$idx->debugmsg("using stop word list: $stop_word_list", 0) if(-f                \
$stop_word_list);

my $t0;
my $t1;

#
# Finding files
#
$idx->debugmsg("using files from: $corpus_root", 0);

$t0 = new Benchmark;

\end{boxedverbatim}

\begin{boxedverbatim}


 find(\&wanted, ( $corpus_root ));
 @files = grep { /$filter/ } @files if($filter);

$idx->debugmsg("total of " . scalar @files . " files retrieved from             \
'$corpus_root'", 0);

$t1 = new Benchmark;
my $timediff_find = timestr(timediff($t1, $t0));


#
# Preparing Index
#
$idx->debugmsg("constructing index object with documents", 0);
$t0 = new Benchmark;

for my $f (@files)
{
	my $gdoc = new Clair::GenericDoc(
		DEBUG => $DEBUG,
		# module_root => $module_root,
		content => $f,
		stem => 1,
		use_parser_module => "shakespear"
	);

	# insert the document into the index object
	$idx->insert($gdoc);
}
$t1 = new Benchmark;
my $timediff_prep = timestr(timediff($t1, $t0));



#
# Building Index
#
$t0 = new Benchmark;
$idx->debugmsg("building index, please wait...", 0);
$idx->clean(); # cleans up any existing index.
my ($invidx, $docidx, $wordidx) = $idx->build();
$t1 = new Benchmark;
my $timediff_build = timestr(timediff($t1, $t0));


#
# Writing Index
#
$t0 = new Benchmark;
$idx->debugmsg("sync-ing (saving) to disk", 0);
$idx->sync();
$t1 = new Benchmark;
my $timediff_sync = timestr(timediff($t1, $t0));


# you can use the methods from the submodules this way
my $hash = $idx->index_read("dirfiles", "caesar");
print Dumper($hash);

# print Dumper($hash);


# my $doc = $idx->index_read($idx->{index_file_format},                         \
"$index_root/document_meta_idx.dbm", 1);
# my $words = $idx->index_read($idx->{index_file_format},                       \
"$index_root/word_idx.dbm", 1);

\end{boxedverbatim}

\begin{boxedverbatim}

my $space = `du -sk $index_root`; 
$space = $1 if($space =~ /(\d+)\s+/);
# my @sorted_words = reverse sort { $words->{$a}->{count} <=>                   \
$words->{$b}->{count} } keys %$words;

# $idx->debugmsg("total documents   : " . scalar keys %$doc, 0);
# $idx->debugmsg("total unique words: " . scalar keys %$words, 0);
$idx->debugmsg("disk space used   : " . $space . " KB", 0);
$idx->debugmsg("file collect took : " . $timediff_find, 0);
$idx->debugmsg("data prep took    : " . $timediff_prep, 0);
$idx->debugmsg("index build took  : " . $timediff_build, 0);
$idx->debugmsg("index write took  : " . $timediff_sync, 0);
# $idx->debugmsg("top 20 words      : list below", 0);

# for my $i (0..19)
# {
	# my $w = $sorted_words[$i];
	# $idx->debugmsg("   $w $words->{$w}->{count}", 0);
# }


# $idx->debugmsg($doc, 1);



# to find all the shakespear html files by scenes
sub wanted
{
	return if(-d $File::Find::name || $File::Find::name =~                         \
/full\.html|index\.html|news\.html|^\./);
	push @files, $File::Find::name;
}


\end{boxedverbatim}

\subsubsection{index\_mldbm\_incremental.pl}
\begin{boxedverbatim}

#!/usr/local/bin/perl

# script: test_index_mldbm_incremental.pl
# functionality: Tests index update using Index/mldbm.pm; requires that
# functionality: index_mldbm.pl was run previously

use strict;
use FindBin;
use vars qw/$DEBUG/;

use Benchmark;
use Clair::GenericDoc;
use Clair::Index;
use Data::Dumper;
use File::Find;


$DEBUG = 0;
my %args;
my @files = ();
my $corpus_root = "$FindBin::Bin/input/index/Shakespear";
my $incremental_root = "$FindBin::Bin/input/index/incremental";
my $index_root = "$FindBin::Bin/produced/index_mldbm",
my $stop_word_list = "$FindBin::Bin/input/index/stopwords.txt";
my $filter = "\.html";

# instantiate the index object
my $idx = new Clair::Index(
	DEBUG => $DEBUG,
	stop_word_list => $stop_word_list,
	index_root => $index_root,
	# rw_modules_root => $rw_module_root,	
);

$idx->debugmsg("using stop word list: $stop_word_list", 0) if(-f                \
$stop_word_list);

my $t0;
my $t1;


# let's try incremental adding of index.
@files = ();
find(\&wanted, ( $incremental_root ));
@files = grep { /$filter/ } @files if($filter);
# print Dumper(\@files);


$t0 = new Benchmark;

# insert, build, and sync
for my $f (@files)
{
	my $gdoc = new Clair::GenericDoc(
		DEBUG => 1,
		# module_root => $module_root,
		content => $f,
		stem => 1,
		use_parser_module => "shakespear"
	);

	# insert the document into the index object
	$idx->insert($gdoc);
}
$idx->build();
$idx->sync();


\end{boxedverbatim}

\begin{boxedverbatim}


$t1 = new Benchmark;
my $timediff = timestr(timediff($t1, $t0));
$idx->debugmsg("incremental index update took : " . $timediff, 0);


my $doc2 = $idx->index_read($idx->{index_file_format},                          \
"$index_root/document_meta_index.dbm", 1);

$idx->debugmsg("total documents   : " . scalar keys %$doc2, 0);
$idx->debugmsg($doc2, 1);


# to find all the shakespear html files by scenes
sub wanted
{
	return if(-d $File::Find::name || $File::Find::name =~                         \
/full\.html|index\.html|news\.html|^\./);
	push @files, $File::Find::name;
}

\end{boxedverbatim}

\subsubsection{index\_mldbm.pl}
\begin{boxedverbatim}

#!/usr/local/bin/perl

# script: test_index_mldbm.pl
# functionality: Tests index creation using Index/mldbm.pm, outputs stats,
# functionality: uses input/index/Shakespear, creates produces/index_mldbm

use strict;
use FindBin;
use lib "$FindBin::Bin/../lib";
use lib "$FindBin::Bin/lib"; # if you are outside of bin path.. just in case
use vars qw/$DEBUG/;

use Benchmark;
use Clair::GenericDoc;
use Clair::Index;
use Data::Dumper;
use File::Find;
use Getopt::Long;
use Pod::Usage;


$DEBUG = 0;
my %args;
my @files = ();
my $corpus_root = "$FindBin::Bin/input/index/Shakespear";
my $incremental_root = "$FindBin::Bin/input/index/incremental";
my $index_root = "$FindBin::Bin/produced/index_mldbm",
my $stop_word_list = "$FindBin::Bin/input/index/stopwords.txt";
my $filter = "\.html";

# Determine the GenericDoc module root here
# my @libpaths = grep { -d $_ && $_ =~ /GenericDoc/ } @INC;
# my $module_root = shift @libpaths;

# @libpaths = grep { -d $_ && $_ =~ /Index/ } @INC;
# my $rw_module_root = shift @libpaths;

GetOptions(\%args, 'help', 'man', 'debug=i', 'datadir=s', 'listfile=s',         \
'filter=s', 'stop_word_list=s') or pod2usage(2);
pod2usage(1) if($args{help});
pod2usage(-exitstatus => 0, -verbose => 2) if($args{man});
$corpus_root = $args{datadir} if($args{datadir});
$DEBUG = $args{debug} if($args{debug});
$stop_word_list = $args{stop_word_list} if($args{stop_word_list});



# instantiate the index object
my $idx = new Clair::Index(
	DEBUG => $DEBUG,
	stop_word_list => $stop_word_list,
	index_root => $index_root,
	# rw_modules_root => $rw_module_root,	
);

$idx->debugmsg("using stop word list: $stop_word_list", 0) if(-f                \
$stop_word_list);

my $t0;
my $t1;

#
# Finding files
#
$idx->debugmsg("using files from: $corpus_root", 0);

$t0 = new Benchmark;

\end{boxedverbatim}

\begin{boxedverbatim}


 find(\&wanted, ( $corpus_root ));
 @files = grep { /$filter/ } @files if($filter);

$idx->debugmsg("total of " . scalar @files . " files retrieved from             \
'$corpus_root'", 0);

$t1 = new Benchmark;
my $timediff_find = timestr(timediff($t1, $t0));


#
# Preparing Index
#
$idx->debugmsg("constructing index object with documents", 0);
$t0 = new Benchmark;

for my $f (@files)
{
	my $gdoc = new Clair::GenericDoc(
		DEBUG => $DEBUG,
		# module_root => $module_root,
		content => $f,
		stem => 1,
		use_parser_module => "shakespear"
	);

	# insert the document into the index object
	$idx->insert($gdoc);
}
$t1 = new Benchmark;
my $timediff_prep = timestr(timediff($t1, $t0));



#
# Building Index
#
$t0 = new Benchmark;
$idx->debugmsg("building index, please wait...", 0);
$idx->clean(); # cleans up any existing index.
my ($invidx, $docidx) = $idx->build();
$t1 = new Benchmark;
my $timediff_build = timestr(timediff($t1, $t0));


#
# Writing Index
#
$t0 = new Benchmark;
$idx->debugmsg("sync-ing (saving) to disk", 0);
$idx->sync();
$t1 = new Benchmark;
my $timediff_sync = timestr(timediff($t1, $t0));


# you can use the methods from the submodules this way
my $modobj = $idx->_load_rw_module("mldbm");
my $hash = $modobj->_mldbm_read("$index_root/document_meta_idx.dbm", $idx);

# print Dumper($hash);


my $doc = $idx->index_read($idx->{index_file_format},                           \
"$index_root/document_meta_idx.dbm", 1);
my $space = `du -sk $index_root`; 
$space = $1 if($space =~ /(\d+)\s+/);

\end{boxedverbatim}

\begin{boxedverbatim}


$idx->debugmsg("total documents   : " . scalar keys %$doc, 0);
# $idx->debugmsg("total unique words: " . scalar keys %$words, 0);
$idx->debugmsg("disk space used   : " . $space . " KB", 0);
$idx->debugmsg("file collect took : " . $timediff_find, 0);
$idx->debugmsg("data prep took    : " . $timediff_prep, 0);
$idx->debugmsg("index build took  : " . $timediff_build, 0);
$idx->debugmsg("index write took  : " . $timediff_sync, 0);

$idx->debugmsg($doc, 1);



# to find all the shakespear html files by scenes
sub wanted
{
	return if(-d $File::Find::name || $File::Find::name =~                         \
/full\.html|index\.html|news\.html|^\./);
	push @files, $File::Find::name;
}



__END__

=head1 NAME

test_index.pl - builds indexes from the corpus

=head1 SYNOPSIS

index.pl [options] 

Options:

 -help            brief help message
 -man             full documentation
 -debug           specify a debug level for verbosity
 -datadir         corpus dir [default: /home/cs6998/hw1/Shakespeare]
 -listfile        file containing a list of data source
 -stop_work_list  provide a list file containing the stop words

=head1 OPTIONS

=over 8

=item B<-help>

Print a brief help message and exits.

=item B<-man>

Prints the manual page and exits.

=back

=head1 DESCRIPTION

B<This program> will slurp in all the files under designated data directory
and create an inverted index for searching.

=cut

\end{boxedverbatim}

\subsubsection{ir.pl}
\begin{boxedverbatim}

#!/usr/local/bin/perl

# script: test_ir.pl
# functionality: Builds a corpus from some text files, then makes an IDF, a
# functionality: TF, and outputs some information from them

# To run this script, you need to have ALECACHE=/tmp (or to some other
# directory) set in your environment.

use warnings;
use strict;
use Clair::Utils::CorpusDownload;
use Clair::Utils::Idf;
use Clair::Utils::Tf;
use DB_File; # This is necessary if running on an NFS drive


my $in_dir = "$FindBin::Bin/input/ir";
my $out_dir = "$FindBin::Bin/produced/ir";
my $corpus_name = "ir_corpus";

# Read the *.txt files from the input directory, taking care to
# prepend the input directory before the filenames.
opendir INPUT, $in_dir or die "Couldn't open $in_dir: $!";
my @files = map { "$in_dir/$_" } grep { /\.txt$/ } readdir(INPUT);
closedir INPUT;

# Make this object so we can get the files into TREC format
my $corpus = Clair::Utils::CorpusDownload->new(
    corpusname => $corpus_name,
    rootdir => $out_dir,
);

# You have to do this because the rootdir and corpus
# parameters passed to the CorpusDownload constructor are ignored.
$corpus->{rootdir} = $out_dir;
$corpus->{corpus} = $corpus_name;

$corpus->buildCorpusFromFiles( filesref => \@files, cleanup => 0 );

# The order of the calls to buildIdf, build_docno_dbm, and buildTf are
# important. It can fail if they are called in a different order.

# Create the idf database file
$corpus->buildIdf( stemmed => 1 );
my $idf = Clair::Utils::Idf->new( rootdir => $out_dir, corpusname =>            \
$corpus_name,
    stemmed => 1 );

# Create the tf database file
$corpus->build_docno_dbm();
$corpus->buildTf( stemmed => 1 );
my $tf = Clair::Utils::Tf->new( rootdir => $out_dir, corpusname =>              \
$corpus_name,
    stemmed => 1 );

# Output some information involving term statistics.
print "nfiles=", scalar @files, "\n";
my @words = qw(the and);
foreach my $word (@words) {
    my $idf_score = $idf->getIdfForWord($word);
    my $tf_score = $tf->getFreq($word);
    my $n_docs = $tf->getNumDocsWithWord($word);
    print "word=$word, idf=$idf_score, tf=$tf_score, n_docs=$n_docs\n";
}

# Output some information involving phrase statistics.

\end{boxedverbatim}

\begin{boxedverbatim}

my @phrase = qw(in the);
my $tf_score = $tf->getPhraseFreq(@phrase);
my $n_docs = $tf->getNumDocsWithPhrase(@phrase);
print "phrase=\"in the\", freq=$tf_score, n_docs=$n_docs\n";

\end{boxedverbatim}

\subsubsection{learn.pl}
\begin{boxedverbatim}

#!/usr/local/bin/perl

# script: test_learn.pl
# functionality: Uses feature vectors in the svm_light format and calculates
# functionality: and saves perceptron parameters; needs features_traintest.pl

use strict;
use FindBin;
# use lib "$FindBin::Bin/../lib";
# use lib "$FindBin::Bin/lib"; # if you are outside of bin path.. just in case
use vars qw/$DEBUG/;

use Benchmark;
use Clair::Learn;
use Data::Dumper;
use File::Find;


$DEBUG = 0;
my %args;
my @train_files = (); # list of train files we will analyze
my @test_files = (); # list of test files we will analyze
my %container = (); # container for our file arrays.

my $results_root = "$FindBin::Bin/produced/features";
mkpath($results_root, 0, 0777) unless(-d $results_root);

my $output = "feature_vectors";
my $train = "$results_root/$output.train";
my $model = "$results_root/model";
my $eta = $args{eta};

unless(-f $train)
{
	print "The train file is required. Make sure features_traintest.pl has been    \
run.\n";
	exit;
}

my $t0;
my $t1;

#
# Finding files
#
$t0 = new Benchmark;


my $lea = new Clair::Learn(DEBUG => $DEBUG, train => $train, model => $model);
my ($w0, $w) = $lea->learn("", $eta); # retrieves the coefficients


$t1 = new Benchmark;
my $timediff = timestr(timediff($t1, $t0));

$lea->debugmsg("learning (perceptron) convergence took: $timediff", 0);
$lea->debugmsg("intercept: $w0\n" . Dumper($w), 1);

# save the output
open M, "> $model" or $lea->errmsg("cannot open file '$model': $!", 1);
print M "intercept $w0\n";
while (my ($feature_id, $weight) = each %$w)
{
	print "id:weight $feature_id:$weight\n";
	print M "$feature_id $weight\n";
}
close M;

\end{boxedverbatim}

\subsubsection{lexrank2.pl}
\begin{boxedverbatim}

#!/usr/local/bin/perl

# script: test_lexrank2.pl
# functionality: Computes lexrank from a stemmed line-based cluster

use strict;
use warnings;
use FindBin;
use Clair::Network;
use Clair::Cluster;
use Clair::Document;
use Clair::Network::Centrality::LexRank;

my $basedir = $FindBin::Bin;
my $input_dir = "$basedir/input/lexrank";

my $c = new Clair::Cluster();

$c->load_lines_from_file("$input_dir/t02_lexrank.input");
$c->stem_all_documents();
my %cos_matrix = $c->compute_cosine_matrix(text_type => 'stem');

my $n = $c->create_network(cosine_matrix => \%cos_matrix);

my $cent = Clair::Network::Centrality::LexRank->new($n);

$cent->centrality();


print "SENT LEXRANK\n";
$cent->print_current_distribution();
print "\n";

\end{boxedverbatim}

\subsubsection{lexrank3.pl}
\begin{boxedverbatim}

#!/usr/local/bin/perl

# script: test_lexrank3.pl
# functionality: Computes lexrank from line-based, stripped and stemmed
# functionality: cluster

use strict;
use warnings;
use FindBin;
use Clair::Network;
use Clair::Cluster;
use Clair::Document;
use Clair::Network::Centrality::LexRank;

my $basedir = $FindBin::Bin;
my $input_dir = "$basedir/input/lexrank";

# Switch to the input directory so that the file list can be
# just filenames without paths (since we don't know absolute path)
chdir "$input_dir";

my $c = new Clair::Cluster();

$c->load_file_list_from_file("filelist.txt", type => 'html', count_id => 1);
$c->strip_all_documents();
$c->stem_all_documents();
my %cos_matrix = $c->compute_cosine_matrix(text_type => 'stem');

my $n = $c->create_network(cosine_matrix => \%cos_matrix);

my $cent = Clair::Network::Centrality::LexRank->new($n);
$cent->centrality();


print "FILE LEXRANK\n";
$cent->print_current_distribution();
print "\n";

\end{boxedverbatim}

\subsubsection{lexrank4.pl}
\begin{boxedverbatim}

#!/usr/local/bin/perl

# script: test_lexrank4.pl
# functionality: Based on an interactive script, this test builds a sentence-
# functionality: based cluster, then a network, computes lexrank, and then
# functionality: runs MMR on it 

use strict;
use warnings;
use FindBin;
use Clair::Config qw( $PRMAIN );
use Clair::Cluster;
use Clair::Document;
use Clair::Network;
use Clair::Network::Centrality::LexRank;
use Clair::Network::Centrality::CPPLexRank;
use Clair::NetworkWrapper;
use File::Spec;
use Getopt::Long;

# This script has been converted from an interactive example script. To
# use it interactively, uncomment the GetOptions part.

# This script is used to run various forms of lexrank with optional MMR 
# reranking.
#
# Each input file must be in the format of one unique meta-data tag and one 
# sentence per line, separated with a tab.
#
# To run an unbiased lexrank on a list of files (uses C++ lexrank):
#     ./lexrank.pl -i myid file1 file2 ... fileN
#
# To run a biased lexrank on a list of files, where each sentence is given
# a boost proportional to its distance from the top of the document (uses
# Perl lexrank):
#     ./lexrank.pl -i myid -b file1 file2 ... fileN
#
# To run a biased lexrank from a file containing query sentences, one per line
# (uses C++ lexrank):
#     ./lexrank.pl -i myid -q bias.txt file1 file2 ... fileN
#
# To use MMR reranking:
#     ./lexrank.pl -i myid -m 0.75
#
# To use generation probabilities instead of cosine similarity:
#     ./lexrank.pl -g
#
# Author: Tony Fader (afader@umich.edu)


# Get command line arguments

my (@files, $id, $rbias, $qbias, $mmr, $size, $clean, $genprob);

my $input_dir = "$FindBin::Bin/input/lexrank4";
opendir INPUT, $input_dir or die "Couldn't open $input_dir: $!";
@files = ("$input_dir/combine1.txt");
closedir INPUT;

$id = "test";
$qbias = "$input_dir/bias.10.1.txt";
$mmr = 0.75;
$genprob = 1;
$clean = 0;

#GetOptions(
#    "i=s" => \$id,

\end{boxedverbatim}

\begin{boxedverbatim}

#    "q=s" => \$qbias,
#    "b"   => \$rbias,
#    "g"   => \$genprob,
#    "m=f" => \$mmr,
#    "s=i" => \$size,
#    "c"   => \$clean
#);
# @files = @ARGV;

#if (@files <= 0 || !defined $id) {
#    print_usage();
#    exit(1);
#} elsif ($rbias && $qbias) {
#    print "Both -b and -q specified\n";
#    exit(1);
#}



# Make a temporary directory to work in to prevent collisions between multiple
# runs
my $out_dir = "$FindBin::Bin/produced/lexrank4/$id";
if (!-e $out_dir) {
    mkdir($out_dir, 0755) or die "Couldn't create directory $id: $!";
    chdir($out_dir) or die "Couldn't chdir to $id: $!";
} elsif (-d $out_dir) {
    chdir($out_dir) or die "Couldn't chdir to $id: $!";
} else {
    die "Unable to create or use directory $id";
}



# Create a sentence cluster from the file list

my @lines = combine_lines(@files);
my $sent_cluster = Clair::Cluster->new();
for (@lines) {
    my @tokens = split /\t/;
    die "Malformed line: $_" unless @tokens == 2;
    my ($meta, $text) = @tokens;
    my $doc = Clair::Document->new(
        string => $text,
        type => "text",
        id => $meta
    );
    $doc->stem();
    $sent_cluster->insert($meta, $doc);
}



# Create a network from the sentence cluster

my $network;

if ($genprob) {
    my %matrix = $sent_cluster->compute_genprob_matrix();
    $network = $sent_cluster->create_genprob_network(
        genprob_matrix => \%matrix,
        include_zeros => 1
    );
} else {
    my %matrix = $sent_cluster->compute_cosine_matrix();
    $network = $sent_cluster->create_network(
        cosine_matrix => \%matrix,
        include_zeros => 1

\end{boxedverbatim}

\begin{boxedverbatim}

    );
}




# Run lexrank
my $cent;
if ($rbias) {
  $cent = Clair::Network::Centrality::LexRank->new($network);

    # Set the order bias
    set_order_bias($network, @files);

    $cent->centrality();

} elsif ($qbias) {

    # Wrap the network to use the CPP implementation of lexrank 
    $network = Clair::NetworkWrapper->new( 
        network => $network,
        prmain => $PRMAIN,
        clean => 1
    );

    # Read the bias files
    my @bias_sents = ();
    open BIAS, $qbias or die "Couldn't read $qbias: $!";
    while (<BIAS>) {
        chomp;
        push @bias_sents, $_;
    }
    close BIAS;

    # Run query-based lexrank
    $cent = Clair::Network::Centrality::CPPLexRank->new($network);
    $cent->compute_lexrank_from_bias_sents(bias_sents => \@bias_sents);
} else {

    # Wrap the network to use the CPP implementation of lexrank 
    $network = Clair::NetworkWrapper->new( 
        network => $network,
        prmain => $PRMAIN,
        clean => 1
    );

    # Run unbiased lexrank
    $cent = Clair::Network::Centrality::CPPLexRank->new($network);
    $cent->centrality();
}


# Run the MMR reranker if necessary

if (defined $mmr) {
    $network->mmr_rerank_lexrank($mmr);
}



# Get the results and print them out

my %scores = %{ get_scores($network) };
my $counter = 0;
foreach my $meta (sort { $scores{$b} cmp $scores{$a} } keys %scores) {
    my $text = $sent_cluster->get($meta)->get_text();
    print "$meta\t$text\t$scores{$meta}\n";

\end{boxedverbatim}

\begin{boxedverbatim}

    $counter++;
    if (defined $size and $counter >= $size) {
        last;
    }
}



# Done

exit(0);


##################
# Some subroutines
##################

sub print_usage {
    print "usage: $0 -i id [options] file1 [file2 ... ]\n" .
          "Options: \n" .
          "  -m value, parameter in [0,1]\n" .
          "  -s size\n" .
          "  -q bias_file,  query-based biased lexrank\n" .
          "  -b, rank-based biased lexrank\n" .
          "  -c, cleanup directory when done\n" .
          "Only one of -q and -b may be specified.\n";
}

sub combine_lines {
    my @files = @_;
    my @lines = ();
    foreach my $file (@files) {
        open FILE, "< $file" or die "Couldn't open $file: $!";
        while(<FILE>) {
            chomp;
            push @lines, $_;
        }
        close FILE;
    }
    return @lines;
}

sub get_scores {
    my $network = shift;
    my $graph = $network->{graph};
    my @verts = $graph->vertices();
    my %scores = ();
    foreach my $vert (@verts) {
        $scores{$vert} = $graph->get_vertex_attribute($vert, "lexrank_value");
    }
    return \%scores;
}

# Given a list of files each containing a list of sents, makes a bias file 
# where each sentence is weighted according to its relative position in the
# file. 
sub set_order_bias {

    my $network = shift;
    my @files = @_;

    # Print the bias file
    open TEMP, "> $out_dir/bias.temp" or die "Couldn't open temp file           \
bias.temp: $!";
    foreach my $file (@files) {
        my @metas;
        open FILE, "< $file" or die "Couldn't open $file for read: $!";

\end{boxedverbatim}

\begin{boxedverbatim}

        while (<FILE>) {
            my ($meta, $text) = split /\t/, $_;
            push @metas, $meta;
        }
        close FILE;

        my $denom = $#metas;
        if ($denom < 0) {
            warn "No sentences in $file";
            next;
        } elsif ($denom == 0) {
            print TEMP "$metas[0] 1\n";
        } else {
            foreach my $i (0 .. $denom) {
                my $weight = ($denom - $i) / $denom;
                print TEMP "$metas[$i] $weight\n";
            }
        }

    }
    close TEMP;

    $network->read_lexrank_bias("$out_dir/bias.temp");
    if ($clean) {
        unlink("bias.temp") or warn "Couldn't remove bias.temp: $!";
    }

}

\end{boxedverbatim}

\subsubsection{lexrank\_large.pl}
\begin{boxedverbatim}

#!/usr/local/bin/perl

# script: test_lexrank_large.pl
# functionality: Builds a cluster from a set of files, computes a cosine matrix
# functionality: and then lexrank, then creates a network and a cluster using
# functionality: a lexrank-based threshold of 0.2

use strict;
use warnings;
use FindBin;
use Clair::Network;
use Clair::Network::Centrality::LexRank;
use Clair::Cluster;
use Clair::Document;

my $basedir = $FindBin::Bin;
my $input_dir = "$basedir/input/lexrank";

# chdir to the input directory so that the filelist can be relative paths
# (since we don't know the absolute path)
chdir $input_dir;

my $c = new Clair::Cluster();

$c->load_file_list_from_file("filelist.txt", type => 'html', count_id => 1);
$c->strip_all_documents();
$c->stem_all_documents();

print "I'm here.  There are ", $c->count_elements, " documents in the           \
cluster.\n";
my $sent_n = $c->create_sentence_based_network;
print "Now I'm here.\n";
print "Sentence based network has: ", $sent_n->num_nodes(), " nodes.\n";

my %cos_matrix = $c->compute_cosine_matrix(text_type => 'stem');

my $n = $c->create_network(cosine_matrix => \%cos_matrix);
my $cent = Clair::Network::Centrality::LexRank->new($n);

$cent->centrality();


print "FILE LEXRANK\n";
$cent->print_current_distribution();
print "\n";

my $lex_network = $n->create_network_from_lexrank(0.2);
print "There are ", $lex_network->num_nodes, " nodes in the network created     \
from lexrank.\n";

my $lex_cluster = $n->create_cluster_from_lexrank(0.2);
print "There are ", $lex_cluster->count_elements(), " documents in the cluster  \
created from lexrank.\nThey have:\n";

my $lex_docs_ref = $lex_cluster->documents();
my %lex_docs = %$lex_docs_ref;

foreach my $doc (values %lex_docs ) {
        print $doc->count_words, " words\n";
}


\end{boxedverbatim}

\subsubsection{lexrank.pl}
\begin{boxedverbatim}

#!/usr/local/bin/perl

# script: test_lexrank.pl
# functionality: Computes lexrank on a small network

use strict;
use warnings;
use FindBin;
use Clair::Network;
use Clair::Network::Centrality::LexRank;

my $basedir = $FindBin::Bin;
my $input_dir = "$basedir/input/lexrank";

my $n = new Clair::Network();

$n->add_node(0, text => 'This is node 0');
$n->add_node(1, text => 'This is node 1');
$n->add_node(2, text => 'This is node 2');
$n->add_node(3, text => 'This is node 3');
$n->add_node(4, text => 'This is node 4');

my $cent = Clair::Network::Centrality::LexRank->new($n);

$cent->read_lexrank_probabilities_from_file("$input_dir/files-sym.cos.ID");
$cent->read_lexrank_initial_distribution("$input_dir/files.uniform");

# Remove following line to remove lexrank bias:
$cent->read_lexrank_bias("$input_dir/files.bias");

print "Initial distribution:\n";
$cent->print_current_distribution();

print "READ PROBABILITIES\n";


$cent->centrality(jump => 0.5);

print "The computed lexrank distribution is:\n";
$cent->print_current_distribution();
print "\n";


\end{boxedverbatim}

\subsubsection{linear\_algebra.pl}
\begin{boxedverbatim}

#!/usr/local/bin/perl

# script: test_linear_algebra.pl
# functionality: A variety of arithmetic tests of the linear algebra module 

use strict;
use warnings;
use FindBin;
use Clair::Utils::LinearAlgebra;

my @v1 = ("1", "2", "3", "4");
my @v2 = ("5", "6", "7", "8");
my @v3 = ("2", "4", "6", "8", "10");
my @v4 = ("1", "3", "5", "7", "9");
my @v5 = ("1", "1", "2", "3", "5");
my @v6 = ("3", "2", "1", "0", "2");

#Test Two -- Inner Product of Vectors One and Two
#Test Two Expected -- 70

print "inner product of ", list_to_string(@v1), " and ", list_to_string(@v2), 
      "\n";
my $test1 = Clair::Utils::LinearAlgebra::innerProduct (\@v1,\@v2);
print "$test1\n";


#Test Seven -- Subtraction of Vectors One and Two
#Test Seven Expected -- (-4, -4, -4, -4)

print "difference of ", list_to_string(@v1), " and ", list_to_string(@v2),      \
"\n";
my @diff = Clair::Utils::LinearAlgebra::subtract (\@v1,\@v2);
print list_to_string(@diff), "\n";

#Test Twelve -- Addition of Vectors One and Two
#Test Twelve Expected -- (6, 8, 10, 12)

print "sum of ", list_to_string(@v1), " and ", list_to_string(@v2), "\n";
my @sum1 = Clair::Utils::LinearAlgebra::add (\@v1,\@v2);
print list_to_string(@sum1), "\n";

#Test Fifteen -- Addition of Vectors Three and Four and Five
#Test Fifteen Expected -- (4, 8, 13, 18, 24)

print "sum of ", list_to_string(@v3), " and ", list_to_string(@v4), " and ", 
    list_to_string(@v5), "\n";
my @sum2 = Clair::Utils::LinearAlgebra::add (\@v3,\@v4,\@v5);
print list_to_string(@sum2), "\n";

#Test Seventeen -- Addition of Vectors One and Two
#Test Seventeen Expected -- (3, 4, 5, 6)

print "mean of ", list_to_string(@v1), " and ", list_to_string(@v2), "\n";
my @mean1 = Clair::Utils::LinearAlgebra::average (\@v1,\@v2);
print list_to_string(@mean1), "\n";

#Test Twenty -- Addition of Vectors Three and Four and Six
#Test Twenty Expected -- (2, 3, 4, 5, 7)

print "mean of ", list_to_string(@v3), " and ", list_to_string(@v4), " and ", 
    list_to_string(@v6), "\n";
my @mean2 = Clair::Utils::LinearAlgebra::average (\@v3,\@v4,\@v6);
print list_to_string(@mean2), "\n";

sub list_to_string {
    return join " ", @_;
}

\end{boxedverbatim}

\subsubsection{mead\_summary.pl}
\begin{boxedverbatim}

#!/usr/local/bin/perl

# script: test_mead_summary.pl
# functionality: Tests MEAD's summarizer on a cluster of two documents,
# functionality: prints features for each sentence of the summary 

use strict;
use warnings;
use FindBin;
use Clair::Cluster;
use Clair::Config;
use Clair::Document;
use Clair::MEAD::Wrapper;
use Clair::MEAD::Summary;

my $out_dir = "$FindBin::Bin/produced/mead_summary";
my $docs = "$FindBin::Bin/input/mead_summary";

my $cluster = Clair::Cluster->new();
my $doc1 = Clair::Document->new( 
    file => "$docs/fed1.txt", 
    id => 1, 
    type => "text"
);
$cluster->insert(1, $doc1);
my $doc2 = Clair::Document->new( 
    file => "$docs/fed2.txt", 
    id => 2, 
    type => "text"
);
$cluster->insert(2, $doc2);

my $mead = Clair::MEAD::Wrapper->new(
    mead_home => $MEAD_HOME,
    cluster => $cluster,
    cluster_dir => $out_dir
);

my $summary = $mead->get_summary();

print "To string:\n";
print $summary->to_string() . "\n\n";

foreach my $i ( 1 .. $summary->size() ) {

    my %sent = $summary->get_sent($i);
    my %feats = $summary->get_features($i);
    my $str = join ",", map { "$_=$feats{$_}" } (keys %feats);

    print "$sent{text} ($sent{did}.$sent{sno}: $str)\n";

}

\end{boxedverbatim}

\subsubsection{mega.pl}
\begin{boxedverbatim}

#!/usr/local/bin/perl

# script: test_mega.pl
# functionality: Downloads documents using CorpusDownload, then makes IDFs,
# functionality: TFs, builds a cluster from them, a network based on a
# functionality: binary cosine, and tests the network for a couple of
# functionality: properties

use strict;
use warnings;
use FindBin;
use Clair::Utils::CorpusDownload;
use Clair::Utils::Idf;
use Clair::Utils::Tf;
use Clair::Document;
use Clair::Cluster;
use Clair::Network;

my $basedir = $FindBin::Bin;
my $gen_dir = "$basedir/produced/mega";

my $corpusref = Clair::Utils::CorpusDownload->new(corpusname => "testhtml",
                rootdir => $gen_dir);

# Get the list of urls that we want to download
my $uref =                                                                      \
$corpusref->poach("http://tangra.si.umich.edu/clair/testhtml/index.html",       \
error_file => "$gen_dir/errors.txt");

my @urls = @$uref;

foreach my $v (@urls) {
	print "URL: $v\n";
}

# Build the corpus using the list of urls
# This will index and convert to TREC format
$corpusref->buildCorpus(urlsref => $uref);


# -------------------------------------------------------------------
#  This is how to build the IDF.  First we build the unstemmed IDF,
#  then the stemmed one.
# -------------------------------------------------------------------
$corpusref->buildIdf(stemmed => 0);
$corpusref->buildIdf(stemmed => 1);

# -------------------------------------------------------------------
#  This is how to build the TF.  First we build the DOCNO/URL
#  database, which is necessary to build the TFs.  Then we build
#  unstemmed and stemmed TFs.
# -------------------------------------------------------------------
$corpusref->build_docno_dbm();
$corpusref->buildTf(stemmed => 0);
$corpusref->buildTf(stemmed => 1);

# -------------------------------------------------------------------
#  Here is how to use a IDF.  The constructor (new) opens the
#  unstemmed IDF.  Then we ask for IDFs for the words "have"
#  "and" and "zimbabwe."
# -------------------------------------------------------------------
my $idfref = Clair::Utils::Idf->new( rootdir => $gen_dir,
                       corpusname => "testhtml" ,
                       stemmed => 0 );

my $result = $idfref->getIdfForWord("have");
print "IDF(have) = $result\n";

\end{boxedverbatim}

\begin{boxedverbatim}

$result = $idfref->getIdfForWord("and");
print "IDF(and) = $result\n";
$result = $idfref->getIdfForWord("zimbabwe");
print "IDF(zimbabwe) = $result\n";

# -------------------------------------------------------------------
#  Here is how to use a TF.  The constructor (new) opens the
#  unstemmed TF.  Then we ask for information about the
#  word "have":
#
#  1 first, we get the number of documents in the corpus with
#    the word "Washington"
#  2 then, we get the total number of occurrences of the word "Washington"
#  3 then, we print a list of URLs of the documents that have the
#    word "Washington"
# -------------------------------------------------------------------
my $tfref = Clair::Utils::Tf->new( rootdir => $gen_dir,
                     corpusname => "testhtml" ,
                     stemmed => 0 );

print "\n---Direct term queries (unstemmed):---\n";
$result = $tfref->getNumDocsWithWord("washington");
my $freq   = $tfref->getFreq("washington");
@urls = $tfref->getDocs("washington");
print "TF(washington) = $freq total in $result docs\n";
print "Documents with \"washington\"\n";
foreach my $url (@urls)  {  print "  $url\n";  }
print "\n";

# -------------------------------------------------------------------
#  Then we do 1-2 with the word "and"
# -------------------------------------------------------------------
$result = $tfref->getNumDocsWithWord("and");
$freq   = $tfref->getFreq("and");
@urls = $tfref->getDocs("and");
print "TF(and) = $freq total in $result docs\n";

# -------------------------------------------------------------------
#  Then we do 1-3 with the word "zimbabwe"
# -------------------------------------------------------------------
$result = $tfref->getNumDocsWithWord("zimbabwe");
$freq   = $tfref->getFreq("zimbabwe");
@urls = $tfref->getDocs("zimbabwe");
print "TF(zimbabwe) = $freq total in $result docs\n";
print "Documents with \"zimbabwe\"\n";
foreach my $url (@urls)  {  print "  $url\n";  }
print "\n";


# -------------------------------------------------------------------
#  Here is how to use a TF for phrase queries.  The constructor (new)
#  opens the stemmed TF.  Then we ask for information about the
#  phrase "result in":
#
#  1 first, we get the number of documents in the corpus with
#    the phrase "result in"
#  2 then, we get the total number of occurrences of the phrase
#    "result in"
#  3 then, we print a list of URLs of the documents that have the
#    word "result in" and the number of times each occurs in the
#    document, as well as the position in the document of the initial
#    term ("result") in each occurrence of the phrase
#  4 finally, using a different method, we print the number of times
#    "result in" occurs in each document in which it occurs (from 3),
#    as well as the position(s) of its occurrence (as in 3)
# -------------------------------------------------------------------
$tfref = Clair::Utils::Tf->new( rootdir => $gen_dir,

\end{boxedverbatim}

\begin{boxedverbatim}

                     corpusname => "testhtml" ,
                     stemmed => 1 );

print "\n---Direct phrase queries (stemmed):---\n";
my @phrase = ("result", "in");
$result = $tfref->getNumDocsWithPhrase(@phrase);
$freq   = $tfref->getPhraseFreq(@phrase);
my $positionsByUrlsRef = $tfref->getDocsWithPhrase(@phrase);
print "freq(\"result in\") = $freq total in $result docs\n";
print "Documents with \"result in\"\n";
foreach my $url (keys %$positionsByUrlsRef)  {
    my $url_freq = scalar keys %{$positionsByUrlsRef->{$url}};
    print "  $url:\n";
    print "      freq      = $url_freq\n";
    print "      positions = " . join(" ", reverse sort keys                    \
%{$positionsByUrlsRef->{$url}}) . "\n";
}
print "\n";

print "The following should be identical to the previous:\n";
foreach my $url (keys %$positionsByUrlsRef) {
    my ($url_freq, $url_positions_ref) =                                        \
$tfref->getPhraseFreqInDocument(\@phrase, url => $url);
    print "  $url:\n";
    print "      freq      = $url_freq\n";
    print "      positions = " . join(" ", reverse sort keys                    \
%$url_positions_ref) . "\n";
}
print "\n\n";


# -------------------------------------------------------------------
#  Then we do 1-4 with the phrase "resulting in"
#  And also print out the number of times "resulting in" is used in each
#  document
#  Because of stemming, the results this time should be the
#  same as those from last time (see directly above)
# -------------------------------------------------------------------

@phrase = ("resulting", "in");
$result = $tfref->getNumDocsWithPhrase(@phrase);
$freq   = $tfref->getPhraseFreq(@phrase);
$positionsByUrlsRef = $tfref->getDocsWithPhrase(@phrase);
print "freq(\"result in\") = $freq total in $result docs\n";
print "Documents with \"resulting in\" (should be the same as for \"result      \
in\")\n";
foreach my $url (keys %$positionsByUrlsRef)  {
    my $url_freq = scalar keys %{$positionsByUrlsRef->{$url}};
    print "  $url:\n";
    print "      freq      = $url_freq\n";
    print "      positions = " . join(" ", reverse sort keys                    \
%{$positionsByUrlsRef->{$url}}) . "\n";
}
print "\n";

print "The following should be identical to the previous:\n";
foreach my $url (keys %$positionsByUrlsRef) {
    my ($url_freq, $url_positions_ref) =                                        \
$tfref->getPhraseFreqInDocument(\@phrase, url => $url);
    print "  $url:\n";
    print "      freq      = $url_freq\n";
    print "      positions = " . join(" ", reverse sort keys                    \
%$url_positions_ref) . "\n";
}
print "\n";



\end{boxedverbatim}

\begin{boxedverbatim}

# -------------------------------------------------------------------
#  Here is how to use a TF for fuzzy OR queries.  We query the
#  (stemmed index of the) corpus as follows:
#
#  1 first, we get the number and scores of documents in the corpus
#    matching a query over the negated term !"thisisnotaword" (# = N),
#    then try the same query formulated as a negated phrase
#  2 then, we get the number and scores of documents in the corpus
#    matching a query over the term "result" (# = A),
#    then try the same query formulated as a phrase
#  3 then, we get the number and scores of documents in the corpus
#    matching a query over the term "in" (# = B)
#  4 then, we get the number and scores of documents in the corpus
#    matching a query over terms "result", "in" (# = C <= A + B)
#  5 then, we get the number and scores of documents in the corpus
#    matching the phrase query "result in" (# = D <= A, B)
#  6 then, we get the number and scores of documents in the corpus
#    matching a query over the negated phrase !"result in" (# = E = N - D)
#  7 finally, we get the number and scores of documents in the corpus
#    matching a query over the phrases "due to", "according to"
# -------------------------------------------------------------------

print "\n---Fuzzy OR Queries (stemmed):---\n";
#1a
    print "Query 1a: !\"thisisnotaword\" (negated term query)\n";
    my ($pTerms, $pNegTerms, $pPhrasePtrs, $pNegPhrasePtrs) = ([],              \
["thisisnotaword"], [], []);
    my %docScores = $tfref->getDocsMatchingFuzzyORQuery($pTerms, $pNegTerms,    \
$pPhrasePtrs, $pNegPhrasePtrs);
    my $N = scalar keys %docScores;
    my @scores = sort {$b <=> $a} values %docScores;
    print "    # docs matching: N = $N\n";
    print "             scores: " . join(" ", @scores) . "\n";
#1b
    print "Query 1b: !\"thisisnotaword\" (negated phrase query)\n";
    ($pTerms, $pNegTerms, $pPhrasePtrs, $pNegPhrasePtrs) = ([], [], [],         \
[["thisisnotaword"]]);
    %docScores = $tfref->getDocsMatchingFuzzyORQuery($pTerms, $pNegTerms,       \
$pPhrasePtrs, $pNegPhrasePtrs);
    $N = scalar keys %docScores;
    @scores = sort {$b <=> $a} values %docScores;
    print "    # docs matching: N = $N\n";
    print "             scores: " . join(" ", @scores) . "\n\n";


#2a
    print "Query 2a: \"result\" (term query)\n";
    ($pTerms, $pNegTerms, $pPhrasePtrs, $pNegPhrasePtrs) = (["result"], [], [], \
[]);
    %docScores = $tfref->getDocsMatchingFuzzyORQuery($pTerms, $pNegTerms,       \
$pPhrasePtrs, $pNegPhrasePtrs);
    my $A = scalar keys %docScores;
    @scores = sort {$b <=> $a} values %docScores;
    print "    # docs matching: A = $A\n";
    print "             scores: " . join(" ", @scores) . "\n";
#2b
    print "Query 2b: \"result\" (phrase query)\n";
    ($pTerms, $pNegTerms, $pPhrasePtrs, $pNegPhrasePtrs) = ([], [],             \
[["result"]], []);
    %docScores = $tfref->getDocsMatchingFuzzyORQuery($pTerms, $pNegTerms,       \
$pPhrasePtrs, $pNegPhrasePtrs);
    $A = scalar keys %docScores;
    @scores = sort {$b <=> $a} values %docScores;
    print "    # docs matching: A = $A\n";
    print "             scores: " . join(" ", @scores) . "\n\n";
#3
    print "Query 3: \"in\"\n";

\end{boxedverbatim}

\begin{boxedverbatim}

    ($pTerms, $pNegTerms, $pPhrasePtrs, $pNegPhrasePtrs) = (["in"], [], [],     \
[]);
    %docScores = $tfref->getDocsMatchingFuzzyORQuery($pTerms, $pNegTerms,       \
$pPhrasePtrs, $pNegPhrasePtrs);
    my $B = scalar keys %docScores;
    @scores = sort {$b <=> $a} values %docScores;
    print "    # docs matching: B = $B\n";
    print "             scores: " . join(" ", @scores) . "\n\n";
#4
    print "Query 4: \"result\", \"in\"\n";
    ($pTerms, $pNegTerms, $pPhrasePtrs, $pNegPhrasePtrs) = (["in"], [], [],     \
[]);
    %docScores = $tfref->getDocsMatchingFuzzyORQuery($pTerms, $pNegTerms,       \
$pPhrasePtrs, $pNegPhrasePtrs);
    my $C = scalar keys %docScores;
    @scores = sort {$b <=> $a} values %docScores;
    print "    # docs matching: C = $C <= A + B = " . ($A + $B) . "\n";
    print "             scores: " . join(" ", @scores) . "\n\n";
#5
    print "Query 5: \"result in\"\n";
    ($pTerms, $pNegTerms, $pPhrasePtrs, $pNegPhrasePtrs) = ([], [], [["result", \
"in"]], []);
    %docScores = $tfref->getDocsMatchingFuzzyORQuery($pTerms, $pNegTerms,       \
$pPhrasePtrs, $pNegPhrasePtrs);
    my $D = scalar keys %docScores;
    @scores = sort {$b <=> $a} values %docScores;
    print "    # docs matching: D = $D <= min{A, B}\n";
    print "             scores: " . join(" ", @scores) . "\n\n";
#6
    print "Query 6: !\"result in\"\n";
    ($pTerms, $pNegTerms, $pPhrasePtrs, $pNegPhrasePtrs) = ([], [], [],         \
[["result", "in"]]);
    %docScores = $tfref->getDocsMatchingFuzzyORQuery($pTerms, $pNegTerms,       \
$pPhrasePtrs, $pNegPhrasePtrs);
    my $E = scalar keys %docScores;
    @scores = sort {$b <=> $a} values %docScores;
    print "    # docs matching: E = $E = N - D = " . ($N - $D) . "\n";
    print "             scores: " . join(" ", @scores) . "\n\n";
#7
    print "Query 7: \"due to\", \"according to\"\n";
    ($pTerms, $pNegTerms, $pPhrasePtrs, $pNegPhrasePtrs) = ([], [],             \
[["due","to"], ["according","to"]], []);
    %docScores = $tfref->getDocsMatchingFuzzyORQuery($pTerms, $pNegTerms,       \
$pPhrasePtrs, $pNegPhrasePtrs);
    my $F = scalar keys %docScores;
    @scores = sort {$b <=> $a} values %docScores;
    print "    # docs matching: F = $F\n";
    print "             scores: " . join(" ", @scores) . "\n\n";



print "\n---Cluster and Network creation:---\n";
# Create a cluster with the documents
my $c = new Clair::Cluster;

$c->load_documents("$gen_dir/download/testhtml/tangra.si.umich.edu/clair/testht \
ml/*", type => 'html');

print "Loaded ", $c->count_elements, " documents.\n";

$c->strip_all_documents;
$c->stem_all_documents;

print "I'm done stripping and stemming\n";


my $count = 0;

\end{boxedverbatim}

\begin{boxedverbatim}

my $c2 = new Clair::Cluster;
foreach my $doc (values %{ $c->documents} ) {
	$count++;

	if ($count <= 40) {
		$c2->insert($doc->get_id, $doc);
	}
}

my %cm = $c2->compute_cosine_matrix();
my %bin_cos = $c2->compute_binary_cosine(0.15);
my $network = $c2->create_network(cosine_matrix => \%bin_cos);

print "Number of documents in network: ", $network->num_documents, "\n";

print "Average diameter: ", $network->diameter(avg => 1), "\n";
print "Maximum diameter: ", $network->diameter(), "\n";

\end{boxedverbatim}

\subsubsection{mmr.pl}
\begin{boxedverbatim}

#!/usr/local/bin/perl

# script: test_mmr.pl
# functionality: Tests the lexrank reranker on a network  

use strict;
use warnings;
use FindBin;
use Clair::Cluster;
use Clair::Network;
use Clair::Network::Centrality::LexRank;
use Clair::Document;


my $input_dir = "$FindBin::Bin/input/mmr";
my $file = "$input_dir/file.txt";
my $bias_file = "$input_dir/bias.txt";
my $lambda = 0.5;

# Split the first document into sentences

open FILE, "< $file" or die "Couldn't open $file: $!";
my $text;
while (<FILE>) {
    $text .= $_;
}
close FILE;
my $document = Clair::Document->new(
    string => $text,
    id => "document",
    type => "text"
);
my @sents = $document->split_into_sentences();


# Split the second document into sentences

open FILE, "< $bias_file" or die "Couldn't open $bias_file: $!";
$text = "";
while (<FILE>) {
    $text .= $_;
}
close FILE;
my $bias_doc = Clair::Document->new(
    string => $text,
    id => "document",
    type => "text"
);
my @bias = $bias_doc->split_into_sentences();


# Make a cluster from the first document's sentences

my $cluster = Clair::Cluster->new();
my $i = 1;
for (@sents) {
    my $doc = Clair::Document->new(
        string => $_,
        type => "text",
        id => $i
    );
    $doc->stem();
    $cluster->insert($i, $doc);
    $i++;
}



\end{boxedverbatim}

\begin{boxedverbatim}

# Turn it into a matrix to run lexrank

my %matrix = $cluster->compute_cosine_matrix();
my $network = $cluster->create_network(
    cosine_matrix => \%matrix,
    include_zeros => 1
);
my $cent = Clair::Network::Centrality::LexRank->new($network);
$cent->compute_lexrank_from_bias_sents( bias_sents => \@bias );


# Run MMR reranker 

$network->mmr_rerank_lexrank($lambda);


# Print out the sentences, ordered by lexrank
my $graph = $network->{graph};
my @verts = $graph->vertices();

my %scores = ();
foreach my $vert (@verts) {
    $scores{$vert} = $graph->get_vertex_attribute($vert, "lexrank_value");
}

foreach my $vert (sort { $scores{$b} cmp $scores{$a} } keys %scores) {
    my $sent = $cluster->get($vert)->get_text();
    print "$sent ($scores{$vert})\n";
}


\end{boxedverbatim}

\subsubsection{networkstat.pl}
\begin{boxedverbatim}

#!/usr/local/bin/perl

# script: test_networkstat.pl
# functionality: Generates a network, then computes and displays a large
# functionality: number of network statistics 

use strict;
use warnings;
use DB_File;
use FindBin;

use Clair::Network;
use Clair::Network::Writer::Edgelist;
use Clair::Network::Writer::Pajek;
use Clair::Cluster;
use Clair::Document;

my $basedir = $FindBin::Bin;
my $input_dir = "input/networkstat";
my $output_dir = "produced/networkstat";

my $old_prefix = "a";
my $threshold = 0.20;

my $prefix = "$output_dir/$old_prefix";

print "prefix: $prefix\n";
print "threshold: $threshold\n";

# Create cluster
my %documents = ();
my $cluster = Clair::Cluster->new(documents => \%documents);
my @files = ();
my @doc_ids = ();

# Open txt file and read in each line, putting it into the cluster as
# a separate document
open (TXT, "<$input_dir/$old_prefix.txt") || die("Could not open                \
$input_dir/$old_prefix.txt.");

my $doc_count = 0;

while (<TXT>)
{
	$doc_count++;
	my $doc = Clair::Document->new(type => 'text', string => "$_",
	                                  id => "$doc_count");
	$cluster->insert($doc_count, $doc);

	print "$doc_count:\t$_\n";
}

my %cos = $cluster->compute_cosine_matrix(text_type => 'text');

## CREATE A.ALL.COS FILE
$cluster->write_cos("$prefix.all.cos", cosine_matrix => \%cos);

# Uncomment to display the cosine matrix:
# foreach my $i (1..$doc_count)
# {
# 	foreach my $j (1..$doc_count)
# 	{
# 		if ($j < $i)
# 		{
# 			print "$j $i $cos{$j}{$i}\n";
# 			print "$i $j $cos{$i}{$j}\n";
# 		}

\end{boxedverbatim}

\begin{boxedverbatim}

# 	}
# }


# Do binary cosine w/ cutoff of 0.15
my %bin_matrix = $cluster->compute_binary_cosine($threshold);


## CREATE A.15.COS FILE
$cluster->write_cos("$prefix$threshold.cos", cosine_matrix => \%bin_matrix,     \
write_zeros => 0);

# Create networks
my $network = $cluster->create_network(cosine_matrix => \%cos, include_zeros => \
1);
my $networkThreshold = $cluster->create_network(cosine_matrix => \%bin_matrix);

# Creating .links files
my $export = Clair::Network::Writer::Edgelist->new();
$export->write_network($network, "$prefix.all.links");
$export->write_network($networkThreshold, "$prefix.links");
$network->write_nodes("$prefix.nodes");
$export->write_network($networkThreshold, "$prefix.linksuniq",
                       skip_duplicates => 1);

### check if the stats file exists
if (not -e "$prefix.stats") {
	print STDERR "creating the .stats file\n";
	`echo statistic: [date] value > $prefix.stats`;
}

my $n1 = $network->num_documents;
my $n2 = $networkThreshold->num_documents;
print_stat("documents", "$n1 vs. $n2");

$n1 = $network->num_pairs;
$n2 = $networkThreshold->num_pairs;
print_stat("pairs", "$n1 vs. $n2");
display_stat("documents");
display_stat("pairs");

my $ext_links = $networkThreshold->num_links(external => 1);
my $int_links = $networkThreshold->num_links(internal => 1);
my $int_links_nm = $networkThreshold->num_links(internal => 1, unique => 1);

print_stat("Number of external links (includes links with multiplicities)",     \
$ext_links);
display_stat("Number of external links");

print_stat("Number of internal links (includes links with multiplicities)",     \
$int_links);
display_stat("Number of internal links (includes links with multiplicities)");

if ($ext_links != 0) {
	print_stat("Ratio of internal to external links", $ext_links/$int_links);
	display_stat("Ratio of internal to external links");
}

print_stat("Number of internal links (no multiplicities allowed)",              \
$int_links_nm);
display_stat("Number of internal links (no multiplicities allowed)");

$networkThreshold->write_db("$prefix.db");
print "PRINTING DB\n";
$networkThreshold->print_db("$prefix.db");
$networkThreshold->write_db("$prefix-xpose.db", transpose => 1);
print "PRINTING TRANSPOSED DB\n";

\end{boxedverbatim}

\begin{boxedverbatim}

$networkThreshold->print_db("$prefix-xpose.db");
$networkThreshold->find_scc("$prefix.db", "$prefix-xpose.db",                   \
"$prefix-scc-db.fin", verbose => 1);
$networkThreshold->get_scc("$prefix-scc-db.fin", "$prefix.link_map",            \
"$prefix.scc");
$export->write_network($networkThreshold, "$prefix-xpose.link",
                       transpose => 1);

print_stat("Average in-degree", "average degree " .                             \
$networkThreshold->avg_in_degree);
display_stat("Average in-degree");
my %in_hist = $networkThreshold->compute_in_link_histogram();
$networkThreshold->write_link_matlab(\%in_hist, $prefix . "_in.m",              \
"$old_prefix-in");
$networkThreshold->write_link_dist(\%in_hist, "$prefix-inLinks");

print_stat("Average out-degree", "average degree " .                            \
$networkThreshold->avg_out_degree);
display_stat("Average out-degree");
my %out_hist = $networkThreshold->compute_out_link_histogram();
$networkThreshold->write_link_matlab(\%out_hist, $prefix . "_out.m",            \
"$old_prefix-out");
$networkThreshold->write_link_dist(\%out_hist, "$prefix-outLinks");

print_stat("Average total-degree", "average degree " .                          \
$networkThreshold->avg_total_degree);
display_stat("Average total-degree");
my %tot_hist = $networkThreshold->compute_total_link_histogram();
$networkThreshold->write_link_matlab(\%tot_hist, $prefix . "_total.m",          \
"$old_prefix-total");
$networkThreshold->write_link_dist(\%tot_hist, "$prefix-totalLinks");

print_stat("Power Law, out-link distribution",                                  \
$networkThreshold->power_law_out_link_distribution);
display_stat("Power Law, out-link distribution");

print_stat("Power Law, in-link distribution",                                   \
$networkThreshold->power_law_in_link_distribution);
display_stat("Power Law, in-link distribution");

print_stat("Power Law, total-link distribution",                                \
$networkThreshold->power_law_total_link_distribution);
display_stat("Power Law, total-link distribution");

my $wscc = $networkThreshold->Watts_Strogatz_clus_coeff(filename =>             \
"$prefix.cc.out");
print_stat("Watts-Strogatz clustering coefficient", $wscc);
display_stat("Watts-Strogatz clustering coefficient");

my $newman_cc = $networkThreshold->newman_clustering_coefficient();
print_stat("Newman clustering coefficient", $newman_cc);
display_stat("Newman clustering coefficient");

my @triangles = $networkThreshold->get_triangles();
print_stat("Network triangles", @triangles);
display_stat("Network triangles");

my $spl = $networkThreshold->get_shortest_path_length("1", "12");
print_stat("Shortest path between node 1 and node 12", $spl);
display_stat("Shortest path between node 1 and node 12");

my %dist = $networkThreshold->get_shortest_paths_lengths("1");
print_stat("Shortest paths between node 1 and reachable nodes", %dist);
display_stat("Shortest paths between node 1 and reachable nodes");


print_stat("Average shortest path",

\end{boxedverbatim}

\begin{boxedverbatim}

	   $networkThreshold->average_shortest_path());
display_stat("Average shortest path");

print_stat("Average directed shortest path", $networkThreshold->diameter(avg => \
1, filename => "$prefix.asp.directed.out", directed => 1) );
display_stat("Average directed shortest path");

print_stat("Average undirected shortest path", $networkThreshold->diameter(avg  \
=> 1, filename => "$prefix.asp.undirected.out", undirected => 1) );
display_stat("Average undirected shortest path");

print_stat("Maximum directed shortest path", $networkThreshold->diameter(max => \
1, filename => "$prefix.diameter.out", directed => 1) );
display_stat("Maximum directed shortest path");

print_stat("Maximum undirected shortest path", $networkThreshold->diameter(max  \
=> 1, filename => "$prefix.diameter.out", undirected => 1) );
display_stat("Maximum undirected shortest path");

write_to_stat("------ COSINE STATISTICS -----------\n");

my ($link_avg_cos, $nl_avg_cos) = $networkThreshold->average_cosines(\%cos);

print_stat("linked average cosine", $link_avg_cos);
display_stat("linked average cosine");

print_stat("not linked average cosine", $nl_avg_cos);
display_stat("not linked average cosine");

my ($link_hist, $nolink_hist) = $networkThreshold->cosine_histograms(\%cos);
$networkThreshold->write_histogram_matlab($link_hist, $nolink_hist, $prefix,    \
$prefix);
my $hist_string = $networkThreshold->get_histogram_as_string($link_hist,        \
$nolink_hist);
write_to_stat($hist_string);
print $hist_string;

print "$prefix\n";

$networkThreshold->create_cosine_dat_files($old_prefix, \%cos, directory =>     \
"produced/networkstat");

print "2\n";

my $dat_stats = $networkThreshold->get_dat_stats("$prefix", "$prefix.links",    \
"$prefix.all.cos");

#produced/networkstat/a/produced/networkstat/a-point-one-all.dat

print "3\n";

write_to_stat($dat_stats);
print $dat_stats;

print "4\n";

$export = Clair::Network::Writer::Pajek->new();
$export->set_name($prefix);
$export->write_network($networkThreshold, "$prefix.net");

#
# Statistics Methods
#
sub print_stat {
  my $name = shift;
	my $value = shift;
	my $date = `date`;

\end{boxedverbatim}

\begin{boxedverbatim}

	chomp($date);
	open (STATS, ">>$prefix.stats");
	print STATS $name,": [$date] $value\n";
	close STATS;
}

sub write_to_stat {
	my $text = shift;
	open (STATS, ">>$prefix.stats");
	print STATS $text;
	close STATS;
}

sub get_stat {
	my $name = shift;
	my $line = `grep "^$name" $prefix.stats`;
	chomp($line);
	my @columns = split (" ", $line);
	return $columns[$#columns];
}

sub display_stat {
	my $name = shift;
	print `grep "^$name" $prefix.stats`;
}

sub not_exists_stat {
	my $name = shift;
	my $st = `grep "^$name" $prefix.stats`;
	return ($st =~ /^\s*$/);
}


\end{boxedverbatim}

\subsubsection{pagerank.pl}
\begin{boxedverbatim}

#!/usr/local/bin/perl

# script: test_pagerank.pl
# functionality: Creates a small cluster and runs pagerank, displaying
# functionality: the pagerank distribution

use strict;
use warnings;
use FindBin;
use Clair::Network;
use Clair::Network::Centrality::PageRank;
use Clair::Cluster;
use Clair::Document;

my $basedir = $FindBin::Bin;
my $input_dir = "$basedir/input/pagerank";

my $c = new Clair::Cluster();
my $doc1 = new Clair::Document(id => 1, type => 'text', string => 'This is      \
document 1');
my $doc2 = new Clair::Document(id => 2, type => 'text', string => 'This is      \
document 2');
my $doc3 = new Clair::Document(id => 3, type => 'text', string => 'This is      \
document 3');
my $doc4 = new Clair::Document(id => 4, type => 'text', string => 'This is      \
document 4');
$c->insert(1, $doc1);
$c->insert(2, $doc2);
$c->insert(3, $doc3);
$c->insert(4, $doc4);

my $n = $c->create_hyperlink_network_from_file("$input_dir/link.txt");

my $cent = Clair::Network::Centrality::PageRank->new($n);
$cent->centrality();

print "NODE PAGERANK\n";
$cent->print_current_distribution();
print "\n";


\end{boxedverbatim}

\subsubsection{query.pl}
\begin{boxedverbatim}

#!/usr/local/bin/perl

# script: query.pl
# functionality: Requires indexes to be built via index_*.pl scripts, shows
# functionality: queries implemented in Clair::Info::Query, single-word and
# functionality: phrase queries, meta-data retrieval methods

use strict;
use FindBin;
# use lib "$FindBin::Bin/../lib";
# use lib "$FindBin::Bin/lib"; # in case you are outside the current dir
use vars qw/$DEBUG/;

use Benchmark;
use Clair::Index;
use Clair::Info::Query;
use Data::Dumper;
use POSIX;


$DEBUG = 0;
my %args;
# my @indexes = qw/word_idx document_idx document_meta_idx/;


$DEBUG = 0;
my $index_root = "$FindBin::Bin/produced/index_mldbm",
my $index_root_dirfiles = "$FindBin::Bin/produced/index_dirfiles",
my $stop_word_list = "$FindBin::Bin/input/index/stopwords.txt";


my $t0;
my $t1;

#
# Initializing index
#
$t0 = new Benchmark;

  # instantiate the index object first.
  my $idx = new Clair::Index(DEBUG => $DEBUG, index_root => $index_root);

  $idx->debugmsg("pre-loading necessary meta indexes.. please wait", 0);

	# and then pass the index object into the query constructor.
  my $q = new Clair::Info::Query(DEBUG => $DEBUG, index_object => $idx, ,       \
stop_word_list => $stop_word_list);

$t1 = new Benchmark;
my $timediff_init = timestr(timediff($t1, $t0));
$idx->debugmsg("index initialization took : " . $timediff_init, 0);


# test some queries
my $output;

$idx->debugmsg("processing query: 'king'", 0);
$output = $q->process_query("king");
print Dumper($output);

$idx->debugmsg('processing query: "julius caesar"', 0);
$output = $q->process_query('"julius caesar"');
print Dumper($output);

$idx->debugmsg('document frequency for: "caesar"', 0);
$output = $q->document_frequency("caesar");
print Dumper($output);

\end{boxedverbatim}

\begin{boxedverbatim}


$idx->debugmsg('term frequency for: "caesar" in doc 76', 0);
$output = $q->term_frequency("76 caesar");
print Dumper($output);

$idx->debugmsg('document_title for doc_id: 37', 0);
$output = $q->document_title("37");
print Dumper($output);

$idx->debugmsg('document_content for doc_id: 37', 0);
$output = $q->document_content("73", 0);
print Dumper($output);


# these results only show up after the incrental index update
$idx->debugmsg("processing query: 'romeo'", 0);
$output = $q->process_query("romeo");
print Dumper($output);

$idx->debugmsg("processing query: 'romeo juliet'", 0);
$output = $q->process_query('"romeo juliet"');
print Dumper($output);



$idx->debugmsg("USING dirfiles formatted index", 0);

undef $idx;
undef $q;

  # instantiate the index object first.
  $idx = new Clair::Index(DEBUG => $DEBUG, index_root => $index_root_dirfiles,  \
index_file_format => "dirfiles"); # NOTE index_file_format param

  $idx->debugmsg("pre-loading necessary meta indexes.. please wait", 0);

	# and then pass the index object into the query constructor.
  $q = new Clair::Info::Query(DEBUG => $DEBUG, index_object => $idx, ,          \
stop_word_list => $stop_word_list);

# test some queries
my $output;

$idx->debugmsg("dirfiles processing query: 'king'", 0);
$output = $q->process_query("king");
print Dumper($output);

$idx->debugmsg('dirfiles processing query: "julius caesar"', 0);
$output = $q->process_query('"julius caesar"');
print Dumper($output);

$idx->debugmsg('dirfiles document frequency for: "caesar"', 0);
$output = $q->document_frequency("caesar");
print Dumper($output);

\end{boxedverbatim}

\subsubsection{random\_walk.pl}
\begin{boxedverbatim}

#!/usr/local/bin/perl

# script: test_random_walk.pl
# functionality: Creates a network, assigns initial probabilities and tests
# functionality: taking single steps and calculating stationary distribution 

use strict;
use warnings;
use FindBin;
use Clair::Network;

my $basedir = $FindBin::Bin;
my $input_dir = "$basedir/input/random_walk";
my $gen_dir = "$basedir/produced/random_walk";

my $n = new Clair::Network();

$n->add_node(1, text => 'Text for node 1');
$n->add_node(2, text => 'Text for node 2');
$n->add_node(3, text => 'More text');

$n->read_transition_probabilities_from_file("$input_dir/t.txt");
$n->read_initial_probability_distribution("$input_dir/i.txt");

print "READ PROBABILITIES\n";

$n->save_transition_probabilities_to_file("$gen_dir/trans_prob.txt");
$n->make_transitions_stochastic();
$n->save_transition_probabilities_to_file("$gen_dir/stoch_trans_prob.txt");
$n->save_current_probability_distribution("$gen_dir/init_prob.txt");

print "WROTE_PROBABILITES BACK\n";

$n->perform_next_random_walk_step();
$n->perform_next_random_walk_step();
$n->perform_next_random_walk_step();

print "PERFORMED RANDOM WALK STEPS\n";

$n->save_current_probability_distribution("$gen_dir/new_prob.txt");

$n->compute_stationary_distribution();

print "COMPUTED STATIONARY DISTRIBUTION\n";
$n->save_current_probability_distribution("$gen_dir/stat_dist.txt");
print "WROTE RESULTS BACK\n";

print "The computed stationary distribution is:\n";
$n->print_current_probability_distribution();
print "\n";


\end{boxedverbatim}

\subsubsection{read\_dirfiles.pl}
\begin{boxedverbatim}

#!/usr/local/bin/perl

# script: test_read_dirfiles.pl
# functionality: Requires index_*.pl scripts to have been run, shows how to
# functionality: access the document_index and the inverted_index, how to
# functionality: use common access API to retrieve information

use strict;
use FindBin;
use lib "$FindBin::Bin/../lib";
use lib "$FindBin::Bin/lib"; # if you are outside of bin path.. just in case
use vars qw/$DEBUG/;

use Benchmark;
use Clair::GenericDoc;
use Clair::Index;
use Data::Dumper;
use File::Find;
use Getopt::Long;
use Pod::Usage;


$DEBUG = 0;
my $index_root = "$FindBin::Bin/produced/index_dirfiles",
my $index_root_mldbm = "$FindBin::Bin/produced/index_mldbm",
my $stop_word_list = "$FindBin::Bin/input/index/stopwords.txt";


# instantiate the index object
my $idx = new Clair::Index(
	DEBUG => 1,
	stop_word_list => $stop_word_list,
	index_root => $index_root,
	index_file_format => "dirfiles",
);

$idx->debugmsg("trying to read the document, positional index hash from:        \
$index_root", 0);

my $hash = {};
my $count = 0;

$hash = $idx->index_read("dirfiles", "caesar");
$count = scalar keys %{$hash->{caesar}};
$idx->debugmsg("total of $count docs contain the word 'caesar'");

$hash = $idx->index_read("dirfiles", "king");
$count = scalar keys %{$hash->{king}};
$idx->debugmsg("total of $count docs contain the word 'king'");

$idx->{index_root} = $index_root_mldbm;
$hash = $idx->index_read("mldbm", "caesar");
$count = scalar keys %{$hash->{caesar}};
$idx->debugmsg("total of $count docs contain the word 'caesar' from mldbm");

$hash = $idx->index_read("mldbm", "king");
$count = scalar keys %{$hash->{caesar}};
$idx->debugmsg("total of $count docs contain the word 'king' from mldbm");

# or access the meta index by supplying the third parameter, with 2nd parameter
# as the meta index name.
$idx->{index_root} = $index_root;
my $dochash = $idx->index_read("dirfiles", "document_meta_index", 2); #         \
document id 2
print Dumper($dochash);

my $dochash2 = $idx->index_read("dirfiles", "document_index", 100); # document  \

\end{boxedverbatim}

\begin{boxedverbatim}

id 100
print Dumper($dochash2);

my $dochash3 = $idx->index_read("dirfiles", "document_index", "all"); # return  \
everything in document_index
$count = scalar keys %{$dochash3};
$idx->debugmsg("retrieved total of $count doc data from document_index");

my $dochash4 = $idx->index_read("dirfiles", "document_meta_index", "all"); #    \
return everything for document_meta_index
$count = scalar keys %{$dochash4};
$idx->debugmsg("retrieved total of $count doc meta data from                    \
document_meta_index");

\end{boxedverbatim}

\subsubsection{sampling.pl}
\begin{boxedverbatim}

#!/usr/local/bin/perl

# script: test_sampling.pl
# functionality: Exercises network sampling using RandomNode and ForestFire 

#!/usr/bin/perl

use strict;
use warnings;
use Clair::Network;
use Clair::Network::Sample::RandomNode;
use Clair::Network::Sample::ForestFire;

my $net = new Clair::Network();

$net->add_node("A");
$net->add_node("B");
$net->add_node("C");
$net->add_node("D");
$net->add_node("E");
$net->add_edge("A", "B");
$net->add_edge("A", "C");
$net->add_edge("A", "D");
$net->add_edge("B", "C");
$net->add_edge("B", "D");

my $sample = Clair::Network::Sample::RandomNode->new($net);

$sample->number_of_nodes(2);

print "Original network: ", $net->{graph}, "\n";
print "Sampling 2 nodes using random node selection\n";
my $new_net = $sample->sample();

print "New network: ", $new_net->{graph}, "\n";

my $fire = new Clair::Network::Sample::ForestFire($net);

print "Sampling 3 nodes using Forest Fire model\n";
$new_net = $fire->sample(3, 0.9);

print "New network: ", $new_net->{graph}, "\n";

\end{boxedverbatim}

\subsubsection{statistics.pl}
\begin{boxedverbatim}

#!/usr/local/bin/perl

# script: test_statistics.pl
# functionality: Tests linear regression and T test code 

use strict;
use warnings;

use Clair::Network;
use Clair::Statistics::Distributions::TDist;

my %hist = (1, 2, 2, 4, 3, 6, 4, 9, 5, 11, 6, 12, 7, 14, 8, 16, 9, 18,
            10, 20, 11, 22
           );
my %bee = (   101.6 => 37,
              240.4 => 39.7,
              180.9 => 40.5,
              390.2 => 42.6,
              360.3 => 42.0,
              120.8 => 39.1,
              180.5 => 40.2,
              330.7 => 37.8,
              395.4 => 43.1,
              194.1 => 40.2,
              135.2 => 38.8,
              210.0 => 41.9,
              240.6 => 39.0,
              145.7 => 39.0,
              168.3 => 38.1,
              192.8 => 40.2,
              305.2 => 43.1,
              378.0 => 39.9,
              165.9 => 39.6,
              303.1 => 40.8
            );



my $net = Clair::Network->new();

my ($coef, $r) = $net->linear_regression(\%bee);

my $n = scalar keys %bee;
my $r_squared = $r**2;

my $df = $n - 2;
my $sr = sqrt((1 - $r_squared) / $df);
my $t = $r / $sr;
my $tdist = Clair::Statistics::Distributions::TDist->new();
my $t_prob = $tdist->get_prob($df, $t) * 2;

print "t_prob: $t_prob\n";

if ($t_prob < 0.05) {
  print "Likely power law relationship (p < 0.05)\n";
}


\end{boxedverbatim}

\subsubsection{stem.pl}
\begin{boxedverbatim}

#!/usr/local/bin/perl

# script: test_stem.pl
# functionality: Tests the Clair::Utils::Stem stemmer

use strict;
use warnings;
use FindBin;
use Clair::Utils::Stem;

my $stemmer = new Clair::Utils::Stem;
my $file = "$FindBin::Bin/input/stem/1.txt";

open FILE, $file or die "Couldn't open $file: $!";
while (<FILE>) {
   chomp $_;
   {
      /^([^a-zA-Z]*)(.*)/ ;
      print $1;
      $_ = $2;
      unless ( /^([a-zA-Z]+)(.*)/ ) { last; }
      my $word = lc $1; # turn to lower case before calling:
      $_ = $2;
      $word = $stemmer->stem($word);
      print $word;
      redo;
   }
   print "\n";
}
close FILE;

\end{boxedverbatim}

\subsubsection{summary.pl}
\begin{boxedverbatim}

#!/usr/local/bin/perl

# script: test_summary.pl
# functionality: Test the cluster summarization ability using various features 

use strict;
use warnings;
use Clair::Document;
use Clair::Cluster;
use Clair::SentenceFeatures qw(:all);
use FindBin;

# Load some documents
my @docs = glob("$FindBin::Bin/input/summary/*");
my $cluster = Clair::Cluster->new();
$cluster->load_file_list_array(\@docs, type => "text", filename_id => 1);

# Create a list of features and assign them uniform weights
my %features = (
    'length' => \&length_feature,
    'position' => \&position_feature,
    'simwithfirst' => \&sim_with_first_feature,
    'centroid' => \&centroid_feature
);
my %weights = map { $_ => 1 } keys %features;

# Compute the features and scale them to [0,1]
$cluster->compute_sentence_features(%features);
$cluster->normalize_sentence_features(keys %features);

# Score the sentences using the weights
$cluster->score_sentences( weights => \%weights );

# Get a ten sentence summary
my @summary = $cluster->get_summary( size => 10 );

foreach my $sent (@summary) {
    my $features = $sent->{features};
    my $score = $sent->{score};
    $sent->{did} =~ /([^\/]+\.txt)/;
    my $did = $1;
    my $sno = $sent->{'index'} + 1;
    print "[$did,$sno,$score]\t$sent->{text}\n";
    foreach my $fname (keys %$features) {
        print "\t$fname $features->{$fname}\n";
    }
}

\end{boxedverbatim}

\subsubsection{wordcount\_dir.pl}
\begin{boxedverbatim}

#!/usr/local/bin/perl

# script: test_wordcount_dir.pl
# functionality: Counts the words in each file of a directory; outputs report 

use strict;
use warnings;
use Clair::Document;
use FindBin;

my $prefix = "$FindBin::Bin/input/wordcount_dir";

#Count words in every Document in a file and return max,min,avergage:
opendir(DIR, $prefix);
my @files = grep { /\.txt$/ } readdir(DIR);
closedir(DIR);

my $doc;

my $num_files = scalar @files;
die "No files in $prefix" if $num_files == 0;

my $file = shift @files;
$file = "$prefix/$file";
$doc = new Clair::Document(type=>'text',file=>$file);
my $max = $doc->count_words();
my $maxFile = $file;
my $min = $doc->count_words();
my $minFile = $file;
my $temp;
my $avg = 0;

foreach $file (@files) {
    $file = "$prefix/$file";
    next unless -f $file;
    $doc = new Clair::Document( type => 'text', file => $file );
    $temp = $doc->count_words();
    $avg = $avg + $temp;
    if($temp > $max){
        $max = $temp;
        $maxFile = $file;
    }
    if($temp < $min){
        $min = $temp;
        $minFile = $file;
    }
}

$avg = $avg / $num_files;
print "The minimum number of words is $min words in file $minFile\n";
print "The maximum number of words is $max words in file $maxFile\n";
print "The average number of words is $avg words\n";

\end{boxedverbatim}

\subsubsection{wordcount.pl}
\begin{boxedverbatim}

#!/usr/local/bin/perl

# script: test_wordcount.pl
# functionality: Using Cluster and Document, counts the words in each file
# functionality: of a directory 

use strict;
use warnings;
use Clair::Cluster;
use Clair::Document;
use FindBin;

my $input_dir = "$FindBin::Bin/input/wordcount";

my $cluster = Clair::Cluster->new();
$cluster->load_documents("$input_dir/*.txt", type => "text", filename_id => 1   \
);
my $docs = $cluster->documents();
print "did\t#words\n";
foreach my $did (keys %$docs) {
    my $doc = $docs->{$did};
    my $words = $doc->count_words();
    print "$did\t$words\n";
}

\end{boxedverbatim}

\subsubsection{xmldoc.pl}
\begin{boxedverbatim}

#!/usr/local/bin/perl

# script: test_xmldoc.pl
# functionality: Tests the XML to text function of Document 

use strict;
use warnings;
use Clair::Document;
use Clair::Cluster;
use FindBin;

my $doc = Clair::Document->new(
			    file => "$FindBin::Bin/input/xmldoc/dow-clean.xml",
			    type => "xml");

$doc->xml_to_text();
my $text = $doc->get_text();
print "Text:\n$text\n";

my @sents = $doc->get_sent();
print "Sentences:\n";
my $i = 1;
for (@sents) {
    print "$i $_\n";
    $i++;
}

\end{boxedverbatim}

\subsubsection{classify\_weka.pl}
\begin{boxedverbatim}

#!/usr/local/bin/perl

# script: test_classify_weka.pl
# functionality: Extracts bag-of-words features from each document
# functionality: in a training corpus of baseball and hockey documents,
# functionality: then trains and evaluates a Weka decision tree classifier,
# functionality: saving its output to files

use strict;
use warnings;
use FindBin;
use lib "$FindBin::Bin/../lib";

use Clair::Document;
use Clair::Cluster;
use Clair::Interface::Weka;;

my $basedir   = $FindBin::Bin;
my $input_dir = "$basedir/input/classify";
my $gen_dir   = "$basedir/produced/classify";



# ---FEATURE EXTRACTION PHASE---
	print "\n---FEATURE EXTRACTION PHASE---";

	# Extract features for training, then for testing
	for my $round (("train", "test")) {
		# Create a cluster
		my $c = new Clair::Cluster;
		$c->set_id("sports");

		# Read every document from the the the training or test directory and insert  \
it into the cluster
		# Convert from HTML to text, then stem as we do so
		while ( <$input_dir/$round/*> ) {
			my $file = $_;

			my $doc = new Clair::Document(type => 'html', file => $file, id => $file);
			$doc->set_class(extract_class($doc->get_html(), $file));  # Set the          \
document's class label
			$doc->strip_html;
			$doc->stem;

			$c->insert($file, $doc);
		}
		# Compute the bag-of-words feature (which actually constitutes a vector) for  \
each document in the cluster
		$c->compute_document_feature(name => "vect", feature =>                       \
\&compute_bag_of_words_vect);

		# Get the number of documents belonging to each class occurring in the        \
cluster
		my %classes = $c->classes();
		print "\nExtracting ", $c->count_elements(), " documents to $round:\n";
		print "    " . $classes{'baseball'} . " baseball documents\n";
		print "    " . $classes{'hockey'} . " hockey documents\n";

		# Write features to ARFF, prepending the specified header
		my $header = "%1. Title: Baseball / Hockey Corpus Dataset ($round)\n" .
					 "%2. Source: 20_newsgroups Corpus\n" .
					 "%      (a) Creator: Ken Lang\n" .
					 "%\n";
		write_ARFF($c, "$gen_dir/$round.arff", $header);
		print "Features written to $gen_dir/$round.arff\n";
	}


\end{boxedverbatim}

\begin{boxedverbatim}


# ---TRAINING PHASE---
	print "\n---TRAINING PHASE---\n";

	# Train a J48 decision tree classifier using 10-fold cross-validation
	print "Training J48 decision tree classifier with 10-fold                      \
cross-validation...\n";
	train_classifier(classifier => 'weka.classifiers.trees.J48',
					 trainfile  => "$gen_dir/train.arff",
					 modelfile  => "$gen_dir/J48.model",
					 logfile    => "$gen_dir/train-10fold-J48.log");
	print "    See $gen_dir/train-crossval-J48.log for log of classifier output    \
from training and 10-fold cross-validation\n";

	# Train a J48 decision tree classifier using cross-validation on the test set
	print "Training J48 decision tree classifier with cross-validation on test     \
set...\n";
	my ($train_10fold_log, $train_test_log) = train_classifier(classifier =>       \
'weka.classifiers.trees.J48',
															   trainfile  => "$gen_dir/train.arff",
															   modelfile  => "$gen_dir/J48.model",
															   testfile   => "$gen_dir/test.arff",
															   logfile    => "$gen_dir/train-test-J48.log");
	print "    See $gen_dir/train-crossval-J48.log for log of classifier output    \
from training and cross-validation on $gen_dir/test.arff\n";


# ---TESTING PHASE---
	print "\n---TESTING PHASE---\n";
	print "Testing classifier predictions...\n";
	# Test the classifier directly on the test set, outputting predictions for     \
individual documents
	my $test_log = test_classifier(classifier => 'weka.classifiers.trees.J48',
								   modelfile => "$gen_dir/J48.model",
								   testfile  => "$gen_dir/test.arff",
								   predfile  => "$gen_dir/test-J48.pred",
								   logfile   => "$gen_dir/test-J48.log");
	print "    See $gen_dir/test-J48.log for log of classifier output from         \
testing\n";
	print "    See $gen_dir/test-J48.pred for log of classifier predictions from   \
testing\n";


# ---DONE---
	print "\nHave a nice day!\n";



# ---AUXILIARY PROCEDURES---

# Extract a document's class
sub extract_class {
    my $html = shift;
    my $file = shift;

    my $label = $1 if ($html =~ m/<DOC GROUP="rec\.sport\.(\w+?)">/);
    die "extract_class - Class label not found in $file" if not defined $label;
    return $label;
}


# Compute the bag-of words feature from 10 pre-selected features (these         \
features were culled earlier
# from the entire set of stemmed terms occurring in the corpus using chi square \
feature selection)
sub compute_bag_of_words_vect {
	my %params = @_;

\end{boxedverbatim}

\begin{boxedverbatim}

	my $docref = $params{document};

	my %tf = $docref->tf(type => "stem");
	my %vect;
	$vect{'hockei'} = $tf{'hockei'}   || 0;
	$vect{'nhl'} = $tf{'nhl'}         || 0;
	$vect{'playoff'} = $tf{'playoff'} || 0;
	$vect{'pitch'} = $tf{'pitch'}     || 0;
	$vect{'basebal'} = $tf{'basebal'} || 0;
	$vect{'goal'} = $tf{'goal'}       || 0;
	$vect{'cup'} = $tf{'cup'}         || 0;
	$vect{'ca'} = $tf{'ca'}           || 0;
	$vect{'bat'} = $tf{'bat'}         || 0;
	$vect{'pitcher'} = $tf{'pitcher'} || 0;

	return \%vect;
}

\end{boxedverbatim}

\subsubsection{lsi.pl}
\begin{boxedverbatim}

#!/usr/local/bin/perl

# script: test_lsi.pl
# functionality: Constructs a latent semantic index from a corpus of
# functionality: baseball and hockey documents, then uses that index
# functionality: to map terms, queries, and documents to latent semantic
# functionality: space. The position vectors of documents in that space
# functionality: are then used to train and evaluate a SVM classifier
# functionality: using the Weka interface provided in Clair::Interface::Weka

use strict;
use warnings;
use FindBin;
use lib "$FindBin::Bin/../lib";

use Clair::Algorithm::LSI;
use Clair::Document;
use Clair::Cluster;
use Clair::Interface::Weka;

use vars qw(@ISA @EXPORT);


my $basedir   = $FindBin::Bin;
my $input_dir = "$basedir/input/lsi";
my $gen_dir   = "$basedir/produced/lsi";


my $index;
# Extract features for training, then for testing
for my $round (("train", "test")) {
	if ($round eq "train") {
		print "\n---LSI TRAINING ROUND---";
	} elsif ($round eq "test") {
		print "\n---LSI TEST ROUND---";
	}
	# Create a cluster
	my $c = new Clair::Cluster;
	$c->set_id("sports");

	# Read every document from the the the training or test directory and insert   \
it into the cluster
	# Convert from HTML to text, then stem as we do so
	while ( <$input_dir/$round/*> ) {
		my $file = $_;

		my $doc = new Clair::Document(file => $file, type => 'html', id => $file);
		$doc->set_class(extract_class($doc->get_html(), $file));  # Set the           \
document's class label
		$doc->strip_html;
		$doc->stem;

		$c->insert($file, $doc);
	}

	# Get the number of documents belonging to each class occurring in the cluster
	my %classes = $c->classes();
	print "\nExtracting ", $c->count_elements(), " documents ($round):\n";
	print "    " . $classes{'baseball'} . " baseball documents\n";
	print "    " . $classes{'hockey'} . " hockey documents\n";

	if ($round eq "train") {
		# On training round, construct document-term matrix and compute SVD           \
(computationally extremely intensive)
		print "\nConstructing document-term matrix and computing its singular value   \
decomposition...\n";
		$index = new Clair::Algorithm::LSI(cluster => $c, type => "stem");

\end{boxedverbatim}

\begin{boxedverbatim}

		$index->build_index();
		print "  Done.\n";
	} elsif ($round eq "test") {
		# On test round, road the previously saved index
		print "\nLoading latent semantic index from file $gen_dir/sports.lsi...\n";
		$index = new Clair::Algorithm::LSI(file => "$gen_dir/sports.lsi",
										   cluster => $c);
		print "  Done.\n";
	}

	# For each document in the cluster, compute the position vector of the         \
document in latent space
	# using the singular value decomposition of the document-term matrix
	$c->compute_document_feature(name => "latent_coord",
								 feature => \&compute_latent_space_position_vect);
	# Write this feature (actually vector of features) to ARFF, prepending the     \
specified header
	my $header = "%1. Title: Baseball / Hockey Corpus Dataset ($round)\n" .
				 "%2. Source: 20_newsgroups Corpus\n" .
				 "%      (a) Creator: Ken Lang\n" .
				 "%\n";
	write_ARFF($c, "$gen_dir/$round.arff", $header);
	print "Features written to $gen_dir/$round.arff\n";

	if ($round eq "train") {
		# Train a support vector machine (SVM) using 10-fold cross-validation
		print "Training support vector machine (SVM) with 10-fold                     \
cross-validation...\n";
		train_classifier(classifier => 'weka.classifiers.functions.SMO',
						 trainfile  => "$gen_dir/train.arff",
						 modelfile  => "$gen_dir/SMO.model",
						 logfile    => "$gen_dir/train-10fold-SMO.log");
		print "  Done.\n";
		print "    See $gen_dir/train-crossval-SMO.log for log of classifier output   \
from training and 10-fold cross-validation\n";

		# Perform various operations on the LSI to illustrate the functionality it    \
provides
		print "\nAssorted LSI Operations:\n";
		my @docids = sort keys %{$c->documents()};
		my $firstdoc = $c->documents()->{$docids[0]};
		# Find documents similar near in latent semantic space to the (arbitrarily)   \
first document in the corpus
		print "\n1.  10 documents most similar to the first \"" .                     \
$firstdoc->get_class() . "\" document:\n";
		my %doc_dists = $index->rank_docs($firstdoc);
		@docids = sort {$doc_dists{$a} <=> $doc_dists{$b} } keys %doc_dists;
		for (my $i=0; $i < 10; $i++) {
			my $class = $c->get($docids[$i])->get_class();
			print "      $docids[$i]\tclass: $class\tdistance:                           \
$doc_dists{$docids[$i]}\n";
		}
		# Find documents far away from that document
		print "\n    10 documents least similar to the first \"" .                    \
$firstdoc->get_class() . "\" document:\n";
		@docids = sort {$doc_dists{$b} <=> $doc_dists{$a} } keys %doc_dists;
		for (my $i=0; $i < 10; $i++) {
			my $class = $c->get($docids[$i])->get_class();
			print "      $docids[$i]\tclass: $class\tdistance:                           \
$doc_dists{$docids[$i]}\n";
		}
		# Find terms near in latent semantic space to the term "hockey"
		print "\n2.  20 terms contextually most related to \"hockey\":\n";
		my %term_dists = $index->rank_terms("hockey");
		my @terms = sort {$term_dists{$a} <=> $term_dists{$b} } keys %term_dists;
		for (my $i=0; $i < 20; $i++) {
			print "        $terms[$i]      \tdistance: $term_dists{$terms[$i]}\n";

\end{boxedverbatim}

\begin{boxedverbatim}

		}
		# Find terms near in latent semantic space to the term "playoff" (which       \
denotes baseball)
		print "\n2.  20 terms contextually most related to \"playoff\":\n";
		%term_dists = $index->rank_terms("playoff");
		@terms = sort {$term_dists{$a} <=> $term_dists{$b} } keys %term_dists;
		for (my $i=0; $i < 20; $i++) {
			print "        $terms[$i]  \tdistance: $term_dists{$terms[$i]}\n";
		}

		# Order the following queries first by nearness in semantic space to the term \
"hockey",
		# then by nearness in semantic space to the term "playoff"
		my @queries = ("goalie stops puck",
					   "pitcher throws to catcher",
					   "extra innings",
					   "overtime");
		print "\n3.  Set of unordered queries:\n";
		print "        " . join("\n        ", @queries);
		print "\n    Ordered by contextual relationship to query \"hockey\":\n";
		my %query_dists = $index->rank_queries("hockey", @queries);
		my @ordered = sort {$query_dists{$a} <=> $query_dists{$b} } keys              \
%query_dists;
		foreach my $query (@ordered) {
			print "        \"$query\"\t\t\tdistance: $query_dists{$query}\n";
		}
		print "    Ordered by contextual relationship to query \"playoff\":\n";
		%query_dists = $index->rank_queries("playoff", @queries);
		@ordered = sort {$query_dists{$a} <=> $query_dists{$b} } keys %query_dists;
		foreach my $query (@ordered) {
			print "        \"$query\"\t\t\tdistance: $query_dists{$query}\n";
		}

		# Save latent semantic index to file
		print "\nSaving latent semantic index to file $gen_dir/sports.lsi...\n";
		$index->save_to_file("$gen_dir/sports.lsi", savecluster => 0);
		print "  Done.\n";
	}
	elsif ($round eq "test") {
		# Test the classifier directly on the test set, outputting predictions for    \
individual documents
		print "Testing SVM predictions...\n";
		my $test_log = test_classifier(classifier =>                                  \
'weka.classifiers.functions.SMO',
									   modelfile => "$gen_dir/SMO.model",
								       testfile  => "$gen_dir/test.arff",
								       predfile  => "$gen_dir/test-SMO.pred",
								       logfile   => "$gen_dir/test-SMO.log");
		print "    See $gen_dir/test-SMO.log for log of classifier output from        \
testing\n";
		print "    See $gen_dir/test-SMO.pred for log of classifier predictions from  \
testing\n";
	}
}

# Delete the latent semantic index from disk (the file is quite large)
unlink "$gen_dir/sports.lsi";

print "\nHave a nice day!\n";



# ---AUXILIARY PROCEDURES---

# Extract a document's class
sub extract_class {
    my $html = shift;

\end{boxedverbatim}

\begin{boxedverbatim}

    my $file = shift;

    my $label = $1 if ($html =~ m/<DOC GROUP="rec\.sport\.(\w+?)">/);
    die "extract_class - Class label not found in $file" if not defined $label;
    return $label;
}

# Compute a document's position in latent semantic space as that space is       \
defined by the singular value
# decomposition (and dimensionality reduction of that decomposition) of the     \
document-term matrix of the
# cluster
sub compute_latent_space_position_vect {
	my %params = @_;
	my $docref = $params{document};

	my $v = $index->doc_to_latent_space($docref);
	my @vect;
	foreach my $elem (list $v) {
		push @vect, $elem;
	}

	return \@vect;
}

\end{boxedverbatim}

\subsubsection{parse.pl}
\begin{boxedverbatim}

#!/usr/local/bin/perl

# script: test_parse.pl
# functionality: Parses an input file and then runs chunklink on it 

use strict;
use warnings;
use FindBin;
use Clair::Utils::Parse;

my $basedir = $FindBin::Bin;
my $input_dir = "$basedir/input/parse";
my $gen_dir = "$basedir/produced/parse";

# Preparing file for parsing
Clair::Utils::Parse::prepare_for_parse("$input_dir/test.txt",                   \
"$gen_dir/parse.txt");

print "PARSING\n";

my $parseout = Clair::Utils::Parse::parse("$gen_dir/parse.txt", output_file =>  \
"$gen_dir/parse_out.txt", options => '-l300');

my $chunkin = Clair::Utils::Parse::forcl("$gen_dir/parse_out.txt", output_file  \
=> "$gen_dir/WSJ_0000.MRG");

print "Now doing chunklink.\n";

my $chunkout = Clair::Utils::Parse::chunklink("$gen_dir/WSJ_0000.MRG",          \
output_file => "$gen_dir/chunk_out.txt", options => '-sph');


\end{boxedverbatim}

\subsection{Utilities}

This section contains different utility scripts that perform common tasks.

\subsubsection{chunk\_document.pl}
\begin{boxedverbatim}

#!/usr/local/bin/perl 
#
# script: chunk_document
#
# functionality: Breaks a text file into multiple files of a given word length
#

use strict;
use warnings;
use Getopt::Long;
use File::Spec;

sub usage;

my $in_file = "";
my $out_dir = "";
my $out_file = "";
my $word_limit = 500;
my $vol = "";
my $dir = "";
my $prefix = "";

my $res = GetOptions("input=s" => \$in_file,
											"output=s" => \$out_dir,
											"words=i" => \$word_limit);

# check for input 
if( $in_file eq "" ){
  usage();
  exit;
}

# check for output directory
if( $out_dir eq "" ){
	usage();
	exit;
} else {
  unless (-d $out_dir) {
    mkdir $out_dir or die "Couldn't create $out_dir: $!";
  }
}

# open infile
open(IN,$in_file) or die "Can't open $in_file: $!";

# get infile name
($vol, $dir, $prefix) = File::Spec->splitpath($in_file);

# read in infile, split into words and print words to outfile till you reach
# word_limit, then start new outfile

my @line = ();
my @bin = ();
my $dump = "";
my $count = 1;
my $word = "";

$out_file = $out_dir.'/'.$prefix.'.'.$word_limit;

while(<IN>){
	#split line into words and move into array
  my @line = split(/ /, $_);

  #add words to array until it's $word_limit long
	foreach $word (@line){
		if($#bin < $word_limit) {
			push (@bin, $word);

\end{boxedverbatim}

\begin{boxedverbatim}

    } else {
      $dump = join(' ', @bin);
			#print "writing: $out_file.$count\n";
			open(OUT, ">$out_file.$count") or die "Can't open $out_file: $!";      
			print OUT $dump;
			close OUT;
			@bin = ($word);
			$count++;
		}
	}
}

#get last words
$dump = join(' ', @bin);
#print "writing: $out_file.$count\n";
open(OUT, ">$out_file.$count") or die "Can't open $out_file: $!";
print OUT $dump;
close OUT;

#
# Print out usage message
#
sub usage
{
  print "usage: $0 --input input_file --output output_dir [--words              \
word_limit]\n\n";
  print "  --input input_file\n";
	print "     Name of the input file\n";
	print "  --output output_dir\n";
	print "     Name of the output directory.\n";
  print "  --words word_limit\n";
	print "     Number of words to include in each file.  Defaults to 500.\n";
	print "\n";
	print "example: $0 --input file.txt --output ./corpus --words 1000\n";
	exit;
}
 

\end{boxedverbatim}

\subsubsection{corpus\_to\_cos.pl}
\begin{boxedverbatim}

#!/usr/bin/perl
# script: corpus_to_cos.pl
# functionality: Calculates cosine similarity for a corpus

use strict;
use warnings;

use Getopt::Long;

use Clair::Cluster;
use Clair::IDF;

sub usage;

my $corpus_name = "";
my $basedir = "produced";
my $out_file = "";
my $sample_size = 0;
my $verbose = 0;
my $stem = 1;

my $res = GetOptions("corpus=s" => \$corpus_name, "base=s" => \$basedir,
		     "output:s" => \$out_file, "sample=i" => \$sample_size,
		     "stem!" => \$stem, "verbose!" => \$verbose);

if (!$res or ($corpus_name eq "") or ($basedir eq "")) {
  usage();
  exit;
}

my $gen_dir = "$basedir";

my $corpus_data_dir = "$gen_dir/corpus-data/$corpus_name";
my $linkfile = "$corpus_data_dir/$corpus_name.links";
my $doc_to_file = "$corpus_data_dir/" . $corpus_name . "-docid-to-file";
my $doc_to_url = "$corpus_data_dir/" . $corpus_name . "-docid-to-url";
my $compress_dbm = "$corpus_data_dir/" . $corpus_name . "-compress-docid";

my $idf_file = "";
if ($stem) {
  my $idf_file = "$corpus_data_dir/" . $corpus_name . "-idf-s";
} else {
  my $idf_file = "$corpus_data_dir/" . $corpus_name . "-idf";
}

if ($verbose) { print "Loading corpus into cluster\n"; }
my $cluster = new Clair::Cluster;
load_corpus($cluster, docid_to_file_dbm => $doc_to_file);

$cluster->strip_all_documents;
if ($stem) {
  $cluster->stem_all_documents;
}

open_nidf($idf_file);

my $text_type = "";
if ($stem) {
  $text_type = "stem";
} else {
  $text_type = "text";
}

my %cos_matrix = $cluster->compute_cosine_matrix(text_type => $text_type);

# default to corpus name + .cos if no output filename given
if ($out_file eq "") {

\end{boxedverbatim}

\begin{boxedverbatim}

  $out_file = $corpus_name . ".cos";
}

my ($vol, $dir, $file);
($vol, $dir, $file) = File::Spec->splitpath($out_file);
if ($dir ne "") {
  unless (-d $dir) {
    mkdir $dir or die "Couldn't create $dir: $!";
  }
}


$cluster->write_cos($out_file, cosine_matrix => \%cos_matrix);

#
# Load a corpus into a cluster
#
sub load_corpus {
  my $self = shift;

  my %parameters = @_;

  my $property = ( defined $parameters{property} ?
		   $parameters{propery} : 'pagerank_transition' );

  my $ignore_EX = ( defined $parameters{ignore_EX} ?
		    $parameters{ignore_EX} : 1 );

  my %docid_to_file = ();

  if (defined $parameters{docid_to_file_dbm}) {
    my $docid_to_file_dbm_file = $parameters{docid_to_file_dbm};
    dbmopen %docid_to_file, $docid_to_file_dbm_file, 0666 or
      die "Cannot open DBM: $docid_to_file_dbm_file\n";
  }

  my %id_hash = ();

  foreach my $id (keys %docid_to_file) {
    if (not exists $id_hash{$id}) {
      if ($id eq "EX") {
	$id_hash{$id} = $id;
      } else {
	my $filename = $docid_to_file{"$id"};
	my ($vol, $dir, $fn) = File::Spec->splitpath($filename);
	my $doc = Clair::Document->new(file => "$filename", id => "$fn",
				       type => 'html');
	$self->insert($doc->get_id, $doc);
	$id_hash{$id} = $doc;
      }
    }
  }

  return $self;
}


#
# Print out usage message
#
sub usage
{
  print "usage: $0 -c corpus_name -o out_file [-b base_dir]\n\n";
  print "  -c corpus_name\n";
  print "       Name of the corpus\n";
  print "  -b base_dir\n";
  print "       Base directory filename.  The corpus is loaded from here\n";

\end{boxedverbatim}

\begin{boxedverbatim}

  print "  -o out_file\n";
  print "       Name of file to write network to\n";
  print "  --sample sample_size\n";
  print "       Instead of computing cosines for the entire corpus, sample      \
sample_size documents uniformly from the document set\n";
  print "  --stem or --no-stem\n";
  print "       Use the stemmed or unstemmed version of the corpus to generate  \
the cosine files\n";

  print "\n";

  print "example: $0 -c bulgaria -o data/bulgaria.graph -b                      \
/data0/projects/lexnets/pipeline/produced\n";

  exit;
}


\end{boxedverbatim}

\subsubsection{corpus\_to\_cos-threaded.pl}
\begin{boxedverbatim}

#!/usr/bin/perl
# script: corpus_to_cos-threaded.pl
# functionality: Calculates cosine similarity using multiple threads
#

use strict;
use warnings;

use Getopt::Long;
use Clair::Cluster;
use MEAD::SimRoutines;
use Clair::IDF;
use threads;
use threads::shared;
use Thread::Queue;
use Storable qw(freeze thaw dclone);

select STDOUT; $| = 1;

sub usage;

my $corpus_name = "";
my $basedir = "produced";
my $output_file = "";
my $sample_size = 0;

my $res = GetOptions("corpus=s" => \$corpus_name, "base=s" => \$basedir,
		     "output:s" => \$output_file, "sample:i" => \$sample_size);

if (!$res or ($corpus_name eq "") or ($basedir eq "")) {
  usage();
  exit;
}

my $gen_dir = "$basedir";
my $verbose = 0;
my $documents : shared;

my $corpus_data_dir = "$gen_dir/corpus-data/$corpus_name";
my $linkfile = "$corpus_data_dir/$corpus_name.links";
my $doc_to_file = "$corpus_data_dir/" . $corpus_name . "-docid-to-file";
my $doc_to_url = "$corpus_data_dir/" . $corpus_name . "-docid-to-url";
my $compress_dbm = "$corpus_data_dir/" . $corpus_name . "-compress-docid";
my $idf_file = "$corpus_data_dir/" . $corpus_name . "-idf-s";

if ($verbose) { print "Loading corpus into cluster\n"; }
my $cluster = new Clair::Cluster;

print "Loading corpus\n";
load_corpus($cluster, $sample_size, docid_to_file_dbm => $doc_to_file);

$cluster->strip_all_documents;
$cluster->stem_all_documents;

my %documents = ();

print "Computing cosine matrix\n";

open_nidf($idf_file);

my %cos_matrix = compute_cosine_matrix($cluster, text_type => 'stem');

# default to corpus name + .cos if no output filename given
if ($output_file eq "") {
  $output_file = $corpus_name . ".cos";
}


\end{boxedverbatim}

\begin{boxedverbatim}

$cluster->write_cos($output_file, cosine_matrix => \%cos_matrix);

#
# Load a corpus into a cluster
#
sub load_corpus {
  my $self = shift;
  my $sample_size = shift;

  my %parameters = @_;

  my $property = ( defined $parameters{property} ?
		   $parameters{propery} : 'pagerank_transition' );

  my $ignore_EX = ( defined $parameters{ignore_EX} ?
		    $parameters{ignore_EX} : 1 );

  my %docid_to_file = ();

  if (defined $parameters{docid_to_file_dbm}) {
    my $docid_to_file_dbm_file = $parameters{docid_to_file_dbm};
    dbmopen %docid_to_file, $docid_to_file_dbm_file, 0666 or
      die "Cannot open DBM: $docid_to_file_dbm_file\n";
  }

  my %id_hash = ();
  my @id_array = ();
  my @sample_array = ();
  my %sample_hash = ();

  foreach my $id (keys %docid_to_file) {
    push @id_array, $id;
  }
  my $id_size = scalar(@id_array);

  if ($sample_size > 0) {
    srand;
    for (my $i = 0; $i < $sample_size; $i++) {
      push @sample_array, $id_array[int(rand($id_size))];
    }
  } else {
    @sample_array = @id_array;
  }

  print "Inserting ", scalar(@sample_array), " documents into cluster\n";
  foreach my $id (@sample_array) {
    if (not exists $id_hash{$id}) {
      if ($id eq "EX") {
	$id_hash{$id} = $id;
      } else {
	my $filename = $docid_to_file{"$id"};
	my $doc = Clair::Document->new(file => "$filename", id => "$id",
				       type => 'html');
	$self->insert($doc->get_id, $doc);
	$id_hash{$id} = $doc;
      }
    }
  }
  print "\n";

  return $self;
}


sub compute_cosine_matrix {
  my $self = shift;


\end{boxedverbatim}

\begin{boxedverbatim}

  my %parameters = @_;

  my $text_type = "stem";
  if (exists $parameters{text_type}) {
    $text_type = $parameters{text_type};
  }

  # deep copy to keep threads::shared happy
  print "Copying documents object\n";
  %documents = %{$self->{documents}};

  my $i = 0;
  my $j = 0;
  my %cos_hash : shared = ();
  my $global_count : shared = 0;

  # Create the document queue
  print "Creating queue\n";
  my $jobs = new Thread::Queue;

  print "Adding ", scalar(keys %documents), " documents to queue\n";
  my $sum = 0;
  foreach my $doc1_key (keys %documents) {
    $i = 0;
    $j++;

    # setup the shared variable
    # must create nested shared data structures by first creating shared
    # leaf nodes (threads::shared docs)
    $cos_hash{$doc1_key} = &share({});

    foreach my $doc2_key (keys %documents) {
      $i++;
      if ($i < $j) {
	my @obj = ($doc1_key, $doc2_key);
# 	$sum++;
# 	if (($sum % 1000) == 0) {
# 	  print $sum / 1000, "\n";
# 	}
	$jobs->enqueue(freeze(\@obj));
      }
    }
  }

  # Create the worker threads
  print "Creating worker threads\n";
  my $x = 0;
  my @threads = ();
  $threads[$x++] = threads->new(\&threaded_cosine, $x, $jobs, \%cos_hash,
				\$global_count, $text_type) for (0..3);

  # wait for them to exit
  $x = 0;
  $threads[$x++]->join() for (0..3);

  $self->{cosine_matrix} = \%cos_hash;
  return %cos_hash;
}

sub threaded_cosine {
  my $num = shift;
  my $jobs = shift;
  my $cos_hash = shift;
  my $global_count = shift;
  my $text_type = shift;

  for (;;) {

\end{boxedverbatim}

\begin{boxedverbatim}

    my $commanddata = thaw($jobs->dequeue_nb);
    return unless $commanddata;
    my ($doc1_key, $doc2_key) = @{$commanddata};
    my $doc1 = $documents{$doc1_key};
    my $doc2 = $documents{$doc2_key};
    my $cos = compute_document_cosine($doc1, $doc2, $text_type);
#    print "thread $num: $doc1_key\n";
    lock ($cos_hash);
    $cos_hash->{$doc1_key}{$doc2_key} = $cos;
    $cos_hash->{$doc2_key}{$doc1_key} = $cos;
#    lock($$global_count);
#    $$global_count++;
#    if (($$global_count % 10) == 0) {
#      print $$global_count / 10, "\n";
#    }
  }
}

#
# Split this out so we can make use of threading
#
sub compute_document_cosine {
  my $document1 = shift;
  my $document2 = shift;
  my $text_type = shift;

  my $text1 = "";
  my $text2 = "";
  if ($text_type eq "stem") {
    $text1 = $document1->get_stem;
    $text2 = $document2->get_stem;
  } elsif ($text_type eq "text") {
    $text1 = $document1->{text};
    $text2 = $document2->{text};
  }

  my $cos = GetLexSim($text1, $text2);

  return $cos;
}


#
# Print out usage message
#
sub usage
{
  print "usage: $0 -c corpus_name -o output_file [-b base_dir]\n\n";
  print "  -c corpus_name\n";
  print "       Name of the corpus\n";
  print "  -b base_dir\n";
  print "       Base directory filename.  The corpus is loaded from here\n";
  print "  -o output_file\n";
  print "       Name of file to write network to\n";
  print "  -s,--sample n\n";
  print "       Take a sample of size n from the documents\n";
  print "\n";

  print "example: $0 -c bulgaria -o data/bulgaria.cos -b                        \
/data0/projects/lexnets/pipeline/produced\n";

  exit;
}


\end{boxedverbatim}

\subsubsection{corpus\_to\_lexical\_network.pl}
\begin{boxedverbatim}

#!/usr/bin/perl
# script: corpus_to_lexical_network.pl
# functionality: Generates a lexical network for a corpus
# In the lexical network, each node is a word, and an edge exists between
# two words if they occur in the same sentences.  Multiple occurences are
# weighted more.
#

use strict;
use warnings;

use Getopt::Long;

use Clair::Cluster;
use Clair::Network::Writer::Edgelist;
#mjschal was here, removing references to Essence.  This doesn't appear to be   \
used:
#use Essence::IDF;

sub usage;

my $corpus_name = "";
my $basedir = "produced";
my $output_file = "";
my $sample_size = 0;
my $verbose = 0;
my $stem = 1;

my $res = GetOptions("corpus=s" => \$corpus_name, "base=s" => \$basedir,
		     "output:s" => \$output_file,
		     "stem!" => \$stem, "verbose!" => \$verbose);

if (!$res or ($corpus_name eq "") or ($basedir eq "")) {
  usage();
  exit;
}

my $gen_dir = "$basedir";

my $corpus_data_dir = "$gen_dir/corpus-data/$corpus_name";
my $doc_to_file = "$corpus_data_dir/" . $corpus_name . "-docid-to-file";

if ($verbose) { print "Loading corpus into cluster\n"; }
my $cluster = new Clair::Cluster;
$cluster->load_corpus($corpus_name, docid_to_file_dbm => $doc_to_file);

$cluster->strip_all_documents;
if ($stem) {
  $cluster->stem_all_documents;
}

my $network = $cluster->create_lexical_network();

if ($output_file ne "") {
  my $export = Clair::Network::Writer::Edgelist->new();
  $export->write_network($network, $output_file, weights => 1);
}


#
# Print out usage message
#
sub usage
{
  print "usage: $0 -c corpus_name -o output_file [-b base_dir]\n\n";
  print "  -c corpus_name\n";
  print "       Name of the corpus\n";

\end{boxedverbatim}

\begin{boxedverbatim}

  print "  -b base_dir\n";
  print "       Base directory filename.  The corpus is loaded from here\n";
  print "  -o output_file\n";
  print "       Name of file to write network to\n";
  print "  --stem or --no-stem\n";
  print "       Use the stemmed or unstemmed version of the corpus to generate  \
the network\n";

  print "\n";

  print "example: $0 -c bulgaria -o bulgaria.graph -b produced\n";

  exit;
}

\end{boxedverbatim}

\subsubsection{corpus\_to\_network.pl}
\begin{boxedverbatim}

#!/usr/bin/perl
# script: corpus_to_network.pl
# functionality: Generates a hyperlink network from corpus HTML files

use strict;
use warnings;

use Getopt::Std;
use vars qw/ %opt /;
use Clair::Network;
use Clair::Network::Writer::Edgelist;
use Clair::Utils::TFIDFUtils;

sub usage;

my $opt_string = "c:b:o:";
getopts("$opt_string", \%opt) or usage();

my $corpus_name = "";
if ($opt{"c"}) {
  $corpus_name = $opt{"c"};
} else {
  usage();
  exit;
}

my $basedir = "produced";
if ($opt{"b"}) {
  $basedir = $opt{"b"};
}
my $gen_dir = "$basedir";

my $output_file = "";
if ($opt{"o"}) {
  $output_file = $opt{"o"};
#  open(OUTFILE, "> $output_file");
} else {
#  *OUTFILE = *STDOUT;
  usage();
  exit;
}

my $verbose = 0;


my $corpus_data_dir = "$gen_dir/corpus-data/$corpus_name";
my $linkfile = "$corpus_data_dir/$corpus_name.links";
my $doc_to_file = "$corpus_data_dir/" . $corpus_name . "-docid-to-file";
my $doc_to_url = "$corpus_data_dir/" . $corpus_name . "-docid-to-url";
my $compress_dbm = "$corpus_data_dir/" . $corpus_name . "-compress-docid";


if ($verbose) { print "Generating hyperlink network\n"; }
my $network = Clair::Network->new_hyperlink_network($linkfile,
						    docid_to_file_dbm =>
						    $doc_to_file,
						    compress_docid =>
						    $compress_dbm);

if ($output_file ne "") {
  write_links($network, $output_file, $doc_to_url);
}


#
# Like write_links in Clair::Network, but print the URL too
#

\end{boxedverbatim}

\begin{boxedverbatim}

sub write_links
{
  my $self = shift;
  my $graph = $self->{graph};

  my $filename = shift;
  my $doc_to_url = shift;

  my %parameters = @_;
  my $skip_duplicates = 0;
  if (exists $parameters{skip_duplicates} &&
      $parameters{skip_duplicates} == 1) {
    $skip_duplicates = 1;
  }

  my $transpose = 0;
  if (exists $parameters{transpose} and $parameters{transpose} == 1) {
    $transpose = 1;
  }

  open(FILE, "> $filename") or die "Could not open file: $filename\n";

  my %seen_edges = ();

  # Open docid to URL database
  my %docid_to_url_dbm = ();
  dbmopen %docid_to_url_dbm, $doc_to_url, 0444 or die;

  foreach my $e ($graph->edges) {
    my $u;
    my $v;

    ($u, $v) = @$e;
    if ($u ne "EX") {
      $u = $docid_to_url_dbm{$u->get_id()};
    }
    if ($v ne "EX") {
      $v = $docid_to_url_dbm{$v->get_id()};
    }
    if ($transpose == 1) {
      my $temp = $u;
      $u = $v;
      $v = $temp;
    }

    if ($skip_duplicates  == 1 || not exists $seen_edges{"$u,$v"}) {
      print(FILE "$u $v\n");
      $seen_edges{"$u,$v"} = 1;
    }
  }

  dbmclose %docid_to_url_dbm;
  close(FILE);
}


#
# Print out usage message
#
sub usage
{
  print "usage: $0 -c corpus_name -o output_file [-b base_dir]\n\n";
  print "  -c corpus_name\n";
  print "       Name of the corpus\n";
  print "  -b base_dir\n";
  print "       Base directory filename.  The corpus is loaded from here\n";
  print "  -o output_file\n";

\end{boxedverbatim}

\begin{boxedverbatim}

  print "       Name of file to write network to\n";
  print "\n";

  print "example: $0 -c bulgaria -o data/bulgaria.graph -b                      \
/data0/projects/lexnets/pipeline/produced\n";

  exit;
}


\end{boxedverbatim}

\subsubsection{cos\_to\_cosplots.pl}
\begin{boxedverbatim}

#!/usr/bin/perl
# script: cos_to_cosplots.pl
# functionality: Generates cosine distribution plots, creating a
# functionality: histogram in log-log space, and a cumulative cosine plot
# functionality: histogram in log-log space
#
# Based on the make_cosiine_plots.pl script by Alex
#

use strict;
use warnings;

use File::Spec;
use Getopt::Long;

sub usage;

my $cos_file = "";
my $num_bins = 100;

my $res = GetOptions("input=s" => \$cos_file, "bins:i" => \$num_bins);

if (!$res || ($cos_file eq "")) {
  usage();
  exit;
}

my ($vol, $dir, $hist_prefix) = File::Spec->splitpath($cos_file);
$hist_prefix =~ s/\.cos//;

my $cosines = "$cos_file";

my @link_bin = ();
$link_bin[$num_bins] = 0;

my $link_total = 0;
my $link_count = 0;
my %cos_hash = ();

my ($doc1,  $doc2,  $cos);
open (COS, $cosines) or die "cannot open $cosines\n";

while(<COS>) {
  chomp;
  ($doc1, $doc2, $cos) = split;
  my $key1 = "$doc1 $doc2";
  my $key2 = "$doc2 $doc1";

  if (($doc1 ne $doc2) &&
      !(exists $cos_hash{$key2}) &&
      !(exists $cos_hash{$key1})) {

    $cos_hash{$key1} = 1;
    my $c = $cos;
    my $d = get_index($c);
    $link_bin[$d]++;
    $link_total += $cos;
    $link_count++;
  }
}
close(COS);

# print final info
print "average cosine is " . $link_total/$link_count . "\n" if 
$link_count>0;

#print "cosine histogram:\n";

\end{boxedverbatim}

\begin{boxedverbatim}


# Commented out  by alex
# Fri Apr 22 23:18:40 EDT 2005
#
# For some reason, matlab decided that today it does not
#  like full paths. So we take them out, and pat matlab
#  on the head.
#
# Just remember that this will produce plots in the
#  current directory now, so CD in to wherever you need
#  to be before piping this stuff into matlab.
#
my $fname = $hist_prefix . "-cosine-hist.m";
my $fname2 = $hist_prefix . "-cosine-cumulative.m";
open(OUT,">$fname") or die ("Cannot write to $fname");
open(OUT2,">$fname2") or die ("Cannot write to $fname2");
print OUT "x = [";
print OUT2 "x = [";
my $cumulative=0;

foreach my $i (0..$#link_bin)
{
   my $out = $link_bin[$i];
   if(not defined $link_bin[$i])
   {
      $out = 0;
   }
   $cumulative+= $out;
   my $thres = $i/100;
#   print "$thres $out\n";
   print OUT "$thres $out\n";
   print OUT2 "$thres $cumulative\n";
}

print OUT "];\n";

my $out_filename = "$hist_prefix"."-cosine-hist";
print OUT "loglog(x(:,1), x(:,2));\n";
print OUT "title(['Number of pairs per cosine in $hist_prefix']);\n";
print OUT "xlabel('Cosine Value');\n";
print OUT "ylabel('Number of pairs');\n";
# Change label font sizes
print OUT "h = get(gca, 'title');\n";
print OUT "set(h, 'FontSize', 16);\n";
print OUT "h = get(gca, 'xlabel');\n";
print OUT "set(h, 'FontSize', 16);\n";
print OUT "h = get(gca, 'ylabel');\n";
print OUT "set(h, 'FontSize', 16);\n";

print OUT "v = axis;\n";
print OUT "v(1) = 0; v(2) = 1;\n";
print OUT "axis(v)\n";
print OUT "print ('-deps', '$out_filename.eps')\n";
print OUT "saveas(gcf, '$out_filename" . ".jpg', 'jpg'); \n";
close OUT;

$out_filename = $hist_prefix . "-cosine-cumulative";
print OUT2 "];\n";
print OUT2 "loglog(x(:,1), x(:,2));\n";
print OUT2 "title(['Number of pairs per cosine in $hist_prefix']);\n";
print OUT2 "xlabel('Cosine Threshold Value');\n";
print OUT2 "ylabel('Number of pairs w/cosine less than or equal to              \
threshold');\n";
# Change label font sizes
print OUT2 "h = get(gca, 'title');\n";
print OUT2 "set(h, 'FontSize', 16);\n";
print OUT2 "h = get(gca, 'xlabel');\n";

\end{boxedverbatim}

\begin{boxedverbatim}

print OUT2 "set(h, 'FontSize', 16);\n";
print OUT2 "h = get(gca, 'ylabel');\n";
print OUT2 "set(h, 'FontSize', 16);\n";

print OUT2 "v = axis;\n";
print OUT2 "v(1) = 0; v(2) = 1;\n";
print OUT2 "axis(v)\n";
print OUT2 "print ('-deps', '$hist_prefix-cosine-cumulative.eps')\n";
print OUT2 "saveas(gcf, '$out_filename" . ".jpg', 'jpg'); \n";
close OUT2;


sub get_index {
  my $d = shift;
  my $c = int($d * $num_bins+0.000001);
#  print "$c $d\n";
  return $c;
}

sub usage {
  print "Usage $0 --input input_file [--bins num_bins]\n\n";
  print "  --input input_file\n";
  print "       Name of the input graph file\n";
  print "  --bins num_bins\n";
  print "       Number of bis\n";
  print "       num_bins is optional, and defaults to 100\n";
  print "\n";
  die;
}


\end{boxedverbatim}

\subsubsection{cos\_to\_histograms.pl}
\begin{boxedverbatim}

#!/usr/bin/perl
# script: cos_to_histograms.pl
# functionality: Generates degree distribution histograms from
# functionality: degree distribution data

use strict;
use warnings;

use File::Spec;
use Getopt::Long;
use Clair::Network;

sub usage;

my $graph_file = "";
my $output_file = "";
my $start = 0.0;
my $end = 1.0;
my $inc = 0.01;
my $hists = 1;
my $verbose = 0;
#my $matlab_script = "/data0/projects/lr/plots/distplots.m";

my $res = GetOptions("input=s" => \$graph_file, "output=s" => \$output_file,
                     "start=f" => \$start, "end=f" => \$end,
                     "step=f" => \$inc,
                     "hists!" => \$hists, "verbose" => \$verbose);

if (!$res or ($graph_file eq "")) {
  usage();
  exit;
}

my ($vol, $dir, $hist_prefix) = File::Spec->splitpath($graph_file);
$hist_prefix =~ s/\.cos//;

if ($verbose) { print STDERR "Loading $graph_file\n"; }
my @edges = load_cos($graph_file);

if ($hists) {
  for (my $i = $start; $i <= $end; $i += $inc) {
    # below is because of some strange rounding bug on the linux machines
    $i = sprintf("%.4f", $i);
    my $cutoff = sprintf("%.2f", $i);
    my @filtered = filter_cosine(\@edges, $cutoff);
    my @hist = link_degree(\@filtered);
    write_hist("hists", $hist_prefix . "." . $cutoff . ".hist", \@hist);
  }
} else {
  if ($verbose) { print STDERR "Skipping writing histogram files\n"; }
}

write_plot("hists", $hist_prefix, $start, $end, $inc);


#
# Write the matlab plot for the cutoff files
#
sub write_plot {
  my $dir = shift;
  my $file = shift;
  my $start = shift;
  my $end = shift;
  my $inc = shift;

  my @hists = ();
  my @cutoffs = ();

\end{boxedverbatim}

\begin{boxedverbatim}

  for (my $i = $end; $i > $start; $i -= $inc) {
    # below is because of some strange rounding bug on the linux machines
    $i = sprintf("%.4f", $i);
    my $cutoff = sprintf("%.2f", $i);
    push (@hists, $dir . "/" . $file . "." . $cutoff . ".hist");
    push (@cutoffs, $cutoff);
  }

  open(MYOUTFILE, ">$file-distplots.m");
  my $file_count = 0;
  my $color_index = 5;
  my $x = "";
  my $y = "";
  my $c = "";

  foreach my $hist (@hists) {
    chomp($hist);
    #$test = "y".$file_count." = load('".$hist."');";
#    print MYOUTFILE $test;
    print MYOUTFILE "y$file_count = load('$hist'); \n";

    print MYOUTFILE "if length(y0) > 75 \n";
    print MYOUTFILE "  y$file_count = y$file_count(1:75) ; \n";

    print MYOUTFILE "else \n";
    print MYOUTFILE "  y$file_count = y$file_count(1:length(y0)); \n";
    print MYOUTFILE "end \n";


    print MYOUTFILE "\n";

    $y = $y."y".$file_count."; ";
    $x = $x."1:1:length(y0); ";
    $c = $c."temp*$color_index;";
    $file_count++;
    $color_index = $color_index + 5;
  }
  print MYOUTFILE "Y = [ $y ]; \n";
  print MYOUTFILE "X = [ $x ]; \n";
  #hard coded to y0
  print MYOUTFILE "temp = ones(1,length(y0) ); \n";

  my $z = "";
  foreach $c (@cutoffs) {
    chomp($c);
    $z = $z."temp*".$c."; ";
  }

  print MYOUTFILE "C = [ $c ]; \n";
  print MYOUTFILE "Z = [ $z ]; \n \n"; # print MYOUTFILE "surf(Z,X,Y); \n";
  print MYOUTFILE "surf(Z,X,Y,C); \n"; # print MYOUTFILE "colormap hsv; \n";

  print MYOUTFILE "xlabel('Cosine similarity threshold');\n";
  print MYOUTFILE "ylabel('Vertex degree');\n";
  print MYOUTFILE "zlabel('Number of nodes');\n";

  print MYOUTFILE "view(-120,37.5); \n";

  my $save = $file . "_" . $start . "_" . $inc . "_" . $end;

  print MYOUTFILE "saveas(gcf,'plots/".$save.".jpg','jpg'); \n";
  print MYOUTFILE "saveas(gcf,'plots/".$save.".eps','eps'); \n";

  close(MYOUTFILE);
}

#

\end{boxedverbatim}

\begin{boxedverbatim}

# Write histogram to file
#
sub write_hist {
  my $dir = shift;
  my $fn = shift;
  my $h = shift;
  my @hist = @{$h};

  unless (-d $dir) {
    mkdir $dir or die "Couldn't create $dir: $!";
  }

  open(OUTFILE, ">", $dir . "/" . $fn) or die "Couldn't open " . $dir . "/" .
    $fn, "\n";

  foreach my $deg (@hist) {
    print OUTFILE "$deg ";
  }
  print OUTFILE "\n";

  close OUTFILE;
}


#
# Load cosine file
#
sub load_cos {
  my $file = shift;

  my @edges = ();

  open(INFILE, $file) or die "Couldn't open $file\n";

  while (<INFILE>) {
    chomp;
    my @array = split(/ /, $_);
    push @edges, \@array;
  }

  close INFILE;

  return @edges;
}


sub link_degree {
  my $vert = shift;
  my @edges = @{$vert};

  my $pagecount = 0;
  my %ct = ();
  my %links = ();
  my %pageswith = ();

  my @hist = ();

  foreach my $e (@edges) {
    my ($from, $to) = @{$e};
    $ct{$from} = 1;
    $ct{$to} = 1;

    if (not exists $links{$to}) {
      $links{$to} = 0;
      $pagecount++;
    }


\end{boxedverbatim}

\begin{boxedverbatim}

    if (not exists $links{$from}) {
      $links{$from} = 0;
      $pagecount++;
    }

    $links{$from}++;
  }

  foreach my $node (grep {$links{$_} == 48} (keys %links)) {
    print "node: $node\n";
  }

  my $total = scalar(keys %ct);

  foreach my $i2 (0..$total-1) {
    $pageswith{$i2} = 0;
  }

  foreach my $node (keys %links) {
    $pageswith{$links{$node}}++;
  }

  foreach my $linkcount (sort {$a <=> $b} keys %pageswith) {
    $hist[$linkcount] = $pageswith{$linkcount};
  }

  return @hist;
}

#
# filter cosine file by cutoff
#
sub filter_cosine {
  my $cref = shift;
  my @cos = @{$cref};
  my $cutoff = shift;

  my @edges = ();

  foreach my $e (@cos) {
    my @links = @{$e};
    my ($l, $r, $c) = @links;
    if ($c >= $cutoff) {
      push @edges, \@links;
    }
  }

  return @edges;
}


#
# Print out usage message
#
sub usage
{
  print "usage: $0 --input input_file [--output output_file] [--start start]    \
[--end end] [--step step]\n\n";
  print "  --input input_file\n";
  print "       Name of the input graph file\n";
  print "  --output output_file\n";
  print "       Name of plot output file\n";
  print "  --start start\n";
  print "       Cutoff value to start at\n";
  print "  --end end\n";
  print "       Cutoff value to end at\n";
  print "  --step step\n";

\end{boxedverbatim}

\begin{boxedverbatim}

  print "       Size of step between cutoff points\n";
  print "\n";

  print "example: $0 --input data/bulgaria.cos --output data/bulgaria.m\n";

  exit;
}

\end{boxedverbatim}

\subsubsection{cos\_to\_networks.pl}
\begin{boxedverbatim}

#!/usr/local/bin/perl
# script: cos_to_networks.pl
# functionality: Generate series of networks by incrementing through cosine     \
cutoffs
#

use strict;
use warnings;
use Getopt::Long;
use File::Spec;
use Clair::Network qw($verbose);
use Clair::Network::Writer::Edgelist;

sub usage;

my $cos_file = "";
my $start = 0.0;
my $end = 1.0;
my $inc = 0.01;
my $graph_dir = "";

my $res = GetOptions("input=s" => \$cos_file, "output=s" => \$graph_dir,
		     "start=f" => \$start, "end=f" => \$end,
		     "step=f" => \$inc);

if ($cos_file eq "") {
  usage();
  exit;
}

my ($vol, $dir, $prefix) = File::Spec->splitpath($cos_file);
$prefix =~ s/\.cos//;
if ($graph_dir eq "") {
  $graph_dir = "graphs/$prefix";
}

unless (-d $graph_dir) {
  `mkdir -p $graph_dir`;
  unless (-d $graph_dir) { die "Couldn't make directory $graph_dir: $!\n"; }
}

my @edges = load_cos($cos_file);

my $test_net = new Clair::Network();
my $net = $test_net->create_cosine_network(\@edges);

for (my $i = $start; $i <= $end; $i += $inc) {
  # below is because of some strange rounding bug on the linux machines
  $i = sprintf("%.4f", $i);
  my $cutoff = sprintf("%.2f", $i);
  my $cos_net = $net->create_network_from_cosines($cutoff);

  my $export = Clair::Network::Writer::Edgelist->new();
  $export->write_network($cos_net,
                         $graph_dir . "/" . $prefix . "-" . $cutoff . ".net");
}

#
# Load cosine file
#
sub load_cos {
  my $file = shift;

  my @edges = ();

  open(INFILE, $file) or die "Couldn't open $file\n";


\end{boxedverbatim}

\begin{boxedverbatim}

  while (<INFILE>) {
    chomp;
    my @array = split(/ /, $_);
    push @edges, \@array;
  }

  close INFILE;

  return @edges;
}

#
# Print out usage message
#
sub usage
{
  print "usage: $0 --input input_file [--output output_directory] [--start      \
start] [--end end] [--step step]\n\n";
  print "  --input input_file\n";
  print "       Name of the input graph file\n";
  print "  --output output_directory\n";
  print "       Name of output directory.  The default is                       \
graphs/input_file_prefix\n";
  print "  --start start\n";
  print "       Cutoff value to start at\n";
  print "  --end end\n";
  print "       Cutoff value to end at\n";
  print "  --step step\n";
  print "       Size of step between cutoff points\n";
  print "\n";

  print "example: $0 --input data/bulgaria.cos --output networks\n";

  exit;
}

\end{boxedverbatim}

\subsubsection{cos\_to\_stats.pl}
\begin{boxedverbatim}

#!/usr/bin/perl
# script: cos_to_stats.pl
# functionality: Generates a table of network statistics for networks by
# functionality: incrementing through cosine cutoffs
#

use strict;
use warnings;
use Getopt::Long;
use File::Spec;
use Clair::Network qw($verbose);
use Clair::Network::Sample::ForestFire;
use Clair::Network::Sample::RandomEdge;
use Clair::Network::Sample::RandomNode;
use Clair::Network::Reader::Edgelist;
use Clair::Network::Writer::Edgelist;
use Clair::Network::Writer::GraphML;

sub usage;

my $delim = "[ \t]+";
my $output_delim = " ";
my $cos_file = "";
my $graphml = 0;
my $threshold;
my $start = 0.0;
my $end = 1.0;
my $inc = 0.01;
my $sample_size = 0;
my $sample_type = "randomnode";
my $out_file = "";
my $graphs = 0;
my $all = 0;
my $stats = 1;
my $single = 0;
my $verbose = 0;

my $res = GetOptions("input=s" => \$cos_file, "output=s" => \$out_file,
                     "delimout=s" => \$output_delim,
                     "graphml" => \$graphml,
                     "threshold=f" => \$threshold, "delim=s" => \$delim,
                     "start=f" => \$start, "end=f" => \$end,
                     "step=f" => \$inc, "graphs:s" => \$graphs,
                     "sample=i" => \$sample_size, "single" => \$single,
                     "sampletype=s" => \$sample_type,
                     "all" => \$all, "stats!" => \$stats,
                     "verbose" => \$verbose);

$Clair::Network::verbose = $verbose;

if ($graphs eq "") {
  # Use default directory graphs if graphs enabled
  $graphs = "graphs";
}

if ($graphs) {
  unless (-d $graphs) {
    mkdir $graphs or die "Couldn't create $graphs: $!";
  }
}

if ($cos_file eq "") {
  usage();
  exit;
}

my $dir;

\end{boxedverbatim}

\begin{boxedverbatim}

my $vol;
my $prefix;
my $file;

($vol, $dir, $prefix) = File::Spec->splitpath($cos_file);
$prefix =~ s/\.cos//;
if ($out_file ne "") {
  ($vol, $dir, $file) = File::Spec->splitpath($out_file);
  if ($dir ne "") {
    unless (-d $dir) {
      mkdir $dir or die "Couldn't create $dir: $!";
    }
  }

  open(OUTFILE, "> $out_file");
  *STDOUT = *OUTFILE;
  select OUTFILE; $| = 1;
}

# make unbuffered
select STDOUT; $| = 1;
select STDERR; $| = 1;
select STDOUT;

my $net;
# Sample network if requested
if ($sample_size > 0) {
  if ($verbose) { print STDERR "Reading in $cos_file\n"; }
  my $reader = new Clair::Network::Reader::Edgelist;
  $net = $reader->read_network($cos_file, undirected => 1, delim => $delim);

  if ($sample_type eq "randomedge") {
    if ($verbose) {
      print STDERR "Sampling $sample_size edges from network using random edge  \
algorithm\n"; }
    my $sample = Clair::Network::Sample::RandomEdge->new($net);
    $net = $sample->sample($sample_size);
  } elsif ($sample_type eq "forestfire") {
    if ($verbose) {
      print STDERR "Sampling $sample_size nodes from network using Forest Fire  \
algorithm\n"; }
    my $sample = Clair::Network::Sample::ForestFire->new($net);
    $net = $sample->sample($sample_size, 0.7);
  } elsif ($sample_type eq "randomnode") {
    if ($verbose) {
      print STDERR "Sampling $sample_size nodes from network using Random Node  \
algorithm\n";
    }
    my $sample = Clair::Network::Sample::RandomNode->new($net);
    $sample->number_of_nodes($sample_size);
    $net = $sample->sample();
  }
} else {
  if ($graphs) {
    # no sampling, just write the graph files
    for (my $i = $start; $i <= $end; $i += $inc) {
      # below is because of some strange rounding bug on the linux machines
      $i = sprintf("%.4f", $i);
      my $cutoff = sprintf("%.2f", $i);
      if ($verbose) {
        print STDERR "Writing graph file for cutoff $cutoff\n";
      };
      open FOUT, ">$graphs/$prefix-$cutoff.graph";

      open(FIN, $cos_file) or die "Couldn't open $cos_file: $!\n";
      while (<FIN>) {
        chomp;

\end{boxedverbatim}

\begin{boxedverbatim}

        my @edge = split(/$delim/);

        my ($u, $v, $w) = @edge;
        if ($w >= $cutoff) {
          print FOUT "$u$output_delim$v$output_delim$w\n";
        }
      }
      close FIN;
      close FOUT;
    }
  }
}

if ($stats) {
  if ($verbose) { print STDERR "Reading in $cos_file\n"; }
  my $reader = new Clair::Network::Reader::Edgelist;
  $net = $reader->read_network($cos_file, undirected => 1,
                               unionfind => 1, delim => $delim);
  if ($single and $threshold) {
    # Run for already generated graph
    print_network($net, $threshold);
  } elsif ($threshold) {
    # Run for just a single cutoff
    run_cutoff($net, $threshold);
  } else {
    # Run for all cutoffs
    if ($net->{directed}) {
#      print "threshold nodes edges diameter lcc avg_short_path                 \
watts_strogatz_cc newman_cc in_link_power in_link_power_rsquared in_link_pscore \
in_link_power_newman in_link_power_newman_error out_link_power                  \
out_link_power_rsquared out_link_pscore out_link_power_newman                   \
out_link_power_newman_error total_link_power total_link_power_rsquared          \
total_link_pscore total_link_power_newman total_link_power_newman_error         \
avg_degree\n";
      print "threshold nodes edges diameter lcc avg_short_path                  \
watts_strogatz_cc hmgd in_link_power in_link_power_rsquared in_link_pscore      \
in_link_power_newman in_link_power_newman_error out_link_power                  \
out_link_power_rsquared out_link_pscore out_link_power_newman                   \
out_link_power_newman_error total_link_power total_link_power_rsquared          \
total_link_pscore total_link_power_newman total_link_power_newman_error         \
avg_degree\n";
    } else {
#      print "threshold nodes edges diameter lcc avg_short_path                 \
watts_strogatz_cc newman_cc power_law power_law_rsquared power_law_pscore       \
power_law_power_newman power_law_newman_error avg_degree\n";
      print "threshold nodes edges diameter lcc avg_short_path                  \
watts_strogatz_cc hmgd power_law power_law_rsquared power_law_pscore            \
power_law_power_newman power_law_newman_error avg_degree\n";
    }
    for (my $i = $start; $i <= $end; $i += $inc) {
      # below is because of some strange rounding bug on the linux machines
      $i = sprintf("%.4f", $i);
      my $cutoff = sprintf("%.2f", $i);

      run_cutoff($net, $cutoff);
    }
  }
}

sub array_to_graphml {
  my $fn = shift;
  my $ed = shift;
  my @edges = @{$ed};

  open(GRAPH, "> $fn") or die "Couldn't open file: $fn\n";

  print GRAPH <<EOH

\end{boxedverbatim}

\begin{boxedverbatim}

<?xml version="1.0" encoding="UTF-8"?>
<graphml xmlns="http://graphml.graphdrawing.org/xmlns"
  xmlns:xsi="http://www.w3.org/2001/XMLSchema-instance"
  xsi:schemaLocation="http://graphml.graphdrawing.org/xmlns
                      http://graphml.graphdrawing.org/xmlns/1.0/graphml.xsd">
EOH
;

  print GRAPH "<key id=\"d1\" for=\"edge\" attr.name=\"weight\"                 \
attr.type=\"double\"/>\n";
  print GRAPH "  <graph id=\"graph\" edgedefault=\"undirected\">\n";

  my %nodes = ();
  foreach my $e (@edges) {
    my ($u, $v, $w) = @{$e};
    $nodes{$u} = 1;
    $nodes{$v} = 1;
  }

  foreach my $v (keys %nodes) {
    print GRAPH "    <node id=\"" . $v . "\"/>\n";
  }

  foreach my $e (@edges) {
    my ($u, $v, $w) = @{$e};
    print GRAPH "    <edge source=\"" . $u . "\" target=\"" . $v . "\">\n";
    print GRAPH "      <data key=\"d1\">" . $w . "</data>\n";
    print GRAPH "    </edge>\n";
  }

  print GRAPH "  </graph>\n";
  print GRAPH "</graphml>\n";

  close(GRAPH);
}



sub run_cutoff {
  my $net = shift;
  my $cutoff = shift;

  if ($verbose) { print STDERR "Creating network for cutoff $cutoff\n"; }
  my $cos_net = $net->create_network_from_cosines($cutoff);

  print_network($cos_net, $cutoff);
  if ($all) {
    # Dump out additional data
    # triangles
    open(FOUT, ">$dir/$prefix-$cutoff.triangles") or die "Couldn't open         \
$dir/$prefix.triangles: $!\n";
    my ($triangles, $triangle_cnt, $triple_cnt) = $net->get_triangles();
    foreach my $triangle (@{$triangles}) {
      print FOUT $triangle, "\n";
    }
    close FOUT;
    # average shortest path matrix
    open(FOUT, ">$dir/$prefix-$cutoff.asp") or die "Couldn't open               \
$dir/$prefix.asp: $!\n";
    # save stdout and redirect it to the file
    *SAVED = *STDOUT;
    *STDOUT = *FOUT;
    $cos_net->print_asp_matrix();
    # restore stdout
    *STDOUT = *SAVED;
    close FOUT;
  }

\end{boxedverbatim}

\begin{boxedverbatim}


#  print "total_size out: ", total_size($cos_net), "\n";

  if ($graphs) {
    write_network($cos_net, $cutoff);
  }
}

sub print_network {
  my $net = shift;
  my $cutoff = shift;

  if ($net->num_nodes > 0) {
    if ($verbose) { print STDERR "Getting network info for cutoff $cutoff\n"; }
    my $stats = $net->get_network_info_as_string();
    print "$cutoff " . $stats . "\n";
  } else {
    print "$cutoff ";
    if ($net->{directed}) {
      print "0 0 0 0 0 0 0 0 0 0 0 0 0 0 0 0 0 0 0 0 0 0 0\n";
    } else {
      print "0 0 0 0 0 0 0 0 0 0 0 0 0\n";
    }
  }
}

sub write_network {
  my $cos_net = shift;
  my $cutoff = shift;

  my $export = Clair::Network::Writer::Edgelist->new();
  $export->write_network($cos_net,
                         "$graphs/$prefix-$cutoff.graph", weights => 1);

  if ($graphml) {
    my $export = Clair::Network::Writer::GraphML->new();
    $export->write_network($cos_net,
                           "$graphs/$prefix-$cutoff.graphml", weights => 1);
  }

  if ($all) {
    # Dump out additional data
    # triangles
    open(FOUT, ">$dir/$prefix-$cutoff.triangles") or die "Couldn't open         \
$dir/$prefix.triangles: $!\n";
    my ($triangles, $triangle_cnt, $triple_cnt) = $net->get_triangles();
    foreach my $triangle (@{$triangles}) {
      print FOUT $triangle, "\n";
    }
    close FOUT;
    # average shortest path matrix
    open(FOUT, ">$dir/$prefix-$cutoff.asp") or die "Couldn't open               \
$dir/$prefix.asp: $!\n";
    # save stdout and redirect it to the file
    *SAVED = *STDOUT;
    *STDOUT = *FOUT;
    $cos_net->print_asp_matrix();
    # restore stdout
    *STDOUT = *SAVED;
    close FOUT;
  }
}


#
# Print out usage message
#

\end{boxedverbatim}

\begin{boxedverbatim}

sub usage
{
  print "usage: $0 --input input_file [--output output_file] [--start start]    \
[--end end] [--step step]\n\n";
  print "  --input input_file\n";
  print "       Name of the input graph file\n";
  print "  --output output_file\n";
  print "       Name of output file.  Dumps the stats to this file\n";
  print "  --start start\n";
  print "       Cutoff value to start at\n";
  print "  --end end\n";
  print "       Cutoff value to end at\n";
  print "  --step step\n";
  print "       Size of step between cutoff points\n";
  print "  --sample sample_size\n";
  print "       Sample from the network\n";
  print "  --sampletype sample_algorithm\n";
  print "       Sampling algorithm to use, can be: randomnode, randomedge,      \
forestfire\n";
  print "  --graphs [directory]\n";
  print "       If set, output a graph file for each cutoff in the specified
directory (defaults to graphs)\n";
  print "  --single\n";
  print "       Generate line for a single threshold.  Must also specify        \
threshold\n";
  print "  --threshold threshold\n";
  print "       Generate network for single threshold and print stats for       \
it.\n";
  print "\n";

  print "example: $0 --input data/bulgaria.cos --output networks\n";

  exit;
}

\end{boxedverbatim}

\subsubsection{crawl\_url.pl}
\begin{boxedverbatim}

#!/usr/bin/perl
# script: crawl_url.pl
# functionality: Crawls from a starting URL, returning a list of URLs
# Output to stdout, or a file

use strict;
use warnings;

use Getopt::Long;
use Clair::Utils::CorpusDownload;
use Clair::Utils::Idf;
use Clair::Utils::Tf;

sub usage;

my $url = "";
my $output_file = "";
my $test = "";
my $verbose = 0;

my $res = GetOptions("url=s" => \$url, "output=s" => \$output_file,
		     "test=s" => \$test, "verbose!" => \$verbose);

if ($url eq "") {
  usage();
  exit;
}

if ($output_file ne "") {
  open(OUTFILE, "> $output_file");
} else {
  *OUTFILE = *STDOUT;
}

# make unbuffered
select STDOUT; $| = 1;
select OUTFILE; $| = 1;


my $corpusref = Clair::Utils::CorpusDownload->new();

if ($verbose) { print "Crawling $url\n"; }
my $uref = 0;
if ($test ne "") {
  $uref = $corpusref->poach($url, error_file => "errors.txt",
			    test => $test);
} else {
  $uref = $corpusref->poach($url, error_file => "errors.txt");
}

foreach my $url (@{$uref}) {
  print OUTFILE $url, "\n";
}

close OUTFILE;

unlink("seen_url", "urls_list");

#
# Print out usage message
#
sub usage
{
  print "usage: $0 -c corpus_name -u url [-b base_dir] [-o output_file]\n\n";
  print "  --url url\n";
  print "       URL to start the crawl from\n";
  print "  --output output filename\n";

\end{boxedverbatim}

\begin{boxedverbatim}

  print "       File to store the URLs in.  If not specified, print them to     \
STDOUT\n";
  print "  --test test regular expression\n";
  print "       Regular expression to test URLs\n";
  print "\n";

  print "example: $0 -c kzoo -b /data0/projects/lexnets/pipeline/produced -u    \
http://www.kzoo.edu/ -o data/kzoo.urls\n";

  exit;
}


\end{boxedverbatim}

\subsubsection{directory\_to\_corpus.pl}
\begin{boxedverbatim}

#!/usr/bin/perl
#
# script: directory_to_corpus.pl
# functionality: Generates a clairlib Corpus from a directory of documents
#

use strict;
use warnings;

use File::Spec;
use Getopt::Long;
use Clair::Utils::CorpusDownload;

sub usage;

my $corpus_name = "";
my $base_dir = "produced";
my $input_dir = "";
my $in_file = "";
my $type = "text";
my $verbose = 0;
my $safe = 0;
my $skipDownload = 0;

my $res = GetOptions("corpus=s" => \$corpus_name, "base=s" => \$base_dir,
		     "directory=s" => \$input_dir, "input=s" => \$in_file,
		     "type=s" => \$type, "verbose" => \$verbose, "skipDownload" =>            \
\$skipDownload);

if (!$res or ($corpus_name eq "")) {
  usage();
  exit;
}

unless (-d $base_dir) {
  mkdir $base_dir or die "Couldn't create $base_dir: $!";
}


my $gen_dir = "$base_dir";

my $corpus_data_dir = "$gen_dir/corpus-data/$corpus_name";

if ($skipDownload)  { 
  $safe = 1;
  print "Skipping download.\n";
}

if ($verbose ) { print "Instantiating corpus $corpus_name in $gen_dir\n"; }

my $corpus = Clair::Utils::CorpusDownload->new(corpusname => "$corpus_name",
				    rootdir => "$gen_dir");

if ($input_dir ne "") {
  $corpus->build_corpus_from_directory(dir => $input_dir, cleanup => 0,
				       safe => $safe, relative => 1, skipCopy => $skipDownload);
} elsif ($in_file ne "") {
  my @files = ($in_file);
  $corpus->buildCorpusFromFiles(filesref => \@files, cleanup => 0, safe =>      \
$safe, skipCopy => $skipDownload);
} else {
  usage();
  exit;
}

sub usage {
  print "Usage $0 --corpus corpus [--input input_file | --directory             \

\end{boxedverbatim}

\begin{boxedverbatim}

input_dir]\n\n";
  print "  --corpus corpus\n";
  print "       Name of the corpus to index\n";
  print "  --base base_dir\n";
  print "       Base directory filename.  The corpus is generated here\n";
  print "  --directory input_dir\n";
  print "       Directory containing files to insert into the corpus\n";
  print "  --input input_file\n";
  print "       File containing filenames of input documents\n";
  print "  --type document_type\n";
  print "       Document type, one of: text, html, stem\n";
  print "  --skipDownload\n";
  print "       Skips copying files into the $base_dir/download folder\n";
  print "  --verbose\n";
  print "       Include verbose output\n";
  print "\n";

  die;
}

\end{boxedverbatim}

\subsubsection{download\_urls.pl}
\begin{boxedverbatim}

#!/usr/bin/perl
# script: download_urls.pl
# functionality: Downloads a set of URLs

use strict;
use warnings;

use Getopt::Std;
use vars qw/ %opt /;
use Clair::Utils::CorpusDownload;

sub usage;

my $opt_string = "b:c:i:";
getopts("$opt_string", \%opt) or usage();

my $corpus_name = "";
#my $corpus_name = "umich2";
if ($opt{"c"}) {
  $corpus_name = $opt{"c"};
} else {
  usage();
  exit;
}

my $url_file = "";
if ($opt{"i"}) {
  $url_file = $opt{"i"};
} else {
  usage();
  exit;
}

my $basedir = "produced";
if ($opt{"b"}) {
  $basedir = $opt{"b"};
}
my $gen_dir = "$basedir";

my $verbose = 0;

if ($verbose ) { print "Instantiating corpus $corpus_name in $gen_dir\n"; }
my $corpus = Clair::Utils::CorpusDownload->new(corpusname => "$corpus_name",
				    rootdir => "$gen_dir");

if ($verbose) { print "Reading URLs\n"; }
my $uref = $corpus->readUrlsFile($url_file);

if ($verbose) { print "Building corpus\n"; }
$corpus->buildCorpus(urlsref => $uref, cleanup => 0);

# write links file
#$corpus->write_links();

#
# Print out usage message
#
sub usage
{
  print "usage: $0 -c corpus_name -i url_file [-b base_dir]\n\n";
  print "  -i url_file\n";
  print "       Name of the file containing a list of URLs from which to build  \
the network\n";
  print "  -c corpus_name\n";
  print "       Name of the corpus\n";
  print "  -b base_dir\n";
  print "       Base directory filename.  The corpus is generated here\n\n";

\end{boxedverbatim}

\begin{boxedverbatim}


  print "example: $0 -c bulgaria -i data/bulgaria.10.urls -b                    \
/data0/projects/lexnets/pipeline/produced\n";

  exit;
}


\end{boxedverbatim}

\subsubsection{generate\_random\_network.pl}
\begin{boxedverbatim}

#!/usr/bin/perl
# script: generate_random_network.pl
# functionality: Generates a random network

use strict;
use warnings;

use Getopt::Long;
use Clair::Network::Generator::ErdosRenyi;
use Clair::Network::Reader::Edgelist;
use Clair::Network::Writer::Edgelist;

sub usage;

my $in_file = "";
my $delim = "[ \t]+";
my $out_file = "";
my $type = "";
my $verbose = 0;
my $undirected = 0;
my $n = 0;
my $m = 0;
my $p = 0;
my $stats = 1;
my $weights = 0;
my $res = GetOptions("input=s" => \$in_file,  "delim=s" => \$delim,
                     "output=s" => \$out_file, "type=s" => \$type,
                     "verbose" => \$verbose, "undirected" => \$undirected,
                     "n=i" => \$n, "m=i" => \$m, "p=f" => \$p,
                     "weights" => \$weights, "stats!" => \$stats);

my $directed = not $undirected;

if (!$res or ($type eq "")) {
  usage();
  exit;
}

my $in_net = 0;
if ($in_file ne "") {
  my $reader = Clair::Network::Reader::Edgelist->new();
  my $in_net = $reader->read_network($in_file,
                                     delim => $delim,
                                     directed => $directed);
  $n = $in_net->num_nodes();
  $m = $in_net->num_links();
}

my $parent_type = "";
my $subtype = "";
if ($type eq "erdos-renyi-gnm") {
  $parent_type = "erdos-renyi";
  $subtype = "gnm";
  if ($m == 0) {
    print "Need m argument for number of edges\n";
    usage();
  }
} elsif ($type eq "erdos-renyi-gnp") {
  $parent_type = "erdos-renyi";
  $subtype = "gnp";
  if ($p == 0) {
    print "Need p argument for probability of edge\n";
    usage();
  }
}

my $net = 0;

\end{boxedverbatim}

\begin{boxedverbatim}

if ($parent_type eq "erdos-renyi") {
  my $generator = Clair::Network::Generator::ErdosRenyi->new(directed =>
                                                             $directed);
  if ($subtype eq "gnm") {
    $net = $generator->generate($n, $m, type => $subtype,
                                weights => $weights,
                                directed => $directed);
  } else {
    $net = $generator->generate($n, $p, type => $subtype,
                                weights => $weights,
                                directed => $directed);
  }
}

if ($out_file ne "") {
  my $export = Clair::Network::Writer::Edgelist->new();
  $export->write_network($net, $out_file, weights => $weights);
}
if ($stats) {
  $net->print_network_info();
}

sub usage {
  print "Usage $0 --output output_file --type type [--verbose]\n\n";
  print "  --input input_file\n";
  print "       Name of the input graph file\n";
  print "  --delim delimiter\n";
  print "          Vertices are delimited by delimter character\n";
  print "  --undirected,  -u\n";
  print "          Treat graph as an undirected graph\n";
  print "  --output output_file\n";
  print "       Name of the output graph file\n";
  print "  --type graph_type\n";
  print "       Type of random graph to generate, can be one of:\n";
  print "            erdos-renyi-gnm: Set number of edges\n";
  print "            erdos-renyi-gnp: Random edge w/ prob p\n";
  print "  -n number_nodes\n";
  print "       Number of nodes\n";
  print "  -m number_edges\n";
  print "       Number of edges\n";
  print "  -p edge_probability\n";
  print "       Probability of edge between two nodes\n";
  print "  --verbose\n";
  print "       Increase verbosity of debugging output\n";
  print "\n";
  die;
}


\end{boxedverbatim}

\subsubsection{idf\_query.pl}
\begin{boxedverbatim}

#!/usr/local/bin/perl
# script: get_idf.pl
# functionality: Looks up idf values for terms in a corpus

use strict;
use warnings;
use Getopt::Long;
use File::Spec;
use Clair::Utils::Idf;

sub usage;

my $base_dir = "";
my $out_file = "";
my $corpus_name = "";
my $query = "";
my $all = '';
my $stemmed = '';
my $dir;
my $vol;
my $file;


my $res = GetOptions("basedir=s" => \$base_dir, "output=s" => \$out_file,
			"corpus=s" => \$corpus_name, 
			"query=s" => \$query,
			"all" => \$all,
			"stemmed" => \$stemmed); 

# check for input dir
if( $base_dir eq "" ){
  usage();
  exit;
}

# check for corpus name
if( $corpus_name eq "" ){
  usage();
  exit;
}

# check for output file
if ($out_file ne "") {
  ($vol, $dir, $file) = File::Spec->splitpath($out_file);
  if ($dir ne "") {
    unless (-d $dir) {
      mkdir $dir or die "Couldn't create $dir: $!";
    }
  }

  open(OUTFILE, "> $out_file");
  *STDOUT = *OUTFILE;
  select OUTFILE; $| = 1;
}

# make unbuffered
select STDOUT; $| = 1;
select STDERR; $| = 1;
select STDOUT;

# check for word query
if( $query eq "" ){
  $all = 1;
} 

# create idf object
my $idf = Clair::Utils::Idf->new(rootdir => "$base_dir", 

\end{boxedverbatim}

\begin{boxedverbatim}

                                 corpusname => "$corpus_name", 
				 stemmed => $stemmed); 

# get idfs
my %idfs = $idf->getIdfs(); 

# print words and idfs to output
if( $all ){
  foreach my $k (keys %idfs) { 
    print "$k: " . $idfs{$k} . "\n"; 
  }
} elsif( $idfs{$query} ) {
  print "$query: " . $idfs{$query} . "\n";
} else {
  print "$query not found\n";
}


#
# Print out usage message
#
sub usage
{
  print "usage: $0 --basedir base_dir --corpus corpus_name [--output            \
output_file] [--query word] [--all] [--stemmed]\n\n";
  print "  --basedir base_dir\n";
  print "      Base directory filename.  The corpus is generated here.\n";
  print "  --corpus corpus_name\n";
  print "      Name of the corpus.\n";
  print "  --output output_file\n";
  print "      Name of output file.  If not given, dumps to stdout.\n";
  print "  --query word\n";
  print "      Term to query.\n";
  print "  --all\n";
  print "      Print out all words and IDF's.  Default.\n";
  print "  --stemmed\n";
  print "      Set whether the input is already stemmed.\n";
  print "\n";
  print "example: $0 --basedir /data0/corpora/sfi/abs/produced --corpus ABS     \
--output ./abs.idf --query hahn --stemmed\n";
  exit;
}

\end{boxedverbatim}

\subsubsection{index\_corpus.pl}
\begin{boxedverbatim}

#!/usr/bin/perl
# script: index_corpus.pl
# functionality: Builds the TF and IDF indices for a corpus 
# functionality: as well as several other support indices
#

use strict;
use warnings;

use File::Spec;
use Getopt::Long;

use Clair::Utils::CorpusDownload;
use Clair::Utils::Tf;
use Clair::Utils::Idf;

sub usage;

my $corpus_name = "";
my $base_dir = "produced";
my $input_dir = "";
my $tf_flag = 1;
my $idf_flag = 1;
my $links_flag = 1;
my $stats_flag = 1;
my $verbose = 0;
my $punc = 0;

my $res = GetOptions("corpus=s" => \$corpus_name, "base=s" => \$base_dir,
		     "tf!" => \$tf_flag, "idf!" => \$idf_flag,
                     "links!" => \$links_flag, "stats!" => \$stats_flag,
                     "verbose" => \$verbose,
                     "punc" => \$punc);

if (!$res or ($corpus_name eq "") or ($base_dir eq "")) {
  usage();
  exit;
}

unless (-d $base_dir) {
  mkdir $base_dir or die "Couldn't create $base_dir: $!";
}


my $gen_dir = "$base_dir";

my $corpus_data_dir = "$gen_dir/corpus-data/$corpus_name";


if ($verbose ) { print "Instantiating corpus $corpus_name in $gen_dir\n"; }
my $corpus = Clair::Utils::CorpusDownload->new(corpusname => "$corpus_name",
                                               rootdir => "$gen_dir");

# index the corpus
print "Indexing the corpus\n";
$corpus->build_docno_dbm();
# Write links file
if ($links_flag) {
  if ($verbose) { print "Building hyperlink database\n"; }
  $corpus->write_links();
}

# Build tf-idf files
if ($idf_flag) {
  if ($verbose) { print "Building IDF database\n"; }
  $corpus->buildIdf(stemmed => 0, punc => $punc);
#  $corpus->buildIdf(stemmed => 1, punc => $punc);

\end{boxedverbatim}

\begin{boxedverbatim}

}
if ($tf_flag) {
  if ($verbose) { print "Building TF database\n"; }
  $corpus->buildTf(stemmed => 0);
  $corpus->buildTf(stemmed => 1);
}

# build document length dist and term counts
if ($stats_flag) {
  if ($verbose) { print "Building document length and term count databases\n";}
  $corpus->build_doc_len(stemmed => 0);
  $corpus->build_term_counts(stemmed => 0);
  $corpus->build_term_counts(stemmed => 1);
}

sub usage {
  print "Usage $0 --corpus corpus\n\n";
  print "  --corpus corpus\n";
  print "       Name of the corpus to index\n";
  print "  --base base_dir\n";
  print "       Base directory filename.  The corpus is located here\n";
  print " --tf, --notf\n";
  print "       Enable or disable building of TF index.  Enabled by default\n";
  print " --idf, --noidf\n";
  print "       Enable or disable building of IDF index.  Enabled by            \
default\n";
  print " --link, --nolinks\n";
  print "       Enable or disable building hyperlink database.  Enabled by      \
default\n";
  print " --stats, --nostats\n";
  print "       Enable or disable building term counts and doc. len. dist.\n";
  print "       Enabled by default\n";
  print " --punc\n";
  print "       Include punctuation in IDF.  Disabled by default.\n";
  print "  --verbose\n";
  print "       Include verbose output\n";
  print "\n";

  die;
}

\end{boxedverbatim}

\subsubsection{link\_synthetic\_collection.pl}
\begin{boxedverbatim}

#!/usr/bin/perl -w
# script: link_synthetic_collection.pl
# functionality: Links a collection using a certain network generator
# Usage: $0
#	 -n <name_of_new_corpus>
#	 -c <input_collection>
#	 -l <link_policy>, any of: {radev, menczer, erdos, watts}
#
# The following arguments are required by the specified policies:
#
# Option and value		Policies		Argument Type
#  -p <link_probability>	erdos, watts		positive float [0,1]
#  -k <k-parameter>		watts			positive integer
#  -w <term_weight_file>	radev			path to term weight file
#  -s <sigmoid_steepness>	radev, menczer		positive float
#  -t <sigmoid_threshold>	radev, menczer		positive float
#  -r <probability_reserve>	radev 			positive float
use strict;

use Getopt::Long;

use Clair::SyntheticCollection;
use Clair::LinkPolicy::WattsStrogatz;
use Clair::LinkPolicy::ErdosRenyi;
use Clair::LinkPolicy::MenczerMacro;
use Clair::LinkPolicy::RadevMicro;

# Default error
sub usage {
  die "
Usage: $0
	 -n <name_of_new_corpus>
         -b <base_directory_of_new_corpus>
	 -c <name_of_input_synthetic_collection>
         -d <base_directory_of_input_collection>
	 -l <link_policy>, any of: {radev, menczer, erdos, watts}

 The following arguments are required by the specified policies:

 Option and value		Policies		Argument Type
  -p <link_probability>		erdos, watts		positive float [0,1]
  -k <num_neighbors>		watts			positive integer
  -w <term_weight_file>		radev			term weight file
  -s <sigmoid_steepness>	radev, menczer		positive float
  -t <sigmoid_threshold>	radev, menczer		positive float
  -r <probability_reserve>	radev 			positive float\n\n";
}

my $corpus_name = "";
my $base_dir = "produced";
my $new_dir = "";
my $new_name = "";
my $link_policy = "";
my $num_neighbors = -1;
my $link_prob = -1;
my $term_weight_file = "";
my $sigmoid_steepness = -1;
my $sigmoid_threshold = -1;
my $prob_reserve = -1;
my $verbose = -1;

my $res = GetOptions("corpus=s" => \$corpus_name, "directory=s" => \$base_dir,
		     "name=s" => \$new_name, "base=s" => \$new_dir,
		     "k=i" => \$num_neighbors,
		     "link=s" => \$link_policy,
		     "probability=f" => \$link_prob,
		     "weight=s" => \$term_weight_file,

\end{boxedverbatim}

\begin{boxedverbatim}

		     "steepness=f" => \$sigmoid_steepness,
		     "threshold=f" => \$sigmoid_threshold,
		     "reserve=f" => \$prob_reserve,
		     "verbose" => \$verbose);

# We need at least -n, -c, -l
unless (($corpus_name ne "") && ($new_name ne "") &&
        ($link_policy ne "")) { usage(); }

# Make sure we can open the existing collection.
# (should croak here if collection does not exist)
my $synthdox = Clair::SyntheticCollection->new (name => $corpus_name,
						base => $base_dir,
						mode => "read_only");

my $new_corpus;

# Verify additional args and create the appropriate corpus.
if ($link_policy eq "radev") {
  # verify args
  unless (($term_weight_file ne "") && ($sigmoid_steepness ne -1) &&
          ($sigmoid_threshold ne -1) && ($prob_reserve ne -1)) { usage() }

  # create corpus
  $new_corpus = Clair::LinkPolicy::RadevMicro->new(base_collection =>
						   $synthdox,
						   base_dir => $new_dir);
  $new_corpus->create_corpus(corpus_name       => $new_name,
			     term_weights      => $term_weight_file,
			     sigmoid_steepness => $sigmoid_steepness,
			     sigmoid_threshold => $sigmoid_threshold,
			     prob_reserve      => $prob_reserve);

} elsif ($link_policy eq "menczer") {
  # verify args
  unless (($sigmoid_steepness ne -1) && ($sigmoid_threshold ne -1)) { usage() }

  # create corpus
  $new_corpus = Clair::LinkPolicy::MenczerMacro->new(base_collection =>
						     $synthdox,
						     base_dir => $new_dir);
  $new_corpus->create_corpus(corpus_name       => $new_name,
			     sigmoid_steepness => $sigmoid_steepness,
			     sigmoid_threshold => $sigmoid_threshold);

} elsif ($link_policy eq "erdos") {
  # verify args
  unless ($link_prob ne -1) { usage() }

  # create corpus
  $new_corpus = Clair::LinkPolicy::ErdosRenyi->new(base_collection =>
						   $synthdox,
						   base_dir => $new_dir);
  $new_corpus->create_corpus(corpus_name => $new_name,
			     link_prob   => $link_prob);

} elsif ($link_policy eq "watts") {
  # verify args
  unless (($link_prob ne -1) && ($num_neighbors ne -1)) { usage() }

  # create corpus
  $new_corpus = Clair::LinkPolicy::WattsStrogatz->new(base_collection =>
						      $synthdox);
  $new_corpus->create_corpus(corpus_name   => $new_name,
			     link_prob     => $link_prob,
			     num_neighbors => $num_neighbors);
} else { usage(); }

\end{boxedverbatim}

\subsubsection{make\_synth\_collection.pl}
\begin{boxedverbatim}

#!/usr/bin/perl
# script: make_synth_collection.pl
# functionality: Makes a synthetic document set
#

use strict;
use warnings;

use File::Spec;
use Getopt::Long;

use Clair::Utils::CorpusDownload;
use Clair::SyntheticCollection;
use Clair::RandomDistribution::Gaussian;
use Clair::RandomDistribution::LogNormal;
use Clair::RandomDistribution::Poisson;
use Clair::RandomDistribution::RandomDistributionFromWeights;
use Clair::RandomDistribution::Zipfian;

sub usage;

my $corpus_name = "";
my $output_name = "";
my $output_dir = "";
my $base_dir = "produced";
my $policy = "";
my $num_docs = 0;
my $verbose = 0;

# Distribution parameters
my $alpha = 0.0;
my $mean = 0.0;
my $variance = 0.0;
my $std_dev = 0.0;
my $lambda = 0.0;

my $res = GetOptions("corpus=s" => \$corpus_name, "base=s" => \$base_dir,
		     "size=i" => \$num_docs, "policy=s" => \$policy,
		     "output=s" => \$output_name,
                     "directory=s" => \$output_dir, "verbose!" => \$verbose,
                     "alpha:f" => \$alpha, "mean:f" => \$mean,
                     "variance:f" => \$variance, "std_dev:f" => \$std_dev,
                     "lambda:f" => \$lambda);

if (!$res or ($corpus_name eq "") or ($num_docs == 0) or
    ($output_name eq "") or ($output_dir eq "") or ($policy eq "")) {
  usage();
  exit;
}

my $gen_dir = "$base_dir";

my $corpus_data_dir = "$gen_dir/corpus-data/$corpus_name";

my $corpus = Clair::Utils::CorpusDownload->new(corpusname => "$corpus_name",
				    rootdir => "$gen_dir");

# index the corpus
my $pwd = `pwd`;
chomp $pwd;

# Get the document length distribution
my %doclen = $corpus->get_doc_len_dist();
# Get term counts
my %tc = $corpus->get_term_counts();

my @doclen_weights = ();

\end{boxedverbatim}

\begin{boxedverbatim}

my @lengths = ();
my @term_weights = ();
my @terms = ();

# Get document length weights
foreach my $k (sort {$doclen{$a} cmp $doclen{$b}} keys %doclen) {
  push @doclen_weights, $doclen{$k};
  push @lengths, ($k, $doclen{$k});
}

# Get term weights
foreach my $k (sort {$tc{$a} cmp $tc{$b}} keys %tc) {
  push @term_weights, $tc{$k};
  push @terms, ($k, $tc{$k});
}

print @term_weights, "\n";
print @doclen_weights, "\n";

my $a;
my $b;

if ($verbose) { print "Reading in term distribution...\n"; }
if ($verbose) { print "Reading in document length distribution...\n"; }
if ($policy eq "randomdistributionfromweights") {
  $a = Clair::RandomDistribution::RandomDistributionFromWeights->new(weights => \
\@term_weights);

  $b = Clair::RandomDistribution::RandomDistributionFromWeights->new(weights => \
\@doclen_weights);
} elsif ($policy eq "gaussian") {
  $a = Clair::RandomDistribution::Gaussian->new(mean => $mean,
                                                variance => $variance,
                                                dist_size => $num_docs);
  $b = Clair::RandomDistribution::Gaussian->new(mean => $mean,
                                                variance => $variance,
                                                dist_size => $num_docs);
} elsif ($policy eq "lognormal") {
  $a = Clair::RandomDistribution::LogNormal->new(mean => $mean,
                                                 std_dev => $std_dev,
                                                 dist_size => $num_docs);
  $b = Clair::RandomDistribution::LogNormal->new(mean => $mean,
                                                 std_dev => $std_dev,
                                                 dist_size => $num_docs);
} elsif ($policy eq "poisson") {
  $a = Clair::RandomDistribution::Poisson->new(lambda => $lambda,
                                               dist_size => $num_docs);
  $b = Clair::RandomDistribution::Poisson->new(lambda => $lambda,
                                               dist_size => $num_docs);
} elsif ($policy eq "zipfian") {
  $a = Clair::RandomDistribution::Zipfian->new(alpha => $alpha,
                                               dist_size => $num_docs);
  $b = Clair::RandomDistribution::Zipfian->new(alpha => $alpha,
                                               dist_size => $num_docs);
}

if ($verbose) { print "Creating collection\n"; }
my $col = Clair::SyntheticCollection->new(name => $output_name,
					  base => $output_dir,
					  mode => "create_new",
					  term_map => \@terms,
					  term_dist => $a,
					  doclen_dist => $b,
					  doclen_map => \@lengths,
					  size => $num_docs);

if ($verbose) { print "Generating documents\n"; }

\end{boxedverbatim}

\begin{boxedverbatim}

$col->create_documents();

chdir $pwd;

#
# Print out usage message
#
sub usage
{
  print "$0\n";
  print "Generate a synthetic corpus\n";
  print "\n";
  print "usage: $0 -c corpus_name [-b base_dir]\n\n";
  print "  --output,-o name\n";
  print "       Name of the generated corpus\n";
  print "  --directory,-d output directory\n";
  print "       Directory to output generated corpus in\n";
  print "  --corpus,-c corpus_name\n";
  print "       Name of the source corpus\n";
  print "  --base,-b base_dir\n";
  print "       Base directory filename.  The corpus is loaded from here\n";
  print "  --policy,-p policy\n";
  print "       Document generation policy: {gaussian, lognormal, poisson,      \
randomdistributionfromweights, zipfian}\n";
  print "  --size, -s number_of_documents\n";
  print "       Number of documents to generate\n";
  print "  --verbose,-v\n";
  print "       Increase debugging verbosity\n";
  print "\n";
  print " The following arguments are required by the spcified policies:\n";
  print "Option and value    Policy               Argument Type\n";
  print "alpha               zipfian              positive float\n";
  print "mean                gaussian,lognormal   positive float\n";
  print "variance            gaussian             positive float\n";
  print "std_dev             lognormal            positive float\n";
  print "lambda              poisson              positive float\n";
  print "\n";
  print "example: $0 -p zipfian --alpha 1.1 -o synthy -d synth_out -c           \
lexrank-sample -b produced -s 10 --verbose\n";

  exit;
}

\end{boxedverbatim}

\subsubsection{network\_growth.pl}
\begin{boxedverbatim}

#!/usr/bin/perl
#
# script: network_growth.pl
# functionality: Generates graphs for queries in web search engine
# functionality: query logs and measures network statistics
# 
# The network edges are updated every time new word (in the ranked word list) 
# is included in measuring the similarities of queries.
# Based on code by Xiaodong Shi
#

use strict;
use warnings;

use File::Path;
use Getopt::Long;
use Clair::Network;
use Clair::Corpus;
use Clair::Cluster;

sub usage;

#my $word_freqs = "sorted_word_freqs_from_50000q.stat";
my $in_file = "sorted_word_freqs_from_1000q.stat";
my $stat_file = "net.stat";
my $delim = "\t\t";
my $sample_size = 1000;
my $corpus_name = "";
my $basedir = "produced";
my $min_freq = 2;
my $verbose = 0;

my $res = GetOptions("corpus=s" => \$corpus_name, "base=s" => \$basedir,
                     "wordfreqs=s" => \$in_file, "delim=s" => \$delim,
                     "sample=i" => \$sample_size, "t=s" => \$stat_file,
		     "minfreq=i" => \$min_freq, "verbose" => \$verbose);

if ($corpus_name eq "") {
  usage();
}

my $corpus = Clair::Corpus->new(corpusname => "$corpus_name",
                                rootdir => "$basedir");

if ($verbose) { print "Loading corpus into cluster\n"; }
my $cluster = new Clair::Cluster;
$cluster->load_corpus($corpus);

#
# 1. Read the corpus file to get the document content
#

my @queries = ();
my %query_hash = ();
my $line_num = 0;

my $docs = $cluster->documents();
foreach my $did (keys %{$docs}) {
  my $doc = $docs->{$did};
  $doc->strip_html();
  my @sents = $doc->get_sentences();
  foreach my $line (@sents) {
    chomp $line;
    # $line = lc($line);
    $line_num++;
    $queries[$line_num-1] = $line;


\end{boxedverbatim}

\begin{boxedverbatim}

    if (not defined $query_hash{$line}) {
      $query_hash{$line} = 1;
    } else {
      $query_hash{$line} = $query_hash{$line} + 1;
    }
  }
}

#
# 2. Read the words and their ranked frequencies from input file.
#

my %freq = $corpus->get_term_counts();

print "Reading finished!\n";
print "Reversing the order of sorted words ...\n";

# Reverse the order of words
my @words = sort { $freq{$a} cmp $freq{$b} } keys %freq;
my %word_rank_hash = ();
my @r_words = ();
my $size = scalar(keys %freq);

for (my $i = 0; $i < $size; $i++) {
  $r_words[$i] = $words[$size - 1 - $i];

  if (exists $word_rank_hash{$r_words[$i]}) {
  } else { 
    $word_rank_hash{$r_words[$i]} = $i + 1;
  }
}

print "Total ", $size, " words. Order reversed!\n";
print "Size of word_rank_hash table: ", scalar(keys %word_rank_hash), "\n";

#
# 3. Take one word each time and build the graph
# 

my $network = Clair::Network->new();
#my $out_file = "$corpus_name.edges";
#if (!(-d $out_file)) {
#  mkpath ($out_file, 1, 0777);
#}
my $out_file = "$corpus_name.wordmodel.nodes";

open (FOUT, ">$out_file") or die "Could not open output file $out_file: $!\n";
print "Writing network nodes to output file $out_file ...\n";

my @qs = keys %query_hash;
foreach (my $i = 0; $i < scalar(@qs); $i++) {
  # add queries to the graph
  $network->add_node($i + 1, $qs[$i]);
  print FOUT (($i+1) .  "\t" . $qs[$i] . "\n");
}

close FOUT;

print "Num. Nodes written: ", $network->num_nodes(), "\n";

# Output network edges into file
#$out_file = $out_dir . "/" . $corpus_name . "/graph";  
#if (!(-d $out_file)) {
#  mkpath ($out_file, 1, 0777);
#}
#$out_file = $out_file . "/edges";
$out_file = "$corpus_name.wordmodel.edges";

\end{boxedverbatim}

\begin{boxedverbatim}

open (FOUT, ">$out_file") or die "Could not open output file $out_file: $!\n";
print "Writing network edges to output file $out_file ...\n";

# Output the network statistics into file
#my $net_stat_file = $out_dir . "/" . $corpus_name . "/stats"; 
#if (!(-d $net_stat_file)) {
#  mkpath ($net_stat_file, 1, 0777);
#}
#$net_stat_file = $net_stat_file . "/net.stat";
my $net_stat_file = "$corpus_name.wordmodel.stats";
open (STAT, ">$net_stat_file") or die "Could not open network stats file        \
$net_stat_file: $!\n"; 

print STAT "threshold nodes edges diameter lcc avg_short_path watts_strogatz_cc \
newman_cc in_link_power in_link_power_rsquared in_link_pscore                   \
in_link_power_newman in_link_power_newman_error out_link_power                  \
out_link_power_rsquared out_link_pscore out_link_power_newman                   \
out_link_power_newman_error total_link_power total_link_power_rsquared          \
total_link_pscore total_link_power_newman total_link_power_newman_error         \
avg_degree\n";

# loop through all distinct queries and add one word at a time; 
# determine if two queries share a common word ranked higher than the added
# word;
for (my $n = 0; $n < $size; $n++) {
  # only if the word appears in more than 1 query, we can measure whether two
  # queries share that same word
  if ((defined $freq{$r_words[$n]}) and ($freq{$r_words[$n]} >= $min_freq)) {
	
  # if there is one of the four conditions, then run the iteration: 
  #    1. the next word has a different frequency from the current one
  #    2. the current word is the first one with frequency equal to min_freq
  #    3. the current word is the first word in the ranked list and its         \
frequency is greater than min_freq (evaluated in the above statement).
  #    4. the current word is the k*50-th in the ranked list. 

  if ((($n < $size - 1) && ($freq{$r_words[$n+1]} ne $freq{$r_words[$n]}))
      || (($n > 0) && ($freq{$r_words[$n - 1]} < $min_freq))
      || ($n % 50 eq 0)) {
    for (my $x = 0; $x < scalar(@qs) - 1; $x++) {
      for (my $y = $x + 1; $y < scalar(@qs); $y++) {
        if (!($network->has_edge($x + 1, $y + 1))) {
          my $k = 0;
          # split the document into word tokens
          my @x_tokens = split(/ /, $qs[$x]);
          my @y_tokens = split(/ /, $qs[$y]);
			
			
          foreach my $x_token (@x_tokens) {
                    
            if ((defined $word_rank_hash{$x_token}) and
                ($word_rank_hash{$x_token} <= $n + 1)) {
              foreach my $y_token (@y_tokens) {
                if ($x_token eq $y_token) {
                  # for simplicity, we don't count the num of
                  # cooccurances of words in them, so we use binary
                  # values instead.
                  $k++;
                  last;
                }
              }
            }
          }

          if ($k > 0) {
            $network->add_edge($x + 1, $y + 1); 
            print FOUT (($x+1) . "\t" . ($y+1) . "\t" . ($n+1) . "\n"); 

\end{boxedverbatim}

\begin{boxedverbatim}

            $network->set_edge_weight($x + 1, $y + 1, 1); 
          }
        }
      }
    }

    print $n + 1 . "\tNum. Edges: " . ($network->num_links()) . "\n";

    my $stat_string = "";
    if ($network->num_links() eq 0) {
      $stat_string = $network->num_nodes() . " " . $network->num_links() . " ";
      $stat_string = $stat_string .
        "0 0 0 0 0 0 0 0 0 0 0 0 0 0 0 0 0 0 0 0\n";
    } else {
      $stat_string = $network->get_network_info_as_string();
    }

    # write network statistics to the file
    print STAT ($n + 1) . " " . $stat_string . "\n";
  }
}
}

close FOUT;
close STAT;


#
# prompt the user about the correct usage of this script
#
sub usage {
  print "usage: $0 --corpus corpus_name [-f query_log_file] [-i                 \
sorted_words_input_file]";
  print "[-s sample_size] [-m min_word_frequency] [-t net_stat_file]\n";
  print "  --corpus, -c corpus_name\n";
  print "          Name of corpus to load\n";
  print "  --sample, -s sample_size\n";
  print "          Calculate statistics for a sample of the network\n";
  print "          By default uses random edge sampling\n";
  print "  --minword, -m min_word_frequency\n";
  print "  -t net_stat_file\n";
  print "\n";
  print "example: $0 -c aol-10000 -f 100000.q ";
  print "-i sorted_word_freqs_from_100000q.stat ";
  print "-s 10000 -m 2 -t aol-10000-query-net.stat\n";
  exit;
}

\end{boxedverbatim}

\subsubsection{network\_to\_plots.pl}
\begin{boxedverbatim}

#!/usr/bin/perl
#
# script: network_to_plots.pl
# functionality: Generates degree distribution plots, creating a
# functionality: histogram in log-log space, and a cumulative degree
# functionality: distribution histogram in log-log space.
#
# Based on the make_cosine_plots.pl script by Alex
#
use strict;
use warnings;

use File::Spec;
use Getopt::Long;

sub usage;

my $cos_file = "";
my $num_bins = 1;

my $res = GetOptions("input=s" => \$cos_file, "bins:i" => \$num_bins);

if (!$res || ($cos_file eq "")) {
  usage();
  exit;
}

my ($vol, $dir, $hist_prefix) = File::Spec->splitpath($cos_file);
$hist_prefix =~ s/\.graph//;

my $cosines = "$cos_file";

my @link_bin = ();
$link_bin[$num_bins] = 0;

my %cos_hash = ();

my ($key1,  $key2);
open (COS, $cosines) or die "cannot open $cosines\n";

while(<COS>) {
  chomp;
  ($key1, $key2) = split;

  if (($key1 ne $key2) &&
      !(exists $cos_hash{$key2}) &&
      !(exists $cos_hash{$key1})) {

    $cos_hash{$key1} = 1;
  }
  if (exists $cos_hash{$key2}) {
    $cos_hash{$key1}++;
  }
}
close(COS);

foreach my $cos (keys %cos_hash) {
  my $deg = $cos_hash{$cos};
  my $d = get_index($deg);
  $link_bin[$d]++;
}

#print "cosine histogram:\n";

# Commented out  by alex
# Fri Apr 22 23:18:40 EDT 2005
#

\end{boxedverbatim}

\begin{boxedverbatim}

# For some reason, matlab decided that today it does not
#  like full paths. So we take them out, and pat matlab
#  on the head.
#
# Just remember that this will produce plots in the
#  current directory now, so CD in to wherever you need
#  to be before piping this stuff into matlab.
#
my $fname = $hist_prefix . "-cosine-hist.m";
my $fname2 = $hist_prefix . "-cosine-cumulative.m";
open(OUT,">$fname") or die ("Cannot write to $fname");
open(OUT2,">$fname2") or die ("Cannot write to $fname2");
print OUT "x = [";
print OUT2 "x = [";
my $cumulative=0;

foreach my $i (0..$#link_bin)
{
   my $out = $link_bin[$i];
   if(not defined $link_bin[$i])
   {
      $out = 0;
   }
   $cumulative += $out;
   my $thres = $i;
   print OUT "$thres $out\n";
   print OUT2 "$thres $cumulative\n";
}

print OUT "];\n";

my $out_filename = "$hist_prefix"."-cosine-hist";
print OUT "loglog(x(:,1), x(:,2));\n";
print OUT "title(['Degree Distribution of $hist_prefix']);\n";
print OUT "xlabel('Degree');\n";
print OUT "ylabel('Number of Nodes');\n";
#print OUT "v = axis;\n";
#print OUT "v(1) = 0; v(2) = 1;\n";
#print OUT "axis(v)\n";
print OUT "print ('-deps', '$out_filename.eps')\n";
print OUT "saveas(gcf, '$out_filename" . ".jpg', 'jpg'); \n";
close OUT;

$out_filename = $hist_prefix . "-cosine-cumulative";
print OUT2 "];\n";
print OUT2 "loglog(x(:,1), x(:,2));\n";
print OUT2 "title(['Degree Distribution of $hist_prefix']);\n";
print OUT2 "xlabel('Degree');\n";
print OUT2 "ylabel('Number of Nodes');\n";
print OUT2 "v = axis;\n";
print OUT2 "v(1) = 0; v(2) = 1\n";
print OUT2 "axis(v)\n";
print OUT2 "print ('-deps', '$hist_prefix-cosine-cumulative.eps')\n";
print OUT2 "saveas(gcf, '$out_filename" . ".jpg', 'jpg'); \n";
close OUT2;


sub get_index {
  my $d = shift;
  my $c = int($d * $num_bins + 0.000001);
#  print "$c $d\n";
  return $c;
}

sub usage {
  print "Usage $0 --input input_file [--bins num_bins]\n\n";
  print "  --input input_file\n";

\end{boxedverbatim}

\begin{boxedverbatim}

  print "       Name of the input graph file\n";
  print "  --bins num_bins\n";
  print "       Number of bins\n";
  print "       num_bins is optional, and defaults to 100\n";
  print "\n";
  die;
}


\end{boxedverbatim}

\subsubsection{print\_network\_stats.pl}
\begin{boxedverbatim}

#!/usr/bin/perl
#
# script: print_network_stats.pl
# functionality: Prints various network statistics
#
use strict;
use warnings;

use Getopt::Long;
use File::Spec;
use Clair::Cluster;
use Clair::Network qw($verbose);
use Clair::Network::Centrality::Betweenness;
use Clair::Network::Centrality::Closeness;
use Clair::Network::Centrality::Degree;
use Clair::Network::Centrality::LexRank;
use Clair::Network::Sample::RandomEdge;
use Clair::Network::Sample::ForestFire;
use Clair::Network::Reader::Edgelist;
use Clair::Network::Writer::Edgelist;
use Clair::Network::Writer::GraphML;
use Clair::Network::Writer::Pajek;

sub usage;

my $delim = "[ \t]+";
my $sample_size = 0;
my $sample_type = "randomedge";
my $fname = "";
my $out_file = "";
my $pajek_file = "";
my $graphml_file = "";
my $extract = 0;
my $stem = 1;
my $undirected = 0;
my $wcc = 0;
my $scc = 0;
my $components = 0;
my $paths = 0;
my $triangles = 0;
my $assortativity = 0;
my $local_cc = 0;
my $all = 0;
my $output_delim = " ";
my $stats = 1;
my $degree_centrality = 0;
my $closeness_centrality = 0;
my $betweenness_centrality = 0;
my $lexrank_centrality = 0;
my $force = 0;
my $graph_class = "";
my $filebased = 0;

my $res = GetOptions("input=s" => \$fname, "delim=s" => \$delim,
                     "delimout=s" => \$output_delim,
                     "output:s" => \$out_file, "pajek:s" => \$pajek_file,
                     "graphml:s" => \$graphml_file,
                     "sample=i" => \$sample_size,
                     "sampletype=s" => \$sample_type,
                     "extract!" => \$extract,
                     "stem!" => \$stem, "undirected" => \$undirected,
                     "components" => \$components, "paths" => \$paths,
                     "wcc" => \$wcc, "scc" => \$scc,
                     "triangles" => \$triangles, "verbose!" => \$verbose,
                     "assortativity" => \$assortativity,
                     "localcc" => \$local_cc, "stats!" => \$stats,
                     "all" => \$all,

\end{boxedverbatim}

\begin{boxedverbatim}

                     "betweenness-centrality" => \$betweenness_centrality,
                     "degree-centrality" => \$degree_centrality,
                     "closeness-centrality" => \$closeness_centrality,
                     "lexrank-centrality" => \$lexrank_centrality,
                     "force" => \$force,
                     "graph-class=s" => \$graph_class,
                     "filebased" => \$filebased);

my $directed = not $undirected;
$Clair::Network::verbose = $verbose;

my $vol;
my $dir;
my $prefix;
($vol, $dir, $prefix) = File::Spec->splitpath($fname);
$prefix =~ s/\.graph//;
if ($all) {
  # Enable all options
  if ($directed) {
    $wcc = 1;
    $scc = 1;
  } else {
    $components = 1;
  }
  $triangles = 1;
  $paths = 1;
  $assortativity = 1;
  $local_cc = 1;
  $betweenness_centrality = 1;
  $degree_centrality = 1;
  $closeness_centrality = 1;
}


if (!$res or ($fname eq "")) {
  usage();
}

my $fh;
my @hyp = ();

# make unbuffered
select STDOUT; $| = 1;

if ($verbose) {
  print "Reading in " . ($directed ? "directed" : "undirected") .
    " graph file\n";
}

my $reader = Clair::Network::Reader::Edgelist->new();
my $net;
my $graph;
if ($graph_class ne "") {
  eval("use $graph_class;");
  $graph = $graph_class->new(directed => $directed);
  $net = $reader->read_network($fname, graph => $graph,
                               delim => $delim,
                               directed => $directed,
                               filebased => $filebased);
} else {
  $net = $reader->read_network($fname,
                               delim => $delim,
                               directed => $directed,
                               filebased => $filebased,
                               edge_property => "lexrank_transition");
}


\end{boxedverbatim}

\begin{boxedverbatim}

# Sample network if requested
if ($sample_size > 0) {
  if ($sample_type eq "randomedge") {
    if ($verbose) {
      print STDERR "Sampling $sample_size edges from network using random edge  \
algorithm\n"; }
    my $sample = Clair::Network::Sample::RandomEdge->new($net);
    $net = $sample->sample($sample_size);
  } elsif ($sample_type eq "forestfire") {
    if ($verbose) {
      print STDERR "Sampling $sample_size nodes from network using Forest Fire  \
algorithm\n"; }
    my $sample = Clair::Network::Sample::ForestFire->new($net);
    $net = $sample->sample($sample_size, 0.7);
  }
}

if ((($net->num_documents > 2000) or ($net->num_links > 4000000)) and
    (!$force) and (!$filebased)) {
  my $error_msg;
  $error_msg .= "Network is too large";
  if ($net->num_documents > 2000) {
    $error_msg .= " (" . $net->num_documents . " > 2000 nodes)";
  }
  if ($net->num_pairs > 4000000) {
    $error_msg .= " (" . $net->num_pairs . " > 4000000 edges)";
  }
  $error_msg .= ", please use sampling\n";
  die $error_msg;
}

# If graphviz dotfile is specified, dump network to that file
#if ($fname ne "") {
#  output_graphviz($net, $out_file);
#}

# If Pajek file is specified, dump network to that file
if ($pajek_file ne "") {
  my $export = Clair::Network::Writer::Pajek->new();
  $export->set_name("pajek");
  $export->write_network($net, "$pajek_file");
}

# If GraphML file is specified, dump network to that file
if ($graphml_file ne "") {
  my $export = Clair::Network::Writer::GraphML->new();
  $export->set_name($fname);
  $export->write_network($net, "$graphml_file");
}

if ($out_file ne "") {
  my $export = Clair::Network::Writer::Edgelist->new();
  $export->write_network($net, $out_file);
}

my $component_net;
if ($extract) {
  # Find the largest connected component
  if ($verbose) { print "Extracting largest connected component\n"; }
  print "Original network info:\n";
  print "  nodes: ", $net->num_nodes(), "\n";
  print "  edges: ", scalar($net->get_edges()), "\n";
  $component_net = $net->find_largest_component("weakly");
} else {
  $component_net = $net;
}


\end{boxedverbatim}

\begin{boxedverbatim}

if ($stats) {
    $component_net->print_network_info(components => $components,
                                       wcc => $wcc, scc => $scc,
                                       paths => $paths,
                                       triangles => $triangles,
                                       assortativity => $assortativity,
                                       localcc => $local_cc,
                                       delim => $output_delim,
                                       verbose => $verbose);
}


# Get centrality measures
if ($degree_centrality) {
  my $degree = Clair::Network::Centrality::Degree->new($component_net);
  my $b = $degree->normalized_centrality();
  open(OUTFILE, "> $prefix.degree-centrality");
  foreach my $v (keys %{$b}) {
    print OUTFILE "$v$output_delim" . $b->{$v} . "\n";
  }
  close OUTFILE;
}
if ($closeness_centrality) {
  my $closeness = Clair::Network::Centrality::Closeness->new($component_net);
  my $b = $closeness->normalized_centrality();
  open(OUTFILE, "> $prefix.closeness-centrality");
  foreach my $v (keys %{$b}) {
    print OUTFILE "$v$output_delim" . $b->{$v} . "\n";
  }
  close OUTFILE;
}
if ($betweenness_centrality) {
  my $betweenness =
    Clair::Network::Centrality::Betweenness->new($component_net);
  my $b = $betweenness->normalized_centrality();
  open(OUTFILE, "> $prefix.betweenness-centrality");
  foreach my $v (keys %{$b}) {
    print OUTFILE "$v$output_delim" . $b->{$v} . "\n";
  }
  close OUTFILE;
}

if ($lexrank_centrality) {
  # Set the cosine value to 1 on the diagonal
  foreach my $v ($component_net->get_vertices) {
    $component_net->set_vertex_attribute($v, "lexrank_transition", 1);
  }

  my $lexrank =
    Clair::Network::Centrality::LexRank->new($component_net);
  my $b = $lexrank->normalized_centrality();
  open(OUTFILE, "> $prefix.lexrank-centrality");
  foreach my $v (keys %{$b}) {
    print OUTFILE "$v$output_delim" . $b->{$v} . "\n";
  }
  close OUTFILE;
}

#
# Print out usage message
#
sub usage
{
  print "usage: $0 [-e] [-d delimiter] -i file [-f dotfile]\n";
  print "or:    $0 [-f dotfile] < file\n";
  print "  --input file\n";
  print "          Input file in edge-edge format\n";

\end{boxedverbatim}

\begin{boxedverbatim}

  print "  --delim delimiter\n";
  print "          Vertices in input are delimited by delimiter character\n";
  print "  --delimout output_delimiter\n";
  print "          Vertices in output are delimited by delimiter (can be printf \
format string)\n";
  print "  --sample sample_size\n";
  print "          Calculate statistics for a sample of the network\n";
  print "          The sample_size parameter is interpreted differently for     \
each\n";
  print "          sampling algorithm\n";
  print "  --sampletype sampletype\n";
  print "          Change the sampling algorithm, one of: randomnode,           \
randomedge,\n";
  print "          forestfire\n";
  print "          randomnode: Pick sample_size nodes randomly from the         \
original network\n";
  print "          randomedge: Pick sample_size edges randomly from the         \
original network\n";
  print "          forestfire: Pick sample_size nodes randomly from the         \
original network\n";
  print "                      using ForestFire sampling (see the tutorial for  \
more\n";
  print "                      information)\n";
  print "          By default uses random edge sampling\n";
  print "  --output out_file\n";
  print "          If the network is modified (sampled, etc.) you can           \
optionally write it\n";
  print "          out to another file\n";
  print "  --pajek pajek_file\n";
  print "          Write output in Pajek compatible format\n";
  print "  --extract,  -e\n";
  print "          Extract largest connected component before analyzing.\n";
  print "  --undirected,  -u\n";
  print "          Treat graph as an undirected graph\n";
  print "  --scc\n";
  print "          Print strongly connected components\n";
  print "  --wcc\n";
  print "          Print weakly connected components\n";
  print "  --components\n";
  print "          Print components (for undirected graph)\n";
  print "  --paths,  -p\n";
  print "          Print shortest path matrix for all vertices\n";
  print "  --triangles,  -t\n";
  print "          Print all triangles in graph\n";
  print "  --assortativity,  -a\n";
  print "          Print the network assortativty coefficient\n";
  print "  --localcc,  -l\n";
  print "          Print the local clustering coefficient of each vertex\n";
  print "  --degree-centrality\n";
  print "          Print the degree centrality of each vertex\n";
  print "  --closeness-centrality\n";
  print "          Print the closeness centrality of each vertex\n";
  print "  --betweenness-centrality\n";
  print "          Print the betweenness centrality of each vertex\n";
  print "  --lexrank-centrality\n";
  print "          Print the LexRank centrality of each vertex\n";
  print "\n";
  print "example: $0 -i test.graph\n";
  print "\n";
  print "Example with sampling: $0 -i test.graph --sample 100 --sampletype      \
randomnode\n\n";

  exit;
}

\end{boxedverbatim}

\subsubsection{sentences\_to\_docs.pl}
\begin{boxedverbatim}

#!/usr/bin/perl
# script: sentences_to_docs.pl
# functionality: Converts a document with sentences into a set of
# functionality: documents with one sentence per document
#
# Make sure a Java interpreter is in your path

use strict;
use warnings;

use File::Spec;
use Getopt::Long;
use Clair::Cluster;
use Clair::Document;

sub usage;

my $in_file = "";
my $dirname = "";
my $basename = "";
my $output_dir = "";
my $single = 0;
my $type = "text";
my $verbose = 0;

my $res = GetOptions("input=s" => \$in_file, "directory=s" => \$dirname,
		     "output=s" => \$output_dir, "singlefile" => \$single,
                     "type=s" => \$type, "verbose" => \$verbose);

if (!$res or ($output_dir eq "")) {
  usage();
  exit;
}

my $vol;
my $dir;
my $prefix;
($vol, $dir, $prefix) = File::Spec->splitpath($output_dir);

if ($dir ne "") {
  unless (-d $dir) {
    mkdir $dir or die "Couldn't create $dir: $!";
  }
}


my $cluster = 0;
if ($dirname ne "") {
  my $pwd = `pwd`;
  chomp $pwd;
  chdir $dirname or die "Couldn't change to $dirname: $!\n";
  if ($verbose) { print "Loading documents from directory $dirname\n"; }
  $cluster = new Clair::Cluster(id => $dirname);
  $cluster->load_documents("*", type => $type, filename_id => 1);
  chdir $pwd or die "Couldn't change back to $pwd: $!\n";
} elsif ($in_file ne "") {
  if ($verbose) { print "Loading documents from file $in_file\n"; }
  my $doc = new Clair::Document(file => $in_file, type => $type,
				id => "document");
  my %docs = ("document", $doc);
  $cluster = new Clair::Cluster(documents => \%docs, id => $in_file);
} else {
  usage();
  exit;
}
if ($verbose) { print "Loaded ", $cluster->count_elements, " documents\n"; }


\end{boxedverbatim}

\begin{boxedverbatim}

if ($type eq "html") {
  if ($verbose) { print "Stripping html\n"; }
  $cluster->strip_all_documents();
}
  

if ($verbose) { print "Creating sentence based cluster\n"; }
my $sentence_cluster = $cluster->create_sentence_based_cluster();

if ((not $single) and (! -d "$output_dir")) {
  if ($verbose) { print "Creating directory $output_dir\n"; }
  mkdir $output_dir;
}

if ($verbose) { print "Saving documents to $output_dir\n"; }
if ($single) {
  # save to single file
  $sentence_cluster->save_documents_to_file($output_dir, 'text');
} else {
  # save to directory
  $sentence_cluster->save_documents_to_directory($output_dir, 'text',           \
name_count => 0);
}

sub usage {
  print "$0: Parse document into sentences and save into a directory or         \
file\n\n";
  print "usage: $0 [--singlefile] --input in_file [--directory directory_name]  \
--output output\n";

  print "  --input in_file\n";
  print "       Input file to parse into sentences\n";
  print "  --directory in_dir\n";
  print "       Input directory to parse into sentences\n";
  print "  --type document_type\n";
  print "       Document type, one of: text, html, stem\n";
  print "  --singlefile\n";
  print "       If true, write output into a single file, one line per          \
sentence\n";
  print "  --output output\n";
  print "       Output filename or directory\n";
  print "\n";

  print "example: $0 -i test/sentences.txt -o sentences\n";
}

\end{boxedverbatim}

\subsubsection{tf\_query.pl}
\begin{boxedverbatim}

#!/usr/local/bin/perl
# script: tf_query.pl
# functionality: Looks up tf values for terms in a corpus
#
# Based on test/test_lookupTFIDF.pl in clairlib

use strict;
use warnings;
use Getopt::Long;
use Data::Dumper;
use Clair::Config;
use Clair::Utils::Tf;
use Clair::Utils::CorpusDownload;

sub usage;

my $corpus_name ="";
my $query = "";
my $stemmed = 0;
my $all = 1;
my $basedir = "";
my $verbose = '';
my @phrase = ();

my $vars = GetOptions("corpus=s" => \$corpus_name,
                     "query=s" => \$query,
                     "basedir=s" => \$basedir,
                     "all" => \$all,
                     "stemmed" => \$stemmed,
                     "verbose" => \$verbose);

if( $corpus_name eq "" ) {
  usage();
  exit;
}

if( $query ne "" ){
	$all = 0;
}

$Clair::Utils::Tf::verbose = $verbose;

if ( $basedir eq "" ) {
	$basedir = "produced";
}
my $gen_dir = "$basedir";

if ($verbose) { print "Loading tf for $corpus_name in $gen_dir\n"; };
my $tf = Clair::Utils::Tf->new(rootdir => "$gen_dir",
                               corpusname => $corpus_name,
                               stemmed => $stemmed);

if( $all ){

	# Use Clair::Utils::CorpusDownload::get_term_counts()
	my $cd = Clair::Utils::CorpusDownload->new(rootdir => "$gen_dir",
                                             corpusname => $corpus_name);
	my %tfs = $cd->get_term_counts(stemmed => $stemmed);
	if( keys(%tfs) == 0 ){
		print "No term counts found.  Perhaps you need to run index_corpus.pl?\n";
	}else{
		foreach my $key (sort keys %tfs) {
			my $freq = $tf->getFreq($key);
    	my $res = $tf->getNumDocsWithWord($key);
			print "$key $freq $res\n";
		}
	}

\end{boxedverbatim}

\begin{boxedverbatim}


}else{

  @phrase = split / /, $query;

	my $res = $tf->getNumDocsWithPhrase(@phrase);

	my $freq = $tf->getPhraseFreq(@phrase);
	my $urls = $tf->getDocsWithPhrase(@phrase);

	if ($verbose) { print "TF($query) = $freq total in $res docs\n"; }
	if ($verbose) { print "Documents with \"$query\"\n"; }

	foreach my $url (keys %{$urls})  {
		my ($url_freq, $match_hash) = $tf->getPhraseFreqInDocument(\@phrase, url =>   \
$url);
		print "  $url: $url_freq\n";
	}
}
sub usage
{
  print "$0: Run TF queries\n";
  print "usage: $0 -c corpus_name -q query [-b base_dir]\n\n";
  print "  --basedir base_dir\n";
  print "       Base directory filename.  The corpus is generated here\n";
  print "  --corpus corpus_name\n";
  print "       Name of the corpus\n";
  print "  --query query\n";
  print "       Term or phrase to query. Enclose phrases in quotes\n";
  print "  --stemmed\n";
  print "       If set, uses stemmed terms.  Default is unstemmed.\n";
	print "  --all\n";
	print "       Prints frequency for all terms(format: term frequency            \
documents)\n";
  print "\n";

  print "example: $0 -c kzoo -q Michigan -b                                     \
/data0/projects/lexnets/pipeline/produced\n";

  exit;
}


\end{boxedverbatim}

\subsubsection{search\_to\_url.pl}
\begin{boxedverbatim}

#!/usr/bin/perl
# script: search_to_url.pl
# functionality: Searches on a Google query and prints a list of URLs

use strict;
use warnings;

use Getopt::Std;
use vars qw/ %opt /;
use Clair::Utils::WebSearch;

sub usage;

my $opt_string = "q:n:";
getopts("$opt_string", \%opt) or usage();

my $num_res = 0;
if ($opt{"n"}) {
  $num_res = $opt{"n"};
} else {
  usage();
  exit;
}

my $query = "";
if ($opt{"q"}) {
  $query = $opt{"q"};
} else {
  usage();
  exit;
}


my @results = @{Clair::Utils::WebSearch::googleGet($query, $num_res)};
foreach my $r (@results)  {
  my ($url, $title, $desc) = split('\t', $r);
  print $url, "\n";
}

sub usage {
  print "usage: $0 -q query -n number_of_results\n";
  print "example: $0 -q pancakes -n 10\n";
}

\end{boxedverbatim}

\subsubsection{wordnet\_to\_network.pl}
\begin{boxedverbatim}

#!/usr/bin/perl
#
# script: wordnet_to_network.pl
# functionality: Generates a synonym network from WordNet
#

use strict;
use warnings;

use WordNet::QueryData;
use Getopt::Long;

sub usage;

my $out_file = "";
my $verbose = 0;
my $res = GetOptions("output=s" => \$out_file, "verbose" => \$verbose);

if (!$res or ($out_file eq "")) {
  usage();
  exit;
}

open(OUTFILE, ">$out_file") or die "Couldn't open $out_file: $!\n";

my $wn = WordNet::QueryData->new;

#my %wn_hash = ();

my @words = $wn->listAllWords("noun");

foreach my $word (@words) {
  foreach my $sense ($wn->querySense($word . "#n")) {
    foreach my $syn ($wn->querySense($sense, "syns")) {
      $syn =~ s/([a-zA-Z]*).*/$1/;
      if ($syn ne "") {
	print OUTFILE "$word $syn\n";
      }
    }
  }
}

close OUTFILE;

sub usage {
  print "Usage $0 --output output_file [--verbose]\n\n";
  print "  --output output_file\n";
  print "       Name of the output graph file\n";
  print "  --verbose\n";
  print "       Increase verbosity of debugging output\n";
  print "\n";
  die;
}


\end{boxedverbatim}

\end{document}